\documentclass[useAMS,usenatbib]{mn2e}

\usepackage{tabularx}
\usepackage{graphicx}
\usepackage{txfonts}
\usepackage{longtable}
\usepackage{hhline}
\usepackage{arydshln}
\usepackage{multirow}
\usepackage{lscape}
\usepackage{array}

\newcommand{\hersc}{{\it Herschel}}
\newcommand{\plk}{{\it Planck}}
\newcommand{\lab}{LABOCA}

\newcommand{\spitz}{{\it  Spitzer}}

\newcommand{\lsun}{$L_\odot$}
\newcommand{\msun}{$M_\odot$}

\newcommand{\mic}{$\mu$m}

\newlength{\pointwidth}
\begin{document}

\title[The submillimeter emission of nearby galaxies]{Dissecting the origin of the submillimeter emission in nearby galaxies with Herschel and \lab}
  
\author[Galametz et al.]
{\parbox{\textwidth}{M. Galametz$^{1,2}$\thanks{e-mail: maud.galametz@eso.org}, 
	       M. Albrecht$^{3}$,
 	       R. Kennicutt$^{1}$,
	       G. Aniano$^{4}$ ,
	       F. Bertoldi$^{3}$,
	       D. Calzetti$^{5}$,
	       K. V. Croxall$^{6}$,
	       D. Dale$^{7}$,
	       B. Draine$^{8}$,
	       C. Engelbracht$^{9,10}$,  
	       K. Gordon$^{11,12}$,     
	       J. Hinz$^{9}$,
	       L. K. Hunt$^{13}$,
	       A. Kirkpatrick$^{5}$,
	       E. Murphy$^{14}$,
	       H. Roussel$^{15}$,
	       R. A. Skibba$^{16}$,
	       F. Walter$^{17}$,
	       A. Weiss$^{18}$
	       and C. D. Wilson$^{19}$}\vspace{0.5cm}\\
\parbox{\textwidth}{$^{1}$Institute of Astronomy, University of Cambridge, Madingley Road, Cambridge CB3 0HA, UK\\
$^{2}$European Southern Observatory, Karl-Schwarzschild-Str. 2, D-85748 Garching-bei-M\"unchen, Germany\\
$^{3}$Argelander-Institut f\"ur Astronomie, Abteilung Radioastronomie, Auf dem H\"ugel 71, D-53121 Bonn, Germany\\
$^{4}$Institut d'Astrophysique Spatiale, Orsay\\
$^{5}$Department of Astronomy, University of Massachusetts, Amherst, MA 01003, USA\\
$^{6}$Department of Astronomy, The Ohio State University, 4051 McPherson Laboratory, 140 W 18th Ave., Columbus, OH, 43210\\
$^{7}$Department of Physics $\&$ Astronomy, University of Wyoming, Laramie, WY 82071, USA\\  
$^{8}$Department of Astrophysical Sciences, Princeton University, Princeton, NJ 08544, USA\\
$^{9}$Steward Observatory, University of Arizona, Tucson, AZ 85721, USA  \\  
$^{10}$Raytheon Company, 1151 East Hermans Road, Tucson, AZ 85756, USA \\
$^{11}$Space Telescope Science Institute, 3700 San Martin Drive, Baltimore, MD, 21218 \\
$^{12}$Sterrenkundig Observatorium, Universiteit Gent, Gent, Belgium\\
$^{13}$INAF Ð Osservatorio Astrofisico di Arcetri, Largo E. Fermi 5, 50125 Firenze, Italy\\
$^{14}$Infrared Processing and Analysis Center, California Institute of Technology, MC 220-6, Pasadena CA, 91125, USA\\
$^{15}$Institut dÕAstrophysique de Paris, Universit\'e Pierre et Marie Curie (UPMC), CNRS (UMR 7095), 75014 Paris, France\\ 
$^{16}$Center for Astrophysics and Space Sciences, Department of Physics, University of California, 9500 Gilman Dr., San Diego, CA 92093, USA\\
$^{17}$Max-Planck-Institut f\"ur Astronomie, K\"onigstuhl 17, D-69117 Heidelberg, Germany\\             
$^{18}$Max-Planck-Institut f\"ur Radioastronomie, Auf dem H\"ugel 69, D-53121 Bonn, Germany\\ 
$^{19}$Department of Physics $\&$ Astronomy, McMaster University, Hamilton, Ontario L8S 4M1, Canada }}

\maketitle{}

 
\begin{abstract}

We model the infrared to submillimeter spectral energy distribution of 11 nearby galaxies of the KINGFISH sample using \spitz\ and \hersc\ data and compare model extrapolations at 870 \mic\ (using different fitting techniques) with LABOCA 870 \mic\ observations. We investigate how the differences between predictions and observations vary with model assumptions or environment. At global scales, we find that modified blackbody models using realistic cold emissivity indices ($\beta$$_c$=2 or 1.5) are able to reproduce the 870 \mic\ observed emission within the uncertainties for most of the sample. Low values ($\beta$$_c$$<$1.3) would be required in NGC0337, NGC1512 and NGC7793. At local scales, we observe a systematic 870 \mic\ excess when using $\beta$$_c$=2.0. The $\beta$$_c$=1.5 or the \citet{Draine_Li_2007} models can reconcile predictions with observations in part of the disks. Some of the remaining ``excesses" occur towards the centres and can be partly or fully accounted for by non-dust contributions such as CO(3-2) or, to a lesser extent, free-free or synchrotron emission. In three non-barred galaxies, the remaining excesses rather occur in the disk outskirts. This could be a sign of a flattening of the submm slope (and decrease of the effective emissivity index) with radius in these objects.
 
\end{abstract}
  


\section{Introduction}

Studying dust emission helps us to probe the star formation activity obscured by dust grains and thus to understand how galaxies evolve through time. Many studies using telescopes such as the Infrared Astronomical Satellite (IRAS) or the {\it Spitzer Space Telescope} (\spitz) have been carried out to better understand the different grain populations contributing to the near-to-far infrared (NIR, FIR) emission of nearby galaxies and investigate their hot and warm dust components. Recent observations of the FIR and submillimeter (submm) using the {\it Herschel Space Observatory} (\hersc) allow us to probe, with increased resolution compared to \spitz, the cold dust emission, study its properties (temperature, grain opacity) within galaxies rather than on integrated scales and analyse how they vary with the Interstellar Medium (ISM) physical conditions \citep[see for instance the analysis on nearby resolved galaxies of][]{Boquien2011, Aniano2012, Bendo2012_2, Foyle2012, Mentuch2012, Smith2012, Draine2013, Galametz2013a}. 

Submm observations of nearby galaxies with \hersc\ and from the ground have suggested that the cold dust properties could vary with the ISM physical conditions, for instance with metallicity. Deriving the dust masses of low-metallicity galaxies using standard dust properties for instance often leads to unphysical dust-to-gas mass ratios (D/G), compared to those expected from their lack of metals \citep{Meixner2010,Galametz2010,Galliano2011}. An excess at submm wavelengths is also often reported in those objects \citep[]{Dumke2004, Galliano2003, Galliano2005, Bendo2006, Marleau2006, Galametz2009,OHalloran2010, Bot2010_2,RemyRuyer2013}. The submm ``excess" is usually defined as the excess emission above that predicted from FIR data (excluding the submm measurement) using a single $\lambda$$^{-\beta}$ emissivity law. (but we will also compare our observations with predictions from a more realistic dust model in the resolved analysis, Section 4). The presence of submm excess in a substantial number of galaxies challenges the models standardly used (more particularly how the submm regime is modelled) and raises the issue of how cold grain properties vary with environment (metallicity, temperature, star formation activity, etc). The origin of the observed excess emission is still an open question. The various hypotheses investigated so far can be classified in two categories:

{\it 1)} On the one hand, the excess could result from an additional dust component not accounted for in the current models. Some studies suggest that this excess could be linked with a reservoir of very cold dust ($<$15K) distributed in dense clumps \citep{Galliano2005,Galametz2011} or magnetic dipole emission from magnetic grains \citep{Draine2012}. Others propose that the excess characterises the so-called anomalous microwave emission generally associated with ``spinning dust" emission \citep{Draine_Lazarian_1998} with a peak frequency which varies with environment or grain properties \citep{Bot2010,Murphy2010,Peel2011,Planck_collabo_2011_SpinningDust,Ysard2010}, although \citet{Draine2012} have argued that rotational emission should not significantly contribute to the 870 \mic\ emission.
 
{\it 2)} On the other hand, the excess could be linked with a variation of dust properties with environment, in particular with temperature or density. Analysis of spatially resolved dust properties in the Galaxy suggest, for instance, that the emissivity of dust grains seems to increase toward the coldest regions of molecular clouds \citep{Juvela2011}, supporting the hypothesis of grain coagulation processes in those environments \citep{Paradis2009}. Moreover, the use of amorphous carbon (AC) instead of standard graphite is sometimes preferred to model carbon dust in the Spectral Energy Distribution (SED) fitting of thermal dust emission in dwarf galaxies  \citep{Meixner2010,OHalloran2010,Galametz2010,Galliano2011}. Indeed, AC grains have a lower emissivity index, resulting in a flatter submm spectrum, and thus require less dust to account for the same emission. This usually leads to lower dust masses and seems in those cases to reconcile the dust-to-gas mass ratios with those expected from the low metallicity of these objects. Many more investigations are necessary to link this excess with the ISM conditions and enable us to disentangle its different possible origins. 

 Even if mostly observed in low-metallicity objects, a submm (850 or 870\mic) excess has also been detected on global scales in solar-metallicity objects \citep{Galametz2011}. Submm excess emission is usually not (or barely) detected at 500 \mic\ in spiral galaxies \citep[][]{Bendo2010,Dale2012,Smith2012,Kirkpatrick2013}, suggesting that observations beyond \hersc\ are necessary to properly probe the presence of excess in these galaxies. We now investigate the potential excess emission beyond 500 \mic\ in 11 nearby galaxies chosen from the KINGFISH programme \citep[Key Insights on Nearby Galaxies: A Far-Infrared Survey with Herschel;][]{Kennicutt2011}, adding 870 \mic\ measurements taken with LABOCA (on APEX) to the NIR-to-FIR dataset. These 11 objects were selected because they are large enough to use the LABOCA mapping mode and their diffuse emission is expected to be bright enough to be detected by the instrument. 
 
 Our \spitz+\hersc+\lab\ dataset is ideal to resolve the potential 870 \mic\ excess and understand its origin. This paper complements the work of \citet{Galametz2012} (hereafter [G12]) that studies the spatially-resolved cold dust properties (temperature, emissivity index, mass) for the same sample. 
The paper is structured as follows. In Section 2, we describe the datasets we use. In Section 3, we study how the 870 \mic\ emission compares with various model extrapolations from the \hersc\ wavebands on global scales. In Section 4, we simulate 870 \mic\ maps using various resolved SED modelling techniques and compare them with our \lab\ observations in order to spatially analyse the differences between these maps. We summarise the analysis in Section 5.

\begin{table*} 
\caption{Galaxy Data}
\label{Galaxy_data}
 \centering
 \begin{tabular}{ccccrccrc}
\hline
\hline
   &&&&&&\\
   Galaxy & Optical  & $\alpha$$_0$ & $\delta$$_0$ & Distance & 12+log(O/H) &  M$_*$ & L$_{TIR}$~~~ & Nuclear\\    
   & Morphology  & (J2000) & (J2000) &  (Mpc)~ &  & (log \msun)  & (log \lsun) & Classification  \\   
   (1) & (2) & (3) & (4) &  (5) & (6)  & (7) & (8) & (9)\\
      &&&&&&\\
     \hline
        &&&&&&\\
     NGC0337   & SBd    & 00h 59m 50.7s & $-$07d 34' 44" & 19.3 	& 8.18 	&9.32& 10.03    &  SF\\
     NGC0628   & SAc    & 01h 36m 41.8s & ~15d 47' 17" &  7.2 		& 8.35 	&9.56&   9.84    &  SF\\
     NGC1097 & SBb    & 02h 46m 18.0s & $-$30d 16' 42"  &  14.2 	& 8.47 	&10.5& 10.62    &  AGN\\
     NGC1291 & SB0/a & 03h 17m 19.1s & $-$41d 06' 32"  &  10.4 	& 8.52 	&10.8&   9.43    &  AGN\\
     NGC1316 & SAB0  & 03h 22m 41.2s & $-$37d 12' 10"  &  21.0 	& 8.77 	&11.5&   9.77    &  AGN\\
     NGC1512 & SBab  & 04h 03m 55.0s & $-$43d 20' 44"  &  11.6 	& 8.56 	&9.92&   9.53    &  AGN\\
     NGC3351 & SBb    & 10h 43m 57.5s & ~11d 42' 19"  &  9.3 	& 8.60 	&10.2&   9.84    &  SF\\
     NGC3621 & SAd    & 11h 18m 18.3s & $-$32d 48' 55"  &  6.5 	& 8.27 	&9.38&   9.83    &  AGN\\
     NGC3627 & SABb & 11h 20m 13.4s & ~12d 59' 27" &  9.4 		& 8.34 	&10.5& 10.40    &  AGN\\
     NGC4826 & SAab & 12h 56m 42.8s & ~21d 40' 50" &  5.3 		& 8.54 	&9.94&   9.56    &  AGN\\
     NGC7793 & SAd    & 23h 57m 50.4s & $-$32d 35' 30"  &  3.9 	& 8.31 	&9.00&   9.22    &  SF\\
        &&&&&&\\
 \hline
\end{tabular}
\begin{list}{}{}
\item[$^{(1)}$] {\small Galaxy Name}
\item[$^{(2)}$] {\small Morphological type from \citet{Kennicutt2011}. NGC0337, NGC1097 and NGC3627 have also been classified as 'peculiar' by \citet{Buta2010}. } 
\item[$^{(3)}$] J2000.0 Right ascension
\item[$^{(4)}$] J2000.0 Declination
\item[$^{(5)}$] Distance in megaparsecs from \citet{Kennicutt2011}
\item[$^{(6)}$] Mean disk oxygen abundance obtained using the calibration of \citet{Pilyugin2005} taken from \citet{Moustakas2010} 
\item[$^{(7)}$] Stellar masses from \citet{Skibba2011} updated for the quoted distances. 
\item[$^{(8)}$] Total infrared luminosities in the 3-to-1100 \mic\ from \citet{Galametz2013b}
\item[$^{(9)}$] Optical spectral classification of the nucleus as proposed by \citet{Moustakas2010}
\end{list}
 \end{table*} 


\section{Observations and data reduction}

This work is based on a combination of \spitz, \hersc\ and \lab\ observations available for a sample of 11 nearby galaxies. A description of the sample is available in the first part of this study ([G12]) and general properties are summarised in Table~\ref{Galaxy_data}. We provide some details of the observations and data reduction processes in this section.

\subsection{Spitzer data}

Warm dust can contribute a non-negligible amount of the 70 \mic\ emission in galaxies. In order to constrain the mid-infrared (MIR) SED, observations from the Multiband Imaging Photometer for Spitzer \citep[MIPS;][]{Rieke2004} at 24 and 70 \mic\ are included in the dataset. The 11 galaxies have been observed with the MIPS as part of the Spitzer Infrared Nearby Galaxies Survey \citep[SINGS;][]{Kennicutt2003}. The full-width maxima (FWHM) of the MIPS 24 and 70 \mic\ PSFs are 6\arcsec\ and 18\arcsec\ respectively. Some galaxies within the sample are also part of the Local Volume Legacy (LVL) survey. All the data were processed with the latest LVL pipeline\footnote{http://irsa.ipac.caltech.edu/data/SPITZER/LVL/LVL$\_$DR5$\_$v5.pdf} \citep{Dale2009}, whether as part of that sample or to make them consistent with it.  
  
\subsection{Herschel data}

The 11 galaxies of the sample have been observed with PACS (Photodetector Array Camera and Spectrometer) and SPIRE (Spectral and Photometric Imaging Receiver) on-board \hersc\ as part of the KINGFISH programme. Those images allow us to sample the peak of the thermal dust emission and to probe the cold dust phases.
PACS observed at 70, 100 and 160 \mic\ with FWHMs of 5\farcs2, 7\farcs7 and 12\arcsec respectively \citep{Poglitsch2010}. The data are processed to Level 1 using the Herschel Interactive Processing Environment (HIPE) software. Scanamorphos was then used to produce the final PACS and SPIRE maps \citep{Roussel2012}. The scanamorphos algorithm was preferred over the standard madmap mapmaking technique because it can subtract the brightness drifts caused by low-frequency noise using the redundancy built in the observations.
PACS maps are calibrated in Jy pixel$^{-1}$ and have a final pixel size of 1\farcs4, 1\farcs7 and 2\farcs85 at 70, 100 and 160 \mic\ respectively. The PACS calibration uncertainties are $\sim$10 $\%$ (see the PACS Observer's Manual\footnote {http://herschel.esac.esa.int/Docs/PACS/html/pacs$\_$om.html}). 
SPIRE produced maps at 250, 350 and 500 \mic, with FWHMs of 18\arcsec, 25\arcsec and 36\arcsec respectively \citep[][]{Griffin2010}\footnote{http://herschel.esac.esa.int/twiki/bin/view/Public/SpirePhotometerBeam ProfileAnalysis}. SPIRE maps are calibrated in Jy beam$^{-1}$. The final pixel size is 6\arcsec, 10\arcsec and 14\arcsec\ at 250, 350 and 500 \mic\ respectively, with calibration uncertainties of $\sim$7\% for the three wave bands (see the SPIRE Observer's Manual\footnote{http://herschel.esac.esa.int/Docs/SPIRE/html/spire$\_$om.html}). We estimate the SPIRE beam sizes using the flux calibration paper of \citet{Griffin2013}. We apply their factors to convert from the point source pipeline to the extended source pipeline. Characterising each band by a solid angle appropriate for a spectrum of the form {\it I$_{\nu}$} $\propto$ $\nu$$^{-1}$, we derive beam sizes of 469.1, 827.2 and 1779.6 arcsec$^2$ for 250, 350 and 500 \mic\ respectively. Fluxes are then converted in MJy~sr$^{-1}$. More details on the data reduction of \hersc\ maps are available in the overview paper of the KINGFISH programme \citep{Kennicutt2011}. 

   \begin{figure*}
    \centering     
    \begin{tabular} {ccc}
  \hspace{-20pt}  {\bf \Large NGC0337} & \hspace{-35pt}{\bf \Large NGC1291} \\
  \hspace{-20pt} \includegraphics[width=10cm]{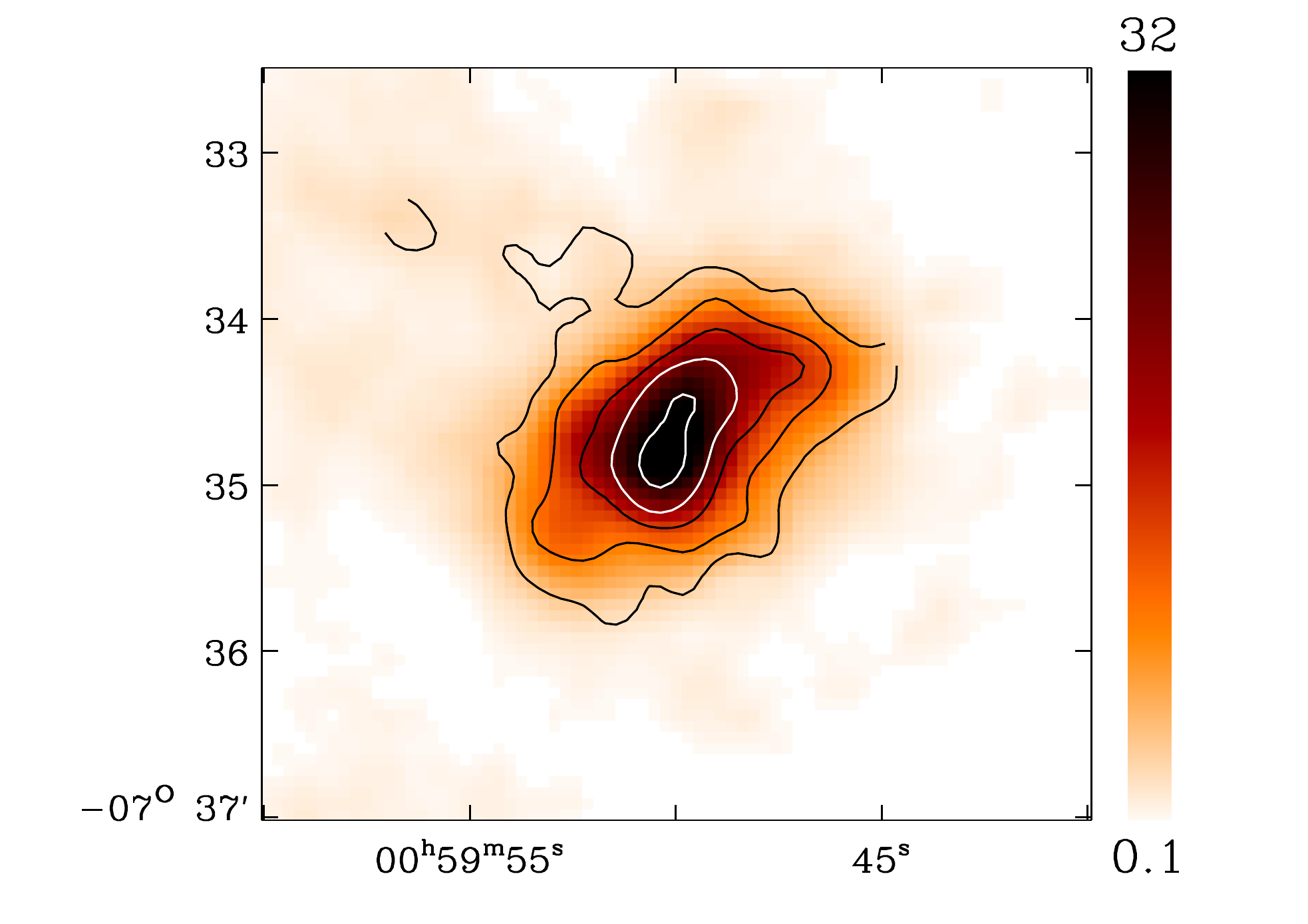} &
\hspace{-35pt}\includegraphics[width=10cm]{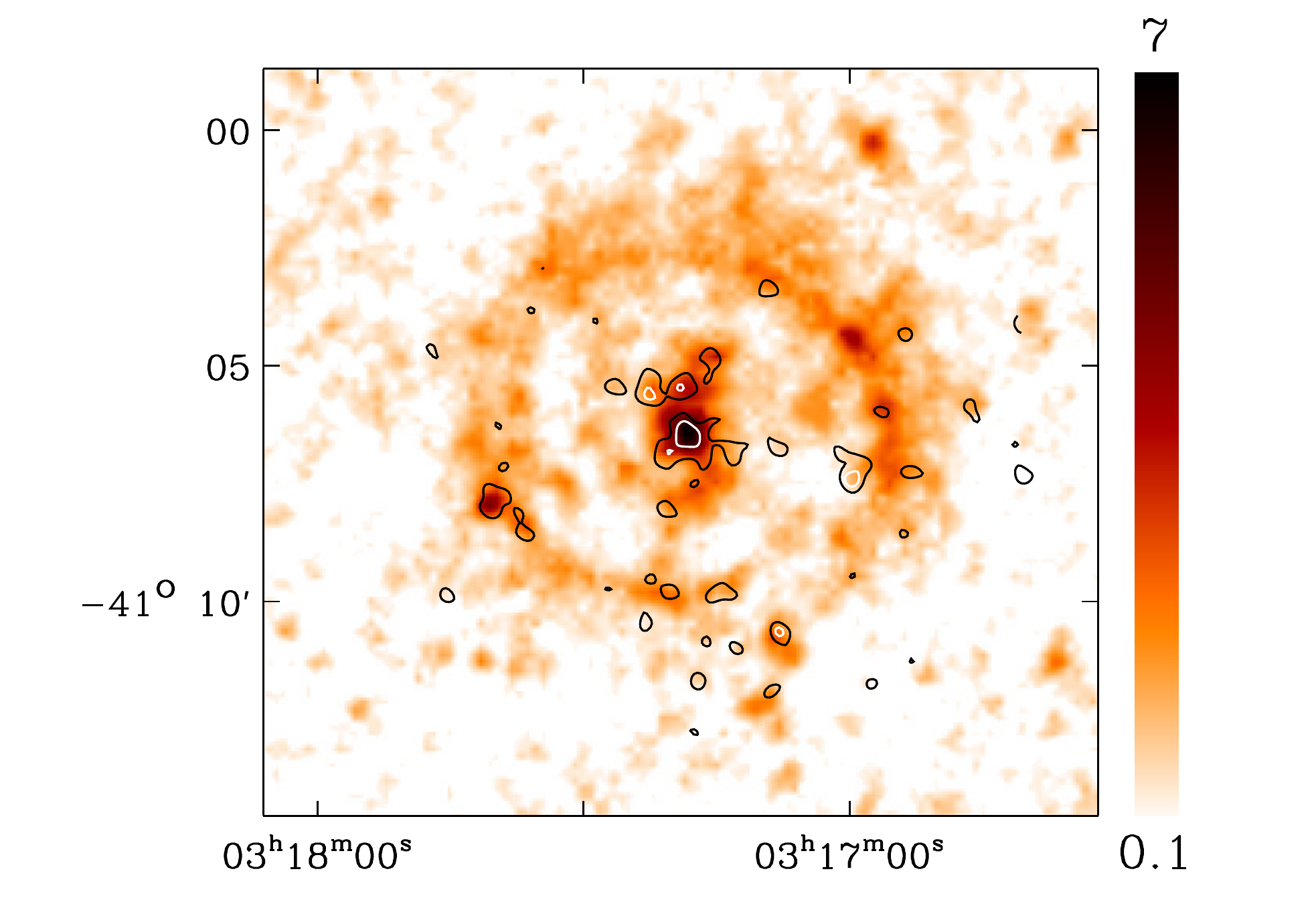} \\
  \hspace{-20pt} {\bf \Large NGC0628} & \hspace{-35pt}{\bf \Large NGC1316 / NGC1317} \\
 & \hspace{-35pt} \multirow{3}{*}{\includegraphics[width=10cm]{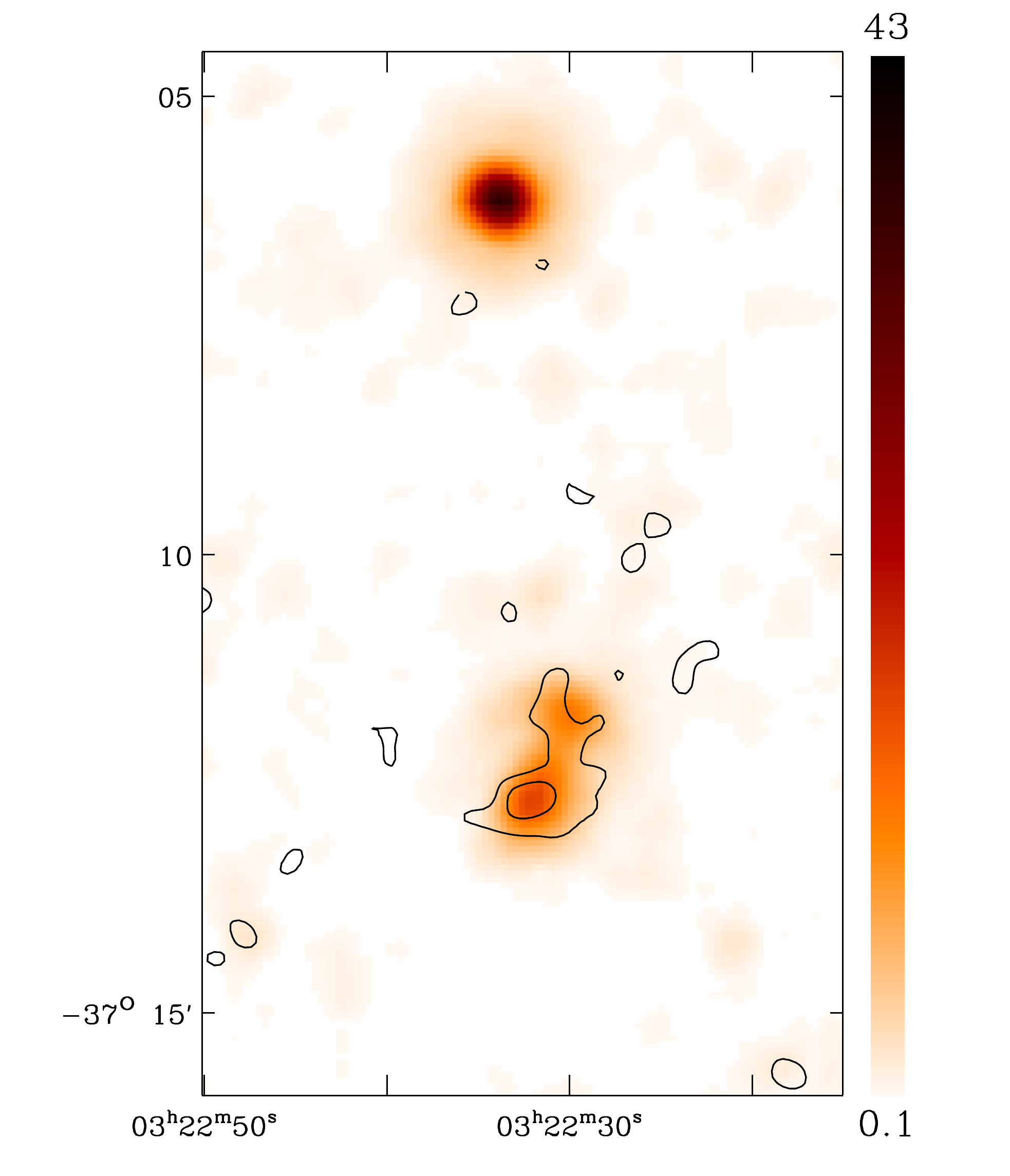}} \\ 
  \hspace{-20pt} \includegraphics[width=10cm]{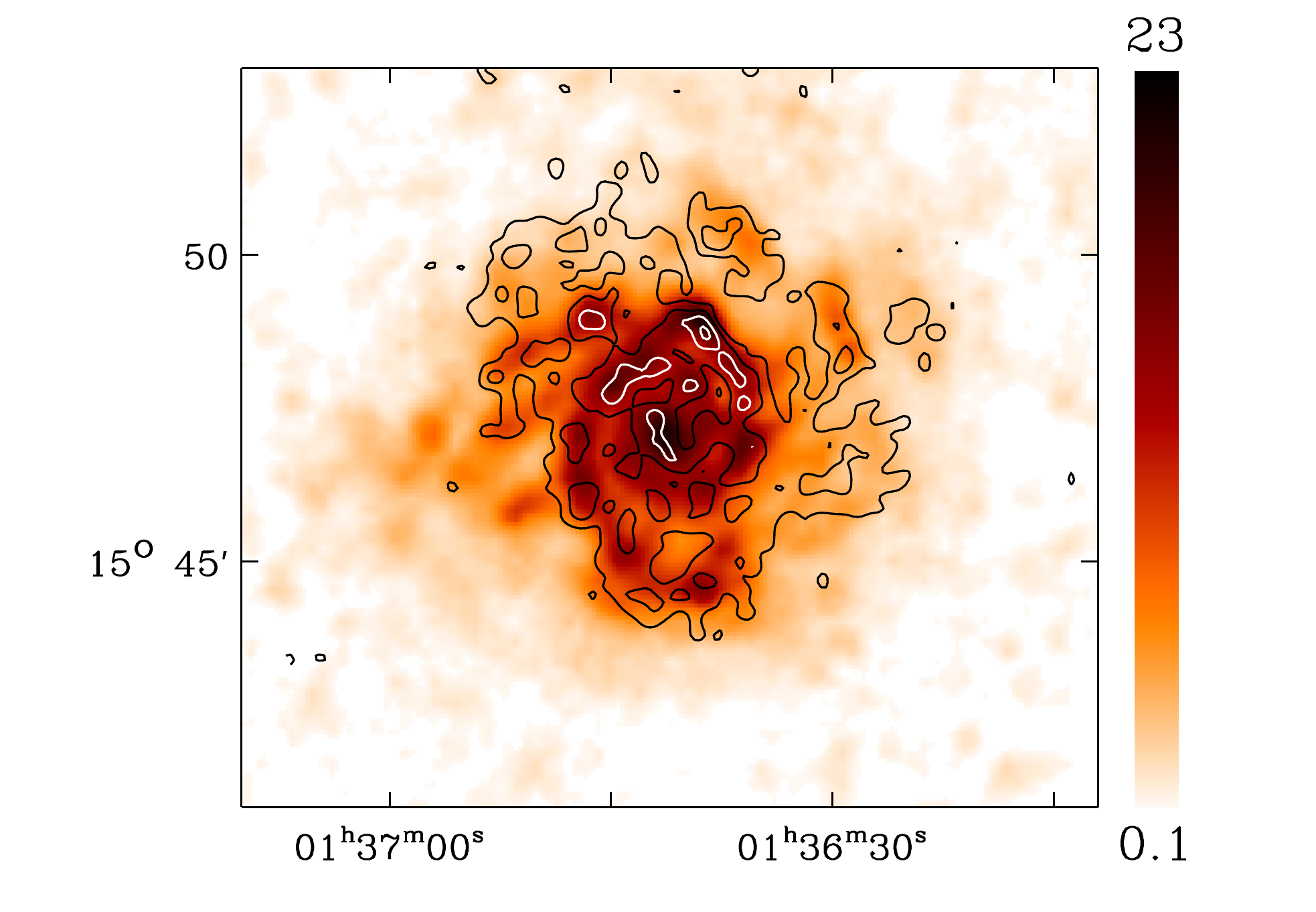} &\\
  \hspace{-20pt} {\bf \Large NGC1097} &  \\  
  \hspace{-20pt} \includegraphics[width=10cm]{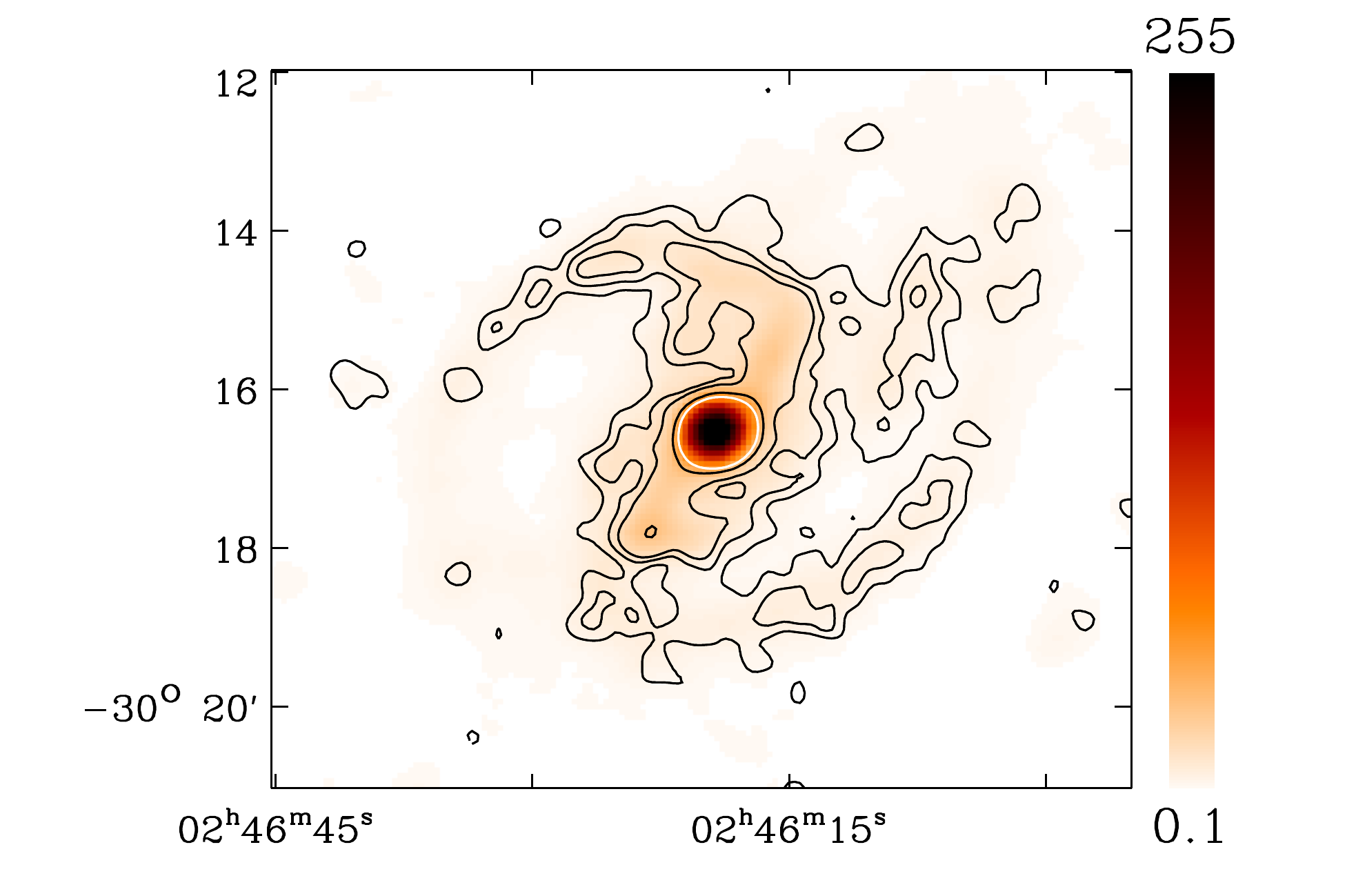} & \\
\end{tabular}
\caption{\hersc/SPIRE 350 \mic\ maps of the sample in MJy~sr$^{-1}$ (color scales are provided for each object in linear scale), with the signal-to-noise of the \lab\ 870 \mic\ maps overlaid as contours. The signal-to-noise maps have been convolved to a resolution of 25\arcsec\ (that of SPIRE 350 \mic) before display. Contours are, for NGC0337: 3, 6, 9, 12 and 17 $\sigma$ where $\sigma$=0.17 MJy~sr$^{-1}$, for NGC0628: 3, 6, 9 and 12 $\sigma$ where $\sigma$=0.19 MJy~sr$^{-1}$, for NGC1097: 3, 6, 9, 20 and 30 $\sigma$ where $\sigma$=0.13 MJy~sr$^{-1}$, for NGC1291: 3 and 6 $\sigma$ where $\sigma$=0.13 MJy~sr$^{-1}$, for NGC1316: 3 and 6 $\sigma$ where $\sigma$=0.20 MJy~sr$^{-1}$. NGC1316 is the southern object in this map, NGC1317 the northern object.} 
\label{350um_Maps}
\end{figure*}

\addtocounter {figure}{-1}
\begin{figure*}
\centering     
\begin{tabular}{ccc}
\hspace{-20pt}{\bf \Large NGC1512} & \hspace{-35pt}{\bf \Large NGC3627} \\   
\hspace{-20pt}\includegraphics[width=10cm]{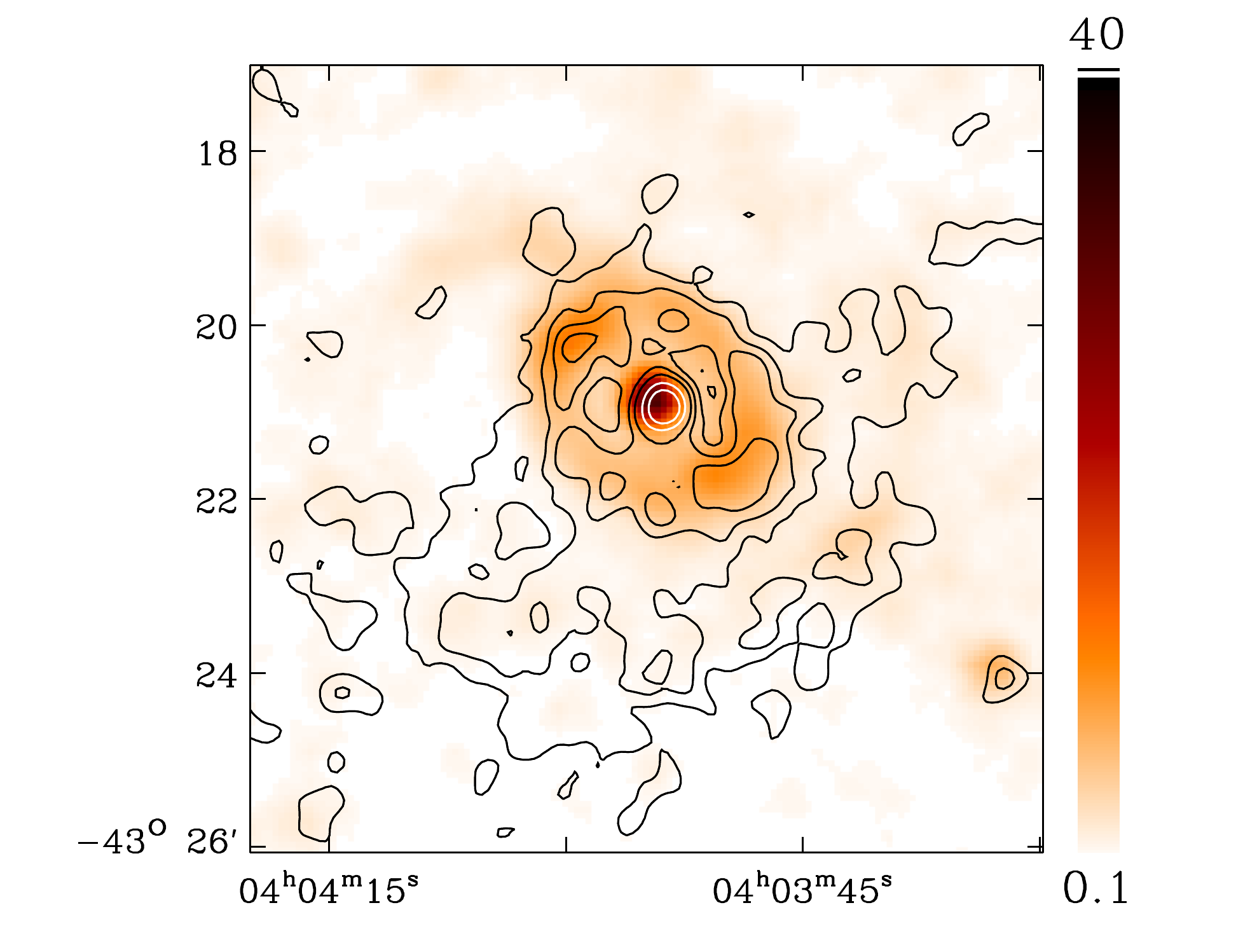} & 
\hspace{-35pt}\includegraphics[width=10cm]{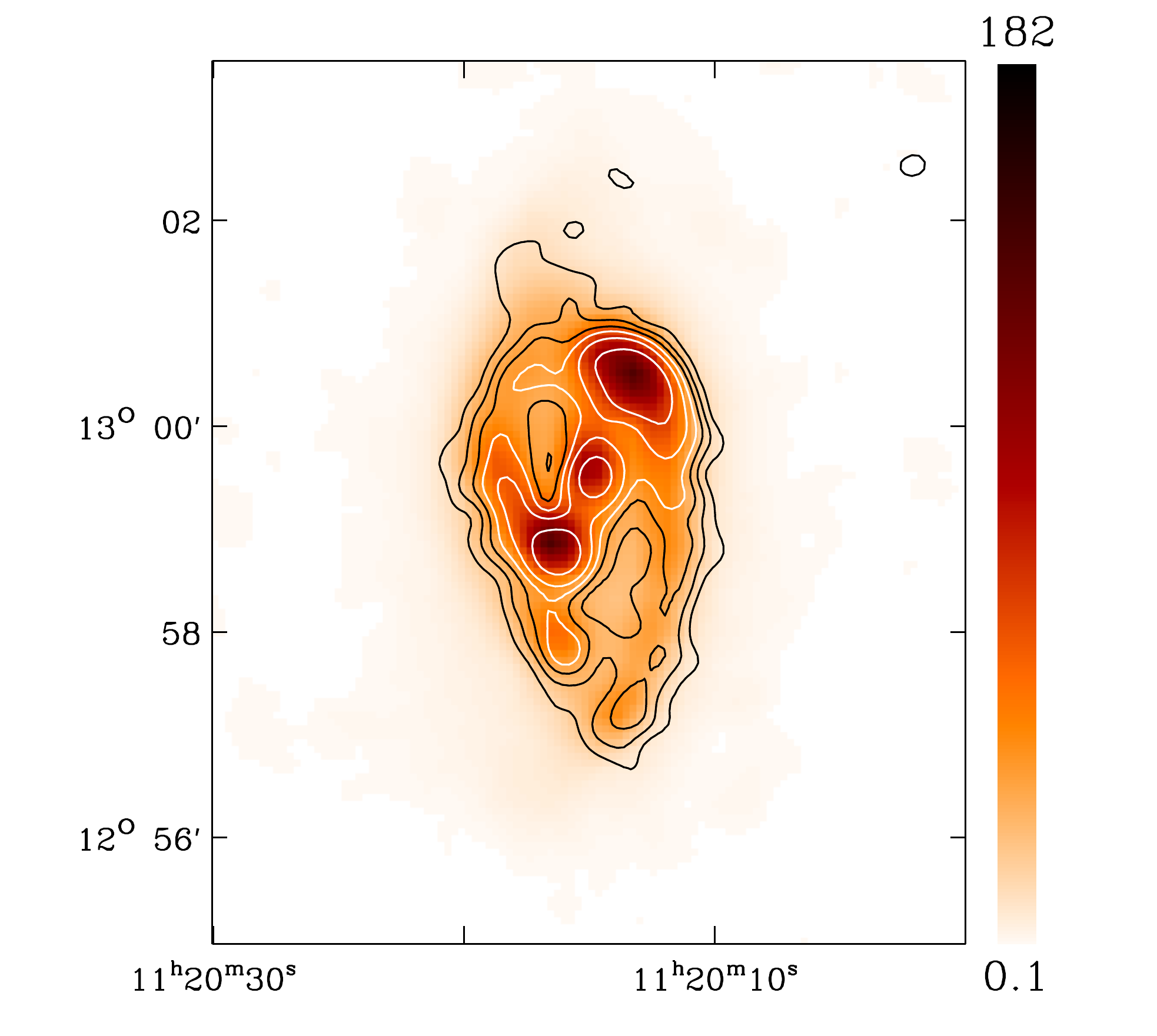} \\
{\bf \Large NGC3351} & {\bf \Large NGC4826} \\   
 \hspace{-20pt}\includegraphics[width=10cm]{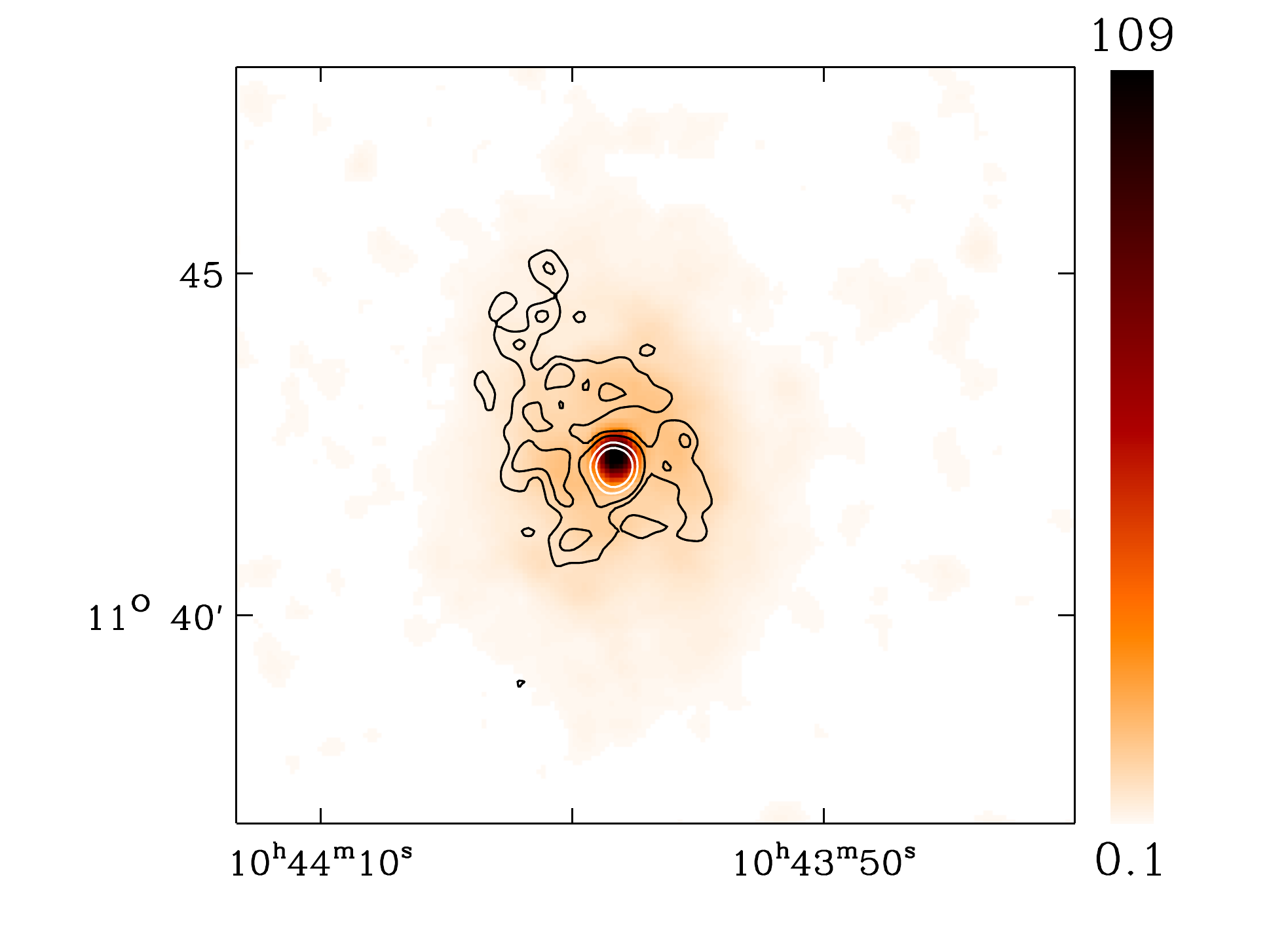} &
 \hspace{-35pt}\includegraphics[width=10cm]{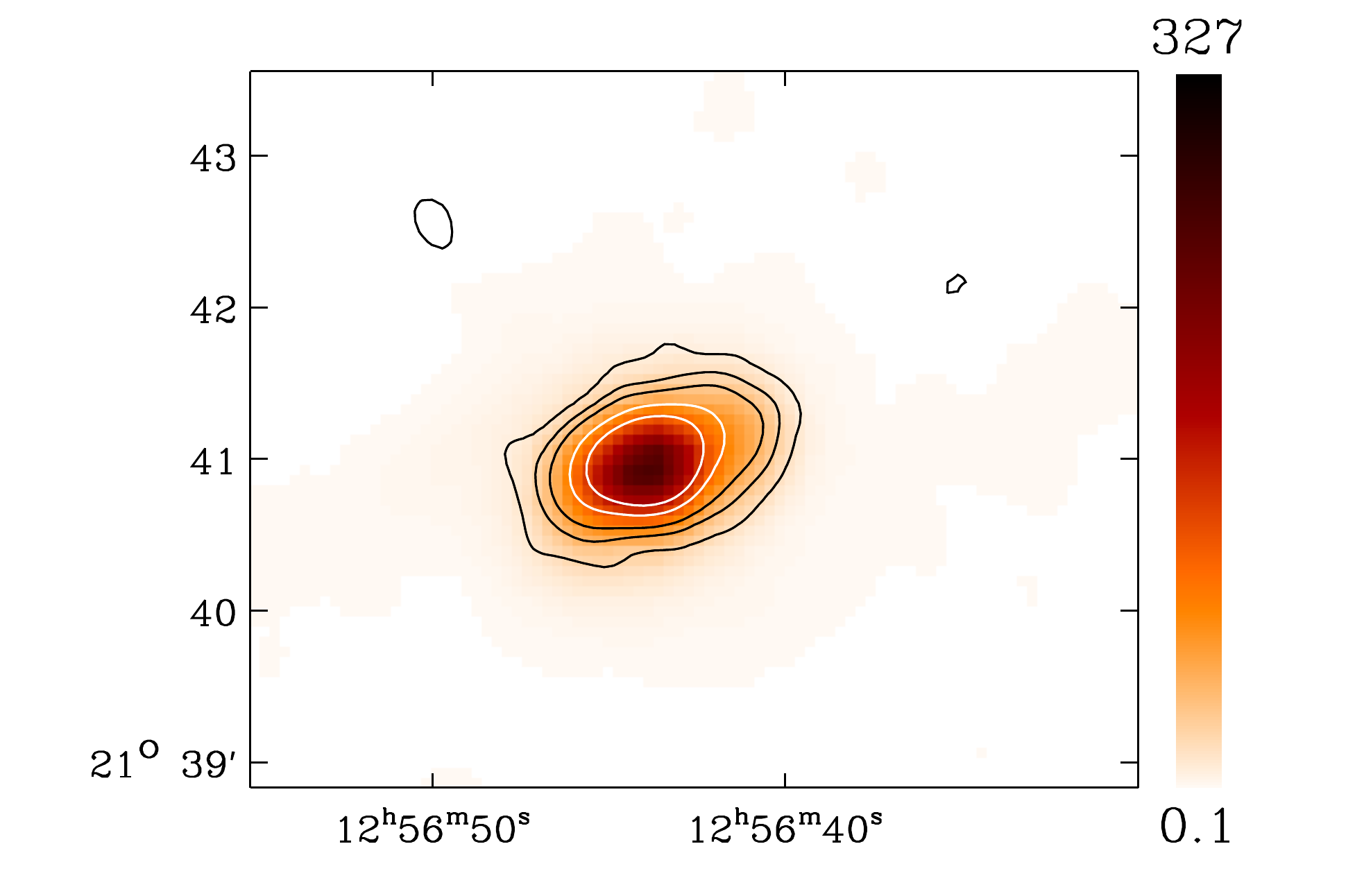} \\
  \hspace{-20pt}{\bf \Large NGC3621} & \hspace{-35pt}{\bf \Large NGC7793} \\   
  \hspace{-20pt}\includegraphics[width=10cm]{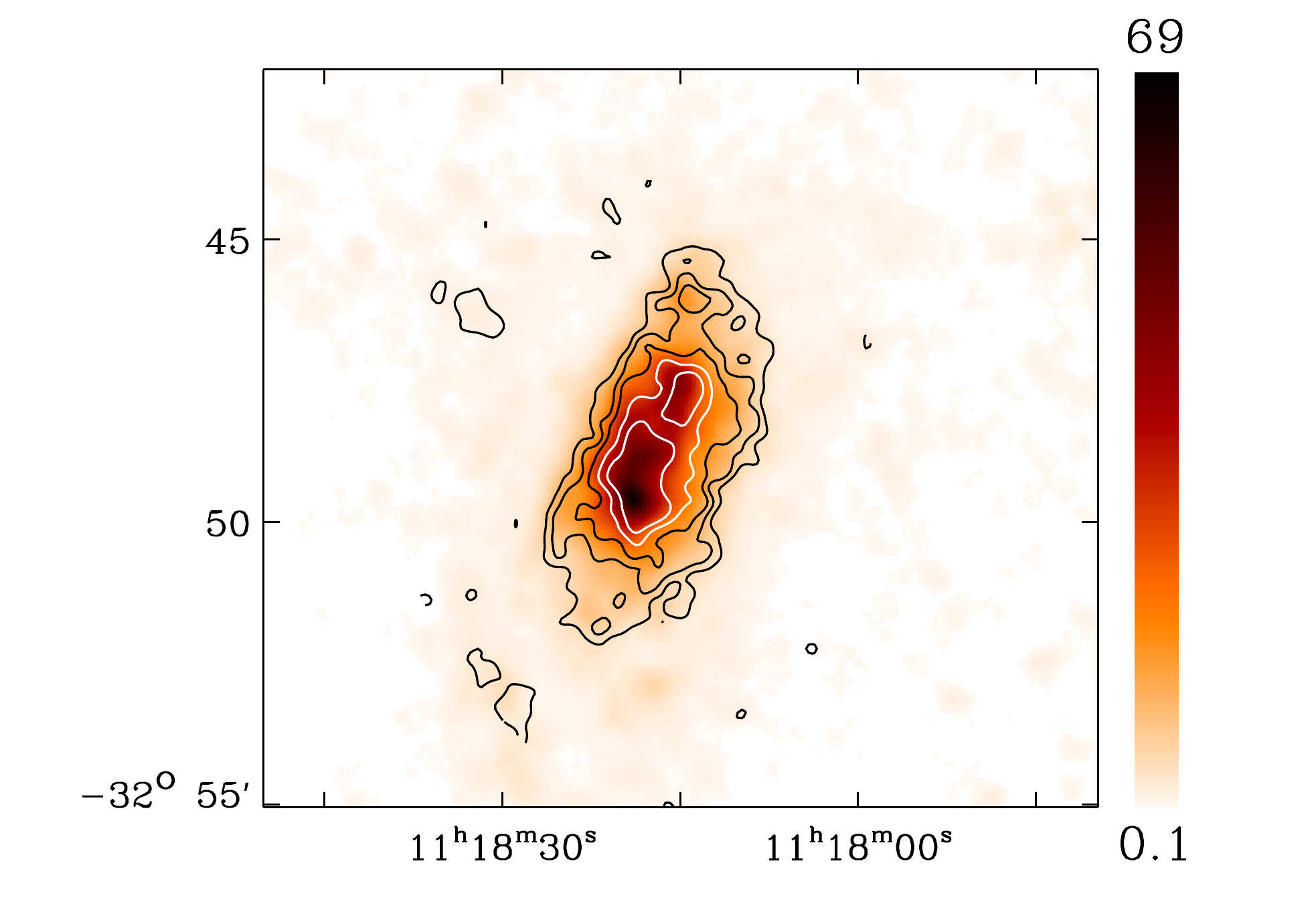} &
\hspace{-35pt}\includegraphics[width=10cm]{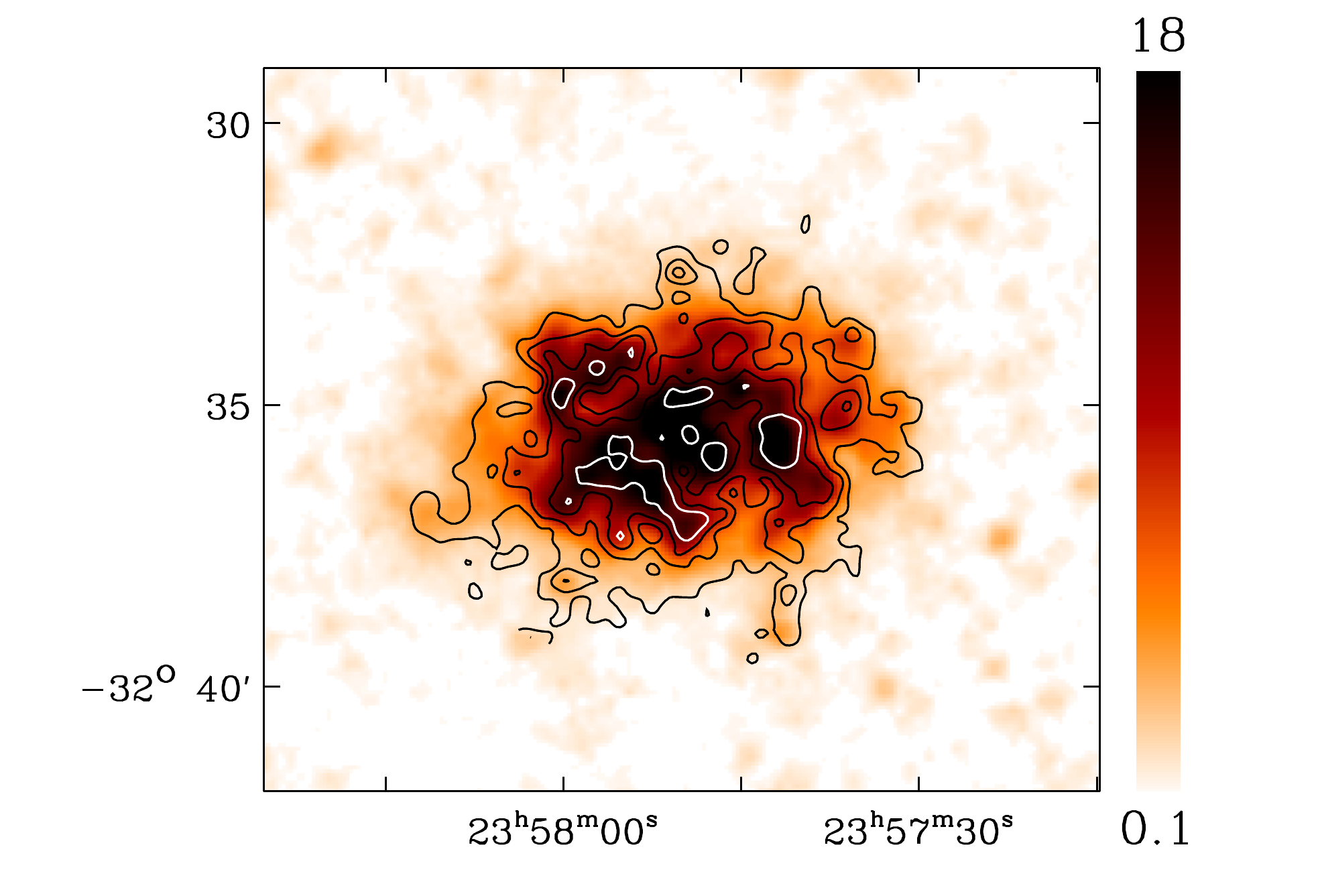} \\ 
\end{tabular}
\caption{continued. Contours are, for NGC1512: 3, 6, 9, 12, 17 and 22 $\sigma$ where $\sigma$=0.09 MJy~sr$^{-1}$, for NGC3351: 3, 5, 12 and 21 $\sigma$ where $\sigma$=0.16 MJy~sr$^{-1}$, for NGC3621: 3, 6, 9, 15 and 21 $\sigma$ where $\sigma$=0.16 MJy~sr$^{-1}$, for NGC3627: 3, 6, 9, 12, 17 and 27 $\sigma$ where $\sigma$=0.22 MJy~sr$^{-1}$, for NGC4826: 3, 9, 15, 27 and 40 $\sigma$ where $\sigma$=0.22 MJy~sr$^{-1}$, for NGC7793: 3, 6, 9, 12 and 17 $\sigma$ where $\sigma$=0.19 MJy~sr$^{-1}$.}
\end{figure*}

\subsection{\lab\ 870 \mic\ maps}

In order to search for potential excess emission beyond 500 \mic, we complement the NIR-to-FIR coverage of the dust emission with \lab\ observations of our galaxies at 870 \mic. \lab\ is a multi-channel bolometer array for continuum observations located on the Atacama Pathfinder EXperiment (APEX) telescope in the Atacama desert, Chile \citep{Siringo2009}. The FWHM of the point spread function (PSF) at 870 \mic\ is $\sim$19\farcs2 \citep{Weiss2009}, thus close to that of SPIRE 250 \mic. 

We refer to \citet{Albrecht2013} for details on the iterative data reduction procedure used to produce the LABOCA maps. A summary is provided here. The mapping is performed in spiral mode (raster pattern). Correction for atmospheric attenuation was enabled through determinations of the zenith opacity via skydips ($\sim$ every two hours). The flux calibration (achieved through observations of Mars, Uranus, Neptune and secondary calibrators) is estimated to be accurate within 9$\%$ rms. The time-ordered data streams are reduced using the BoA software\footnote{BoA was developed at MPIfR (Max-Planck-Institut fŸr Radio Astronomy, Bonn, Germany), AIfA (Argelander-Institut f{\"u}r Astronomie, Bonn, Germany), AIRUB (Astronomisches Institut der Ruhr-Universit{\"a}t, Bochum, Germany), and IAS (Institut d'Astrophysique Spatiale, Orsay, France)}, following steps of calibration, opacity and temperature drifts correction, flat fielding, conversion to Jy,  flagging of dead or bad channels, removal of the correlated noise (on the global array or induced in groups sharing the same electronics), flagging of stationary points or data taken above a given scanning velocity and acceleration threshold, $^3$He temperature drift correction, median baseline removal and despiking. A final intensity map is obtained by co-adding the individual subscans (pixel size of 4\arcsec\ $\times$ 4\arcsec) along with a weight map. The steps of correlated noise or median baseline removal can lead to a subtraction of real flux in the maps. This can be solved by using an iterative approach. From the first map produced using the technique previously described, a Ôsource modelÕ is created by isolating the pixels superior to a given signal-to-noise ratio (areas of apparent source emission defined with polygons) and the data points corresponding to the model are excluded from the baseline determination steps while re-running the reduction pipeline. The new map produced is used to build a new model and serves as an input for the next iteration. The process is repeated until the intensity map converges. Rms and signal-to-noise maps are finally produced using a jackknife technique. 

The \hersc/350 \mic\ maps are shown in Fig.~\ref{350um_Maps} with the signal-to-noise of the \lab\ 870 \mic\ maps overlaid as contours. Before display, the \lab\ maps are convolved to the resolution of the SPIRE 350 \mic\ maps (i.e.25\arcsec\ in Fig.~\ref{350um_Maps}) using the convolution kernels\footnote{http://www.astro.princeton.edu/$\sim$ganiano/Kernels.html} developed by G. Aniano \citep[see][for a description of the technique]{Aniano2011}. We restrict the black and white contours to the apertures used by \cite{Dale2012} to perform the global \hersc\ photometry of the galaxies. Therefore, the 870 \mic\ contours of NGC1317, companion of NGC1316 and also detected with LABOCA are not shown in Fig.~\ref{350um_Maps}.

\section{The 870 \mic\ emission on global scales}

\subsection{Global flux densities}
 
Along with a detailed description of the \lab\ observations and data reduction, \citet{Albrecht2013} provide the global photometry of the sample. Their elliptical apertures are defined to cover the extended emission as observed in the \spitz/MIPS 160 \mic\ maps.

Global radio emission, namely thermal bremsstrahlung emission (free-free) from electrons in the hot ionised gas or synchrotron emission from relativistic electrons moving in a magnetic field, is a possible contribution to the 870 \mic\ continuum emission. \citet{Albrecht2013} estimate the free-free contributions from H$\alpha$ measurements (with which free-free emission directly scales) using integrated H$\alpha$ + [N{\sc ii}] measurements corrected for [N{\sc ii}] contribution within the H$\alpha$ filters \citep{Kennicutt2009}. Assuming a thermal electron temperature of 10$^{4}K$ and the prescription from \citet{Niklas1997} (their eq. 2), they convert the H$\alpha$ fluxes to a free-free emission at 870 \mic\ of:

\begin{equation}
\left ( \frac{S_{ff}}{mJy}\right )=5.329 \times 10^{11}~K~\left ( \frac{S_{H\alpha}}{erg~cm^{-2}~s^{-1}} \right )
\end{equation}

\noindent where K is determined to be $\sim$2 in the sample of \citet{Niklas1997}. The free-free contribution is then subtracted from global radio continuum fluxes from the literature to determine the synchrotron emission. They determine the radio spectral index $\alpha$ (defined as {\it S$_{\nu}$} $\propto$ $\nu$$^{-\alpha}$) and extrapolate towards the LABOCA frequency. They also used a composite model combining a thermal and non-thermal fraction (with 0.1 as the thermal spectral index and a free non-thermal spectral index), model that leads to similar contributions, namely a total free-free+synchrotron contribution at 870 \mic\ of less than 2$\%$ on average for the sample. We refer to their study for further details on the technique and results for individual objects. 

CO line emission is another possible non-dust contribution to the 870 \mic\ emission since the CO(3-2) transition occurs at 867 \mic. When available, \citet{Albrecht2013} determine the CO(3-2) line emission from CO(3-2) observations, deriving the flux densities from:

\begin{equation}
S_{CO(3-2)}=\frac{2k\nu}{c^3\Delta\nu}\int{I_{CO(3-2)}d\Omega}
\end{equation}

\noindent with k the Boltzmann constant, $\nu$ the line frequency, c the speed of light, $\Delta$$\nu$ the bolometer bandwidth and I$_{CO(3-2)}$ the integral of the velocity-integrated intensity over the solid angle of the source $\Omega$ ($\Omega$ is derived from the main beam-size in the case of single-point CO observations). CO(3-2) estimates are otherwise derived from CO(1-0) or CO(2-1) observations (see Section 4.4). The global CO(3-2) contributions are ranging from less than 3$\%$ in NGC0337 and NGC0628 to up to 14$\%$ in NGC4826 \citep{Albrecht2013}. We use the global 870 \mic\ values corrected for free-free + synchrotron and CO line contribution to the 870 \mic\ flux in the following section. 
Because local contributions from CO, free-free and synchrotron emission can vary across the galaxies, we return to the estimates of the resolved non-dust contributions in the individual galaxies in Section 4.4.\\

We compare the integrated \lab\ flux densities with 870 \mic\ estimates extrapolated from observations at shorter IR/submm wavelengths. The global \spitz\ and \hersc\ flux densities are taken from \cite{Dale2007} and \citet{Dale2012} respectively. Their photometric elliptical apertures (provided in [G12]) are chosen to encompass the total emission at each \spitz\ and \hersc\ wavelength. Uncertainties on \spitz\ and \hersc\ fluxes were computed as a combination in quadrature of the calibration uncertainty and the measurement uncertainty. The SPIRE beam sizes have been updated since the publication of \citet{Dale2012} and [G12]. The SPIRE global flux densities are thus rescaled to the new beam size estimates (see Section 2.2 for values), leading to a 7-9$\%$ correction compared to [G12]. 
Discrepancies can be observed between MIPS 70 \mic\ and \hersc/PACS 70 \mic\ fluxes \citep{Aniano2012} because PACS is less sensitive than MIPS to diffuse emission and because the instruments possess different filter profiles. We thus decide to use both observations in our modelling, to be conservative. Color corrections will be applied during the SED model fitting.
The contribution of C+ to the 160 \mic\ emission is considered to be minor (C{\sc ii}/L$_{TIR}$$\sim$0.15$\%$ from \citet{Brauher2008} or \citet{Malhotra2001} and L$_{TIR}$/$\nu$F$_{\nu}$(160)$\sim$2.5 from \citet{Dale2009} so C{\sc ii}/$\nu$F$_{\nu}$(160)$\sim$0.4$\%$).
The SPIRE flux densities are also not corrected for line contribution but potential line features (such as [N\,{\sc ii}]205\mic\ or CO) are considered to be minor compared to continuum emission.  \\

{\it Comparison with \plk\ data -} We can compare the global \hersc\ 350 and 500 \mic\ flux densities and the LABOCA 870 \mic\ flux densities with the \plk\ global flux densities at 350 \mic, 550 \mic, 850 \mic\ (FWHMs of 4.3\arcmin, 4.7\arcmin, 4.8\arcmin\ respectively). The \plk\ flux densities are provided by the Planck Catalogue of Compact Sources \citep[PCCS][]{Planck_collabo_PCCS_2013} and obtained through the IPAC Infrared Science Archive\footnote{http://irsa.ipac.caltech.edu/applications/planck/}. Most of our sources are resolved by \plk. We choose the flux densities estimated using the APERFLUX technique (fitting of a variable circular aperture centred at the position of the source). The \plk\ global flux densities at 350 \mic, 550 \mic, 850 \mic\ and 1.3mm are overlaid on the global SEDs of the sample (see Section 3.2 for the description of the modelling) in Fig.~\ref{GlobalSEDs}. We observe a very good agreement between the global flux densities in the [350-850] \mic\ range. The mean value of the \plk350/SPIRE350 \mic\ flux density ratios is 0.96$\pm$0.19 and the mean value of the \plk850/LABOCA870 \mic\ flux density ratios is 0.87$\pm$0.28. The mean value of the \plk550/SPIRE500 \mic\ flux density ratios is 0.64$\pm$0.15, but becomes 0.94$\pm$0.23 when \plk\ 550 \mic\ flux densities are converted to 500 \mic\ flux densities using a (550/500)$^4$ factor between the two (i.e. assuming dust with a $\nu$$^2$ opacity and a Rayleigh-Jeans behaviour). Using the GAUFLUX values from the PCCS (fitting of a variable Gaussian model to the source), ratios are equal to 1.20$\pm$0.21 (when converted as above) at 500 \mic\ and 1.07$\pm$0.30 at 870 \mic, which is also in very good agreement as well. In particular for the galaxy NGC1512, the GAUFLUX estimates are more consistent with the \hersc\ and LABOCA data than the APERFLUX estimates. 
In some galaxies, the dispersion in the \plk-to-SPIRE or \plk-to-LABOCA ratios are relatively high compared to the calibration uncertainties, i.e. 1\% at 850\mic\ and 10\% at 350 and 550\mic\ \citep{Planck_collabo_PCCS_2013}, 7\% for SPIRE (see Section 2.2 for references), and 9\% for LABOCA. The dispersions in the Planck/SPIRE ratios is however consistent with what has been reported in other studies \citep[see the work of ][for instance]{Herranz2013}. Some contamination of the \plk\ fluxes could originate from the large beam sizes of the telescope, in particular for NGC0337 which is smaller than \plk's beam. This could explain the large differences  we observe for this particular galaxy. Differences between the spectral responses of the \plk, LABOCA and SPIRE instruments are not taken into account in this comparison either, which could also explain the dispersions.\\

\subsection{Global SED modelling technique}

As in [G12], we modelled the global SED of thermal dust emission of our galaxies from 24 to 500 \mic\ using modified blackbodies (MBB). We refer to this study for a discussion of the model, the choice of parameters, and the possible degeneracies. We remind some of the potential issues of the use of MBBs at the end of this section. Since we want to investigate potential excess emission at 870\mic, the \lab\ observations are not included in the fitting procedure. In [G12], we show that a non-negligible fraction of the 70 \mic\ fluxes can be attributed to warm dust. We thus choose to account for this contribution by using a two-temperature (warm + cold dust) fitting technique, each component being modelled as a modified blackbody:

\begin{equation}
L_\nu(\lambda) = A_w \lambda^{-\beta_w} B_\nu(\lambda,T_w) +
         A_c \lambda^{-\beta_c} B_\nu(\lambda,T_c)
\end{equation}

\noindent {\it B$_\nu$} is the Planck function. {\it T$_w$} and {\it T$_c$} are the temperature of the warm and cold component respectively while $\beta$$_w$ and $\beta$$_c$ are the emissivity index of the warm and cold component respectively. {\it A$_w$} and {\it A$_c$} are scaling coefficients, functions of the optical depths contributed by the cold and warm dust components at a given reference wavelength. We determine the color correction for each band by convolving our 2MBB spectra with the instrumental response function using the different \spitz\ and \hersc\ user's manual conventions\footnote{The link to PACS and SPIRE observer manuals in which these conventions can be found are provided earlier in this paper. The reader can access the MIPS Instrument handbook here: 

http://irsa.ipac.caltech.edu/data/SPITZER/docs/mips/mipsinstrumenthandbook/}. The curve fitting procedure applies a Levenberg-Marquardt $\chi$$^2$ minimisation and weights flux densities by the inverse squared of the uncertainty. 
The emissivity index of the warm component is fixed to a value $\beta$$_w$ = 2.0, an approximation of the opacity in the standard \citep{Li_Draine_2001} dust models. Discussion on the influence of the warm dust component and of the 24-to-100 \mic\ flux variations in the sample can be found in [G12] (their Section 3.4). We fix the emissivity index of the cold dust component to two different values: $\beta$$_c$=2.0 and $\beta$$_c$=1.5, in order to study how the two models match the observations at 870 \mic. 
Degeneracies between the temperature and the emissivity index are observed when the two parameters are allowed to vary at the same time. This can lead to an artificial anti-correlation between the emissivity index and the temperature due to noise effects \citep[][among others]{Shetty2009}. An ``apparent" flattening of the slope of the SED could be created as well from temperature mixing effects. On global scales as well as on local scales (our pixels cover large ISM elements), various dust grain populations with different temperatures could be mixed, leading to biases in the emissivity estimates. For both reasons, [G12] caution the use of a variable emissivity to model SEDs at the local scale of consideration. We will not apply this technique on the resolved analysis (Section 4). We nevertheless model the free $\beta$$_c$ assumption at global scales in order to test which values are invoked to explain the SPIRE slopes without {\it a priori} assumption the dust emissivity index value and detect potential strong flattening of the submm slope in our objects. 

While \citet{Klaas2001} or \citet{Dunne_Eales_2001} showed that the SEDs of nearby galaxies could be modelled with either one modified blackbody with a variable $\beta$ or multiple modified blackbodies with $\beta$=2.0, others showed that the bulk of the far-infrared emission from dust grains could be fitted with dust heated by a single radiation field \citep{Draine2007}. Recently, other studies have shown that dust emitting in the 100-to-500 \mic\ range may be heated by both the star forming regions and the older stellar populations \citep{Boquien2011, Bendo2012}, which questions the single MBB fitting of that full wavelength range. Isothermal or 2-temperature fits are not fully able to model the range of grain temperatures observed in each of our resolved ISM elements. We remind the reader that what we call ``emissivity index'' in the following analysis is thus, in fact, the ``effective emissivity index'' (i.e. the power-law index resulting from the different grain populations) rather than the ``intrinsic emissivity index" (i.e. the power-law index of the MBB modelling the emission of a dust grain at a given temperature). The following analysis thus primarily investigates submm slope trends with environment rather than directly probing the intrinsic emissivity properties of cold grains. \\

\begin{figure*}
    \centering 
	\vspace{-40pt}  
	\begin{tabular}{c}
	\hspace{-30pt}
\includegraphics[width=19.5cm]{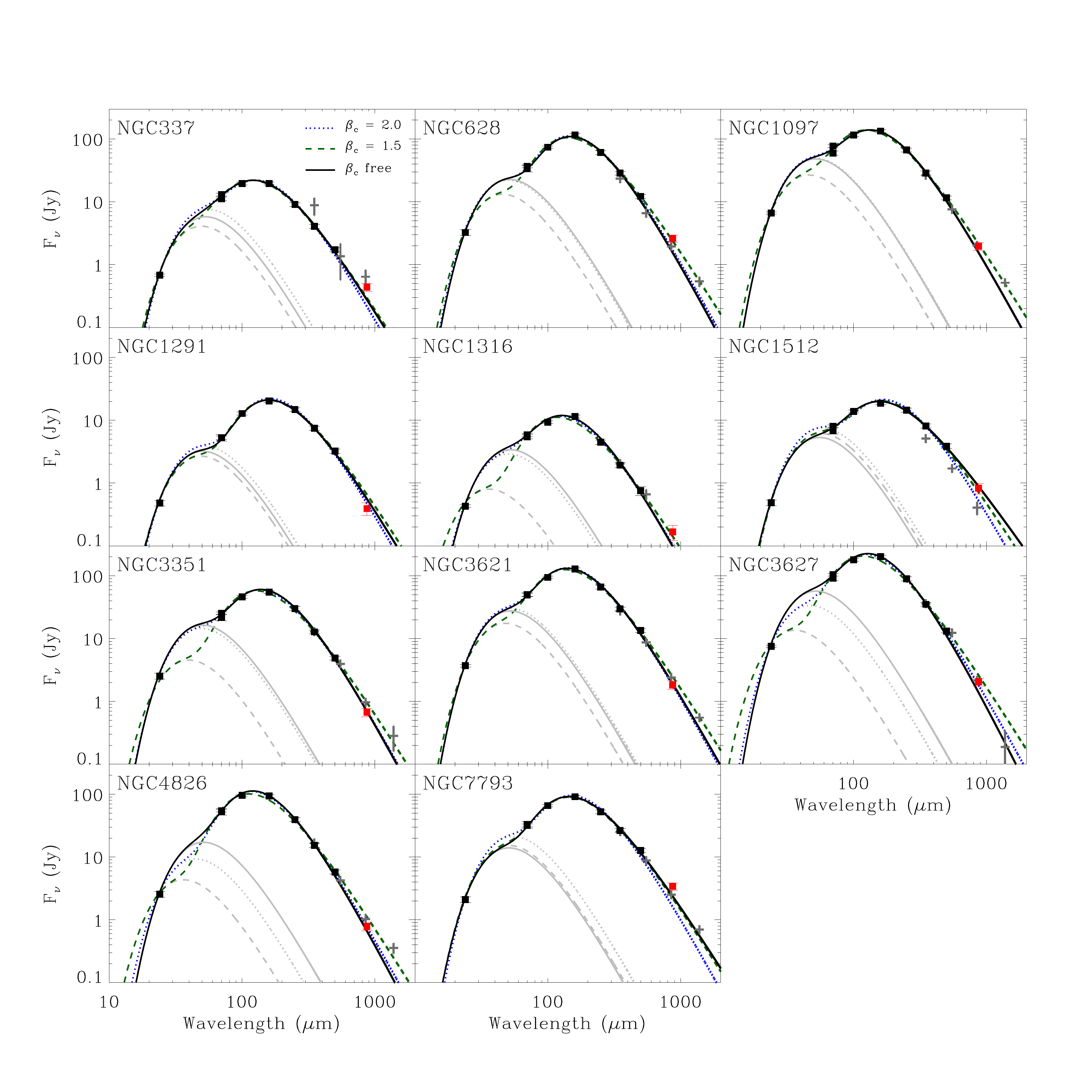} \\ 
 \end{tabular}
 \vspace{-30pt}
\caption{Global SEDs obtained with a two-temperature fitting technique. The model using an emissivity index of the cold component $\beta$$_c$ fixed to 2.0 is shown by a dotted blue line, with $\beta$$_c$ fixed to 1.5 by a long dashed green line and with a free $\beta$$_c$ by a black solid line. The corresponding warm components (with $\beta$$_w$ fixed to 2.0 in the three cases) are overlaid in grey. MIPS, PACS and SPIRE data are overlaid with black rectangles. Observed \lab\ flux densities (not included in the fit) are overlaid in red and have been corrected for CO and radio continuum (see Section 3.1). We also indicate, when available, \plk\ global flux densities at 350, 550 \mic, 850 \mic\ and 1.3 mm (grey crosses) but do not include them in the fit. The height of the grey crosses represents the uncertainties of the \plk\ fluxes.}

    \label{GlobalSEDs}
\end{figure*}

We present the global SEDs obtained using the three different hypotheses: $\beta$$_c$ fixed to 2.0 (dotted blue line), fixed to 1.5 (dashed green line) or free (plain black line) in Fig.~\ref{GlobalSEDs}. We overlay the warm dust component for each model in grey, PACS and SPIRE data with black rectangles and the LABOCA 870 \mic\ flux densities corrected for free-free and synchrotron and CO line contribution \citep[taken from][]{Albrecht2013} with red rectangles. 
As mentioned before, we also overlay (when available) the \plk\ global flux densities at 350 \mic, 550 \mic, 850 \mic\ as well as the \plk\ 1.3mm flux density but did not use them in the fit.

\begin{table*}
\centering 
\caption{\large Comparison of the modelled and observed integrated 870 \mic\ flux densities}
\begin{tabular}{cccccc}
\hline
\hline
\\
 Galaxy & {\bf 870~obs~corrected~$^{a}$} & \multicolumn{2}{c}{$\beta$$_c$} &  ~~870~modelled~~ &  ~~$\frac{\textrm{870~obs~corrected}~^{a}}{\textrm{870~modelled}}$~~	 \\
 \cline{3-4} 
&	{\bf (Jy)} &	~~Method~~ & ~~Value~~		&	(Jy) & \\
\\
\hline
\\
NGC0337&
{\bf 0.44$\pm$0.07} 	&fixed & 2.0 & 		0.22 $\pm$ 0.02 &    2.03 $\pm$ 0.47\\
	 			&&fixed & 1.5  & 		0.30 $\pm$ 0.03 &    1.47 $\pm$ 0.35\\
				&&free &  1.65 $\pm$ 0.15 &  0.26 $\pm$ 0.04  &  1.66 $\pm$ 0.48\\
&\\
NGC0628&
{\bf 2.61$\pm$0.34} 	&fixed & 2.0 & 	1.69 $\pm$ 0.15 &    1.55 $\pm$ 0.34\\
	 			&&fixed & 1.5  & 	2.42 $\pm$ 0.19 &    1.08 $\pm$ 0.23\\
	 			&&free &  2.12 $\pm$ 0.20 &   1.58 $\pm$ 0.27 &    1.65 $\pm$ 0.50\\
&\\
NGC1097&
{\bf 1.96$\pm$0.28} 	&fixed & 2.0 & 	1.58 $\pm$ 0.14 &    1.24 $\pm$ 0.28\\
	 			&&fixed & 1.5  & 	2.20 $\pm$ 0.18 &    0.89 $\pm$ 0.20\\
	 			&&free &  2.02 $\pm$ 0.19 &  1.57 $\pm$ 0.25 &    1.25 $\pm$ 0.38\\
&\\
NGC1291&
{\bf 0.39$\pm$0.09}	         &fixed & 2.0 & 	0.47 $\pm$ 0.05 &    0.83 $\pm$ 0.27\\
	 			&&fixed & 1.5  & 	0.65 $\pm$ 0.05 &    0.60 $\pm$ 0.19\\
	 			&&free &  1.72 $\pm$ 0.18 &  0.56 $\pm$ 0.08 &    0.70 $\pm$ 0.26\\
&\\
NGC1316&
{\bf 0.17$\pm$0.05} 	&fixed & 2.0 & 	0.10 $\pm$ 0.02 &    1.61 $\pm$ 0.68\\
	 			&&fixed & 1.5  & 	0.14 $\pm$ 0.02 &    1.18 $\pm$ 0.46\\
	 			&&free &  2.15 $\pm$ 0.28 &  0.10 $\pm$ 0.03 &    1.68 $\pm$ 0.92\\
&\\
NGC1512&
{\bf 0.83$\pm$0.16} 	&fixed & 2.0 & 	0.53 $\pm$ 0.05 &    1.57 $\pm$ 0.47\\
	 			&&fixed & 1.5  & 	0.72 $\pm$ 0.06 &    1.15 $\pm$ 0.33\\
	 			&&free &  1.16 $\pm$ 0.19 &  0.90 $\pm$ 0.14  &  0.92 $\pm$ 0.32\\
&\\
NGC3351&
{\bf 0.67$\pm$0.10} 	&fixed & 2.0 & 	0.71 $\pm$ 0.06 &    0.94 $\pm$ 0.22\\
	 			&&fixed & 1.5  & 	0.95 $\pm$ 0.07 &   0.70 $\pm$ 0.16\\
	 			&&free &  2.10 $\pm$ 0.21 &  0.66 $\pm$ 0.11 &    1.02 $\pm$ 0.32\\
&\\
NGC3621&
{\bf 1.83$\pm$0.27} 	&fixed & 2.0 & 	1.81 $\pm$ 0.15 &    1.01 $\pm$ 0.23\\
	 			&&fixed & 1.5  & 	2.48 $\pm$ 0.18 &    0.74 $\pm$ 0.16\\
	 			&&free &  1.91 $\pm$ 0.16 &  1.87 $\pm$ 0.28 &   0.98 $\pm$ 0.29\\
&\\
NGC3627&
{\bf 2.05$\pm$0.29} 	&fixed & 2.0 & 	1.81 $\pm$ 0.13 &   1.13 $\pm$ 0.24\\
	 			&&fixed & 1.5  & 	2.61 $\pm$ 0.17 &    0.78 $\pm$ 0.16\\
	 			&&free &  2.39 $\pm$ 0.19 &  1.45 $\pm$ 0.24 &    1.41 $\pm$ 0.44\\
&\\
NGC4826&
{\bf 0.78$\pm$0.11} 	&fixed & 2.0 & 	0.76 $\pm$ 0.07 &    1.02 $\pm$ 0.24\\
	 			&&fixed & 1.5  & 	1.11 $\pm$ 0.07 &   0.70 $\pm$ 0.15\\
	 			&&free &  2.21 $\pm$ 0.17 &  0.67 $\pm$ 0.11 &   1.16 $\pm$ 0.36\\
&\\
NGC7793&
{\bf 3.40$\pm$0.46} 	&fixed & 2.0 & 	1.62 $\pm$ 0.13 &   2.10 $\pm$ 0.45\\
	 			&&fixed & 1.5  & 	2.22 $\pm$ 0.19 &    1.53 $\pm$ 0.34\\
	 			&&free &  1.44 $\pm$ 0.16 & 2.36 $\pm$ 0.35 &   1.44 $\pm$ 0.41\\
&\\
\hline
\end{tabular}
\label{Predictions_global}
\begin{list}{}{}
\item $^a$ These fluxes are taken from \citet{Albrecht2013}. The observed 870 \mic\ flux densities have been corrected for CO and radio continuum. 
 \end{list}
\end{table*}

\subsection{870 \mic\ predictions and comparison with observations}

We deduce global flux estimates at 870 \mic\ from the different 2MBB models and estimate the uncertainties using a Monte-Carlo technique: we generate 100 sets of modified 24-to-500 \mic\ constraints, with fluxes varying randomly within the individual error bars (following a normal distribution around their nominal value). The absolute calibration is the dominating source of uncertainty and is highly correlated across the SPIRE bands (SPIRE's observers' manual). SPIRE measurements are thus randomly modified but in a similar correlated way. We then apply our fitting technique to the 100 modified datasets. The global 870 \mic\ estimates we extrapolate are the medians of these distributions; we take the standard deviation of each distribution as the uncertainty. \\

Table~\ref{Predictions_global} summarises for each galaxy the global 870 \mic\ estimate and 1-$\sigma$ uncertainty obtained using our 3 different MBB fitting techniques ($\beta$$_c$=2.0, 1.5, and $\beta$$_c$ free). The ratios between the observed 870 \mic\ flux densities of \citet{Albrecht2013}, corrected for free-free and synchrotron emission and line contribution (called ``870~obs~corrected" in the table), and our extrapolations from the different models ($\beta$$_c$=2.0 or 1.5 and $\beta$$_c$ free) are also provided. The update of the SPIRE beam sizes discussed in Section 3.1 does not significantly change the $\beta$ values derived when this parameter is free in the models compared to the results of [G12] (variations by less than 5$\%$ on average). 

For 7 galaxies (NGC1097, NGC1291, NGC1316, NGC3351, NGC3621, NGC3627 and NGC4826), the observed 870 \mic\ flux densities match the values estimated using $\beta$$_c$=2.0, within the uncertainties. The global 870 \mic\ emission in those objects could thus be explained by thermal emission modelled using dust grains with standard properties.  However, we note that for NGC1316 and NGC1097, using $\beta$$_c$=1.5 does provide a better prediction of the observed 870 \mic\ (observed-to-modelled ratio closer to unity) than the ``$\beta$$_c$=2.0" case. 

The 4 remaining galaxies (NGC0337, NGC0628, NGC1512 and NGC7793) show a 870 \mic\ excess when $\beta$$_c$=2.0 is used. For NGC0628 and NGC1512, using a flatter emissivity index $\beta$$_c$=1.5 can reconcile the observed and modelled 870 \mic\ global emission within the uncertainties, as shown by their observed-to-modelled ratio close to unity. For the two remaining galaxies NGC0337 and NGC7793, the submm slope constrained from \spitz+\hersc\ data is already flat ($\beta$$_c$ $<$1.7 when free to vary) and we observe an excess of the 870 \mic\ emission compared to the extrapolations, even when an emissivity $\beta$$_c$=1.5 is used ($\sim$50\% above the model predictions on average), even if this excess is statistically weak (inferior to a 2$\sigma$ level). A cold dust component with an even lower emissivity index could be necessary in these galaxies to match the observations. Including the 870 \mic\ data in the fitting procedure leads to $\beta$$_c$ values of 1.33$\pm$0.15 and 1.19$\pm$0.14 for NGC0337 and NGC7793 respectively, thus at the lower end of values typically used in the literature. This questions the origin of the excess detected at 870 \mic\ and the use of a simple isothermal component to fit the cold dust population up to 870 \mic\ in those two objects. 

As mentioned before, using $\beta$$_c$=1.5 reconciles the observed and modelled 870 \mic\ global emission for NGC1512, within the 870 \mic\ uncertainty. Nevertheless, when $\beta$$_c$ is allowed to vary, the model fit to \spitz\ and \hersc\ data favours very low $\beta$$_c$ values ($\beta$$_c$=1.16). Including the 870 \mic\ data in the fitting procedure, we find $\beta$$_c$=1.22$\pm$0.15 for NGC1512. The 870 \mic\ observation thus confirms the shallow slope. The 870 \mic\ observation reveals an extended structure in the southeast of NGC 1512 that is not observed in the SPIRE maps (Fig.\ref{350um_Maps}) and accounts for 30-40\% of the global flux. \citet{Albrecht2013} could not firmly determine the nature of this extended emission that contaminates the map and therefore the global 870 \mic\ flux estimates. They note that the global flux density is only a factor 1.3 higher than the Planck value reported for this object in the Early Release Compact Source Catalogue \citep[][GAUFLUX technique]{Planck_collabo_2011_ERCSC}. This is coherent with the contamination we estimate.

   \begin{figure*}
    \centering   
 \includegraphics[width=18cm]{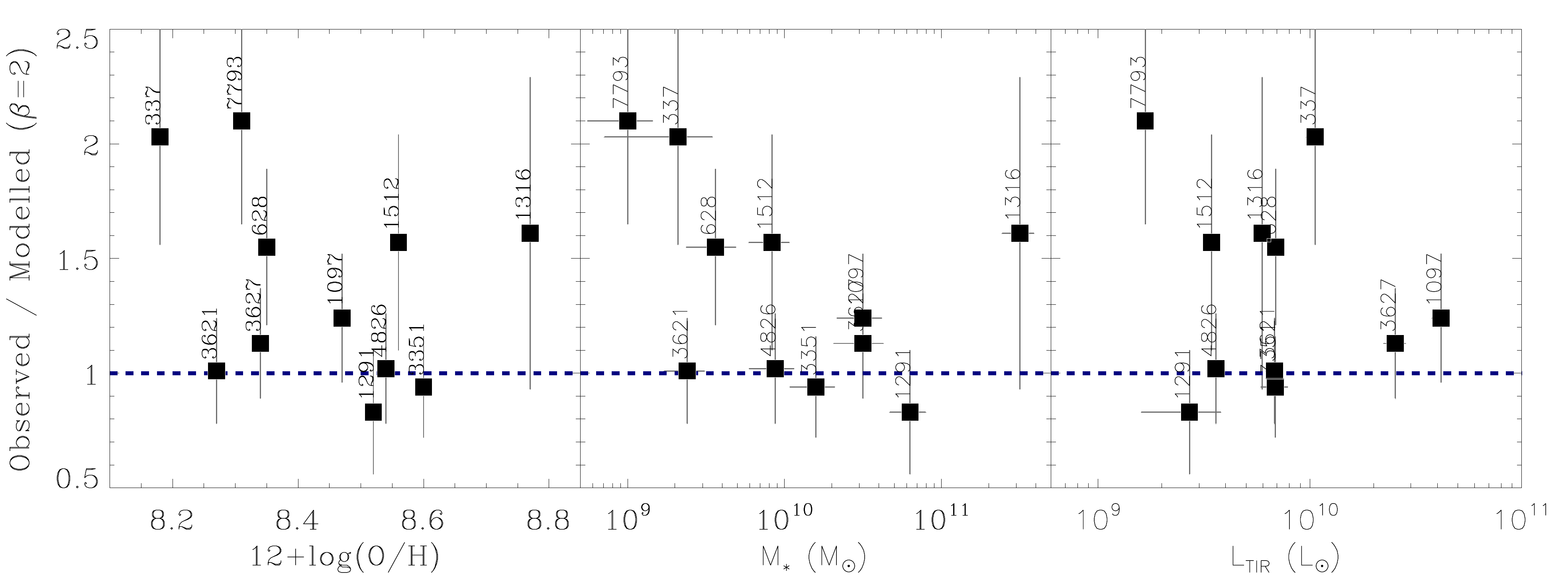}     \\ 
 \includegraphics[width=18cm]{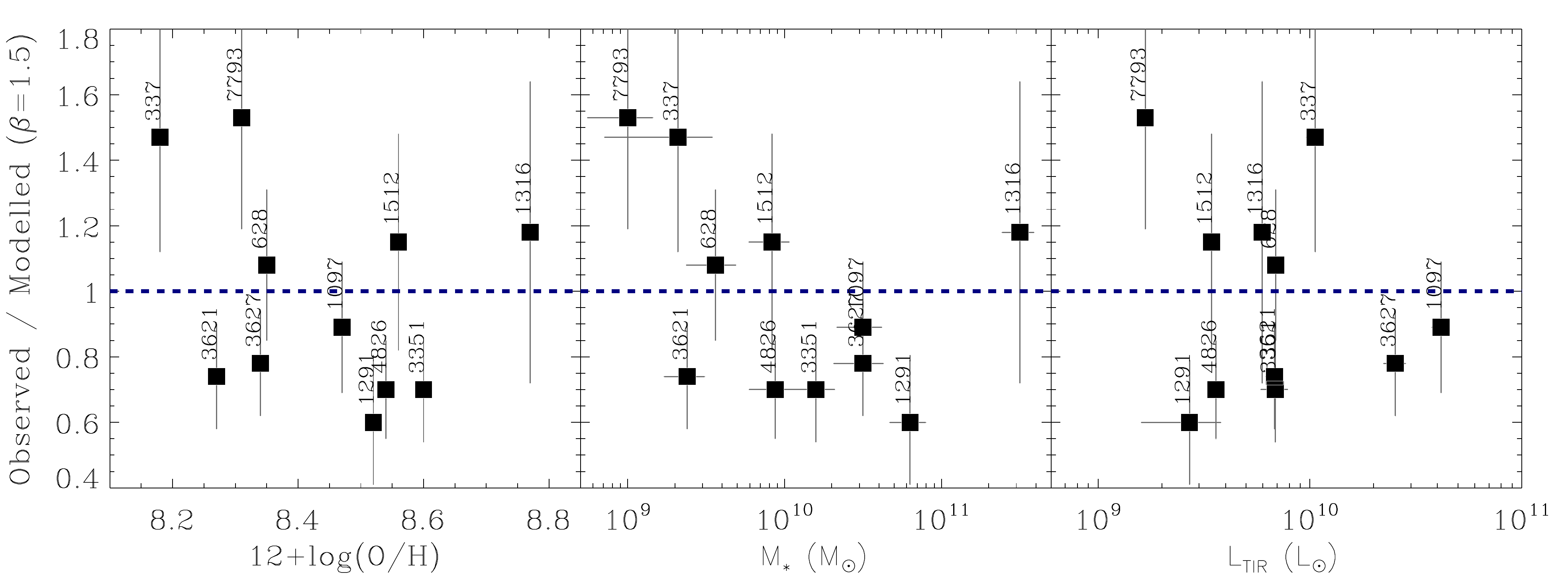} \\ 
\caption{Ratio between the observed 870 \mic\ flux densities and the 870 \mic\ estimates predicted from the $\beta$ = 2 (top) and $\beta$ = 1.5 (bottom) model as a function of, from left to right, metallicity (expressed as oxygen abundances), stellar mass in (\msun) and total infrared luminosity (in \lsun). The observed 870 \mic\ flux densities are corrected for free-free and synchrotron emission and line contribution (see Section 3.1). Labels indicate the galaxies 'NGC' numbers. The horizontal line indicates unity.}
    \label{ObsoverExtrap}
\end{figure*}

\subsection{Dependence with global properties}

We now analyse how the excess (or deficit) of the observed 870 \mic\ emission compared to our extrapolations varies with the global characteristics of our objects. In Fig.~\ref{ObsoverExtrap}, we plot the ratios between the observed 870 \mic\ flux densities and the 870 \mic\ estimates extrapolated from the $\beta$=2 (top panel) and $\beta$=1.5 (bottom panel) models as a function of the average metallicity expressed as oxygen abundances (left) and as a function of the total stellar mass in \msun\ (middle). Metallicities, listed in Table~\ref{Galaxy_data}, are taken from \citet{Kennicutt2011} and obtained using the calibration of \citet{Pilyugin2005}. We use the stellar masses derived by \citet{Skibba2011}. The observed 870 \mic\ flux densities are corrected for free-free and synchrotron emission and CO(3-2) line contribution. In this figure, the ratio of the observed-to-modelled flux can be used as a ``proxy" for submm slope variations to a certain extent, since it characterises how the observed slope varies compared to a theoretical slope of $\beta$=2 or $\beta$=1.5. 
 
The difference we derive between the observed and the modelled 870 \mic\ flux densities varies from one object to the other but does not seem to show a particular trend with metallicity, at least within the average oxygen abundance range probed by our sample. We however find significantly higher observed fluxes than those extrapolated from our models in NGC0337 and NGC7793, two of the 3 galaxies with the lowest metallicities. This could potentially be a sign of a flattening of the submm slope (or decrease of the effective emissivity index $\beta$) or a presence of ``submm excess" with metallicity. We observe a weak trend with the global stellar mass of the galaxies as well as with galaxy morphology (not plotted in this figure). The galaxy with the highest metallicity (and the highest stellar mass) NGC1316 seems to be an outlier to this general trend but the uncertainty on the ratio is significant for this object. 

Figure~\ref{ObsoverExtrap} (right) shows the ratios plotted as a function of the total infrared luminosities of the galaxies, a quantity often used to derive star formation rates \citep{Kennicutt1998}. The total infrared luminosities of the KINGFISH sample are taken from \citet{Galametz2013b} and reported in Table~\ref{Galaxy_data}. No clear dependence is observed with this parameter. 


\section{The 870 \mic\ emission on local scales}

In order to understand the range of emissivity index values we observe on global scales, we now try to characterise variations of the submm slope at 870 \mic\ within our galaxies.

\subsection{Extrapolations using 2MBB techniques}

We convolve our \spitz\ and \hersc\ images to the lowest resolution of the dataset, namely that of SPIRE 500 \mic, using the convolution kernels generated by \citet{Aniano2011}. Maps are then projected to a common sample grid with a pixel size of 14\arcsec\ (the standard pixel size of the SPIRE 500 \mic\ images). This size corresponds to ISM elements ranging from $\sim$265 pc for our closest galaxy NGC7793 to $\sim$1.4 kpc for NGC1316. We then use our two-temperature fitting technique to model the SED in each pixel. We only perform the SED fitting for ISM elements above the 2$\sigma$ detection at 870 \mic\ \citep[according to the signal-to-noise maps produced by][]{Albrecht2013} but the LABOCA data are not included in the fitting process. We refer to [G12] for a detailed discussion of the various cold dust temperature and emissivity index maps derived using MBB fitting. The update of the SPIRE beam sizes mentioned in Section 3 does not affect our conclusions on the distribution of these cold dust properties. 

As previously mentioned in Section 3.2, we choose to fix $\beta$$_c$ in the following analysis. This will {\it i)} prevent biases (T-$\beta$ degeneracy, temperature mixing) from affecting our maps and {\it ii)} enable us to study the behaviour of the submm slope on resolved scales compared to models for which dust properties are perfectly constrained. Beyond the 870 \mic\ excess issue, we thus also probe potential variations of the submm slope in our objects. 

We produce 870 \mic\ maps of our 11 galaxies for 2 different cases, $\beta$$_{c}$ fixed to 2.0 and 1.5. We also convolve the \lab\ images to the SPIRE 500 \mic\ resolution using the convolution kernels developed by \citet{Aniano2011} (see Section 2.2 for web reference) and regrid the maps to our common pixel grid (pixel size: 14\arcsec). For this study, the 870 \mic\ maps have not been corrected for CO or radio contamination. We discuss their potential contribution to the observed 870 \mic\ fluxes in Section 4.4. Figure~\ref{Excess_maps_NGC3627} gives an example of the observed 870 \mic\ map of NGC3627 (top line) along with the 870 \mic\ maps extrapolated from the models with $\beta$$_{c}$=2.0 (second line, first panel) or 1.5 (second line, middle panel). The maps of the full sample derived using the 2MBB techniques are given in appendix in Fig.~\ref{Excess_maps}. We use the same color scale for the observed and modelled 870 \mic\ maps to allow a direct comparison.

\subsection{Extrapolations using the \citet{Draine_Li_2007} model}

We want to compare the 870 \mic\ estimates obtained from the 2MBB resolved techniques with those predicted by a more physical (as far as dust composition and ISM physics are concerned) SED model. We thus also build pixel-by-pixel 870 \mic\ maps using the resolved SED modelling of the KINGFISH sample provided in \citet{Aniano2012} and \citet{Aniano2013} where \spitz\ + \hersc\ data up to 500 \mic\ are modelled using the \citet{Draine_Li_2007} dust models (hereafter [DL07]) on a local basis. Maps are derived at the resolution of SPIRE 500 \mic, for a 14\arcsec $\times$ 14\arcsec\ pixel grid. We refer to the two previously mentioned studies for details on the SED modelling technique. We will remind the reader of a few principles in this section. In addition to the KINGFISH sample, this SED fitting technique has also been used on M31, modelled out to 1.5 optical radius in \citet{Draine2013}. They show that deviations from the [DL07] model appear in the centre (galactocentric distances lower than 6 kpc), suggesting steeper submm opacity spectral index but that the DL07 model can fit the emission out to 500 \mic\ elsewhere, within the uncertainties.

In the [DL07] dust models, the dust size distribution and its composition are considered to be uniform in each resolution element for which the model is run. 
Sources of the infrared emission are old stars, Polycyclic Aromatic Hydrocarbons, graphite and silicate grains. The stellar emission ($>$3 \mic) is mostly constrained by the two first \spitz/IRAC bands and modelled by scaling a blackbody function, using a photospheric temperature of 5000K \citep{Bendo2006}. In each pixel, the dust is assumed to be heated by a distribution of starlight intensities, but with most of the dust heated by a single starlight intensity that is interpreted as the starlight intensity in the diffuse ISM. The spectral shape of the radiation field is assumed to be that of the Galactic diffuse ISM estimated by \citet{Mathis1983}. The modelling procedure uses the bandpasses of the instruments to apply color-corrections and convergence to a preferred model is reached using a $\chi$$^2$ minimisation technique. 

We extrapolate 870 \mic\ maps of the sample from the [DL07] resolved SED models of \citet{Aniano2013}. Figure~\ref{Excess_maps_NGC3627} shows an example of the modelled 870 \mic\ map for NGC3627 (second line, third panel). We provide the 870 \mic\ maps of the full sample derived with this SED modelling technique in appendix in Figure~\ref{Excess_maps} (bottom panels). Here, again, the same color scale was applied between observed and modelled 870 \mic\ maps to allow a direct comparison.

The modelled 870 \mic\ maps obtained using the [DL07] formalism are quantitatively similar to those obtained using a 2MBB model with $\beta$$_c$=1.5. In the wavelength range covered by SPIRE, the SED extrapolated from the [DL07] models is in fact similar to a single temperature modified blackbody with $\beta$$_c$=2.0 but the model can reach colder temperatures due to the mixture of temperatures it includes. The [DL07] dust models also incorporate small modifications to the amorphous silicate opacity, especially for $\lambda$ $>$ 250 \mic, in order to better match the average high Galactic latitude dust emission spectrum measured by COBE-FIRAS \citep{Wright1991,Reach1995, Finkbeiner1999}. Thus, [DL07] models already include a ``Galactic submm excess" (modifications inferior to 12 $\%$ for the 250 \mic $<$ $\lambda$ $<$ 1100 \mic\ range). Emission in excess of the [DL07] models thus means an excess compared to what could be explained by Galactic-like dust only.

 \subsection{Absolute and relative difference maps at 870 \mic}

In order to compare the modelled luminosity L$_{\nu}$$^{\textrm{\small mod}}$ with the observed luminosity L$_{\nu}$$^{\textrm{\small obs}}$ at 870 \mic, we derive maps of the absolute difference defined as:

\begin{equation} 
\textrm{Absolute~Difference}~=~L_{\nu}^{\textrm{\small obs}}(870~\mu m) - L_{\nu}^{\textrm{\small mod}}(870~\mu m)
\end{equation}
\vspace{10pt} 

\noindent and maps of the relative difference defined as:

\begin{equation}
\textrm{Relative~Difference}~=~\frac{L_{\nu}^{\textrm{\small obs}}(870~\mu m) - L_{\nu}^{\textrm{\small mod}}(870~\mu m)}{ L_{\nu}^{\textrm{\small mod}}(870~\mu m) } 
\end{equation}
\vspace{10pt}

\begin{figure*}
\centering
\vspace{-10pt}
\begin{tabular}  { m{0cm} m{5.1cm} m{5.1cm} m{5.1cm}  m{0.7cm}}      
&\hspace{5cm}\rotatebox{90}{\Large 870 \mic\ Observed} & 
\includegraphics[width=5.7cm]{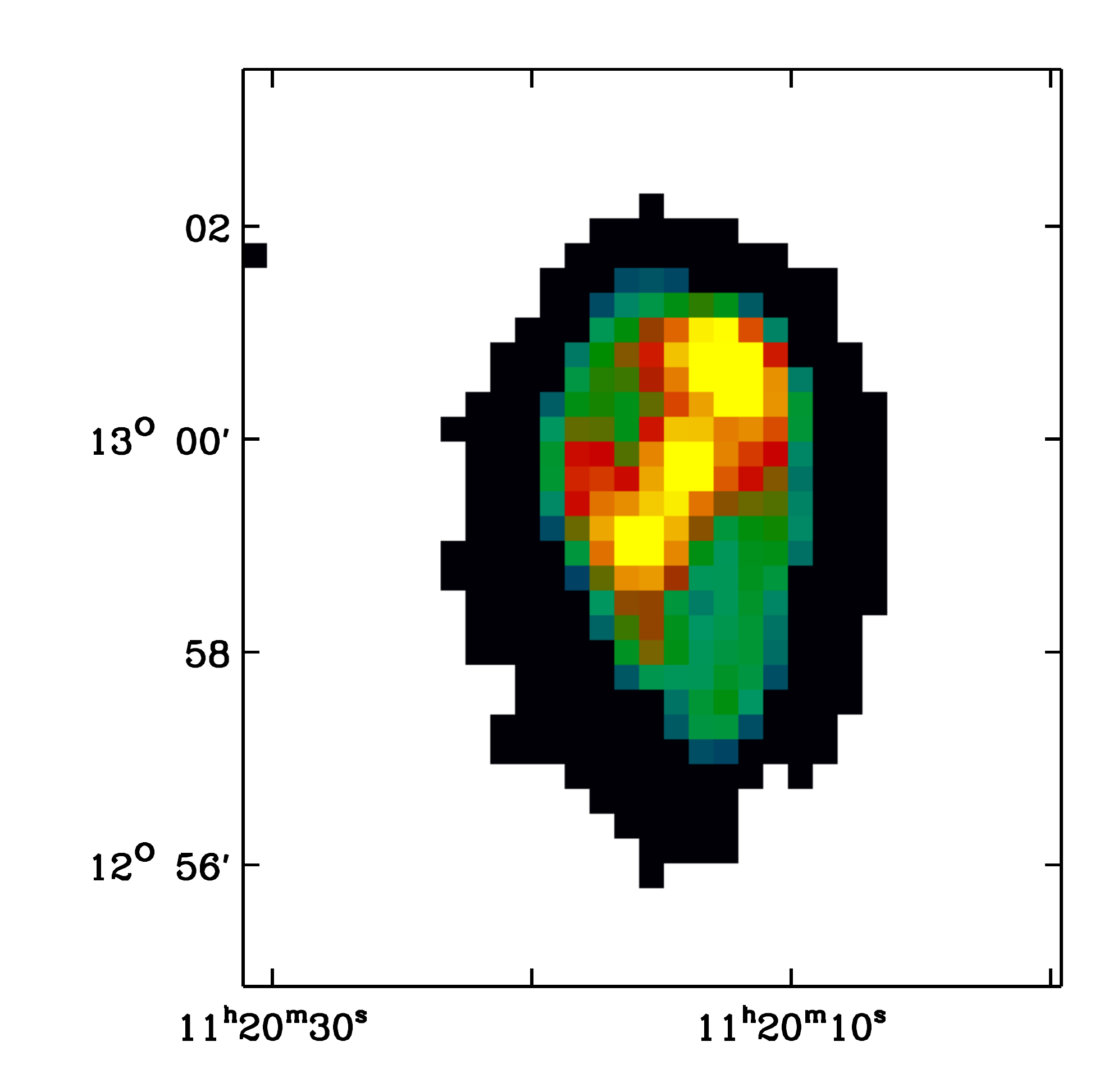} &&
\rotatebox{90}{\includegraphics[width=4cm, height=0.9cm]{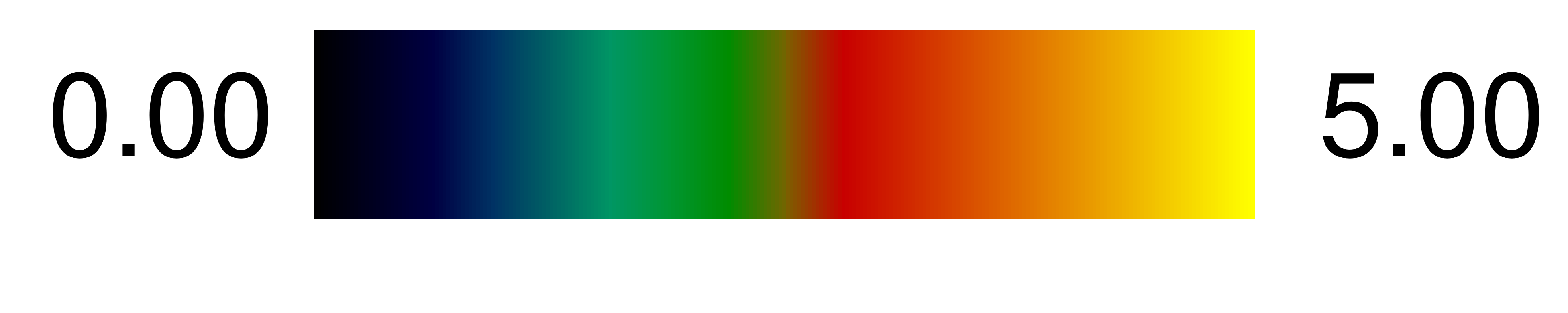}}  \\
& {\Large \hspace{2.2cm}$\beta$$_c$ = 2.0 model} & {\Large \hspace{2.2cm}$\beta$$_c$ = 1.5 model}  & {\Large \hspace{2.2cm}[DL07] model} & \\

\rotatebox{90}{\Large 870 \mic\ Modelled} & 
\includegraphics[width=5.7cm]{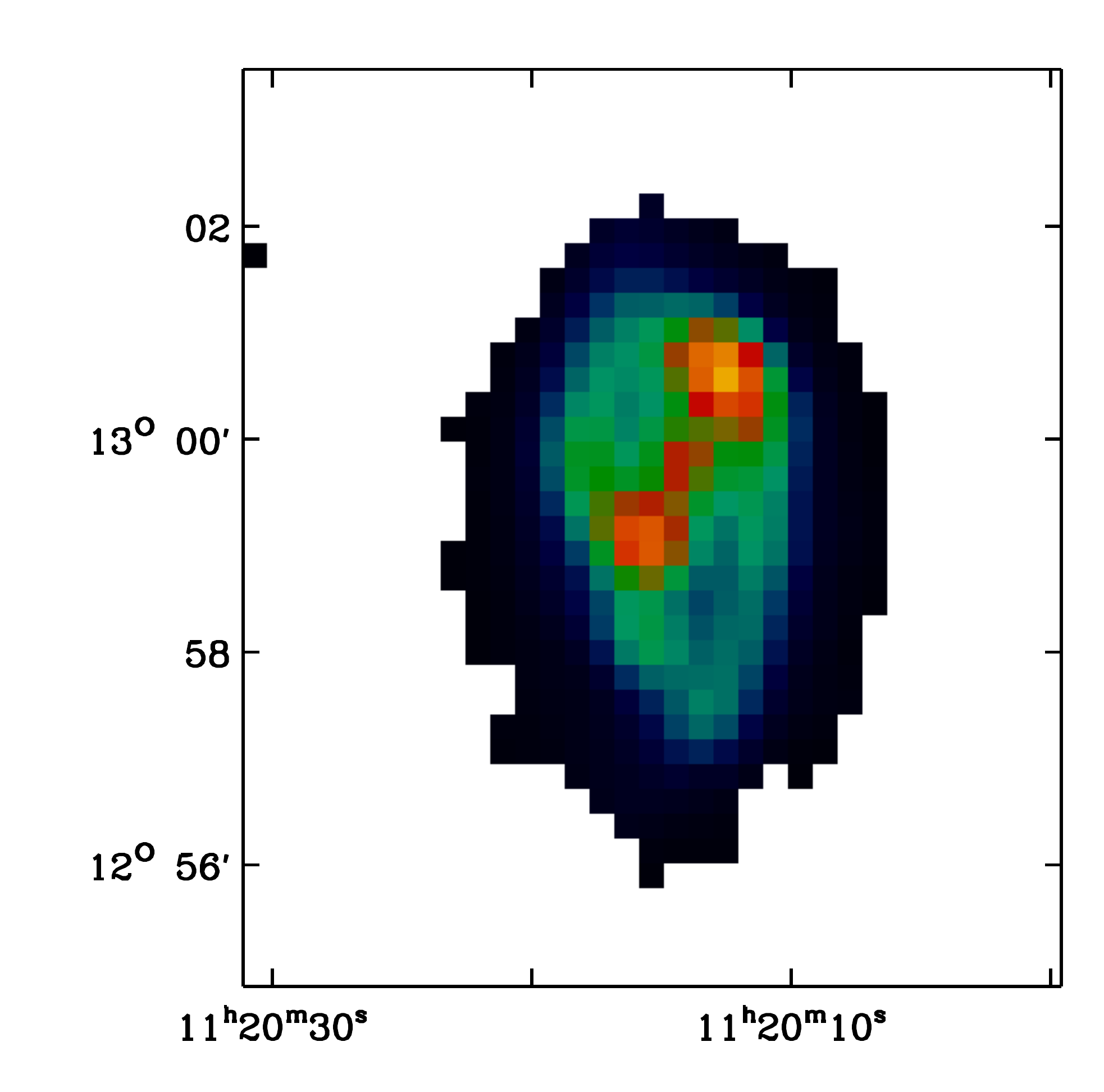} &
\includegraphics[width=5.7cm]{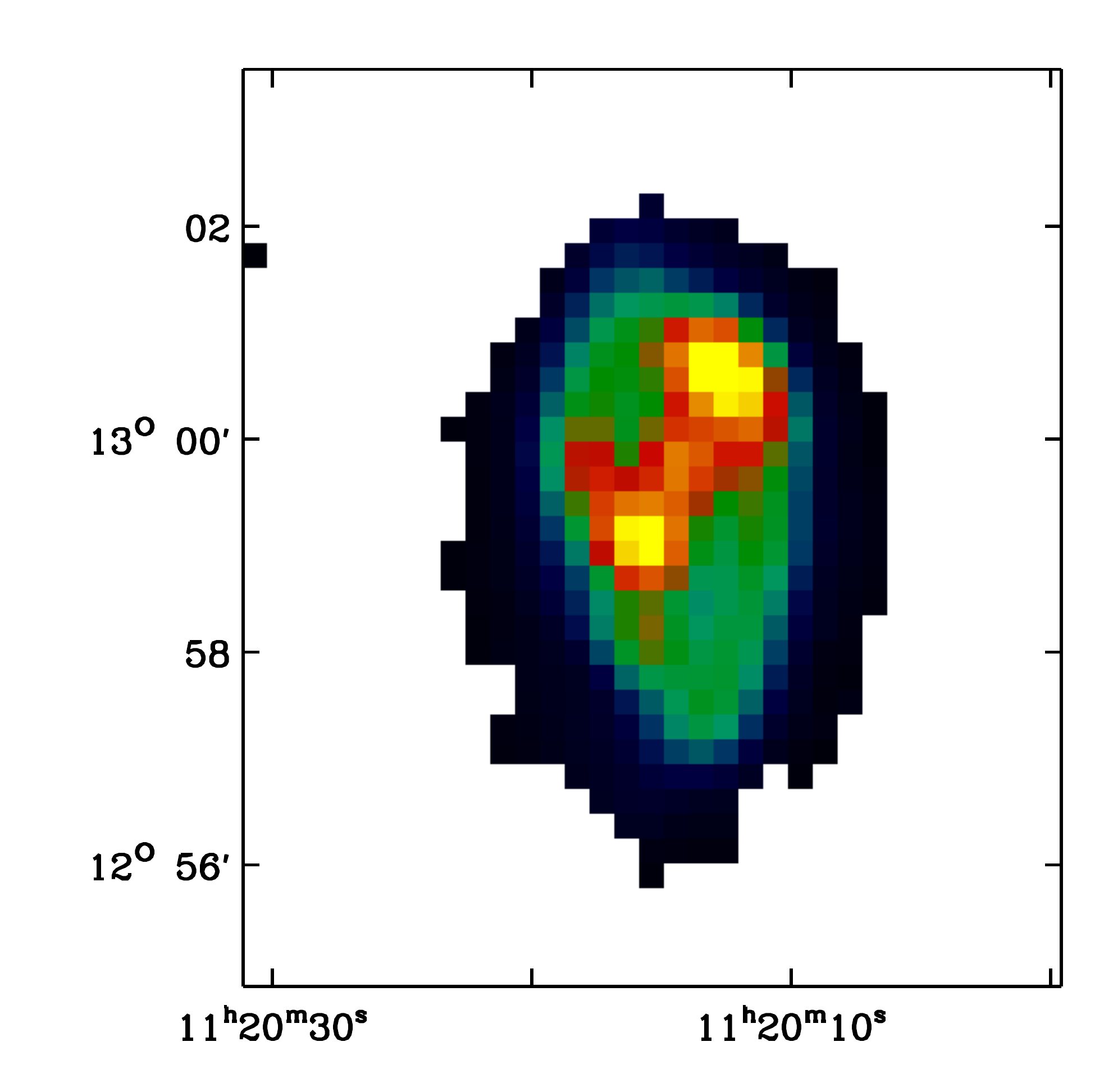} &
\includegraphics[width=5.7cm]{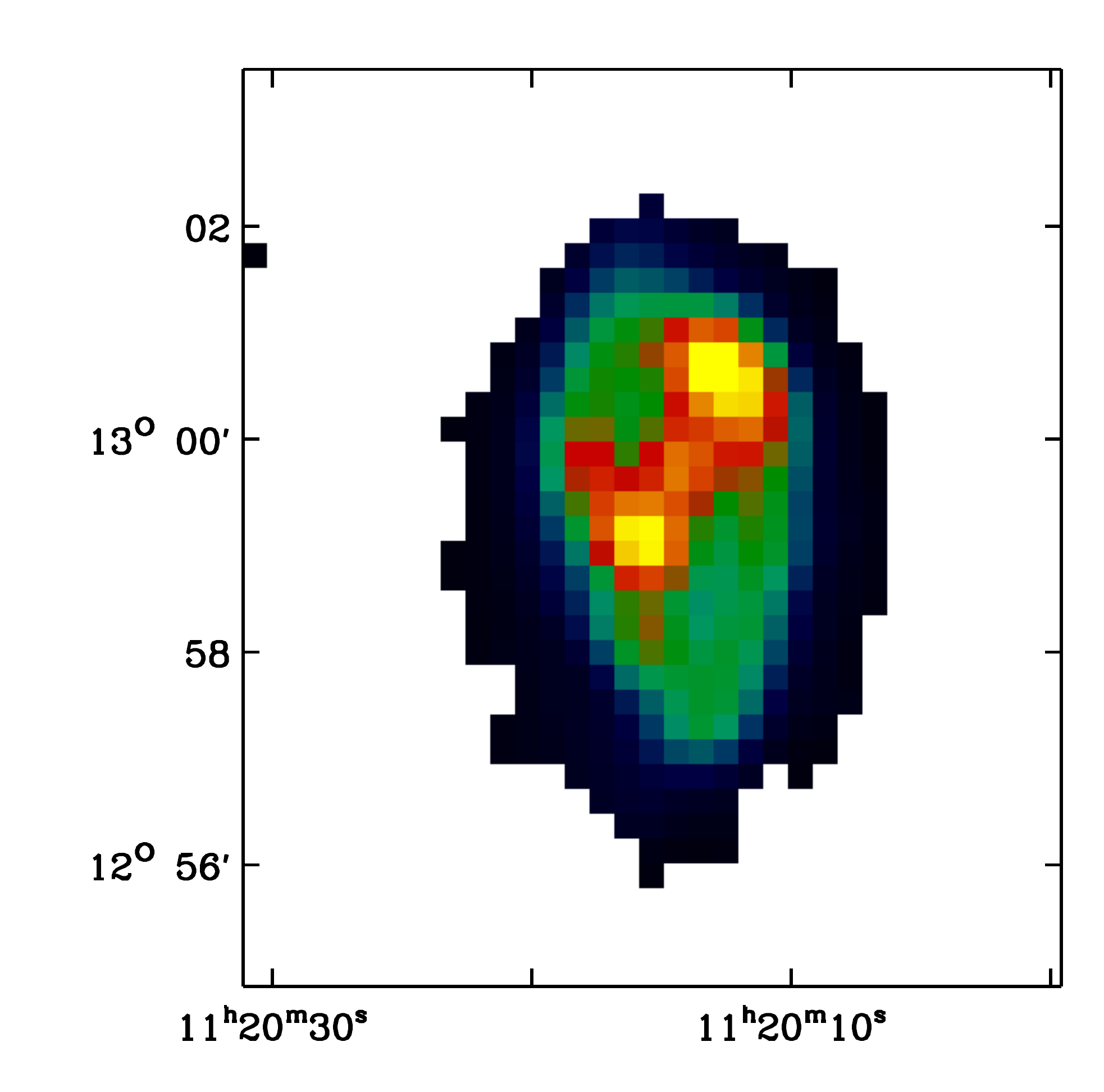}  &
\rotatebox{90}{\includegraphics[width=4cm, height=0.9cm]{NGC3627_Extrap870_ColorBars}}  \\
	
\rotatebox{90}{\Large Absolute Difference} &
\includegraphics[width=5.7cm]{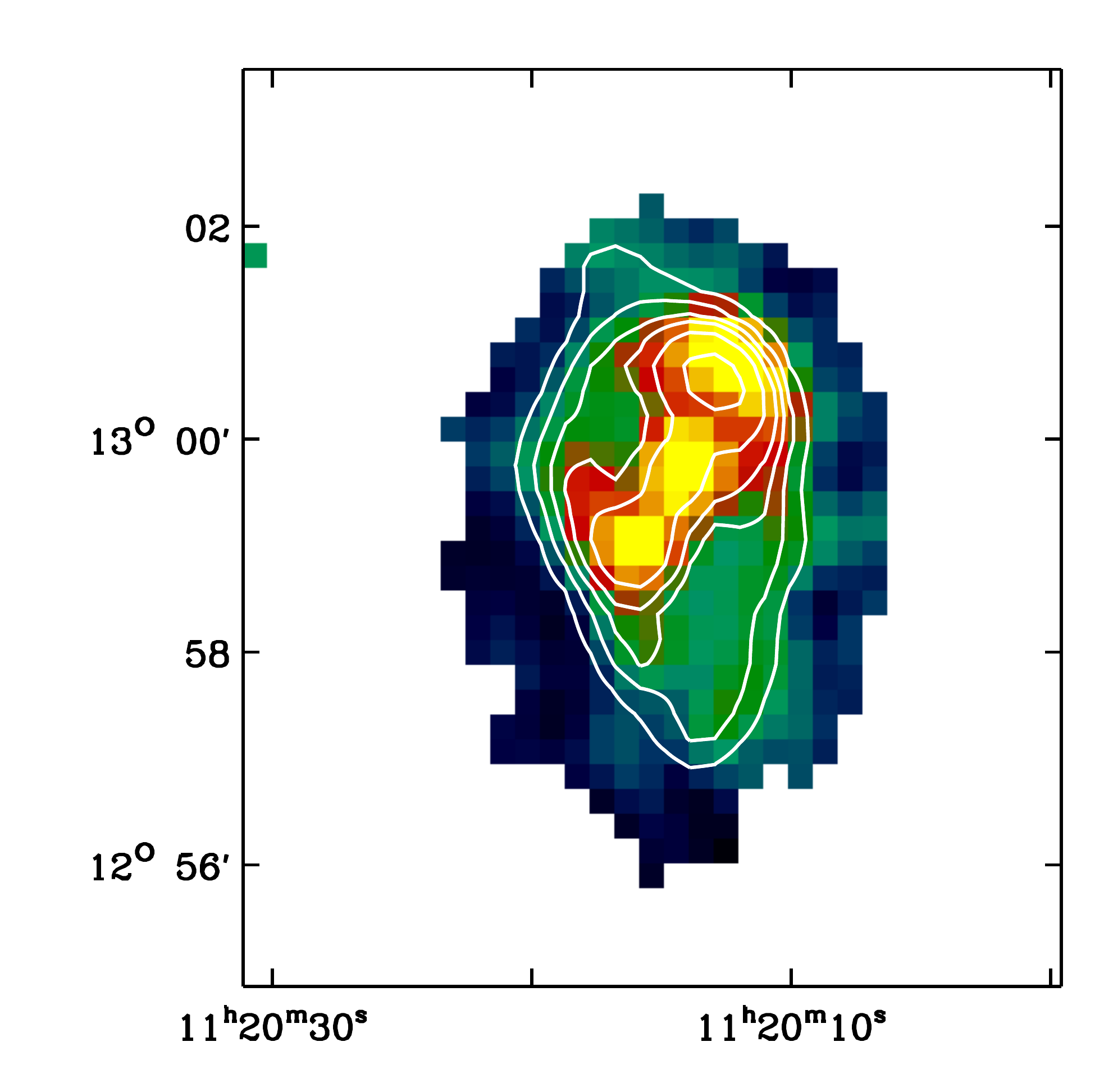} & 
\includegraphics[width=5.7cm]{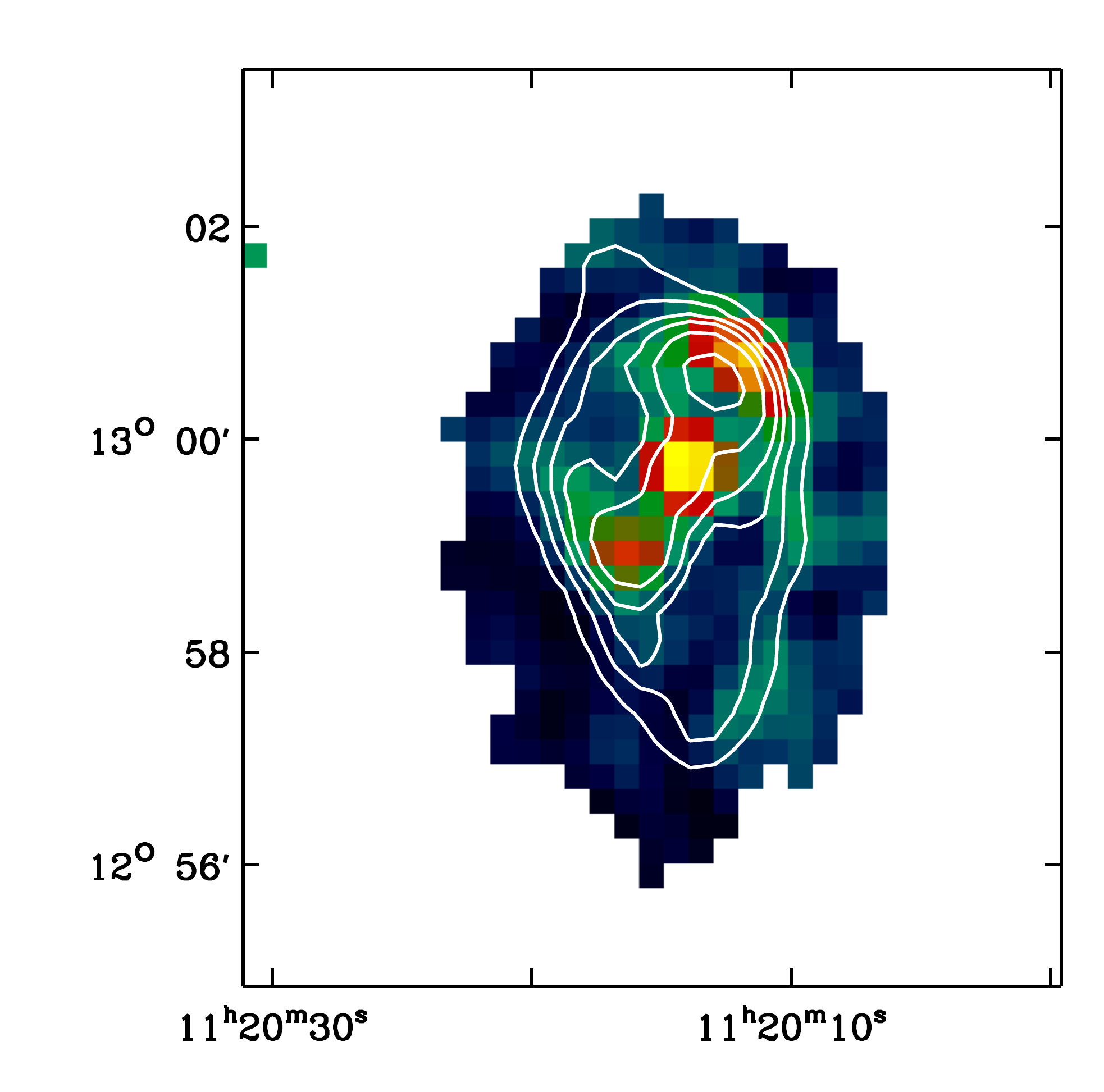} &
\includegraphics[width=5.7cm]{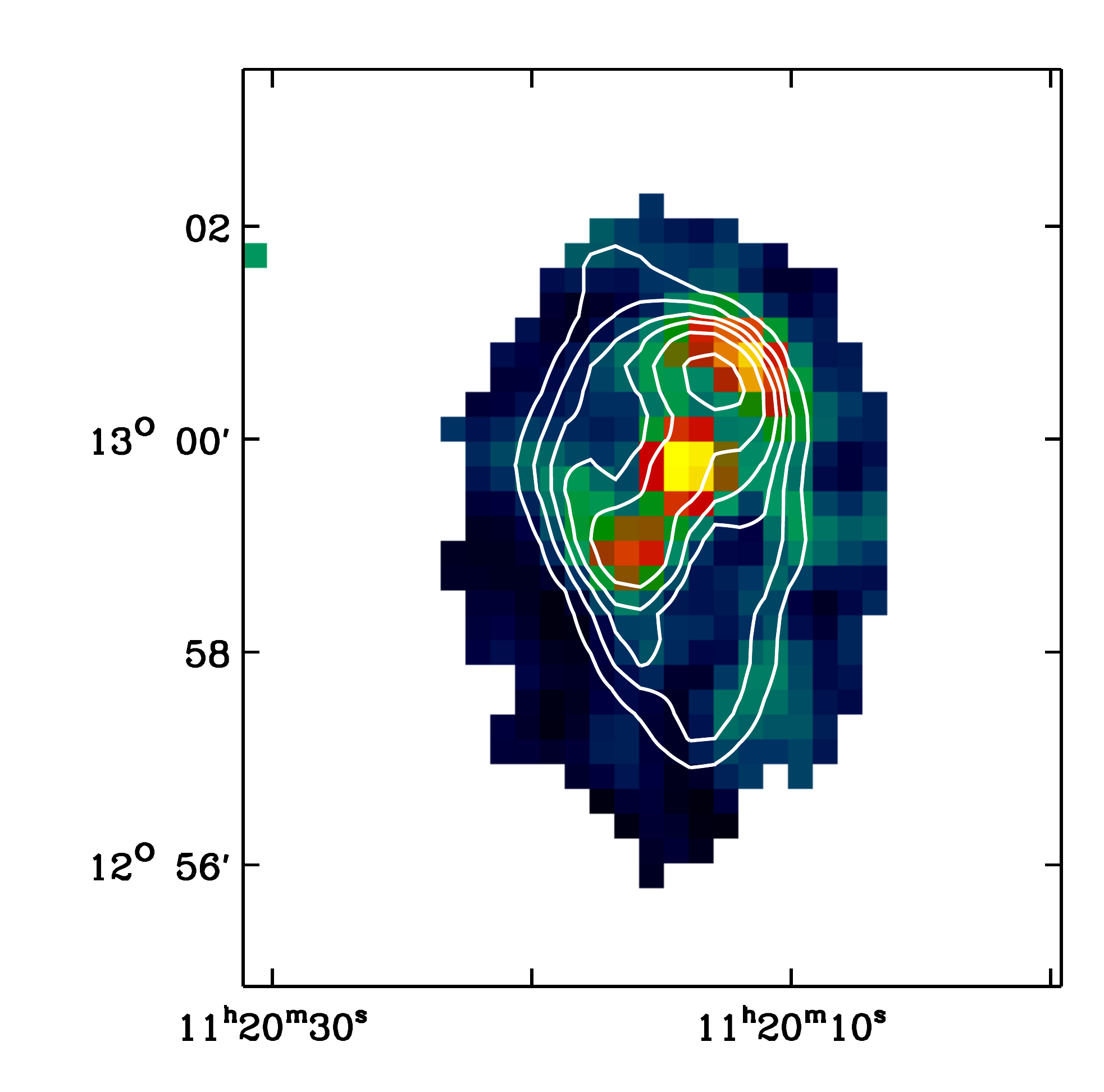}  &
\rotatebox{90}{\includegraphics[width=4cm, height=0.9cm]{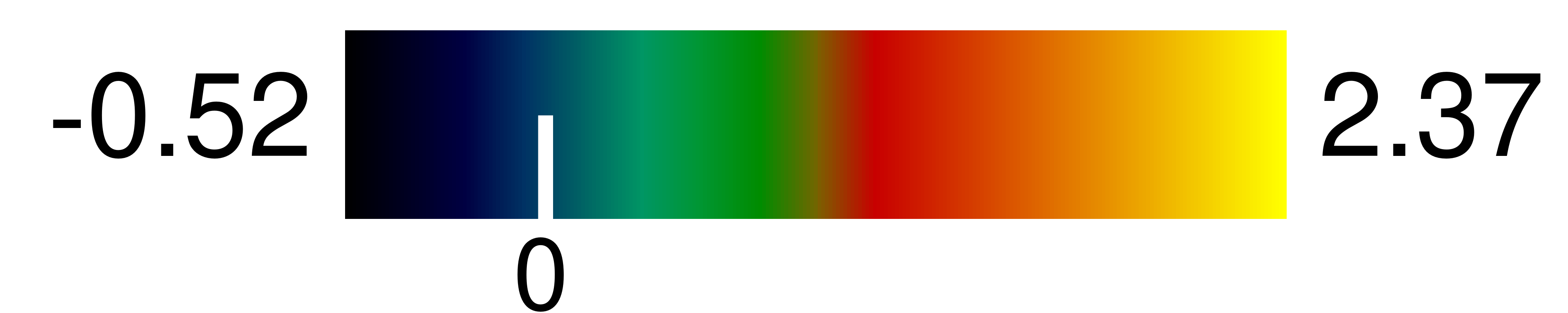}}  \\
	 
\rotatebox{90}{\Large Relative Difference} &
\includegraphics[width=5.7cm]{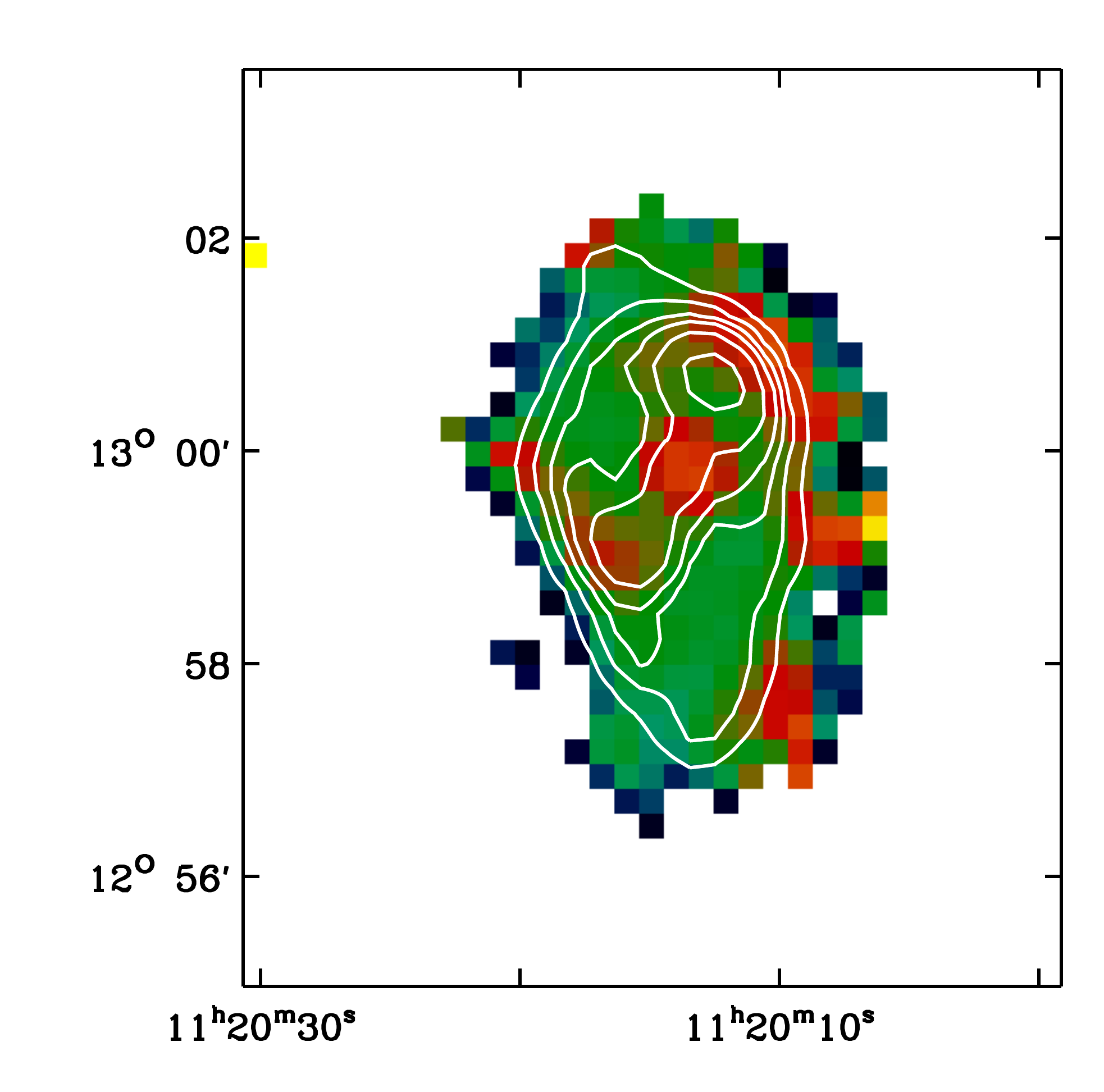} &
\includegraphics[width=5.7cm]{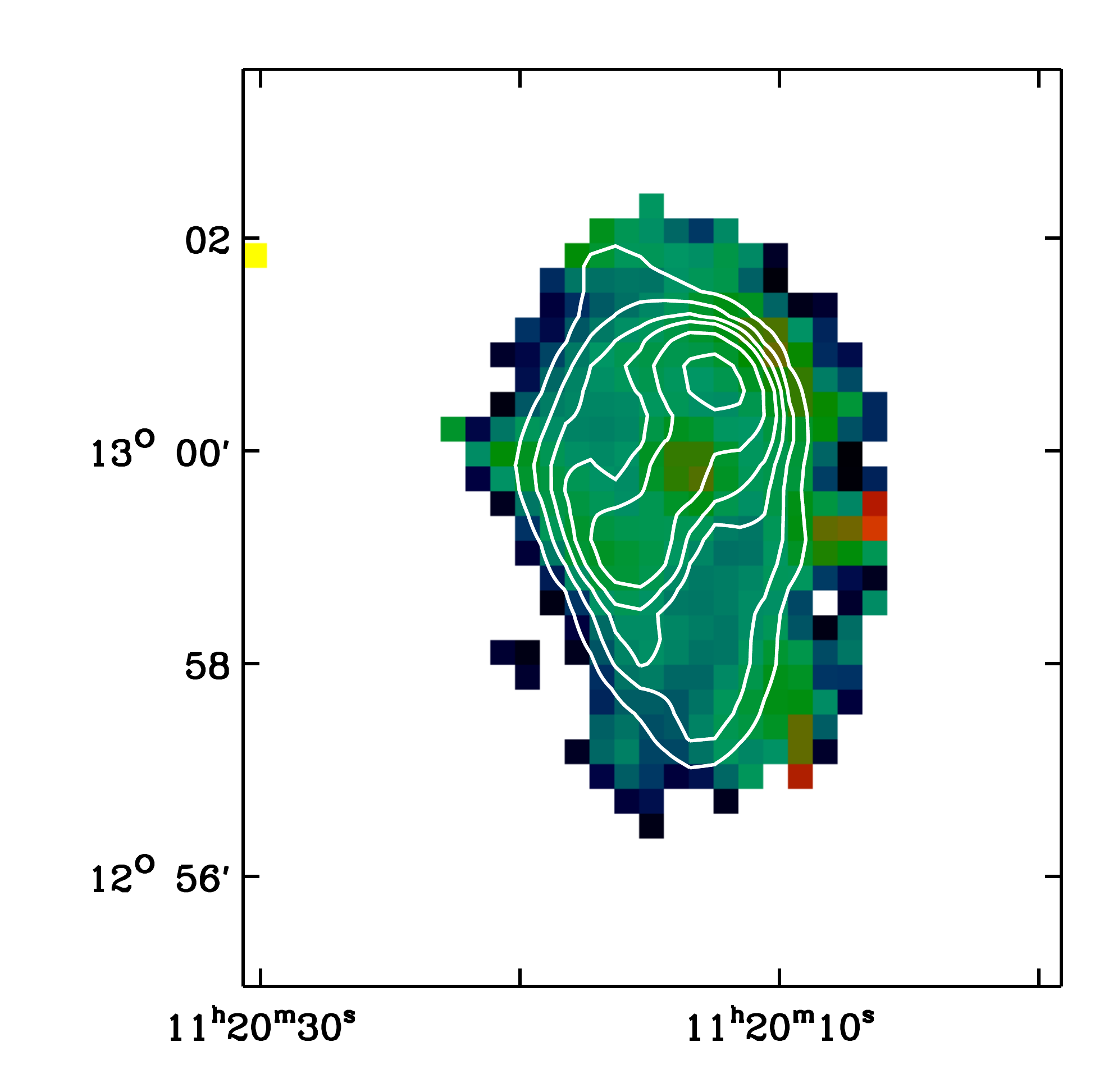} &
\includegraphics[width=5.7cm]{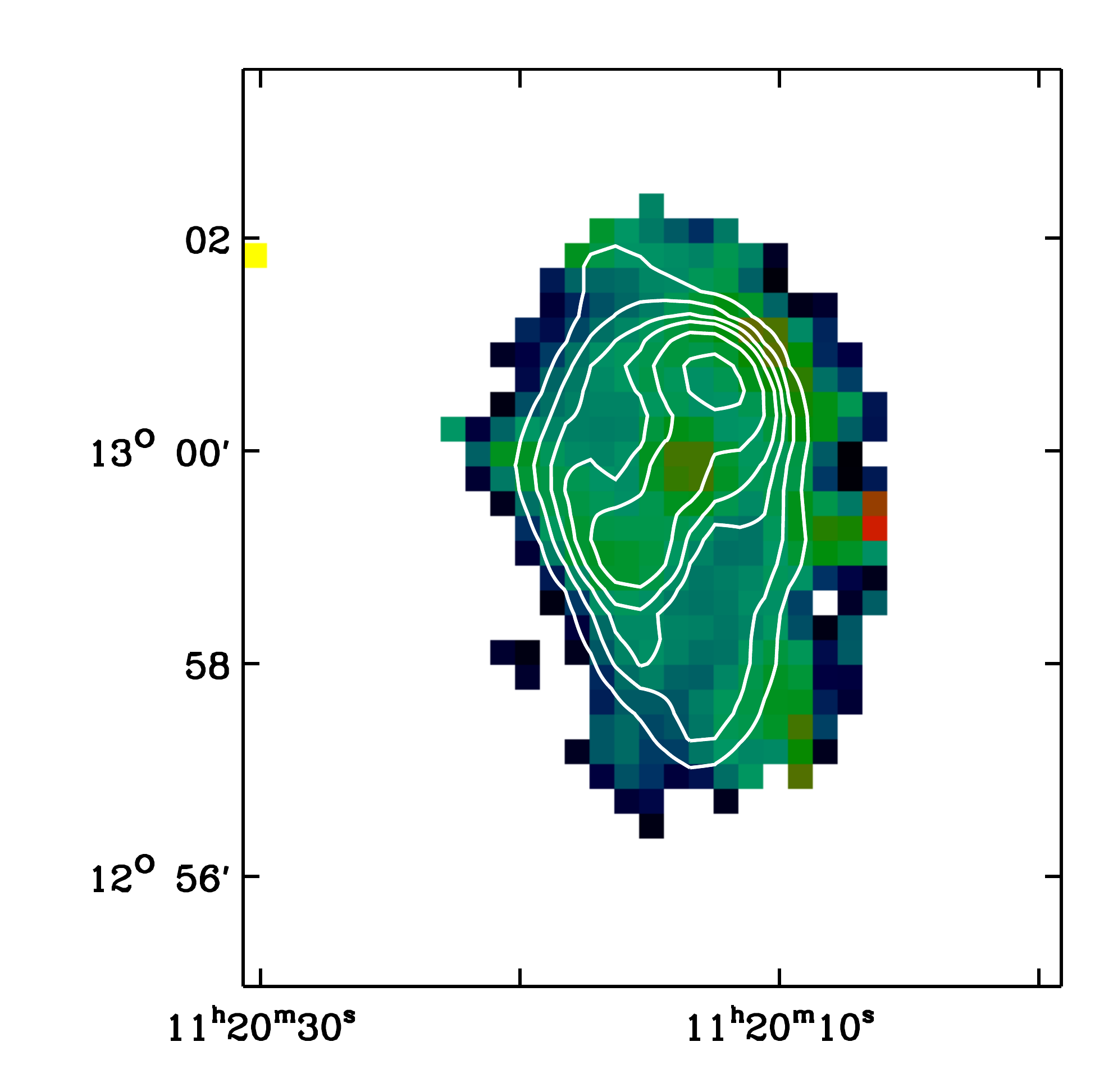}  &
\rotatebox{90}{\includegraphics[width=4cm, height=0.9cm]{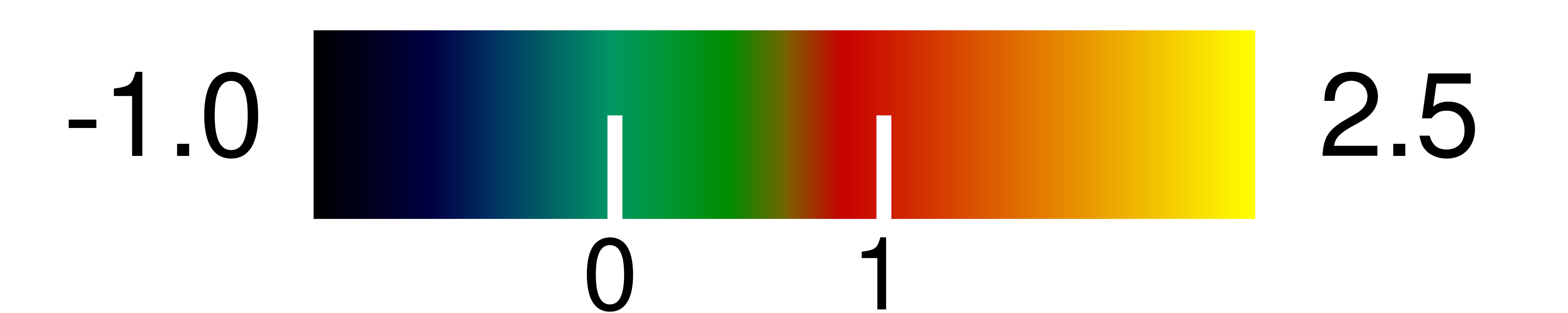}}  \\	   
	
\end{tabular}  
\caption{{\it Top panel:} Flux density at 870 \mic\ map observed with \lab\ (in MJy~sr$^{-1}$) for the galaxy NGC3627 not corrected for CO and radio continuum. {\it Second line panels:} Modelled 870 \mic\ maps (in MJy~sr$^{-1}$) derived, from left to right, using our two-modified blackbody procedure with $\beta$$_c$ fixed to 2, fixed to 1.5 or using the [DL07] formalism. {\it Third line panels:} Absolute difference between the observed and the modelled 870 \mic\ map in MJy~sr$^{-1}$ (defined as observed flux - modelled flux) for the different SED models (see Section 4.3). {\it Bottom line panels:} Relative difference defined as (observed flux - modelled flux) / (modelled flux). We provide the color scales for each line on the right-hand side. Zeros are indicated with a white marker on the color bars. The signal-to-noise of the LABOCA 870 \mic\ map is overlaid as contours on the absolute and relative difference maps (same levels than in Fig.~\ref{350um_Maps}}).
    \label{Excess_maps_NGC3627}
\end{figure*}

Figure~\ref{Excess_maps_NGC3627} shows an example of the 870 \mic\ absolute difference maps (third line panels) and relative difference maps (bottom line panels) obtained from our 3 resolved modelling techniques for the galaxy NGC3627. We provide the absolute and relative difference maps of the full sample in appendix in Fig.~\ref{Excess_maps}. We indicate zeros in the color scales with a white marker to allow a visual separation between 870 \mic\ excess or deficit compared to the different SED models.  The signal-to-noise of the LABOCA 870 \mic\ map is overlaid as contours on the absolute and relative difference maps. The first contour shown corresponds to a 3$\sigma$ 870 \mic\ detection. As mentioned in the previous section, the modelled 870 \mic\ maps obtained using the [DL07] formalism and those obtained using a modified blackbody with $\beta$$_c$=1.5 are quantitatively similar so our conclusions are the same for these two methods. \\

When $\beta$$_c$ is fixed to a standard value of 2, we systematically observe regions with a positive difference between the observed and the modelled 870 \mic\ emission in our maps. This positive difference can be localised in the centre of the galaxy, like in NGC1097, or extended across the object (NGC0337 or NGC7793 for instance). The 870 \mic\ maps extrapolated from the $\beta$$_c$=1.5 model are brighter than those with $\beta$$_c$=2.0 due to the flatter submm slope induced by this SED model that leads to higher extrapolations at 870 \mic. As a direct consequence, the absolute difference derived using $\beta$$_c$=1.5 is systematically smaller than that derived using $\beta$$_c$=2.0. 

We now qualitatively and quantitatively describe the trends observed for each individual galaxy. Part of the excess we detect can be linked to contamination by free-free, synchrotron, or CO emission. We quantify this contribution in Section 4.4. \\

 {\it NGC0337.---} The distributions of the observed and modelled 870 \mic\ emission are very similar. However, quantitatively, the observed 870 \mic\ is significantly higher than our model predictions, regardless of the model we use. The absolute excess above the model prediction at 870 \mic\ is radially decreasing, following the distribution of the star formation activity of the galaxy as well as that of the dust mass surface density mapped in [G12]. The relative excess seems to be non homogeneous across the galaxy. The relative excess in regions detected with LABOCA at a 5-$\sigma$ level is $\sim$70$\%$ on average when $\beta$$_c$=1.5, with a standard deviation of 25\%. This is slightly higher than what is found on global scales for this object (but we remind the reader that we are excluding the lowest-surface brightness regions here). Unfortunately, because this is one of most distant galaxies of the sample, our poor resolution does not allow us a very detailed study of the 870 \mic\ residual distribution.

{\it NGC0628.---} As one can see from Fig.~\ref{350um_Maps}, part of the emission in the southeast side of the galaxy is not detected in the \lab\ map. The modelled 870 \mic\ maps are thus brighter than the emission actually observed with \lab, whatever the model, and the absolute difference thus shows a deficit of emission in that southeast region. For the rest of the galaxy, the distributions of the observed and modelled maps are similar, with the 870 \mic\ emission (modelled and observed) following the distribution of the emission traced by SPIRE wavebands. We observe a strong 870 \mic\ excess above the $\beta$$_c$=2.0 model across the galaxy. The difference between the observed and modelled 870 \mic\ emission partly but not fully disappears when we use a lower $\beta$$_c$=1.5 or the [DL07] dust model. This excess emission is not homogeneously distributed across the disk of the galaxy and does not follow the galaxy structure. Indeed, the absolute and relative difference maps of NGC0628 do not show a statistically significant excess emission in the centre but peaks in the low-surface brightness regions of the galaxy where cold dust at 870 \mic\ emission is detected by \lab\ at a reasonable 3$\sigma$ level. 

{\it NGC1097.---} The observed and modelled 870 \mic\ maps are similar, with a very strong 870 \mic\ emission in the centre of the galaxy (the galaxy possesses a moderate AGN) and along the bar, and two fainter spiral arms. A positive difference between the observed and modelled map is seen across the galaxy when using the $\beta$$_c$=2.0 MBB model. The 870 \mic\ emission seems to be better explained when a $\beta$$_c$=1.5 MBB model or the [DL07] formalism are used. An excess above the model prediction remains nonetheless in the central region when $\beta$$_c$=1.5 (we obtain an average of $\sim$54$^{\pm35}$$\%$ in regions with a 20-$\sigma$ detection) and in the few inter-arm resolved elements that are modelled.

{\it NGC1291.---} The three modelled 870 \mic\ maps follow the distribution of the SPIRE maps, namely a bright centre elongated in the north-south direction and a diffuse cold dust ring (prominent longward of 160 \mic) with localised bright spots in the ring in the northwest and southeast arcs \citep[see][for a detailed analysis of this cold dust ring with \hersc]{Hinz2012}. We analyse the excess in resolved elements detected at 870 \mic\ to a 2$\sigma$ level, which, unfortunately, restricts the study to the centre of the galaxy where a strong excess above the $\beta$$_c$=1.5 model prediction is detected ($\sim$59$\%$ on average in the regions we modelled, with a standard deviation of 30\%). The major part of the cold dust ring is not detected with \lab\ (see Fig.~\ref{350um_Maps}), which prevents us to conclude about any trend in the ring of NGC1291. It also implies that the global 870 \mic\ flux density provided for this galaxy in \citet{Albrecht2013} should be considered as a lower limit. A higher global flux density would reconcile the 870 \mic\ data with the integrated SED models derived in Section 3 for this object.

{\it NGC1316.---} The modelled and observed maps of the galaxy follow the distribution of the warm or cold dust emission traced by \hersc\ longward of 70 \mic, with two peaks located on each side (northwest and southeast) of the star forming centre (indicated by the MIPS 24 \mic\ central peak). We observe a positive difference between the observed and the modelled 870 \mic\ emission throughout the galaxy when the $\beta$$_c$ =2.0 MBB model is used. The ``excess" follows the distribution of star forming regions (namely a central peak co-spatial with the 24 \mic\ peak) rather than that of the cold dust emission or the dust mass surface density as mapped in [G12]. When using the [DL07] model prediction, the very central relative excess is about $\sim$70$\%$ above the modelled 870 \mic\ emission on average. We note that this galaxy possesses a moderate AGN. We discuss potential non-dust contamination (synchrotron emission in the centre in that case) in Section 4.4.1. 

{\it NGC1512.---} The \lab\ emission follows the same distribution but is systematically brighter than our model predictions ($\sim$40$\%$ brighter in the northern part of the galaxy when using the $\beta$=1.5 or the [DL07] models). We detect a significant excess emission above our model predictions in the southeast part of NGC1512, due to a faint extended structure detected at 870 \mic\ (Fig.~\ref{350um_Maps}). Faint blobs are detected in the SPIRE maps in that part of the galaxy but it is difficult to determine if the SPIRE emission belongs to the galaxy due to confusion in the SPIRE maps. NGC1512 is in interaction with NGC1510 located in the southwest of the map in Fig~\ref{350um_Maps}. We refer to the work of \citet{Koribalski2009} where the H{\sc i} emission of the pair is analysed. One might argue that the diffuse 870 \mic\ emission we observe to the south could be a continuation of the NW-SE diffuse 500 \mic\ tidal arm observed between the two galaxies and result from very cold dust in compressed gas. Unfortunately, \citet{Albrecht2013} could not firmly conclude on whether or not the emission is real or an artefact in the 870 \mic\ map itself.

{\it NGC3351.---} The modelled 870 \mic\ maps show a central peak in the emission surrounded by a diffuse disk emission but only part of the emission in the disk is detected with the \lab\ instrument (see Fig.~\ref{350um_Maps}). For this galaxy as well, the non-detection of disk structure with \lab\ indicates that the global 870 \mic\ flux density provided for this galaxy in \citet{Albrecht2013} should be considered as a lower limit. An emission in excess is systematically detected in the centre of the galaxy, whatever the model, but the relative 870 \mic\ difference is small ($<$30$\%$ for regions with a 3-$\sigma$ LABOCA detection). An 870 \mic\ excess above the $\beta$$_c$=1.5 model is also observed and radially increasing to a factor of 2 in the northeast region of the galaxy, i.e. in the only region of the disk detected with \lab\ at a 3$\sigma$ level. 

{\it NGC3621.---} Our modelled maps qualitatively match the observed distribution. They show a radial decrease of the 870 \mic\ emission and two peaks in the centre of the galaxy, the southern being the strongest. However, the observed 870 \mic\ emission is brighter than our $\beta$$_c$=2.0 model prediction, with a distribution of the absolute difference map following the galaxy structure and the dust mass surface density mapped in [G12]. This trend remains when the $\beta$$_c$=1.5 model or the [DL07] model are used. The excess is, however, statistically weak in these two cases, with a 870 \mic\ relative excess of 19$^{\pm 12}$$\%$ on average in regions with a 5-$\sigma$ LABOCA detection. Peaks of the relative excess emission (barely visible for the $\beta$$_c$=1.5 or [DL07] models) are distributed in low-surface brightness regions rather than in the centre of the galaxy.

{\it NGC3627.---} The 870 \mic\ map follows the distribution of the SPIRE observations, namely a strong emission in the centre and peaks at the bar ends (bar in the NW-SE direction) where molecular hydrogen \citep{Wilson2012} and large dust reservoirs ([G12]) are located. The star formation primarily occurs in these two knots where the SF efficiency is enhanced \citep{Watanabe2011}. For all models, the absolute 870 \mic\ excess follows the spiral structure of the galaxy (and the dust mass surface density), with excess maxima corresponding to 870 \mic\ peaks and spiral arms still distinguishable. The relative 870 \mic\ excess above the $\beta$$_c$=1.5 or [DL07] models shows moderate peaks in the centre and each end of the bar. We obtain an average relative excess of 30\% for these regions (detected by LABOCA at a 20-$\sigma$ level), with a standard deviation of 19\%.

{\it NGC4826.---} The map of the 870 \mic\ absolute difference between observation and predictions, like that of the 870 \mic\ emission itself, shows a smooth radial decrease of the excess in this object. The relative difference is non homogeneous and statistically weak on average across the whole object in the $\beta$$_c$=1.5 or [DL07] cases. This is consistent with what is found at global scale for NGC4826.

{\it NGC7793.---} The modelled maps derived for the galaxy follow the distribution of the cold dust emission traced by SPIRE and the 870 \mic\ observation. NGC7793 is a flocculent galaxy with peaks of star formation spread out in the disk (but only partly resolved at the resolution we are working at in this study) and most of the disk is detected with \lab\ at a level $>$3$\sigma$. Like for the galaxy NGC0337 (a galaxy that shows a similar flattening of the submm slope at global scale), we observe a strong 870 \mic\ absolute excess throughout the galaxy. It however does not follow the spiral structure in this object and peaks are observed within the disk rather than in the centre (the relative excess is lower than 30$\%$ in the nucleus). The trend is confirmed by the relative difference maps in which the 870 \mic\ difference between the observations and the models seems to radially increase (by more than a factor of 2) toward the south of the galaxy.\\

For half of the sample, the global fit favours steeper slopes ($\beta$$_c$=2.0) compared to the resolved study. This could primarily be linked with the fact that we are restricting our resolved study to pixels with good detections in the LABOCA data, excluding the faint outskirts where the 870 \mic\ emission is weak. These global versus resolved discrepancies could also partly be due to effects of non-linearity in the SED models we are using and reported for instance in \citet{Galliano2011} or \citet{Galametz2012}. Comparing estimates obtained on global and local scales, they show that the dust mass obtained from integrated fluxes could be much lower than that obtained on a resolved basis and attribute this effect to a possible dilution of the cold dust regions in hotter regions. Integrated models could consequently be unable to account for the cold dust populations, biasing the SEDs toward warmer dust (or steeper submm slope) in MBB models or higher minimum heating intensities for the [DL07] models. The global versus resolved discrepancies we observe re-inforce the importance of systematically confronting integrated results (obtained for instance for high-redshift sources that are lacking spatial resolution) and resolved properties derived in the nearby Universe.

\subsection{Non-dust contribution to the 870 \mic\ emission}

\subsubsection{Free-free and synchrotron contribution}

As mentioned in Section 3, free-free and synchrotron emission can be a source of contamination of our 870 \mic\ observations but does not significantly ($<$2$\%$) contribute, at global scale at least, to the 870 \mic\ continuum emission in most of our objects. In the galaxy centers, contribution to the 870 \mic\  by synchrotron emission could increase in case the galaxy hosts an AGN. Two galaxies, NGC1097 and NGC1316 possess moderate AGNs. Quantifying the contribution of synchrotron to the 1.3 mm emission in a large sample of galaxies (they neglect the free-free emission), \citet{Albrecht2007} showed that the mean contribution of AGN is on average of $\sim$8$\%$ at this wavelength while less than 2$\%$ in normal galaxies. We thus expect a more significant non-dust contribution from the synchrotron emission in the centre of NGC1097 and NGC1316 than for the rest of the sample. Radio contamination could also moderately affect the central resolved elements of other galaxies whose nucleus is classified as ``AGN" in the \citet{Moustakas2010} classification. Finally, the properties of the dust populations in the centre of NGC1097 could also be affected by the presence of a prominent starbursting ring \citep{Hummel1987}.

\subsubsection{Local CO contribution}

The $^{12}$CO(3-2) line emission can also contribute to the 870 \mic\ emission in some galaxies of the sample. The 870 \mic\ maps we use to derive the difference maps are not corrected for the CO(3-2) contribution. \citet{Albrecht2013} provide details of the CO observations available for our sample from which they derive the global estimates of the CO line contribution at 870 \mic\ we quote in this section. At present, there are no CO observations available for the galaxies NGC1512 and NGC3621.\\

\begin{figure*}
\centering
\begin{tabular}  { m{0cm} m{5.5cm} m{5.5cm} m{5.5cm}  }    
& {\Large \hspace{1cm}CO(3-2) contribution} & {\Large \hspace{1.5cm}Absolute Difference}  & {\Large \hspace{1.5cm}Relative Difference}  \\
& {\Large \hspace{2.3cm} ($\%$)} & {\Large \hspace{1cm}corrected for CO(3-2)}  & {\Large \hspace{1cm}corrected for CO(3-2)}  \\
&\\
\rotatebox{90}{\Large NGC0337} & 
\includegraphics[width=5.2cm]{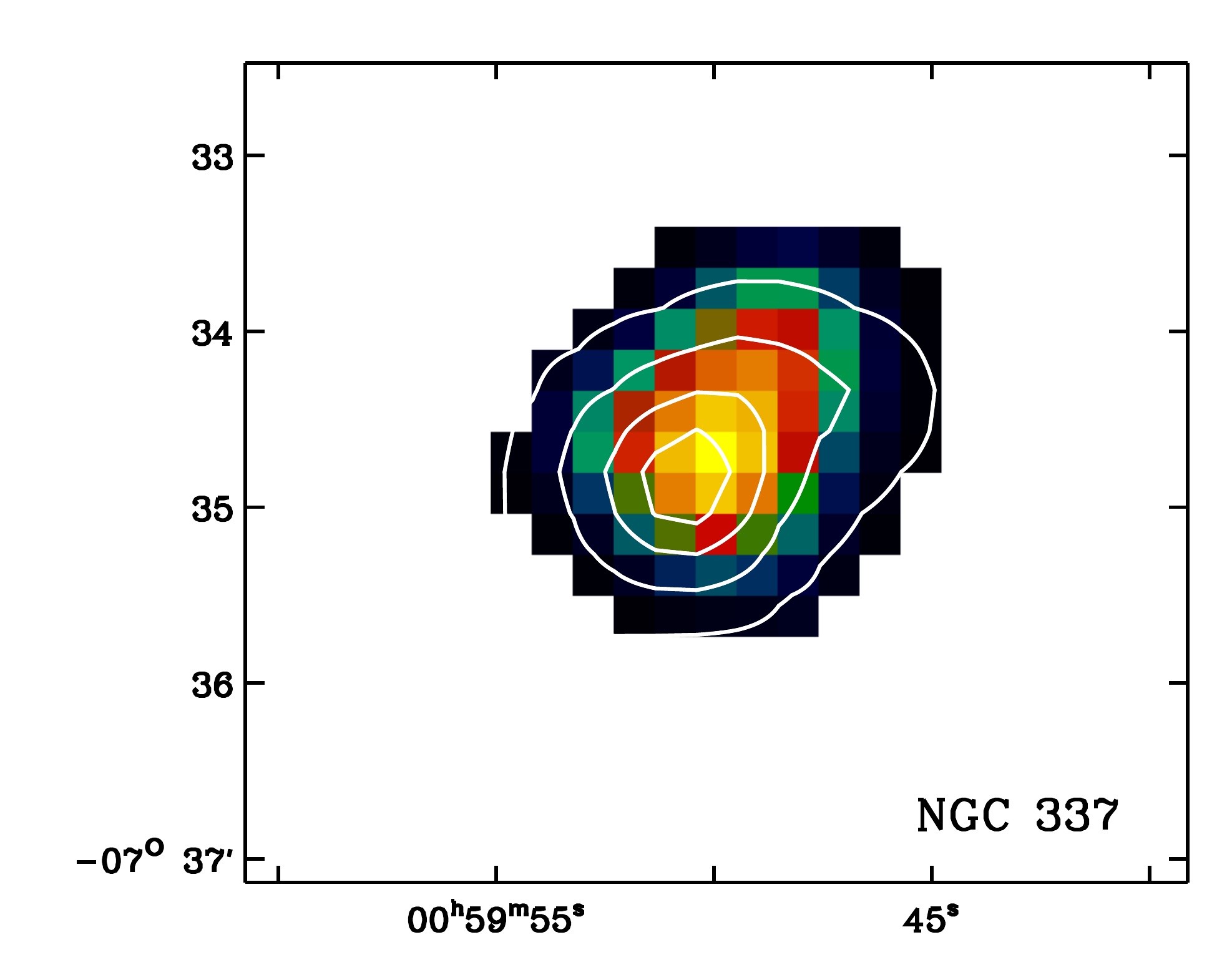} &
\includegraphics[width=5.2cm]{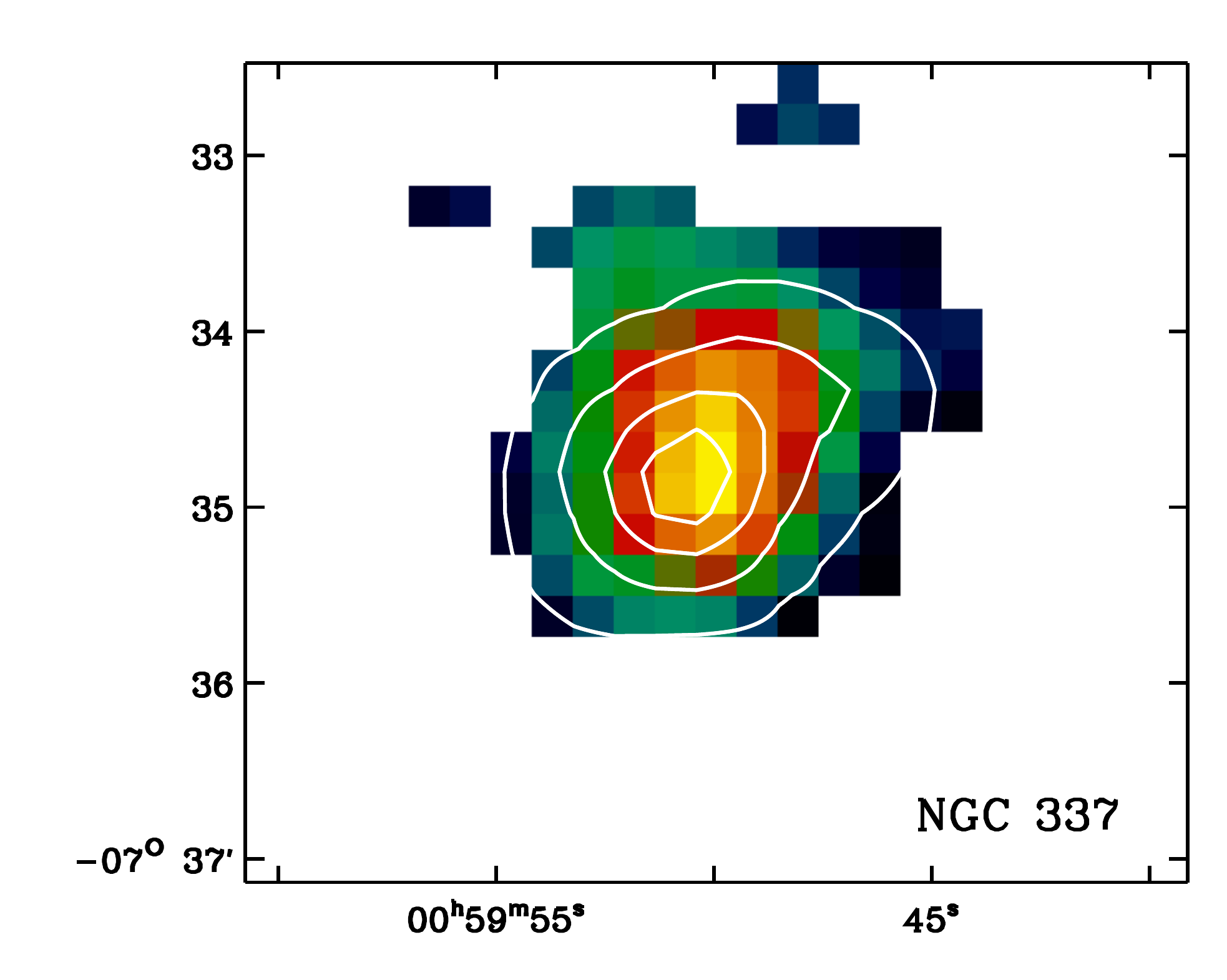} &
\includegraphics[width=5.2cm]{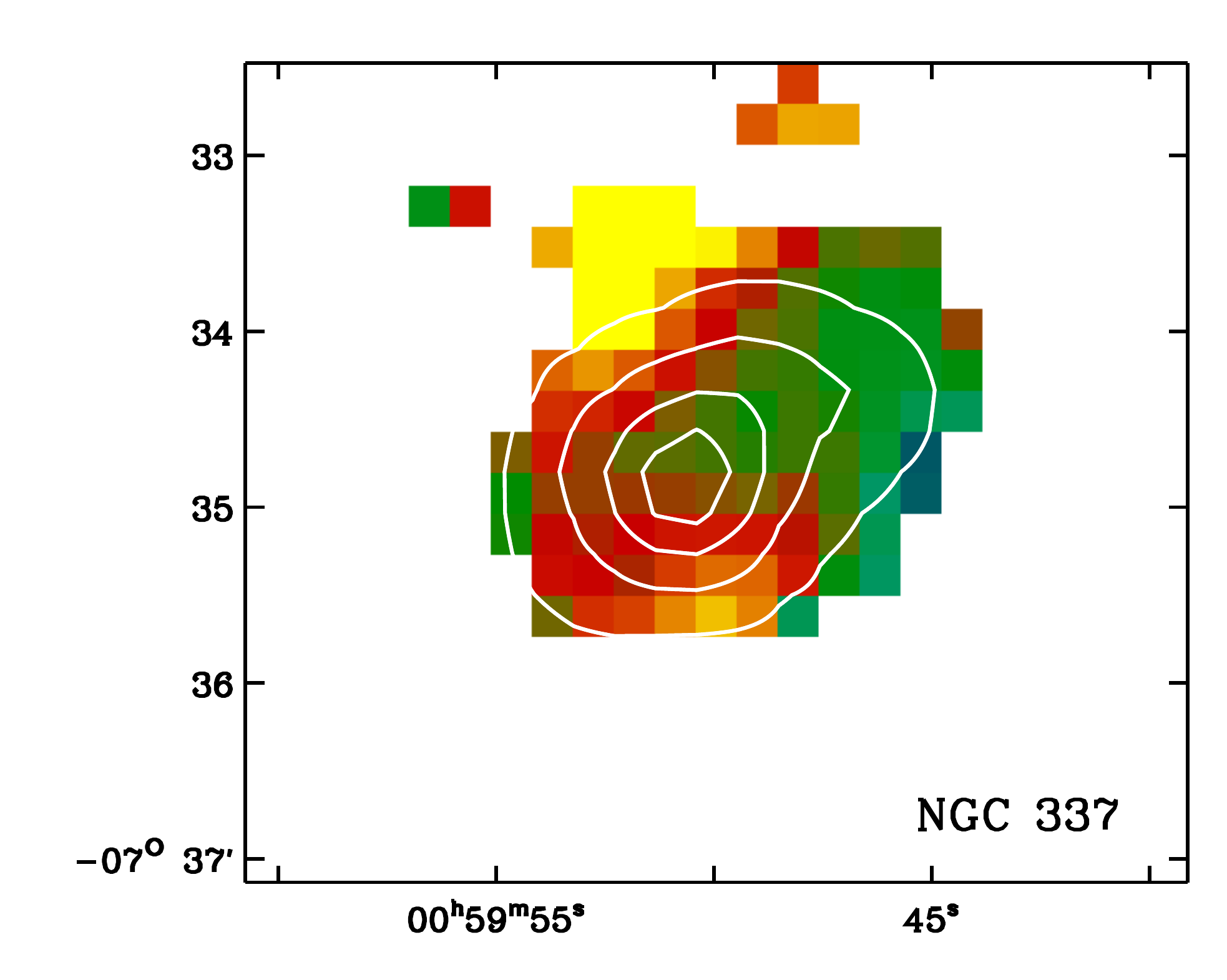} \\

& \hspace{0.8cm}\includegraphics[width=4.5cm, height=0.7cm]{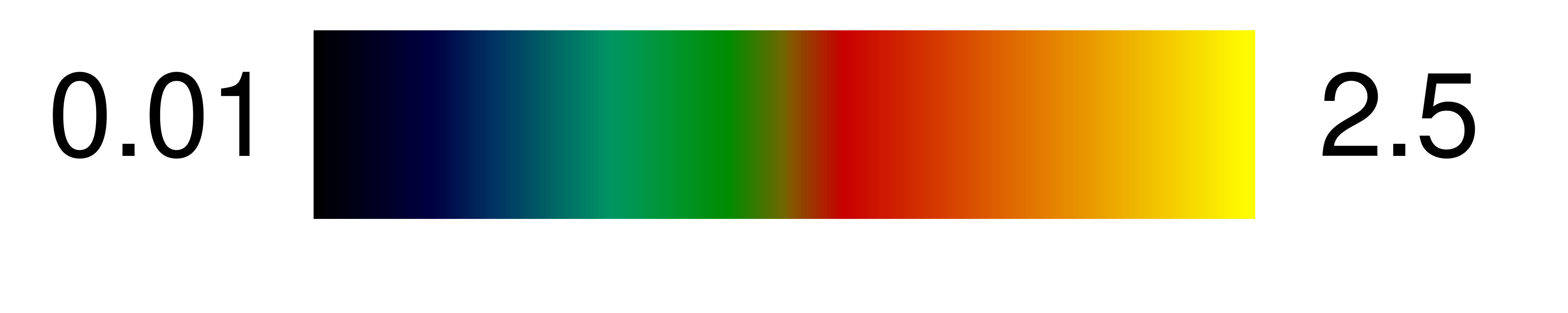} &
\hspace{0.8cm}\includegraphics[width=4.5cm, height=0.7cm]{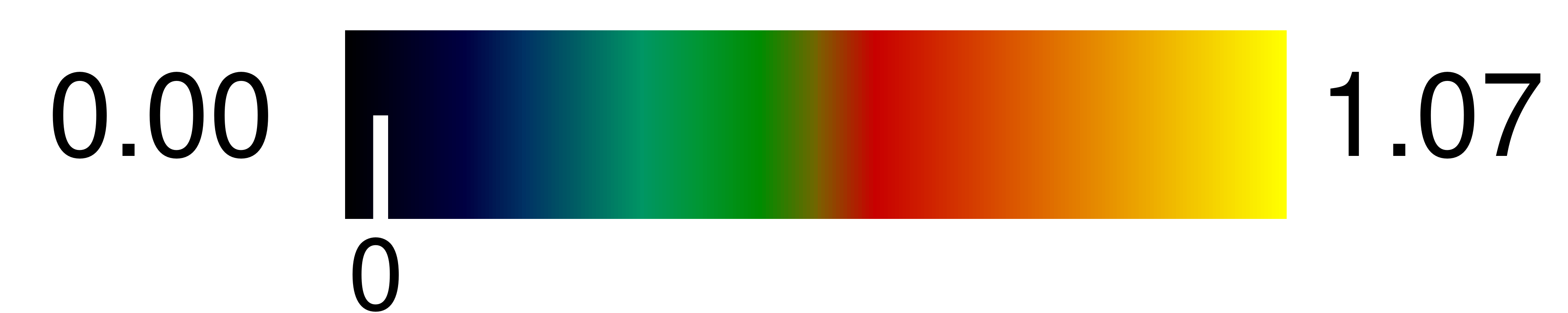} &
\hspace{0.8cm}\includegraphics[width=4.5cm, height=0.7cm]{RelativeExcess_ColorBars} \\
	
\rotatebox{90}{\Large NGC0628} & 
\includegraphics[width=5.2cm]{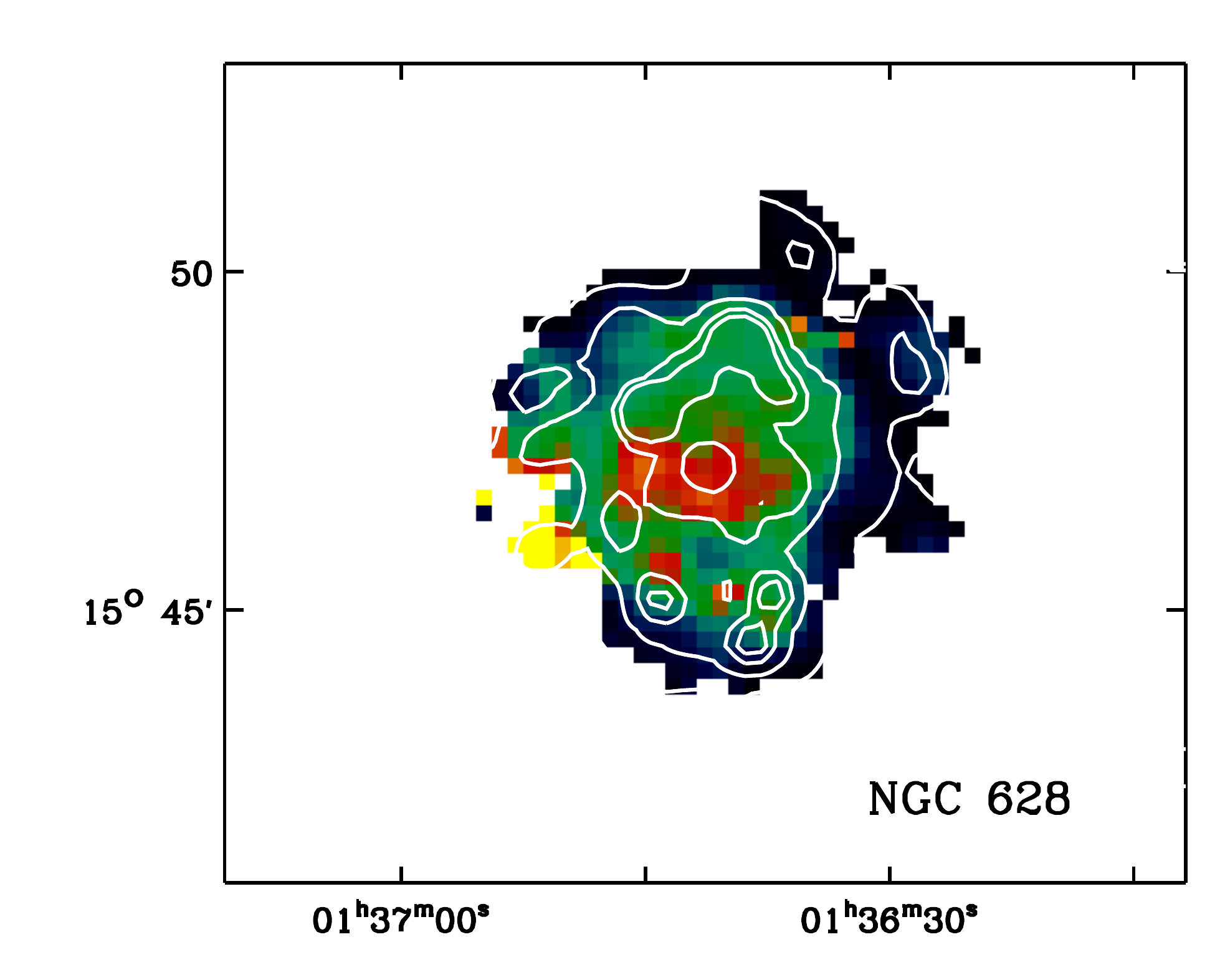} &
\includegraphics[width=5.2cm]{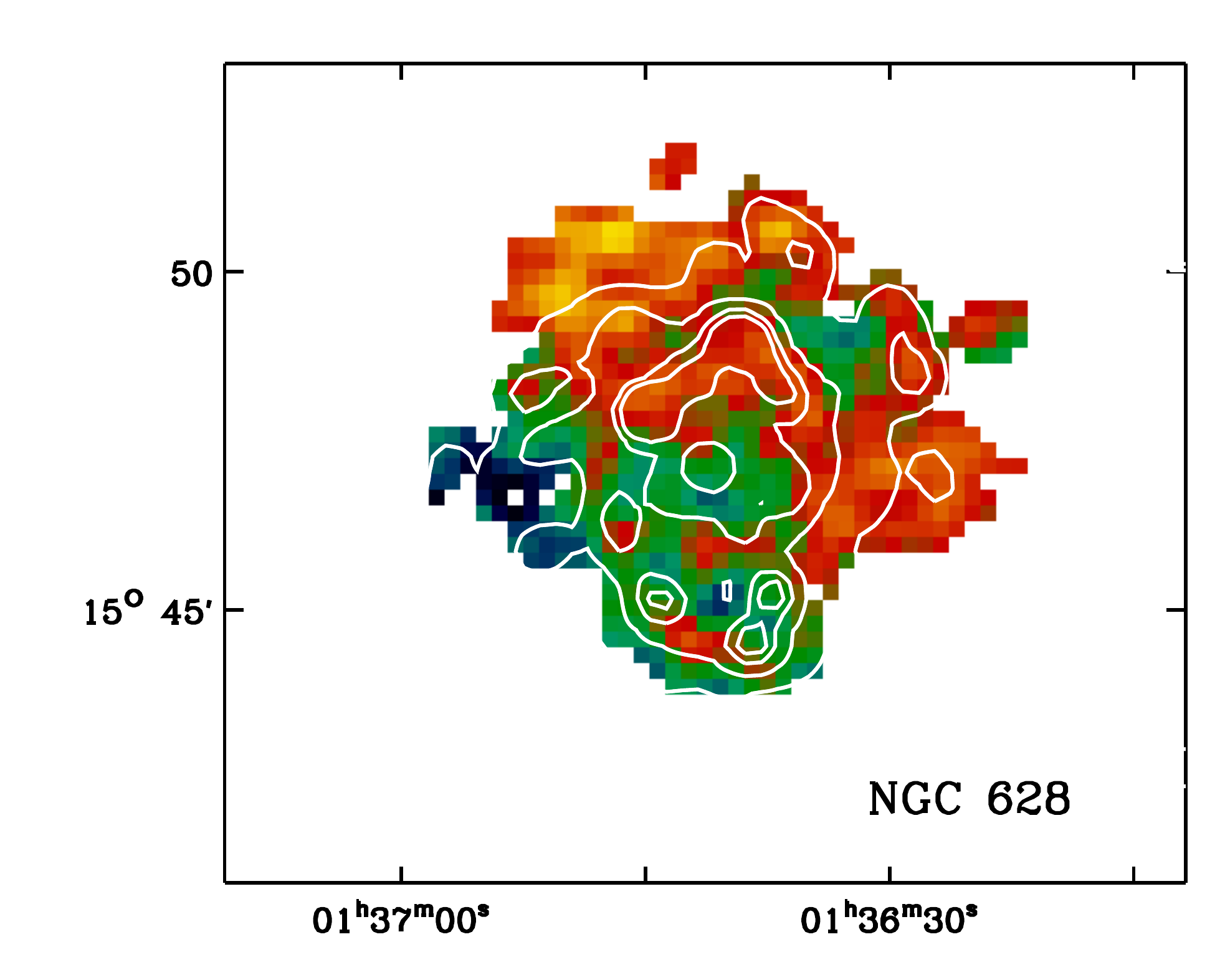} &
\includegraphics[width=5.2cm]{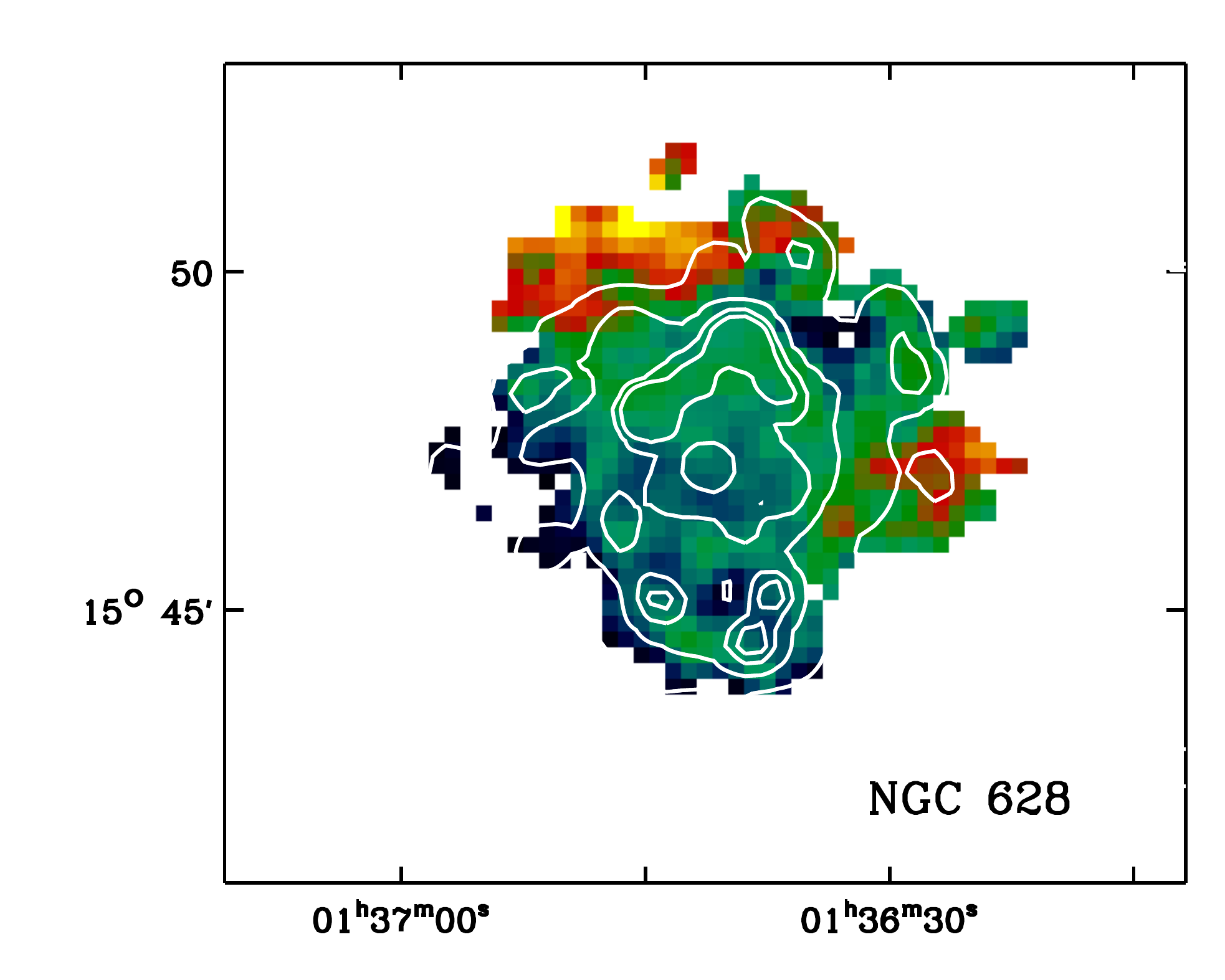}\\
& \hspace{0.8cm}\includegraphics[width=4.5cm, height=0.7cm]{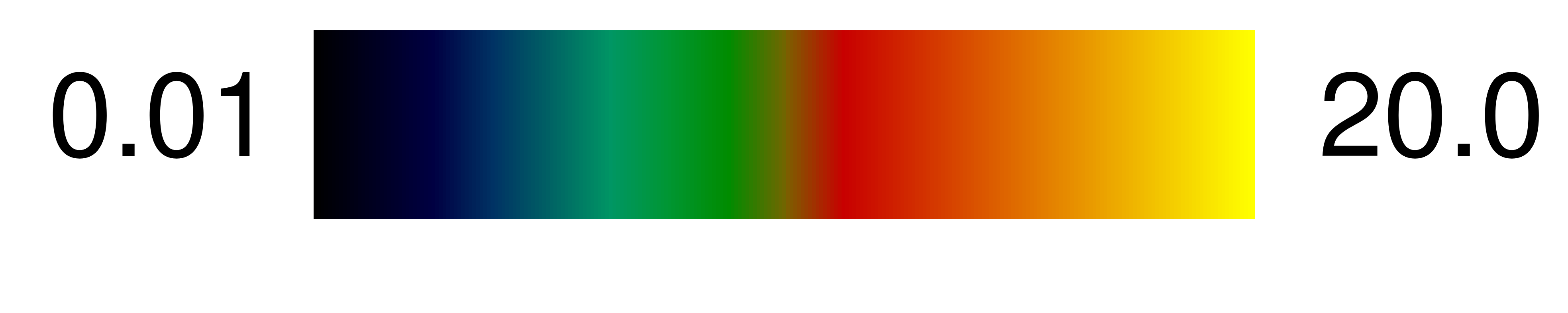} &
\hspace{0.8cm}\includegraphics[width=4.5cm, height=0.7cm]{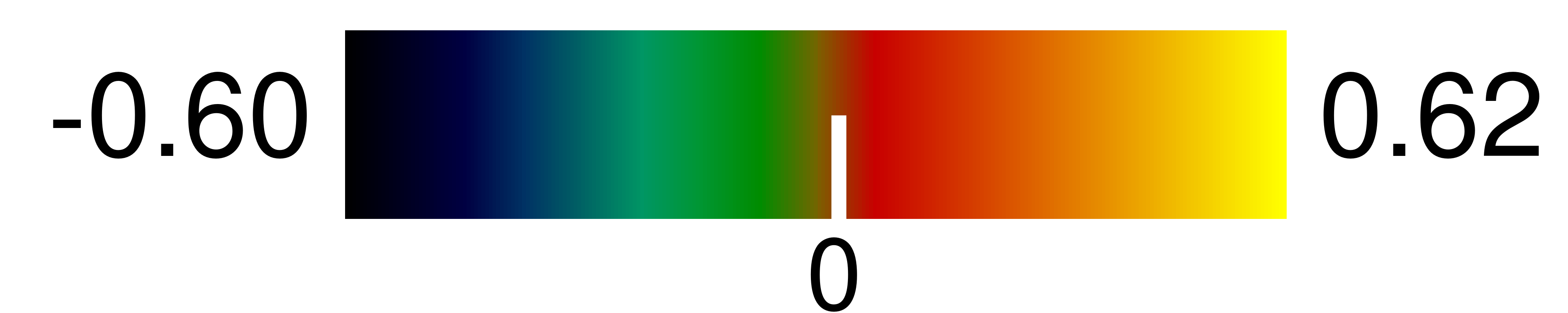} &
\hspace{0.8cm}\includegraphics[width=4.5cm, height=0.7cm]{RelativeExcess_ColorBars} \\
	 
\rotatebox{90}{\Large NGC3351} & 
\includegraphics[width=5.2cm]{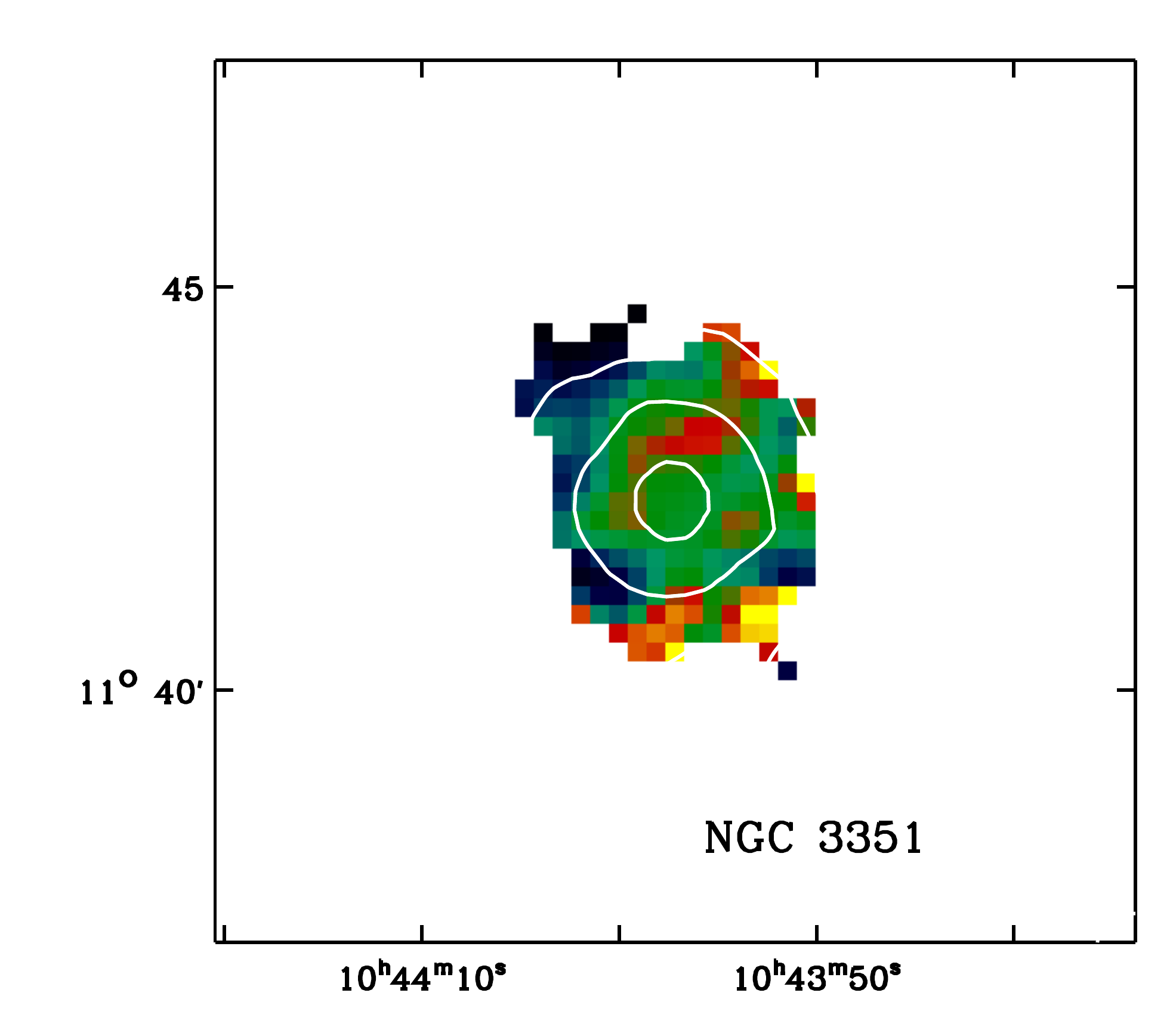} &
\includegraphics[width=5.2cm]{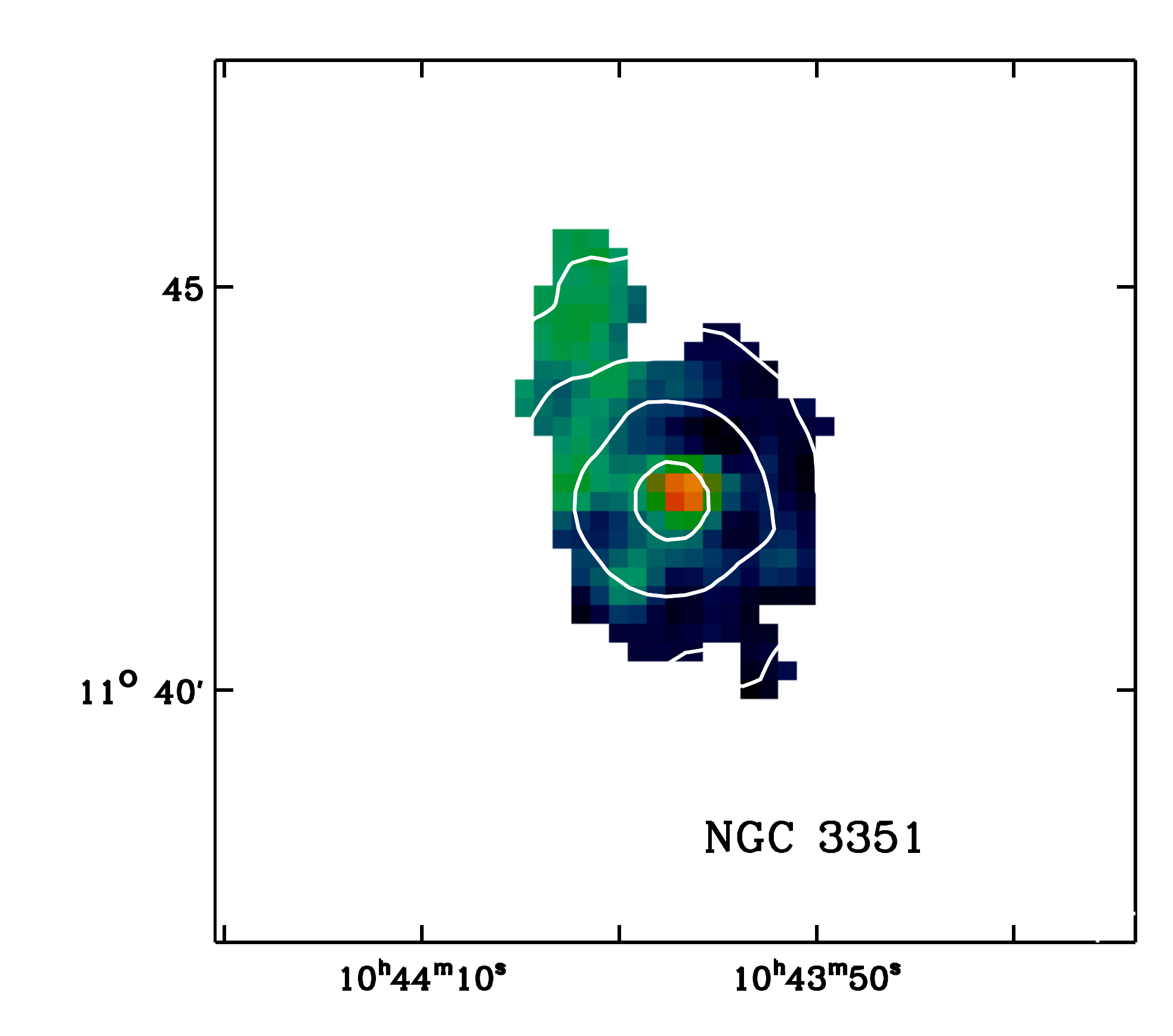} &
\includegraphics[width=5.2cm]{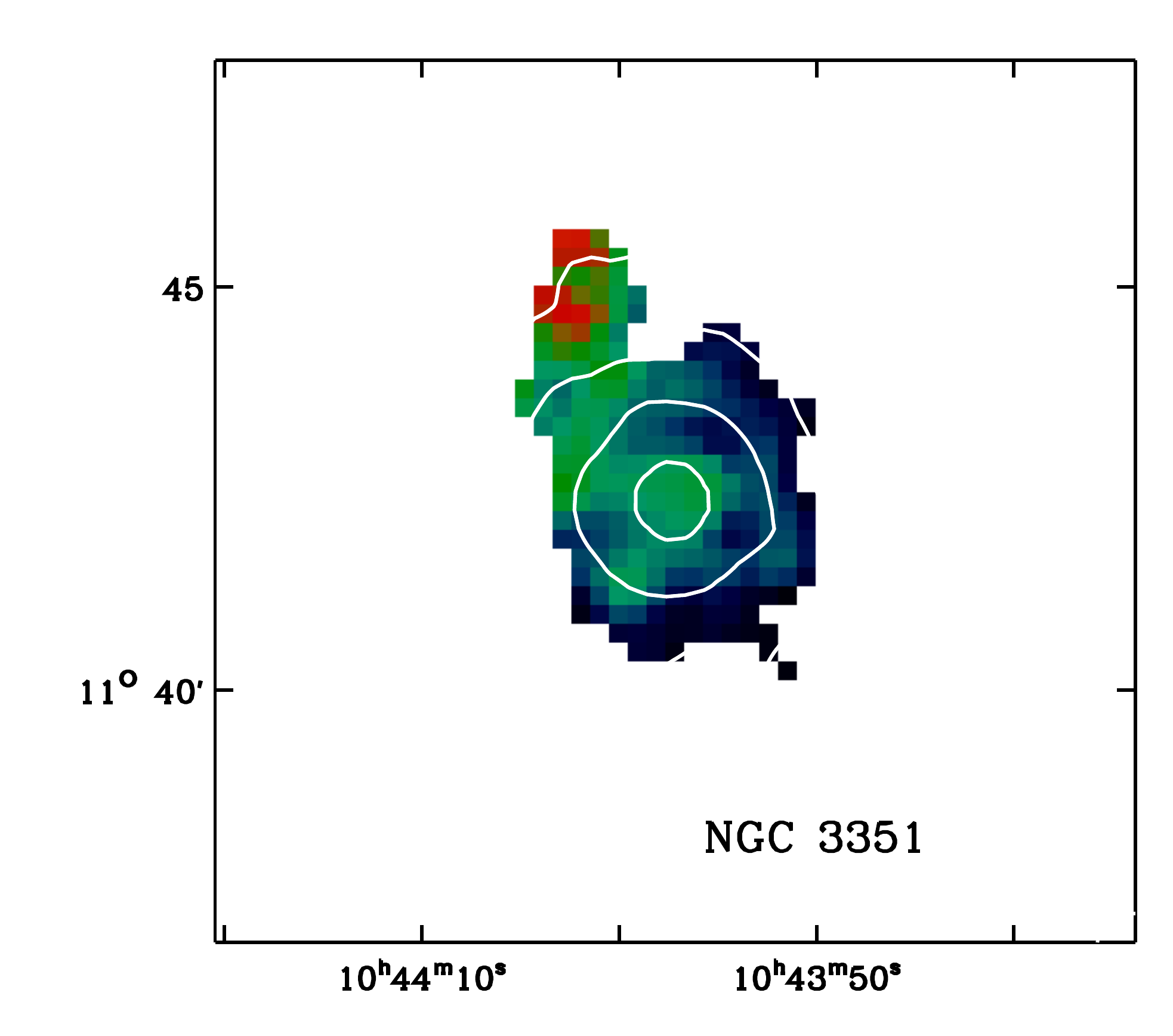} \\
& \hspace{0.8cm}\includegraphics[width=4.5cm, height=0.7cm]{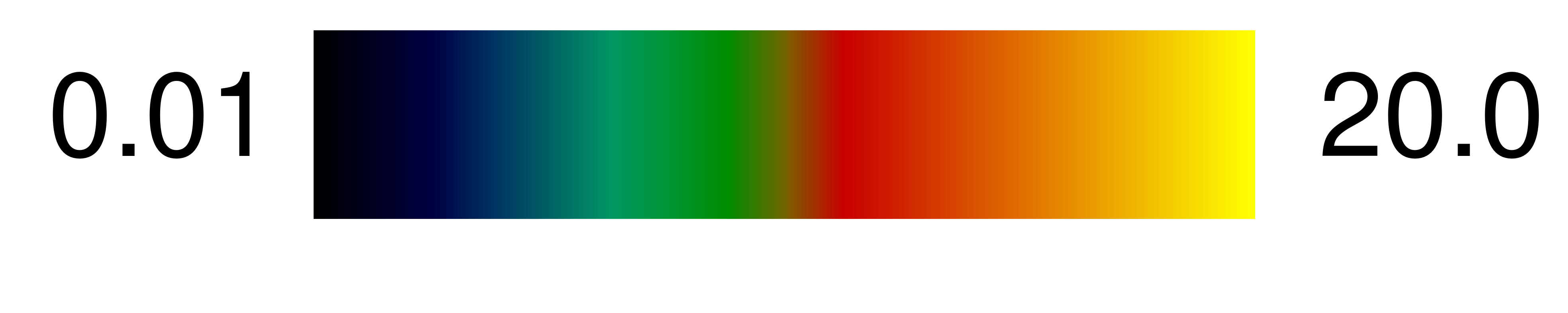} &
\hspace{0.8cm}\includegraphics[width=4.5cm, height=0.7cm]{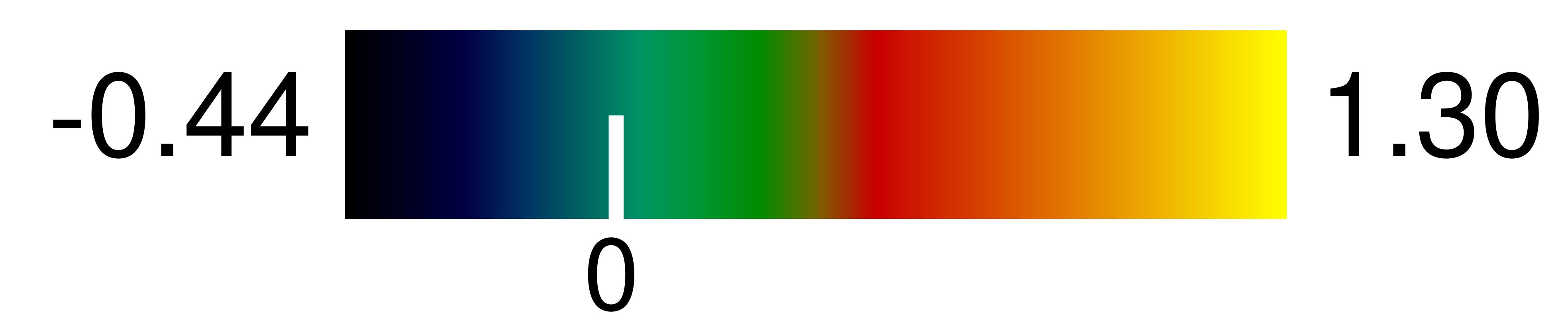} &
\hspace{0.8cm}\includegraphics[width=4.5cm, height=0.7cm]{RelativeExcess_ColorBars} \\
	 
\rotatebox{90}{\Large NGC3627} & 
\includegraphics[width=5.2cm]{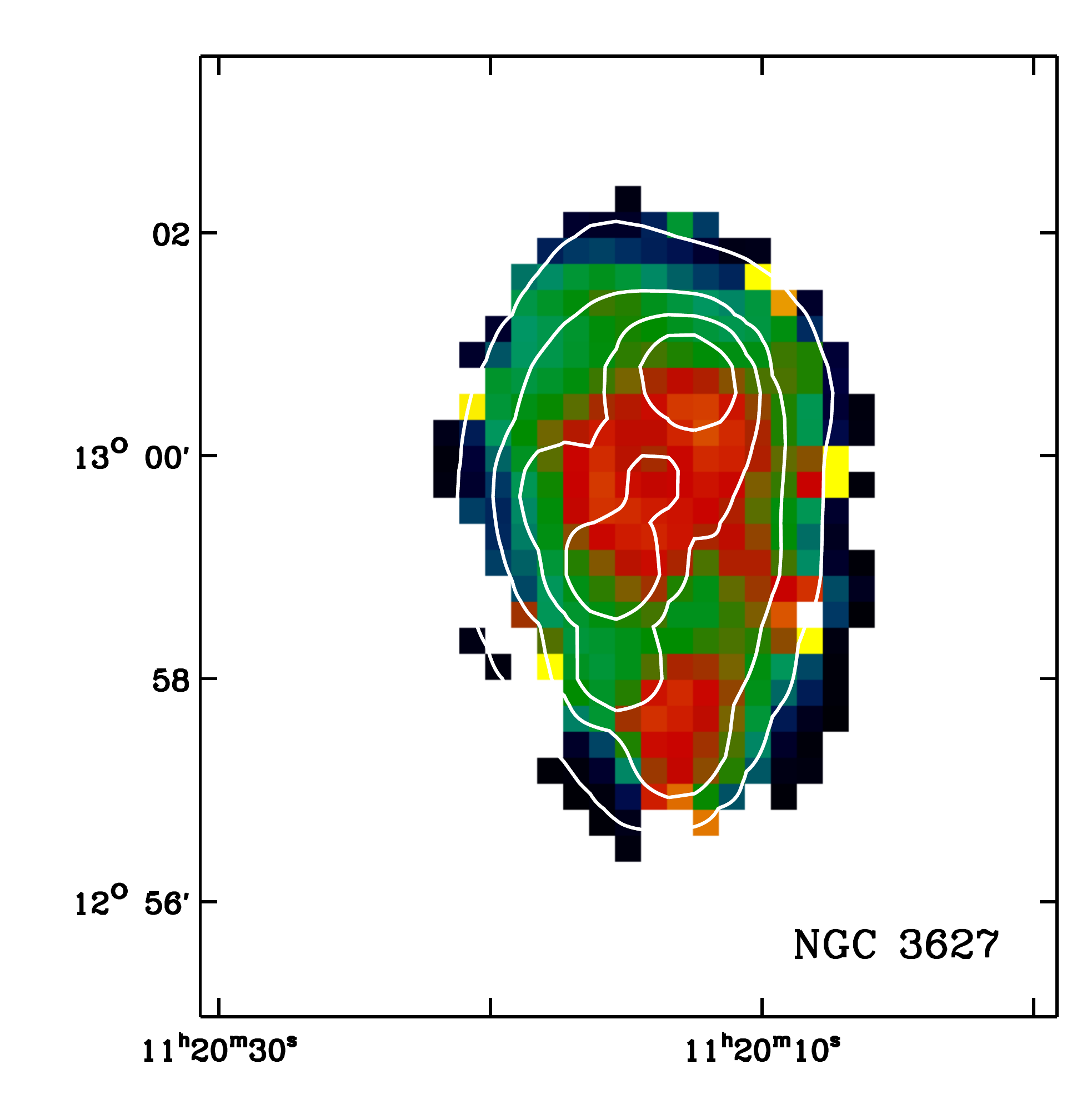} &
\includegraphics[width=5.2cm]{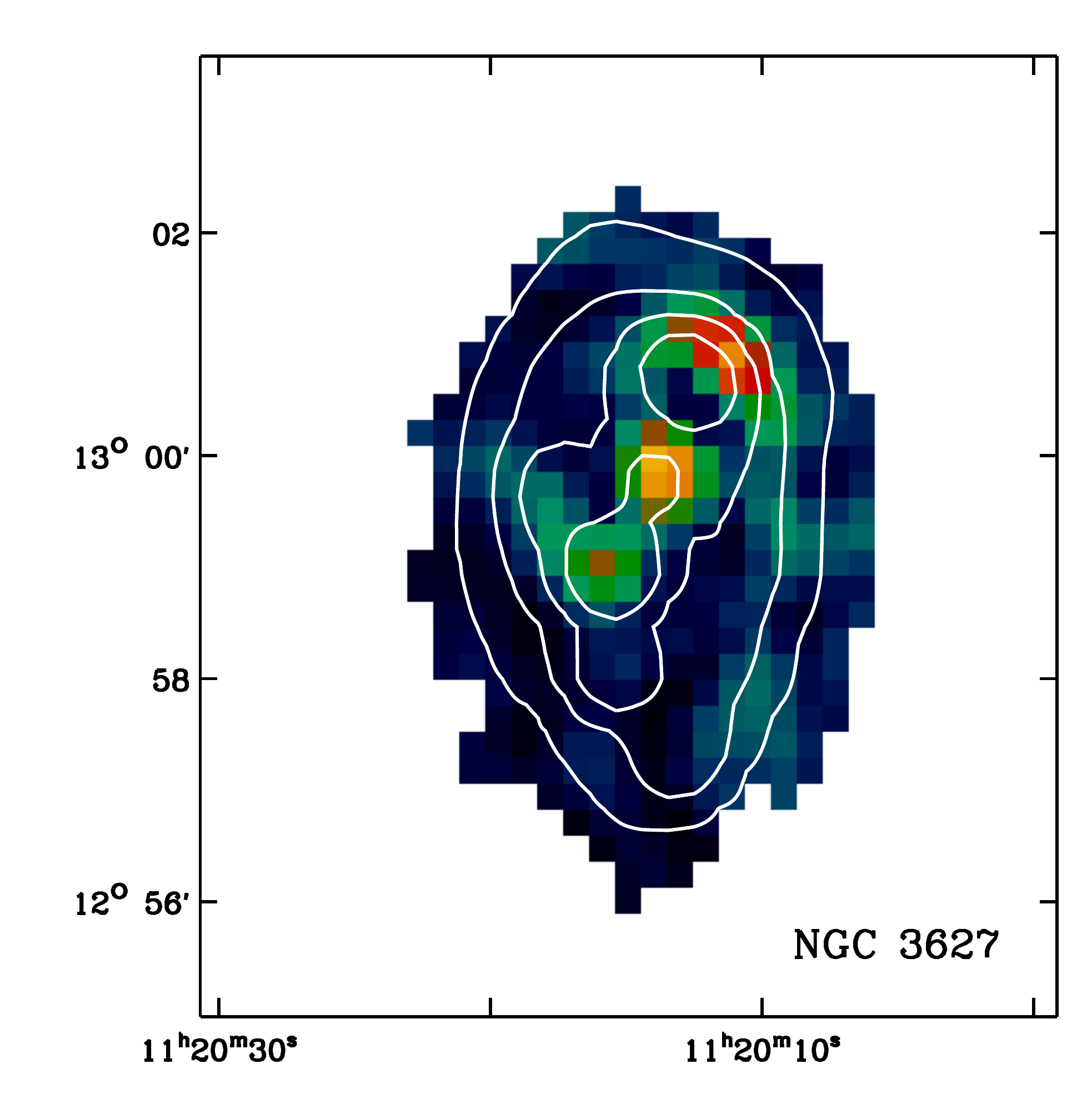} &
\includegraphics[width=5.2cm]{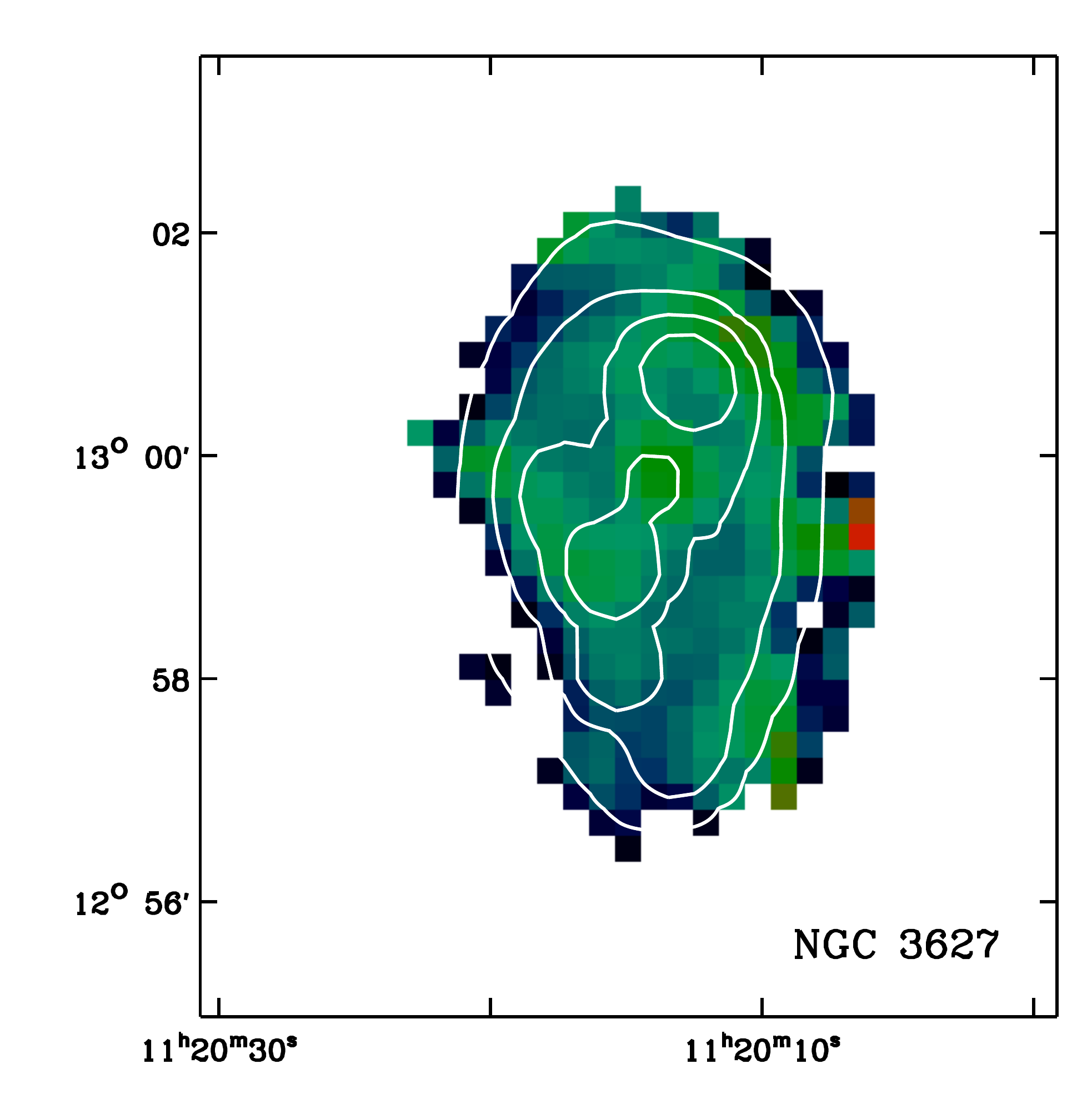} \\
& \hspace{0.8cm}\includegraphics[width=4.5cm, height=0.7cm]{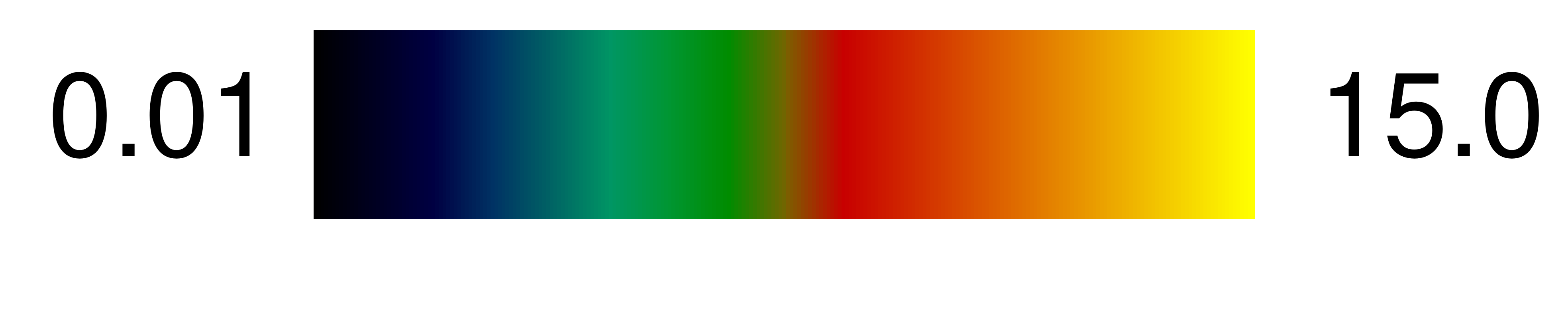} &
\hspace{0.8cm}\includegraphics[width=4.5cm, height=0.7cm]{NGC3627_Excess_ColorBars} &
\hspace{0.8cm}\includegraphics[width=4.5cm, height=0.7cm]{RelativeExcess_ColorBars} \\

\end{tabular}  
\caption{{\it Left column:} CO(3-2) contribution to the 870 \mic\ emission (in $\%$, see Section 4.4.2). {\it Middle column:} Excess emission on absolute scales (in MJy~sr$^{-1}$) corrected for CO(3-2). The modelled 870 \mic\ map is obtained using the [DL07] formalism. {\it Right column:} Corresponding relative excess corrected for CO(3-2). MIPS 24 \mic\ contours are overlaid: for NGC0337: 0.4, 1.5, 4 and 7 MJy sr$^{-1}$); for NGC0628: 0.25, 0.7, 1.5 and 2.1 MJy sr$^{-1}$; for NGC3351: 0.1, 0.3, 1.1 and 10 MJy sr$^{-1}$; for NGC3627: 0.6, 2.2, 6 and 15 MJy~sr$^{-1}$.}
    \label{Excess_maps_minusCO}
\end{figure*}

{\it From single-point measurements - }
Direct CO(3-2) single-point measurements ($\sim$22\arcsec\ beam) were taken toward the centre of NGC1097 \citep{Petitpas_Wilson_2003} and NGC4826 \citep{Mao2010}. For NGC1097, the contribution of the 3-2 line to the measured 870 \mic\ emission in this beam is S$_{CO(3-2)}$$\sim$101 mJy. This represents 20$\%$ of the total \lab\ flux contained in the 56\arcsec\ central region. Added to the synchrotron contribution to the 870 \mic\ (8$\%$), this could mostly explain the central excess (40$\%$ on average) we observe. For NGC4826, a contributing S$_{CO(3-2)}$ flux density of 122 mJy is estimated, representing 14$\%$ of the total flux, here again a non-negligible contribution to the central 870 \mic\ excess emission. 
A CO(2-1) single-point measurement (22\arcsec\ beam) was taken toward the centre of NGC1316 \citep{Horellou2001}. Using a brightness temperature ratio R$_{3-2,2-1}$=0.36 \citep[mean value of the ratio derived in][]{Wilson2012}, \citet{Albrecht2013} estimated a contribution of S$_{CO(3-2)}$$\sim$12 mJy, which represents, like in NGC1097, up to 20$\%$ of the central 56\arcsec\ central region. Here, again, added to the radio contribution to the 870 \mic, we can explain half of the emission in excess in the centre. 
Finally, only CO(1-0) single point measurements (43\arcsec\ beam) are available toward the centre of NGC1291 \citep{Tacconi1991} and NGC7793. Using a brightness temperature ratio R$_{3-2,1-0}$=0.18 \citep[mean value of the ratio derived in][]{Wilson2012}, \citet{Albrecht2013} estimate the S$_{CO(3-2)}$ flux density to reach 28 mJy at most in NGC1291, minor compared to the excess emission we observe, and 240 mJy in NGC7793. This represents in both cases $\sim$6.5$\%$ of the 870 \mic\ emission, not sufficient to explain the excess we observe. We note that in NGC7793, the single point observation is, anyhow, not sufficient to probe CO contribution in the low-surface brightness regions where the excess emission is mostly detected.

{\it From CO maps - }
CO(2-1) maps of NGC0337, NGC0628, NGC3351 and NGC3627 were obtained as part of the the HERA CO-Line Extragalactic Survey \citep[HERACLES;][]{Leroy2009}. We derive maps of the CO(3-2) line contribution to the 870 \mic\ maps, using a brightness temperature ratio R$_{3-2,2-1}$=0.36 to convert the CO(2-1) to CO(3-2) emission. Figure~\ref{Excess_maps_minusCO} (left column) shows the CO(3-2)-to-870 \mic\ flux density ratios for these four galaxies. 
MIPS 24 \mic\ contours are overlaid to indicate the distribution of the star-forming regions across the galaxies.
We also show the absolute difference maps (Fig.~\ref{Excess_maps_minusCO}, middle column) and the relative difference maps (Fig.~\ref{Excess_maps_minusCO}, right column) now corrected for CO(3-2) emission. The modelled 870 \mic\ maps used in this analysis are those obtained using the [DL07] model. For NGC0337 and NGC0628, the CO(3-2) line contamination is estimated to represent less than 3$\%$ of the 870 \mic\ emission on average across the galaxy, with a peak at ~12$\%$ in the centre for NGC0628. This contribution is not sufficient to explain the 870 \mic\ excesses we observe above our model extrapolations for these two objects. For NGC3351, an 8$\%$ CO(3-2) line contamination to the 870 \mic\ emission is derived in the centre, which can fully explain the weak central excess we detect for this object. Finally, a 11$\%$ line contamination is derived on average for NGC3627, but with a peak of the CO contribution to the 870 \mic\ in the centre of the galaxy. After correction, a weak relative excess ($<$0.5 so 50$\%$ above the modelled flux) remains in the centre and on each side of the bar of the galaxy. Directly using the CO(3-2) observation of NGC3627 taken from the JCMT Nearby Galaxy Legacy Survey \citep[NGLS;][]{Wilson2012} lowers the relative excess to ~40$\%$ in the centre and at each end of the bar.

   \begin{figure*}
    \centering  
	\begin{tabular}{c}
	\hspace{-30pt}
\includegraphics[width=19cm]{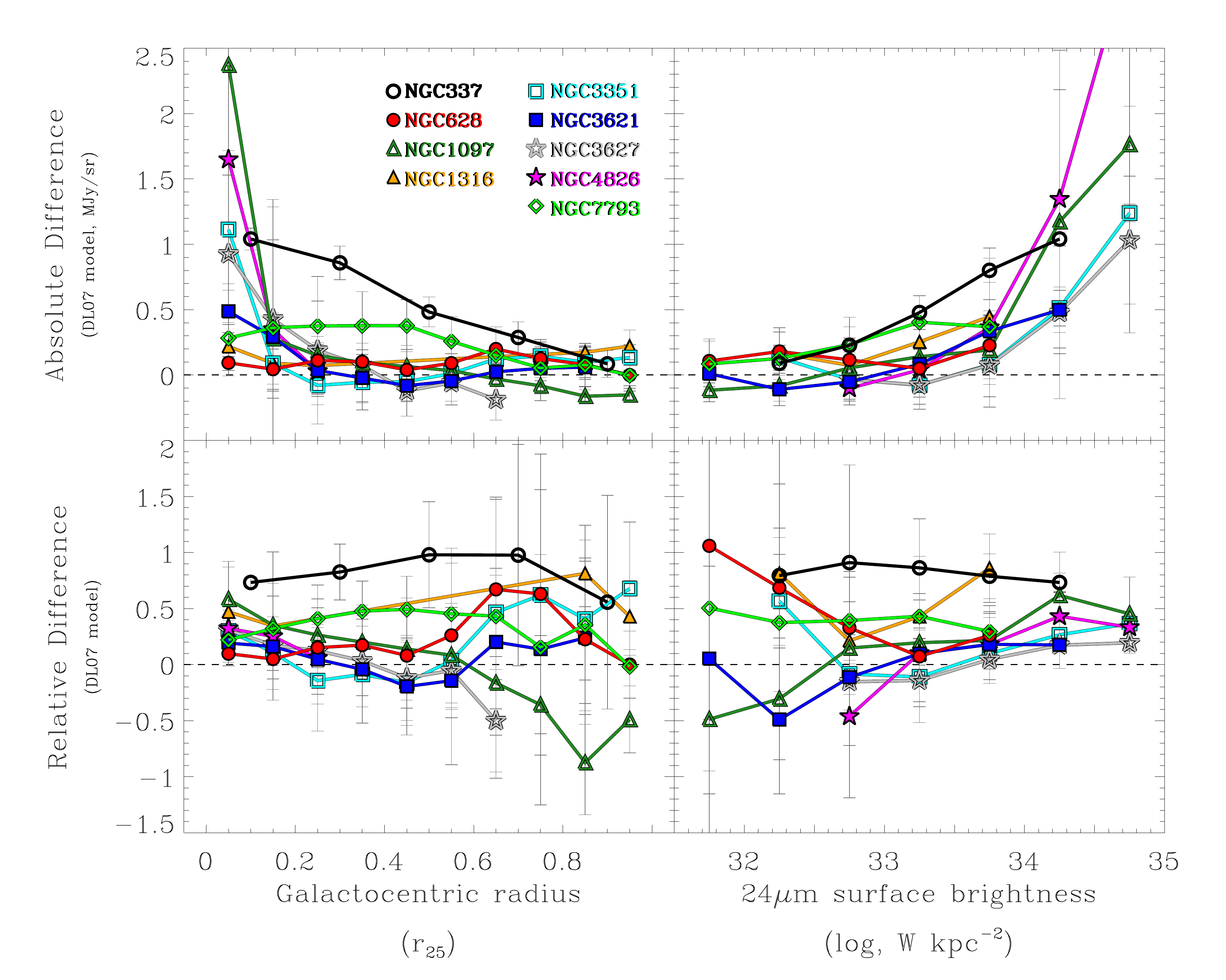} \\ 
 \end{tabular}
\caption{{\it Top:} Absolute difference between the observed 870 \mic\ flux densities (not corrected for CO and radio continuum) and the 870 \mic\ estimates extrapolated from the [DL07] dust model as a function of galactocentric radius (left, see Section 4.5.1) and 24 \mic\ surface brightness in units of W~kpc$^{-2}$ (right, see Section 4.5.2). The different galaxies of the sample are color-coded. The horizontal dashed line indicates when the observation matches the model. {\it Bottom:} Same with relative differences.}
\label{Excess_dependence_radius_mips24}
\end{figure*}

 \subsection{Analysis}
 
 \subsubsection{Average radial dependence}
 
Resolved analysis of the IR emission of nearby galaxies have showed that dust properties are particularly expected to vary from the nuclei to the outskirts of galaxies (see for instance the radial profiles of various dust properties like the TIR-to-UV ratio, the dust-to-gas mass ratio or the Polycyclic Aromatic Hydrocarbons fraction in \citet{Munoz2009_2}, of dust abundance gradients in \citet{Mattsson2012} or the bulge versus disk comparison of \citet{Engelbracht2010}, among others). In order to study the radial dependence of our results, we plot in Fig.~\ref{Excess_dependence_radius_mips24} the absolute (top left) and relative (bottom left) difference between the observed and modelled 870 \mic\ emission (using the [DL07] formalism) as a function of galactocentric radius. For each object, we normalise the radius to the approximate major radius of the galaxy. These major radii are taken from \citet{Kennicutt2011} and approximately match the r$_{25}$ radii of \citet{deVaucouleurs1991}. We exclude the galaxies NGC1291 and NGC1512 from this analysis because of the low statistics we have in the first case and the unknown nature of the 870 \mic\ {southern structure} in the second case. For the remaining objects, we restrict the analysis to resolved elements with a signal-to-noise superior to 5 in the 870 \mic\ band. We bin our resolved elements in 10 annuli from the centre to r$_{25}$ (radii are not corrected for inclination) and take the medians in these annuli. The optical radius of NGC0337 is only 1.5\arcmin, which corresponds to 6 resolved elements of 14\arcsec. Elements are thus binned in 5 annuli rather than 10 for this object. We calculate the median when 3 resolved elements at least fulfil our criteria to work on better statistics. Uncertainties are the standard deviations in each annulus. 

The absolute differences decrease with radius in most cases, with a systematic 870 \mic\ excess (positive difference) in the centre (typically regions located within 0.3 r$_{25}$). Some galaxies like NGC0337 show a smooth decrease with radius. Others have a sharp 870 \mic\ excess profile in the very centre such as for NGC1097, NGC3627, or NGC4826, essentially linked with the contribution of CO and radio to the 870 \mic\ emission, as discussed in the previous section. The galaxy NGC1316 hosts a low-luminosity X-ray AGN \citep{KimD2003} but the possible radio contamination of the 870 \mic\ emission does not lead to a similar broken profile. We observe a sharp decrease of the absolute difference in NGC3351 as well but this could be due to the non-detection of the 870 \mic\ emission previously mentioned, leading to an underestimate of the observed 870 \mic\ flux density across most of the disk of the galaxy. The two non-barred objects NGC0628 and NGC7793 have very flat profiles, i.e. almost no dependence of the absolute excess compared to model predictions on radial distance. The results of [G12] also suggest differences in the submm spectra between barred and non-barred galaxies (when $\beta$$_c$ is allowed to vary). We discuss this further in Section 4.5.3. 

The radial profiles of most of the galaxies are confined to the $\pm$50$\%$ region, which suggests that the [DL07] model is, on average per radial bin, sufficient to explain the 870 \mic\ emission we observe, within uncertainties. We observe a noticeable relative difference in NGC0337 for which the 870 \mic\ difference is systematically above the $\pm$50$\%$ threshold. Even if the relative 870 \mic\ excess of NGC0337 and NGC7793 seems to increase with radius across the disks of the objects before dropping in the outskirts, the trend is, within the uncertainties, consistent with a constant relative (and positive) difference. Some of the decreases obtained in the outskirt regions could be linked with the low statistics (and low 870 \mic\ surface brightnesses) toward r$_{25}$ radii.

\subsubsection{Dependence on the star formation activity}

Several studies have tried to correlate the submm excess with the IR emission. Studying sub-regions along the radius of NGC4631, \citet{Bendo2006} showed, for instance, that their excess 850 \mic\ emission significantly varies in the galaxy and that low excess values were predominantly found in faint IR regions, even if the trend is statistically weak. In Fig.~\ref{Excess_dependence_radius_mips24}, we plot the absolute (top right) and relative (bottom right) difference between the observed and the modelled 870 \mic\ emission (using the [DL07] formalism) as a function of the 24 \mic\ surface brightness. The 24 \mic\ binning differs from the radial binning, the 24 \mic\ emission being indeed sensitive to the warm dust populations and a good tracer for active star formation \citep[see][among others]{Calzetti2007_2}. We use the same detection criteria and require the same number of resolved elements to calculate the median in each bin of surface brightness than those applied for the radius profiles. Uncertainties are the standard deviations in each annulus. 

For all galaxies, the absolute differences increase with the 24 \mic\ surface brightness, with an 870 \mic\ emission consistent with the [DL07] model prediction in low 24 \mic\ surface brightnesses and an excess at high surface brightnesses. More particularly, we observe a systematic 870 \mic\ excess emission when $\nu$L$_{\nu}$(24 \mic) $>$ 5$\times$10$^{33}$ W~kpc$^{-2}$, with a significant 870 \mic\ absolute excess emission in the bright star forming regions ($\nu$L$_{\nu}$(24 \mic) $>$ 5$\times$10$^{34}$ W~kpc$^{-2}$) of NGC1097, NGC3351, NGC3627, NGC4826. This is consistent with the results of Section 4.5.1, the bright star forming regions being located in the galaxy centres (or close in the case of NGC3627) and here again the central excess is partly (if not mostly) linked with contribution of CO and radio to the 870 \mic\ emission. The relative difference profiles show a variety of behaviours within the sample, with large uncertainties, especially in regions with low 24 \mic\ surface brightness. The 870 \mic\ relative differences for the galaxies NGC0337 and NGC7793 show rather flat profiles, so no dependence with the star formation activity either. This absence of clear dependence makes it difficult to understand what could be driving the excess emission we detect at global scales. The galaxies NGC0628, NGC3351 and NGC3621 show an increase toward star forming regions, even if the radial averages have large uncertainties (partly due to non detections in the disk for NGC0628 and NGC3351). We discuss these galaxies in more detail in the following section.

 \subsubsection{Decrease of the effective emissivity index?}

In Fig.~\ref{Excess_dependence_radius_mips24}, we observe from the profiles of resolved elements with a 870 \mic\ detection at 5$\sigma$ averaged in annuli that the relative excess does not show a particular dependence with radius. However, we can see from Fig.~\ref{Excess_maps} that the relative differences between observed and modelled 870 \mic\ emission preferentially peak in the disk or the outskirts for the galaxies NGC0628 and NGC7793 (in NGC3621 as well to a lesser extent) when probed in low-surface brightness regions. Whatever causes the excess emission in these objects seems to have a predominant effect in these regions rather than in the centre. 
Modelling the thermal dust emission in a strip of the Large Magellanic Cloud, \citet{Galliano2011} also derived a submm excess map of the region (excess at 500 \mic\ in their study) and found that the excess is inversely correlated with the dust mass surface density, thus increasing toward low-surface brightnesses as well. In [G12], we showed that when the temperature and the emissivity are allowed to vary in the resolved 2MBB fitting process, the ``effective" emissivity index seems to radially decrease toward low surface brightness regions. The high 870 \mic\ excess emission observed in the northeast part of the disk in NGC3351 (unfortunately the only region of the disk detected with \lab) also corresponds to a region where low values of the emissivity index $\beta$$_c$ were derived in [G12]. The two independent results (870 \mic\ excess detected in the disk or outskirts when $\beta$$_c$ is fixed versus radial flattening of the [250-500\mic] slope when $\beta$$_c$ is a free parameter) could both be explained by a radial flattening of the submm slope in those particular objects.

NGC0628, NGC3621 and NGC7793 are non-barred galaxies, and three of the four late-type objects of the sample (SAc, SAd and SAd morphology respectively). The fourth late-type galaxy is NGC0337 in which we also detect an excess predominant in the outskirts, even if the compactness of the source does not allow us to properly analyse the excess distribution. The radial decrease of the effective emissivity index was also observed in the Scd galaxy M33 \citep{Tabatabaei2011} or the SA(s)b galaxy Andromeda \citep[][]{Smith2012} but not in the SAB(s)c galaxy M83 \citep{Foyle2012} or the other barred spiral galaxies of our KINGFISH/LABOCA sample \citep{Galametz2012}. The submm radial flattening thus seems to be dependent on the galaxy morphology and favoured in late-type non-barred spirals. 

At our working resolution unfortunately, many physical parameters possess a radial dependence. As shown in \citet{Moustakas2010}, NGC0628, NGC3621 and NGC7793 possess, for instance, strong radial metallicity gradients and metallicity is, as discussed in Section 3.4, a driver of variations in the properties of cold dust \citep[][R\'{e}my-Ruyer et al., in prep]{Sandstrom2012_2}. We also know that the dust temperature slowly decreases with radius or that the molecular gas distribution varies and could rapidly become patchy toward the outskirts of spirals. The low resolution of our data prevents us from studying where the excess is produced on sub-kpc scales. The lack of coverage (gap between 500 and 870 \mic\ and no constraint longward of 870\mic) also does not allow us to know if the excess can purely be explained by a smooth flattening (so by variations in the dust opacity or by a smooth dust temperature range for instance) or if the 870 \mic\ is produced by a separate population (cold dust, spinning dust, magnetic grains) that could translate into a break in the submm slope.


\section{Conclusions}

We combine \spitz, \hersc\ and \lab\ data to model with MBBs the global IR-to-submm emission of 11 nearby galaxies. We investigate how the potential excess emission at submm wavelengths varies with various model assumptions (MBB with $\beta$$_c$ fixed or free) and show that:

\begin{itemize}
 \item For 7 galaxies of the sample (NGC1097, NGC1291, NGC1316, NGC3351, NGC3621, NGC3627 and NGC4826), the global 870 \mic\ emission can be explained, within the uncertainties, by thermal emission using an emissivity index $\beta$$_c$=2.0.
 
 \item For the 4 other objects (NGC0337, NGC0628, NGC1512 and NGC7793), the integrated 870 \mic\ emission is in excess compared to that expected from cold dust with an emissivity index $\beta$$_c$=2.0. Using an emissivity index $\beta$$_c$ of 1.5 reproduces the global excess we observe in NGC0628, and also lead to better predictions of the 870 \mic\ emission in the previously mentioned two galaxies NGC1097 and NGC1316. For NGC1512, the observed 870 \mic\ flux and the 870 \mic\ modelled using $\beta$$_c$=1.5 are consistent within the error bars but the model fitted to \spitz+\hersc\ data favours very low $\beta$$_c$ values when allowed to vary ($\beta$$_c$=1.16). Lower values ($\beta$$_c$$<$1.3) would also be required to fit the 870 \mic\ data in NGC0337 and NGC7793, which questions the use of an isothermal component to fit the cold dust population up to 870 \mic\ in those objects.  

\end{itemize}

We apply the same methodology on local scales and produce 870 \mic\ maps using 2MBBs models with fixed $\beta$$_c$ or the [DL07] dust models in order to compare them to the observed emission:

\begin{itemize}

\item We observe a systematic 870 \mic\ excess when the emissivity index of $\beta$$_c$=2.0 is used to model the ISM elements we selected (above a given signal-to-noise criterion). Using $\beta$$_c$=1.5 can, in many cases, reconcile the observed and the modelled 870 \mic\ emission. The (absolute and relative) difference maps derived using the [DL07] formalism are similar to those obtained using $\beta$$_c$=1.5. 

\item Maps of the absolute difference between observed and modelled 870 \mic\ emission show, in many cases, a decrease of this quantity with radius and an increase with star formation (traced using the 24 \mic\ surface brightness). In NGC1097, NGC1316, NGC3627 and NGC4826, this excess can be partly or fully explained by contributions from CO(3-2) emission and/or free-free and synchrotron radiation.

\item The relative excess profiles have larger uncertainties with various behaviours from one galaxy to another and no clear dependence in the case of NGC0337 and NGC7793 with either parameter (radius or star formation). This raises the issue of what could be the drivers of the major excesses we detect in these two objects. This study should be extended to closer (and thus more resolved) objects to increase our statistics.

\item In the non-barred spirals NGC0628, NGC3621 and NGC7793 however, the relative excess maps show peaks in the disk or toward low surface brightnesses. A radial flattening of the submm slope (so a decrease of the ``effective" emissivity index with radius) as suggested by the studies of [G12], could explain the distribution of the 870 \mic\ excess we observe.

\end{itemize}


\section*{Acknowledgments}
We would like to thank the referee for a very constructive report that helped to improve the clarity/robustness of the results. The research of C.D.W. is supported by grants from the Natural Sciences and Engineering Research Council of Canada.
PACS has been developed by MPE (Germany); UVIE (Austria); KU Leuven, CSL, IMEC (Belgium); CEA, LAM (France); MPIA (Germany); INAF-IFSI/OAA/OAP/OAT, LENS, SISSA (Italy); IAC (Spain). This development has been supported by BMVIT (Austria), ESA-PRODEX (Belgium), CEA/CNES (France), DLR (Germany), ASI/INAF (Italy), and CICYT/MCYT (Spain). 
SPIRE has been developed by a consortium of institutes led by Cardiff Univ. (UK) and including: Univ. Lethbridge (Canada); NAOC (China); CEA, LAM (France); IFSI, Univ. Padua (Italy);IAC (Spain); Stockholm Observatory (Sweden); Imperial College London, RAL, UCL-MSSL, UKATC, Univ. Sussex (UK); and Caltech, JPL, NHSC, Univ. Colorado (USA). This development has been supported by national funding agencies: CSA (Canada); NAOC (China); CEA, CNES, CNRS (France); ASI (Italy); MCINN (Spain); SNSB (Sweden); STFC, UKSA (UK); and NASA (USA).


\bibliographystyle{mn2e}
\bibliography{/Users/maudgalametz/Documents/Work/Papers/mybiblio.bib}

 
\newpage
 \appendix
 \section{870 \mic\ maps predicted from our different SED models and 870 \mic\ absolute and relative difference maps}
   
\begin{figure*}
\centering
\begin{tabular}  { m{0cm} m{5.1cm} m{5.1cm} m{5.1cm}  m{0.7cm}}    
{\Large \bf~~~~~~~~~~NGC0337} &&&\\  
&\hspace{5cm}\rotatebox{90}{\Large 870 \mic\ Observed} & 
\includegraphics[width=5.7cm]{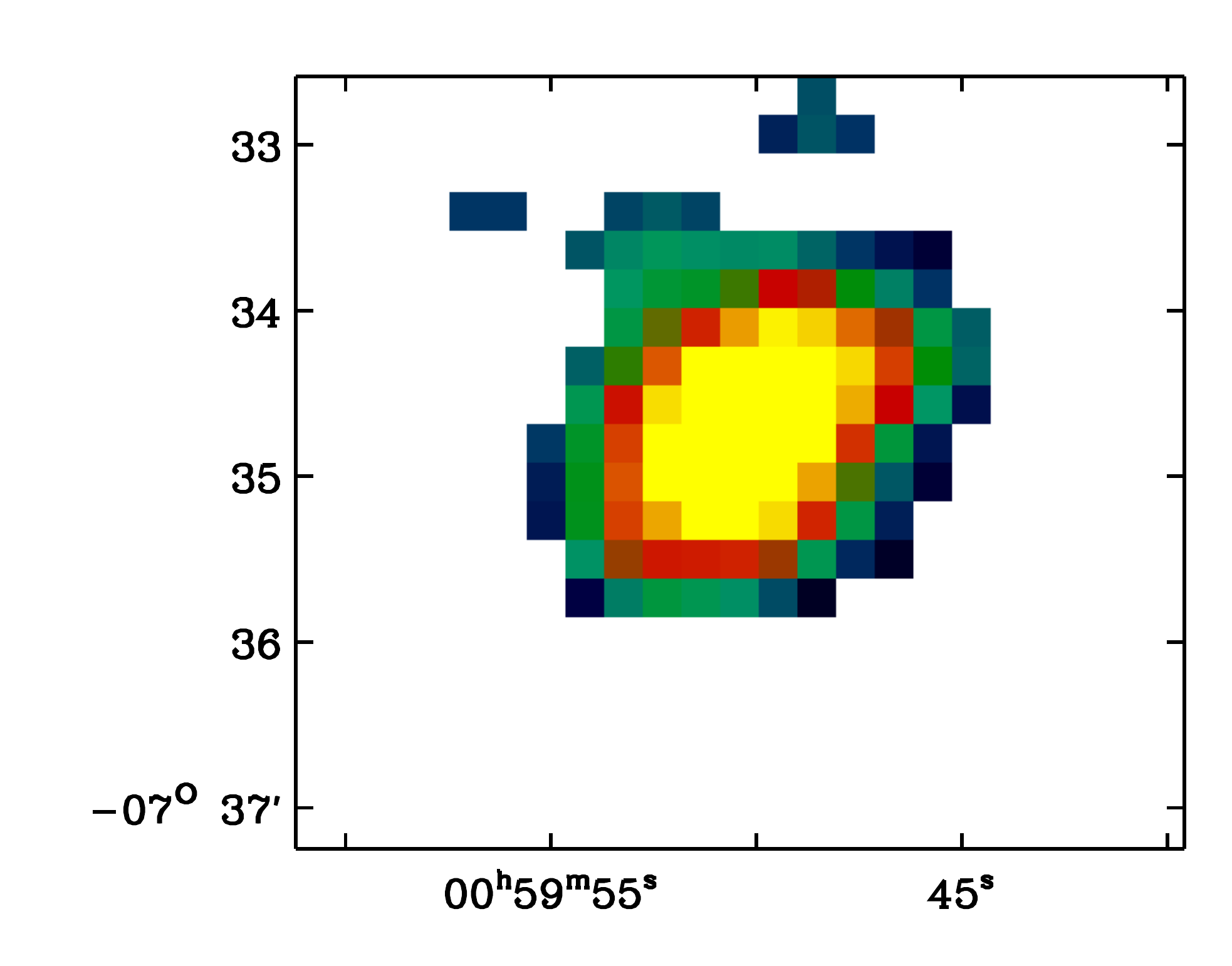} &&
\rotatebox{90}{\includegraphics[width=4cm, height=0.9cm]{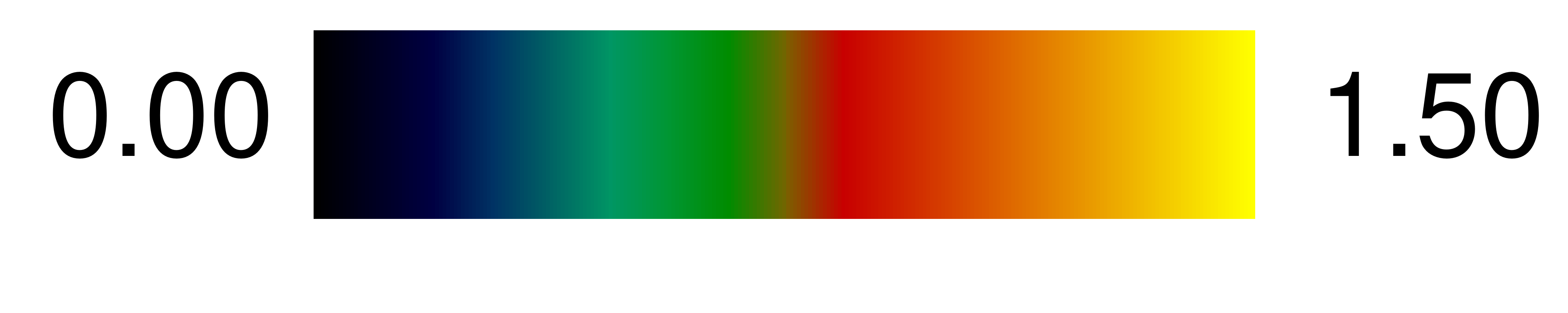}}  \\
&&\\
& {\Large \hspace{2.2cm}$\beta$$_c$ = 2.0 model} & {\Large \hspace{2.2cm}$\beta$$_c$ = 1.5 model}  & {\Large \hspace{2.2cm}[DL07] model} & \\

\rotatebox{90}{\Large 870 \mic\ Modelled} & 
\includegraphics[width=5.7cm]{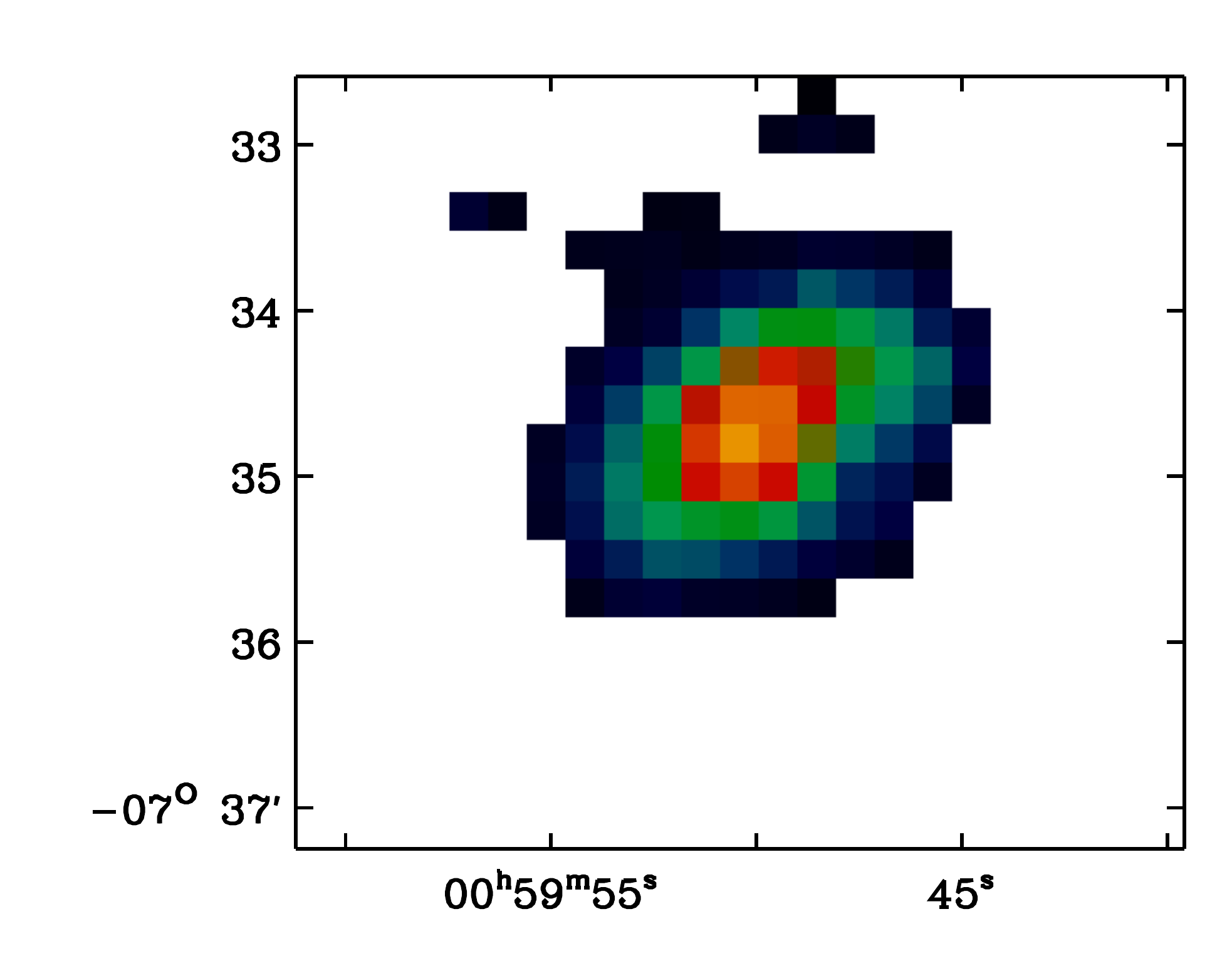} &
\includegraphics[width=5.7cm]{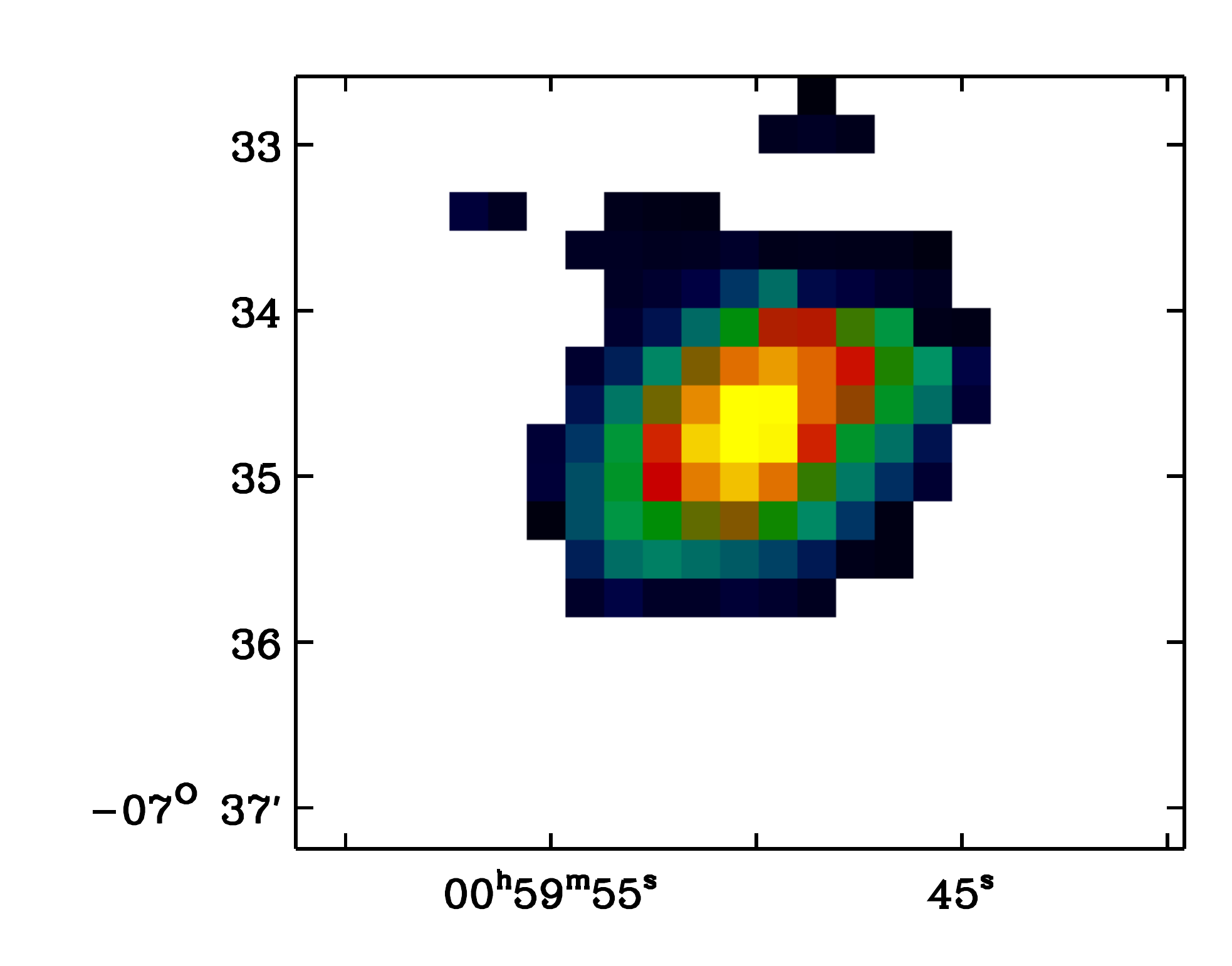} &
\includegraphics[width=5.7cm]{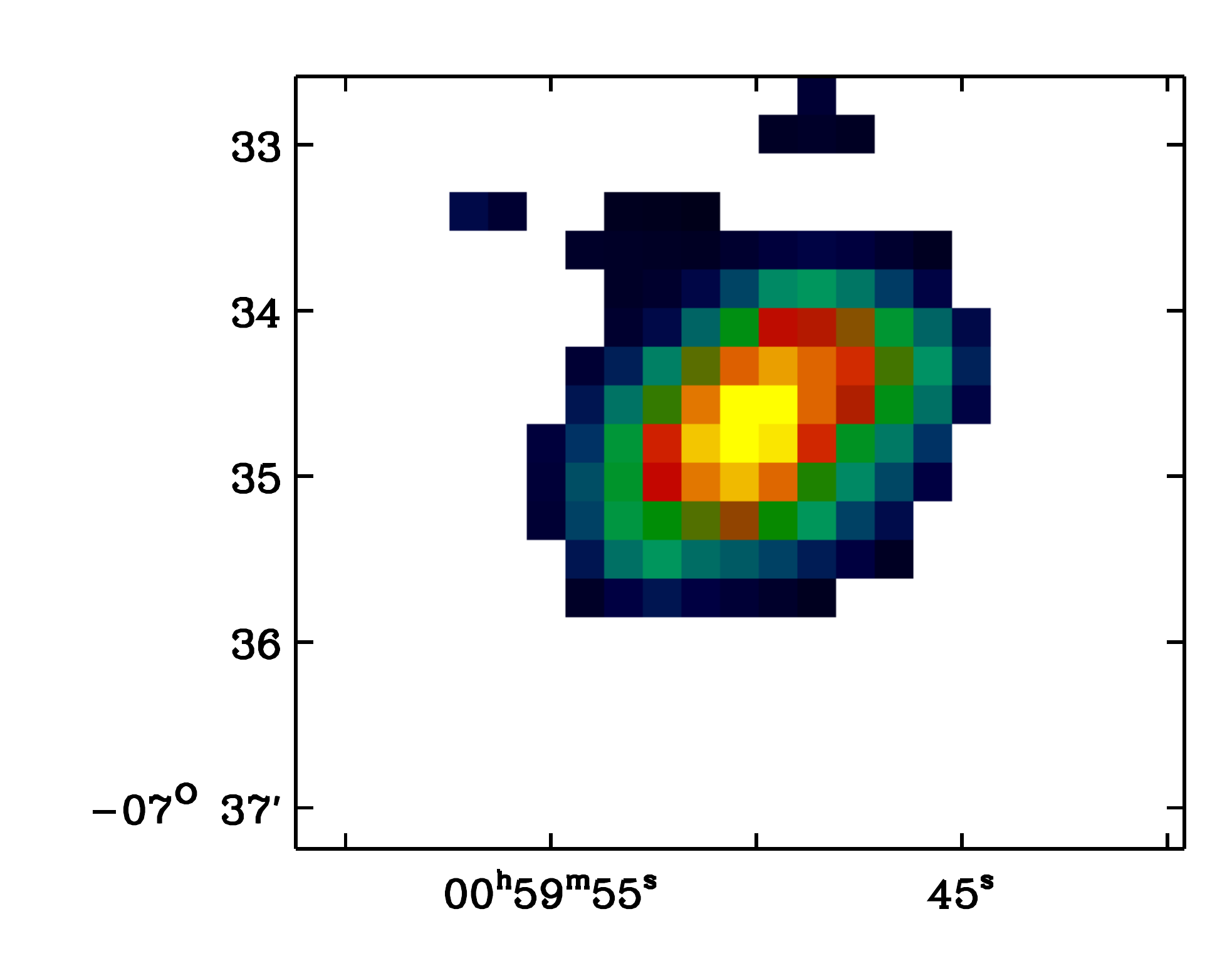}  &
\rotatebox{90}{\includegraphics[width=4cm, height=0.9cm]{NGC0337_Extrap870_ColorBars}}  \\
	
\rotatebox{90}{\Large Absolute Difference} &
\includegraphics[width=5.7cm]{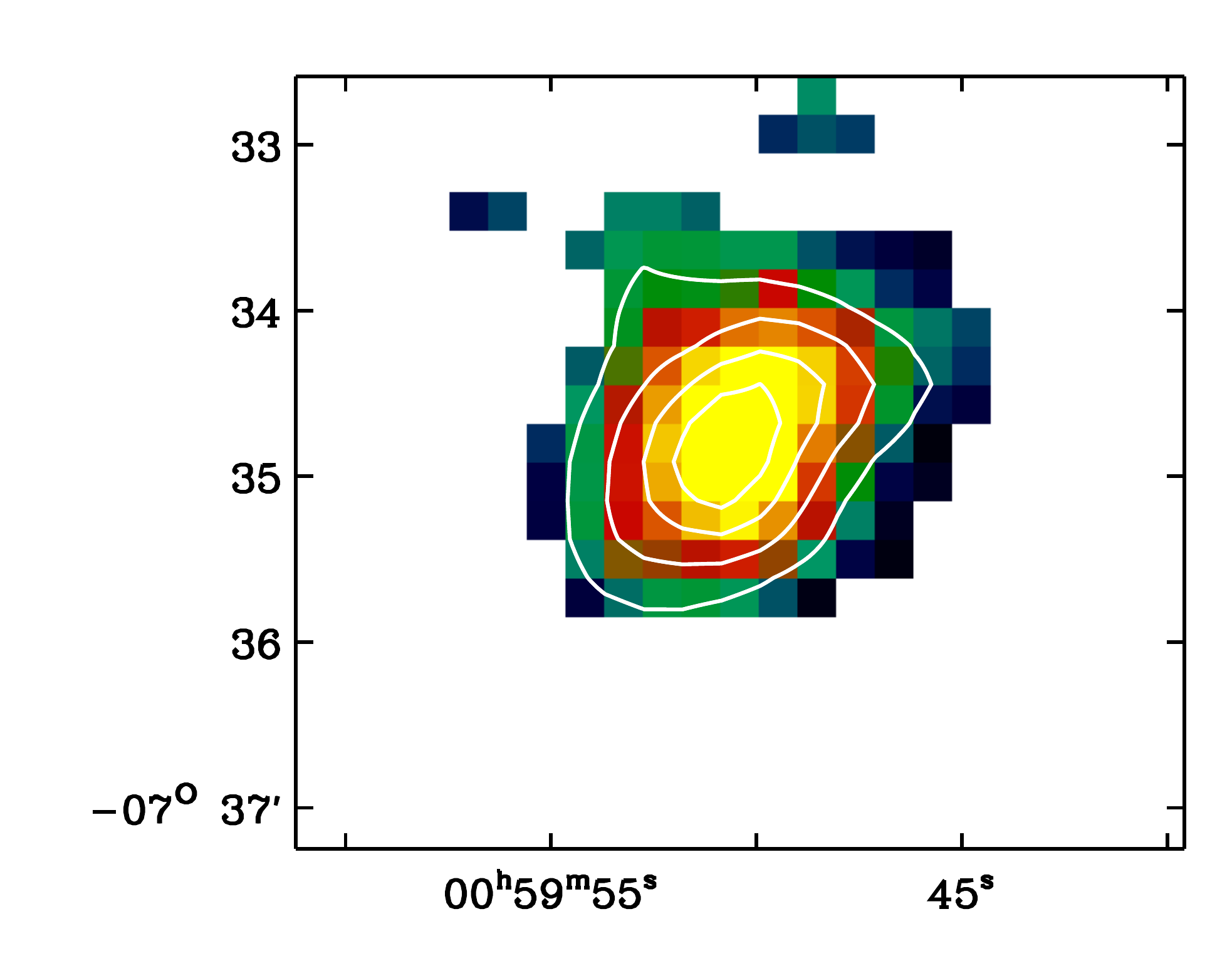} & 
\includegraphics[width=5.7cm]{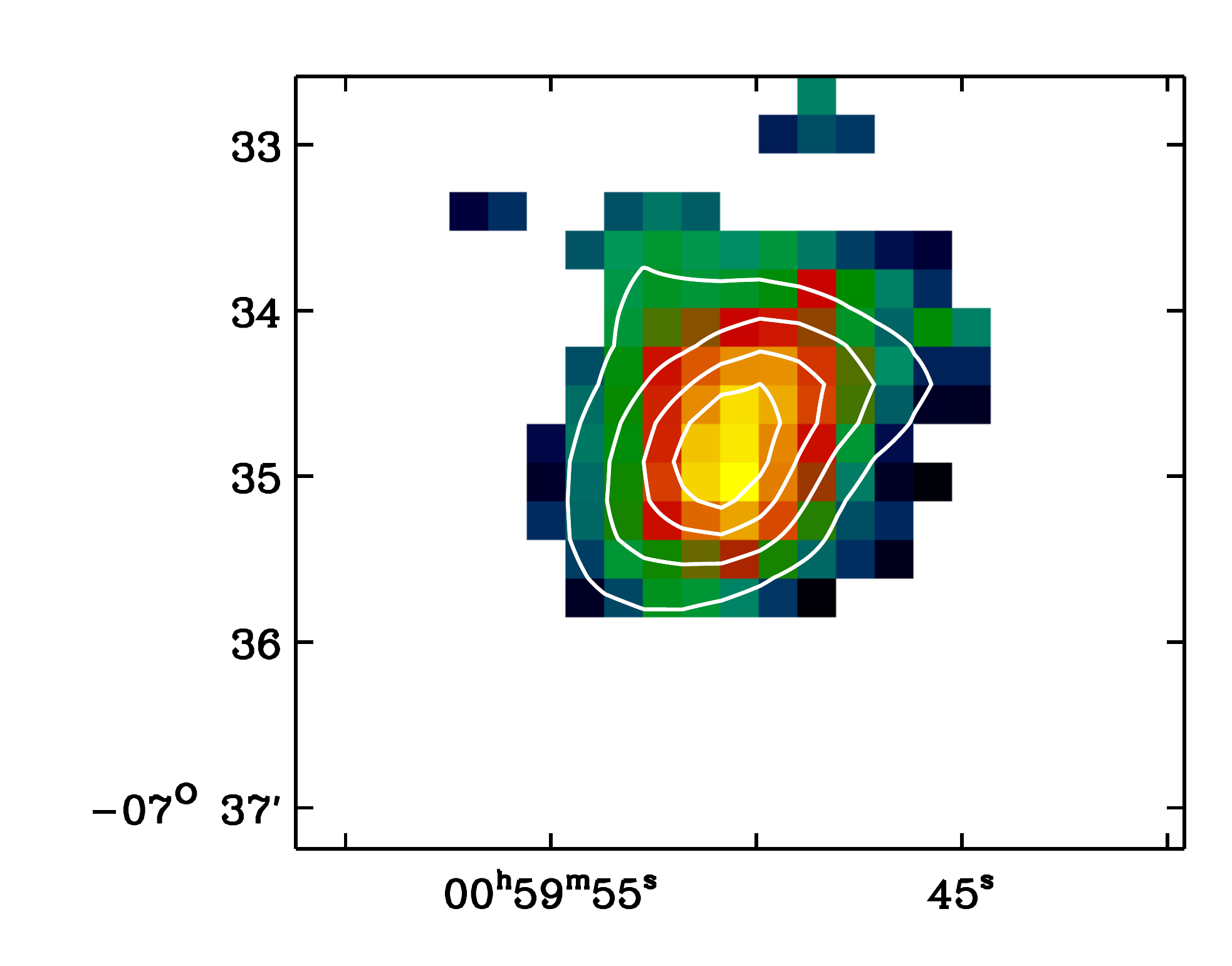} &
\includegraphics[width=5.7cm]{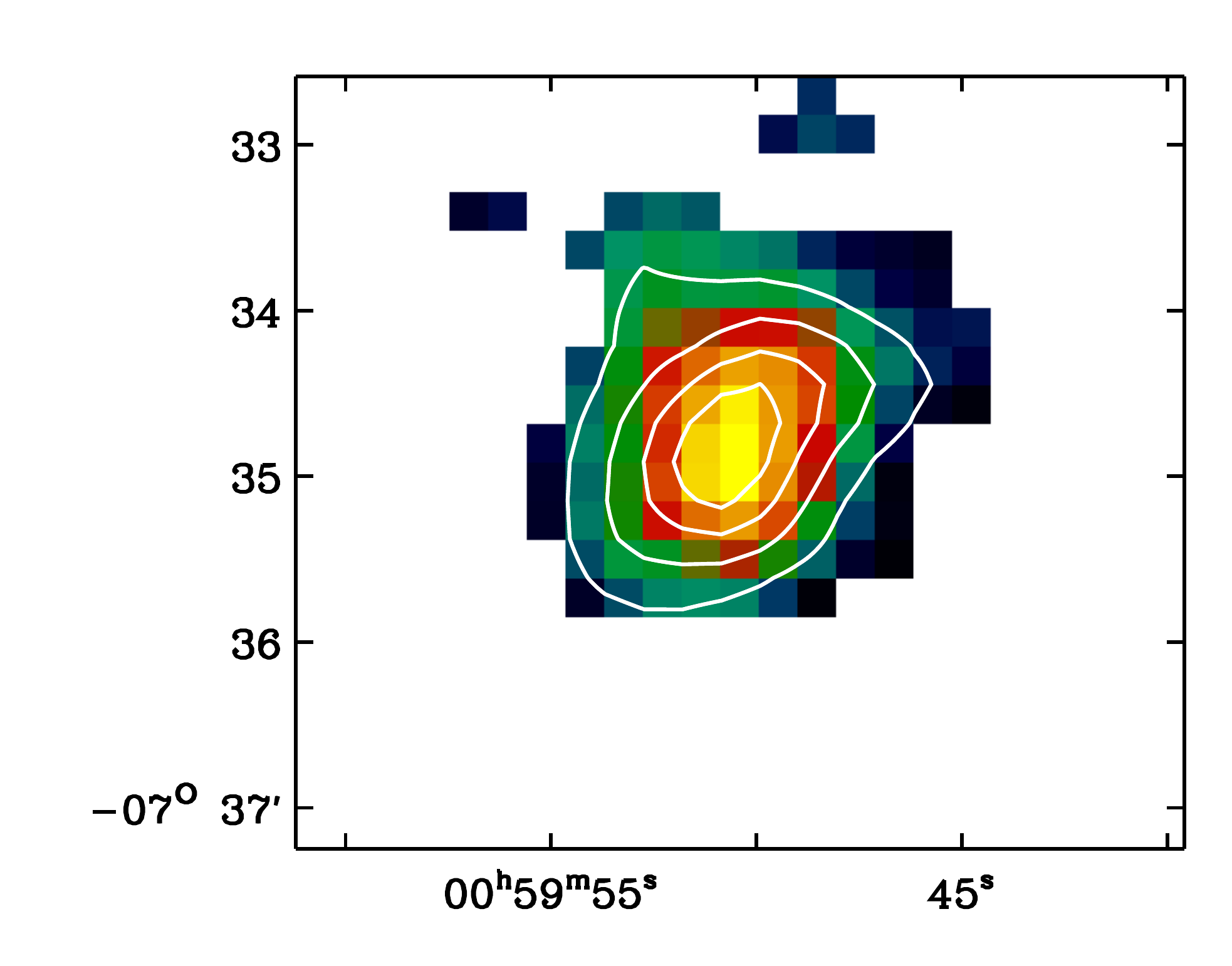}  &
\rotatebox{90}{\includegraphics[width=4cm, height=0.9cm]{NGC0337_Excess_ColorBars}}  \\
	 
\rotatebox{90}{\Large Relative Difference} &
\includegraphics[width=5.7cm]{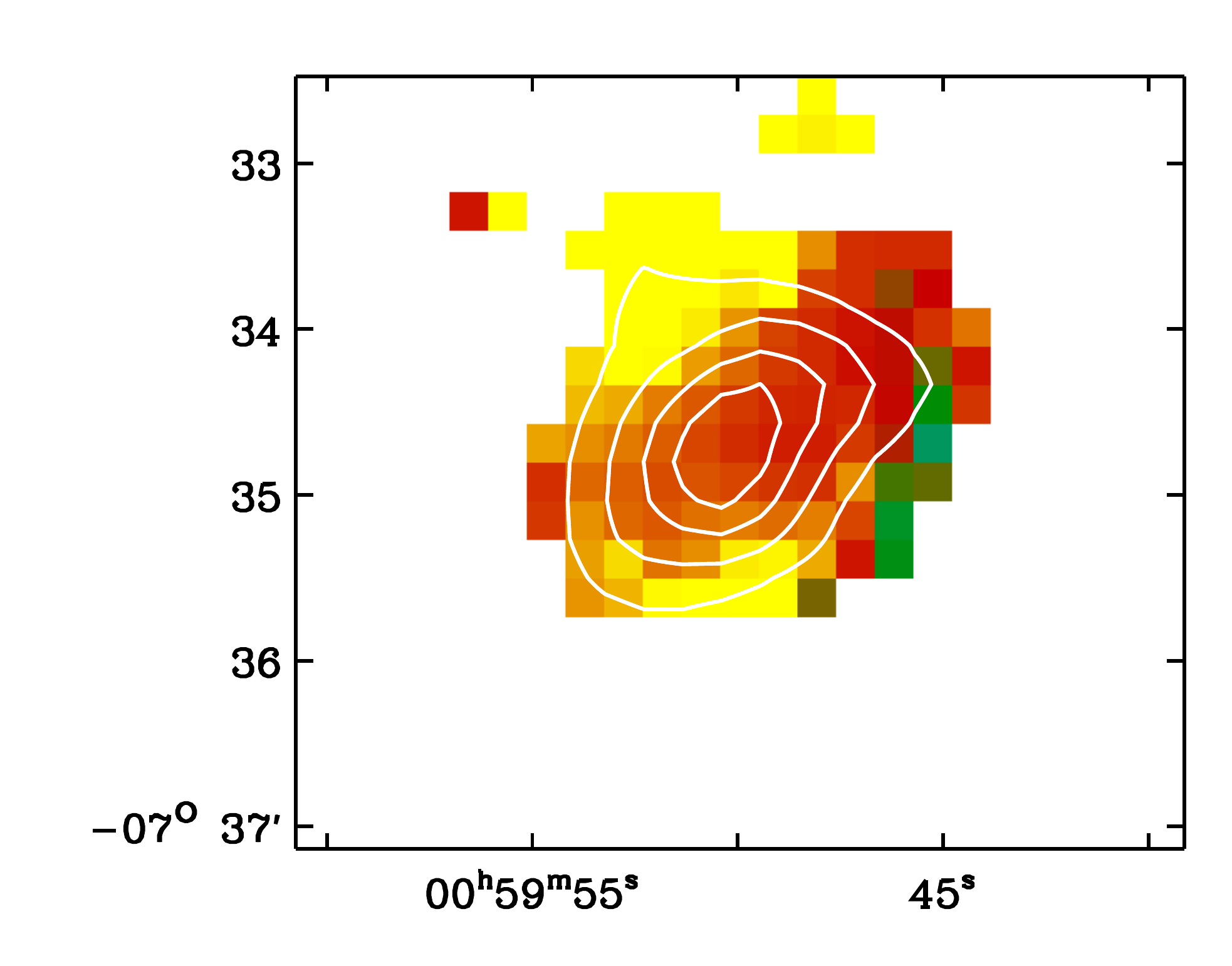} &
\includegraphics[width=5.7cm]{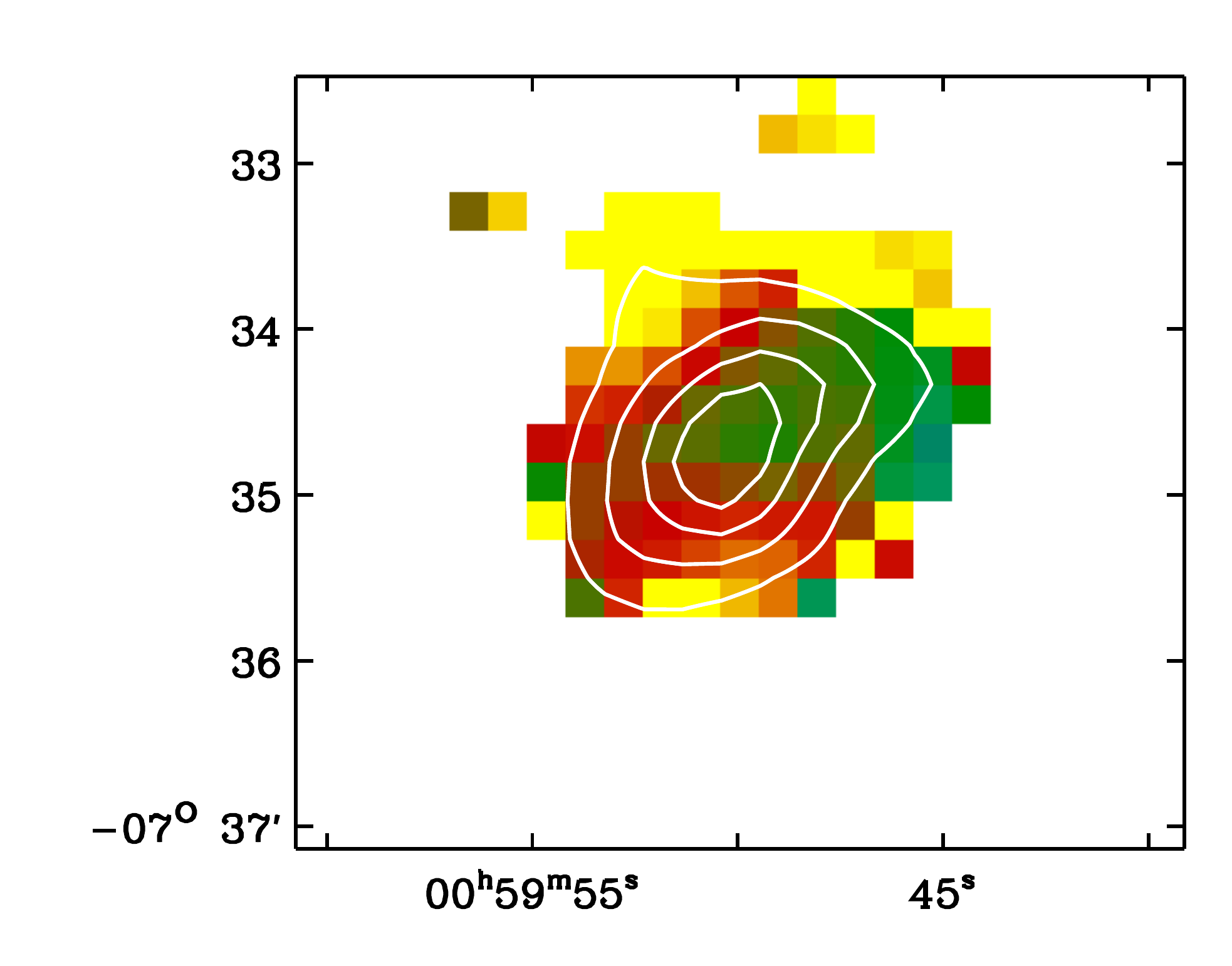} &
\includegraphics[width=5.7cm]{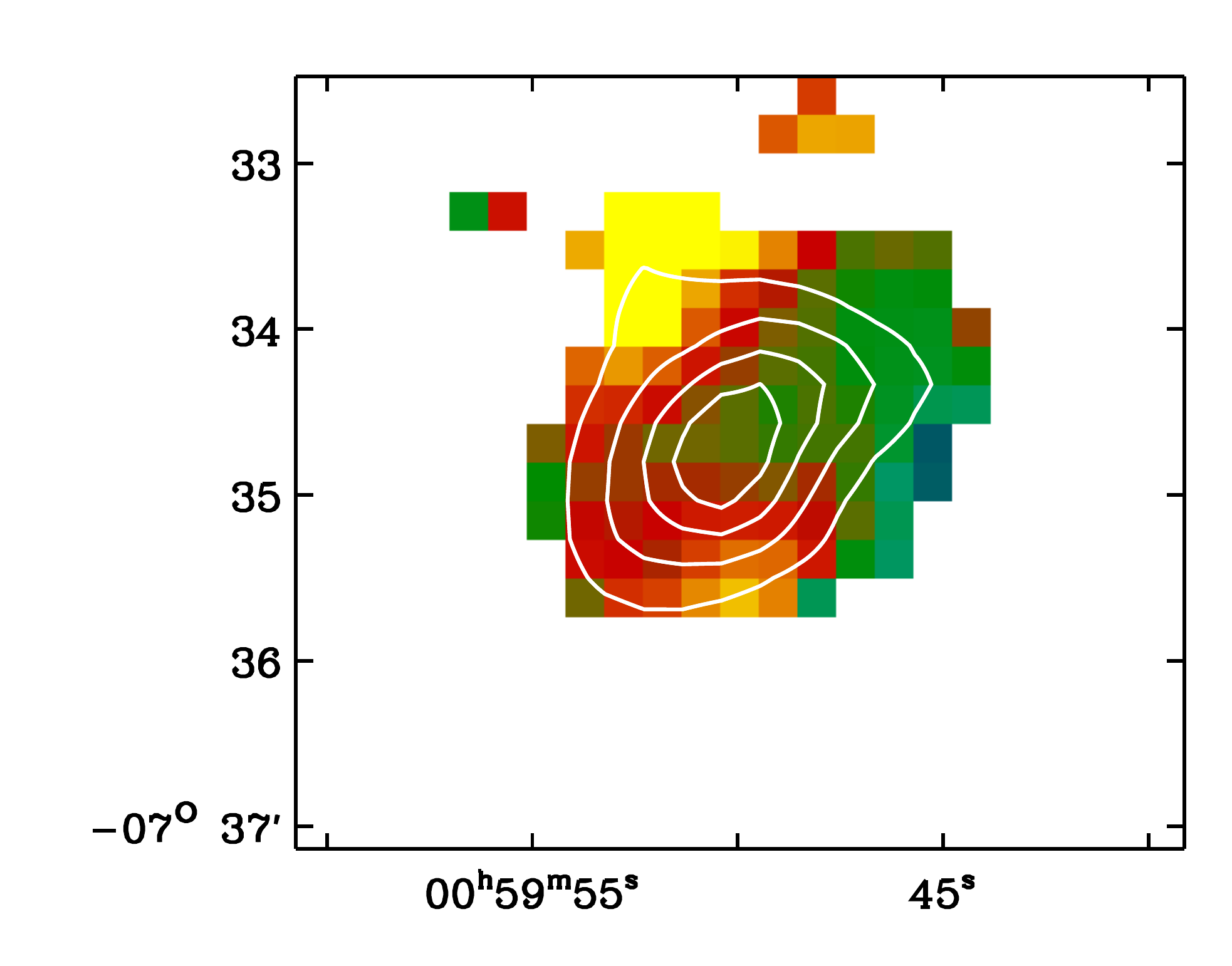}  &
\rotatebox{90}{\includegraphics[width=4cm, height=0.9cm]{RelativeExcess_ColorBars}}  \\	   
\end{tabular}  
\caption{For each galaxy: {\it Top panel:} Flux density at 870 \mic\ map observed with \lab\ (in MJy~sr$^{-1}$) not corrected for CO and radio continuum. {\it Second line panels:} modelled 870 \mic\ maps (in MJy~sr$^{-1}$) derived, from left to right, using our two-modified blackbody procedure with $\beta$$_c$ fixed to 2, fixed to 1.5 or using the [DL07] formalism. {\it Third line panels:} Absolute difference between the observed and the modelled 870 \mic\ map in MJy~sr$^{-1}$ (defined as observed flux - modelled flux) for the different SED models. {\it Bottom line panels:} Relative excess defined as (observed flux - modelled flux) / (modelled flux). We provide the color scales for each line on the right-hand side. Zeros are indicated with a white marker on the color bars. The signal-to-noise of the LABOCA 870 \mic\ map is overlaid as contours on the absolute and relative difference maps (same levels than in Fig.~\ref{350um_Maps}).}
    \label{Excess_maps}
\end{figure*}

\newpage
\addtocounter {figure}{-1}
\begin{figure*}
\centering
\begin{tabular}  { m{0cm} m{5.1cm} m{5.1cm} m{5.1cm}  m{0.7cm}}    
{\Large \bf~~~~~~~~~~NGC0628} &&&\\  
&\hspace{5cm}\rotatebox{90}{\Large 870 \mic\ Observed} & 
\includegraphics[width=5.7cm]{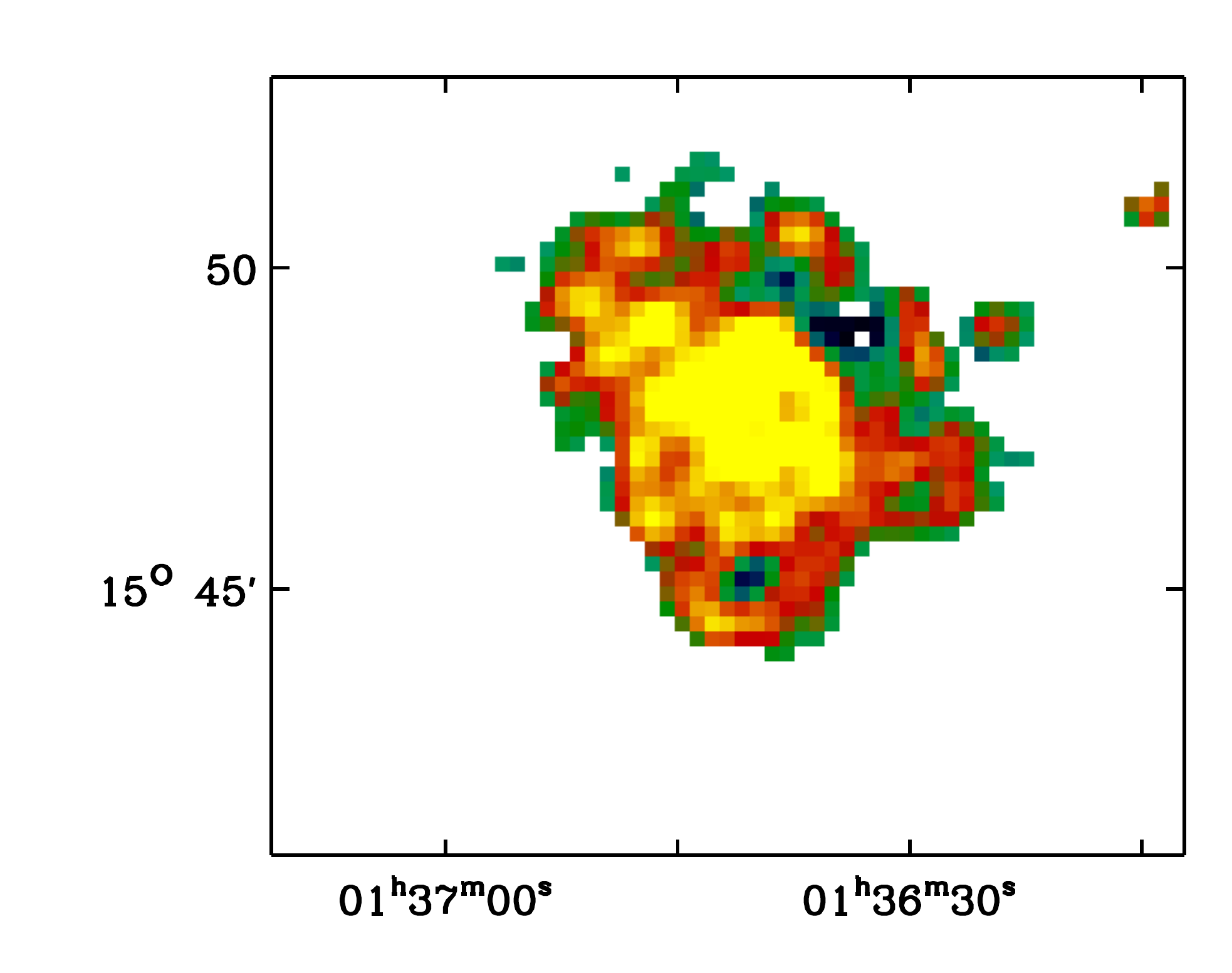} &&
\rotatebox{90}{\includegraphics[width=4cm, height=0.9cm]{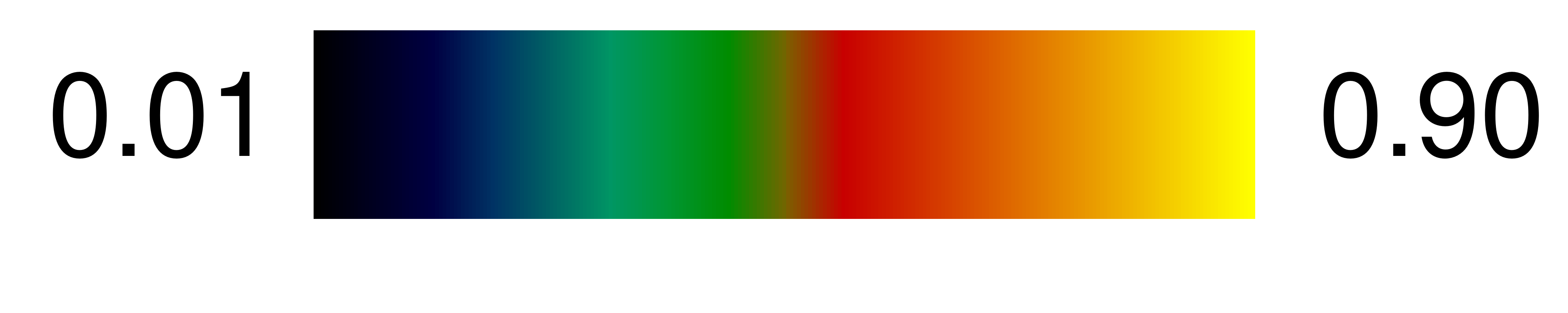}}  \\
&&\\
& {\Large \hspace{2.2cm}$\beta$$_c$ = 2.0 model} & {\Large \hspace{2.2cm}$\beta$$_c$ = 1.5 model}  & {\Large \hspace{2.2cm}[DL07] model} & \\

\rotatebox{90}{\Large 870 \mic\ Modelled} & 
\includegraphics[width=5.7cm]{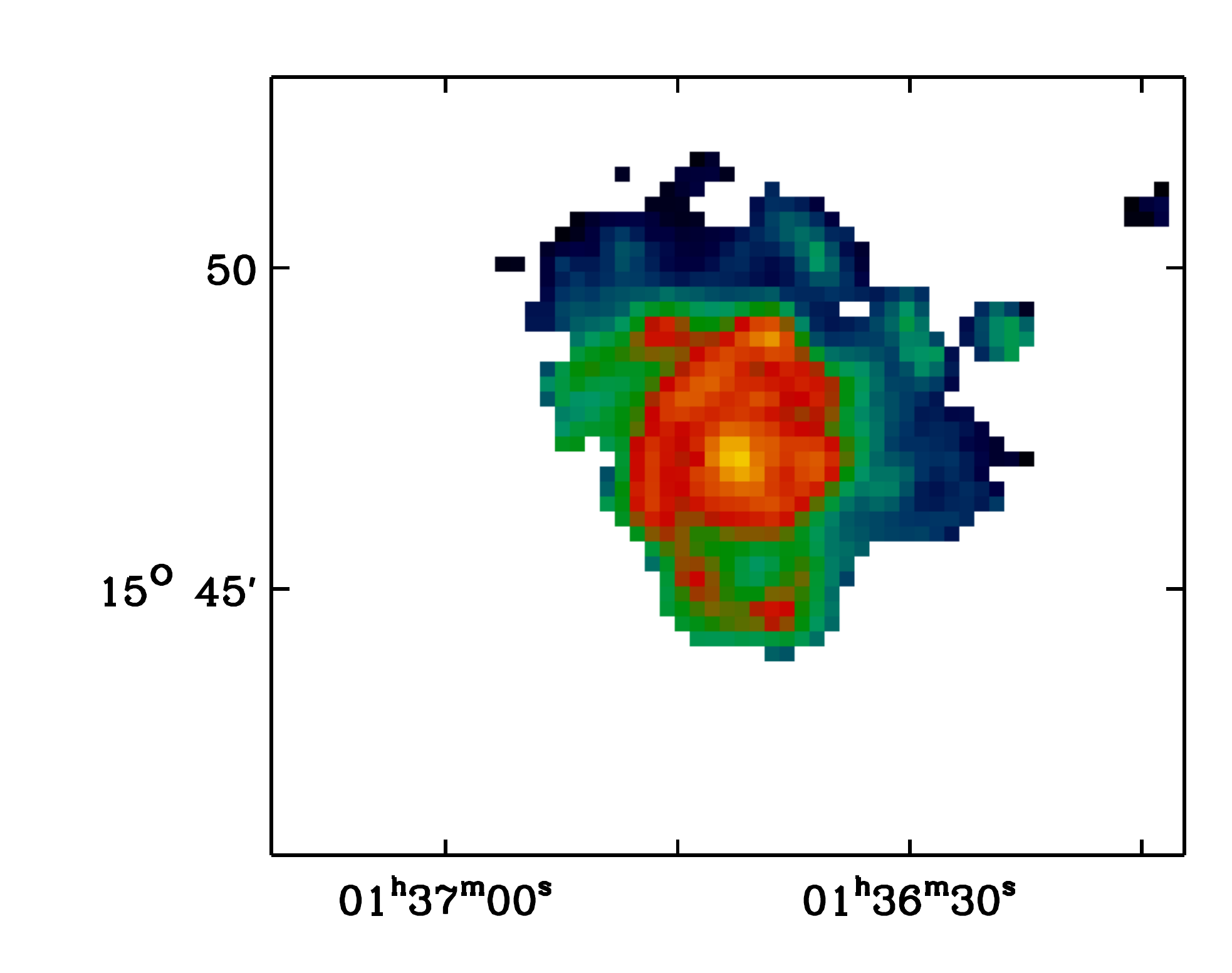} &
\includegraphics[width=5.7cm]{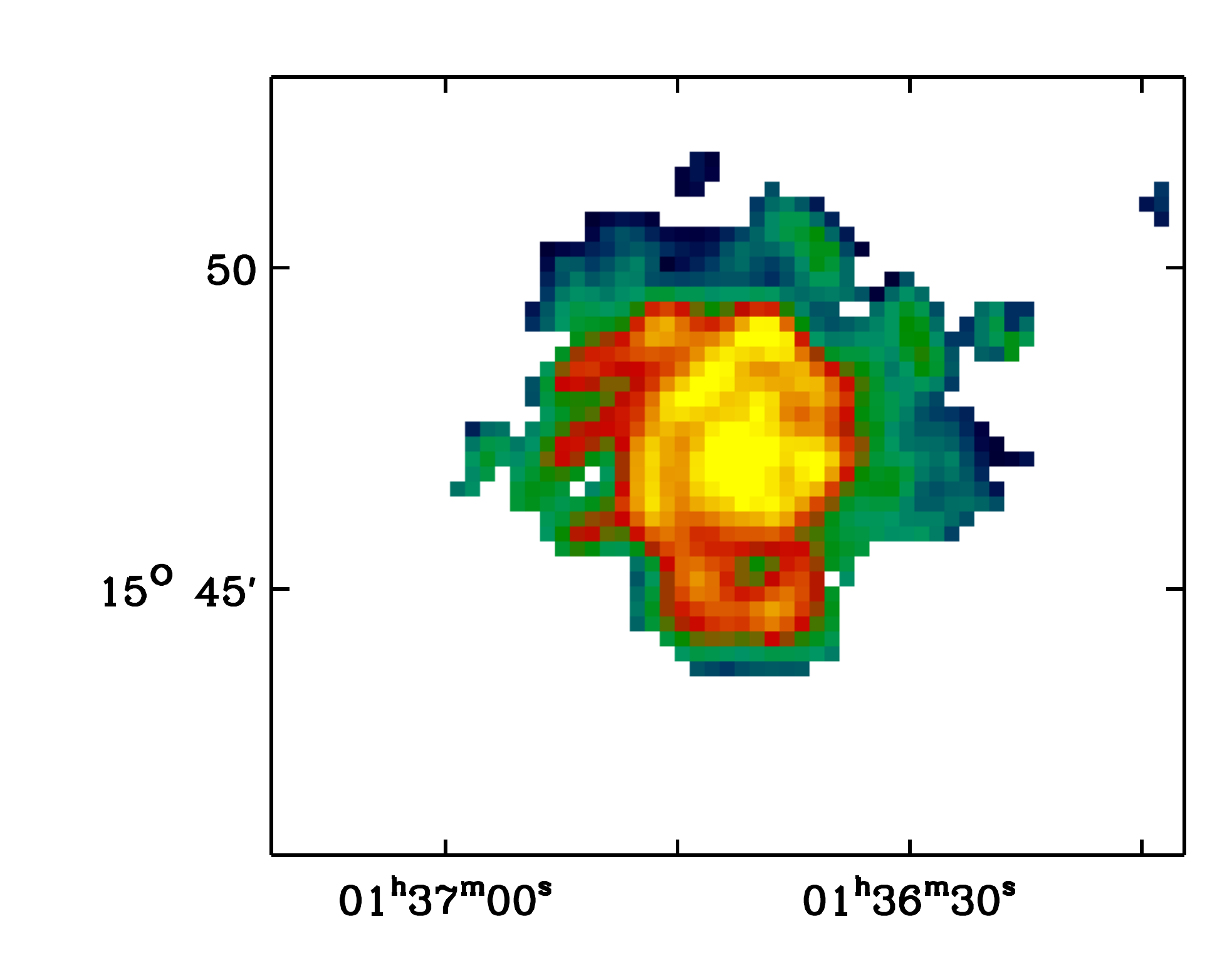} &
\includegraphics[width=5.7cm]{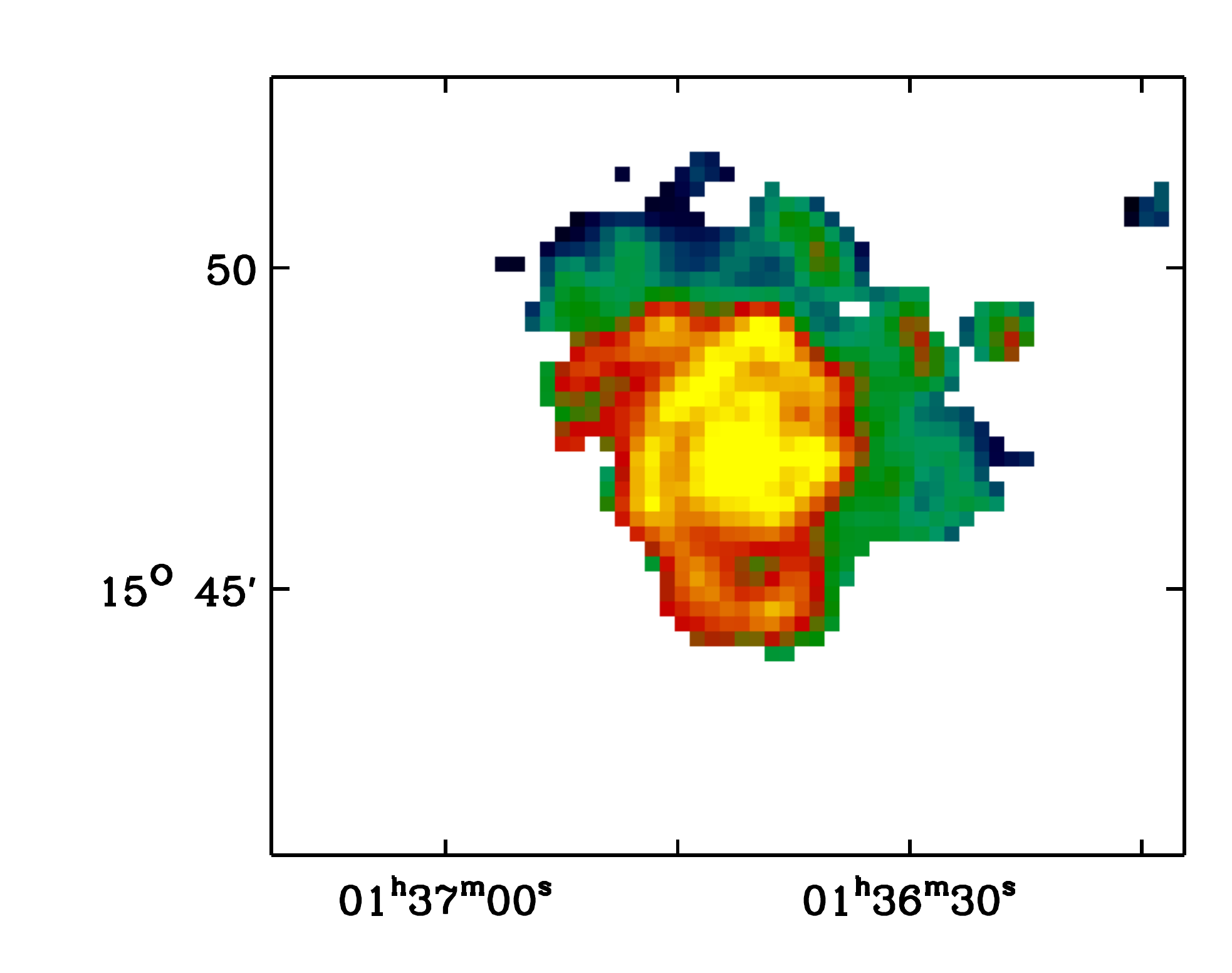}  &
\rotatebox{90}{\includegraphics[width=4cm, height=0.9cm]{NGC0628_Extrap870_ColorBars}}  \\
	
\rotatebox{90}{\Large Absolute Difference} &
\includegraphics[width=5.7cm]{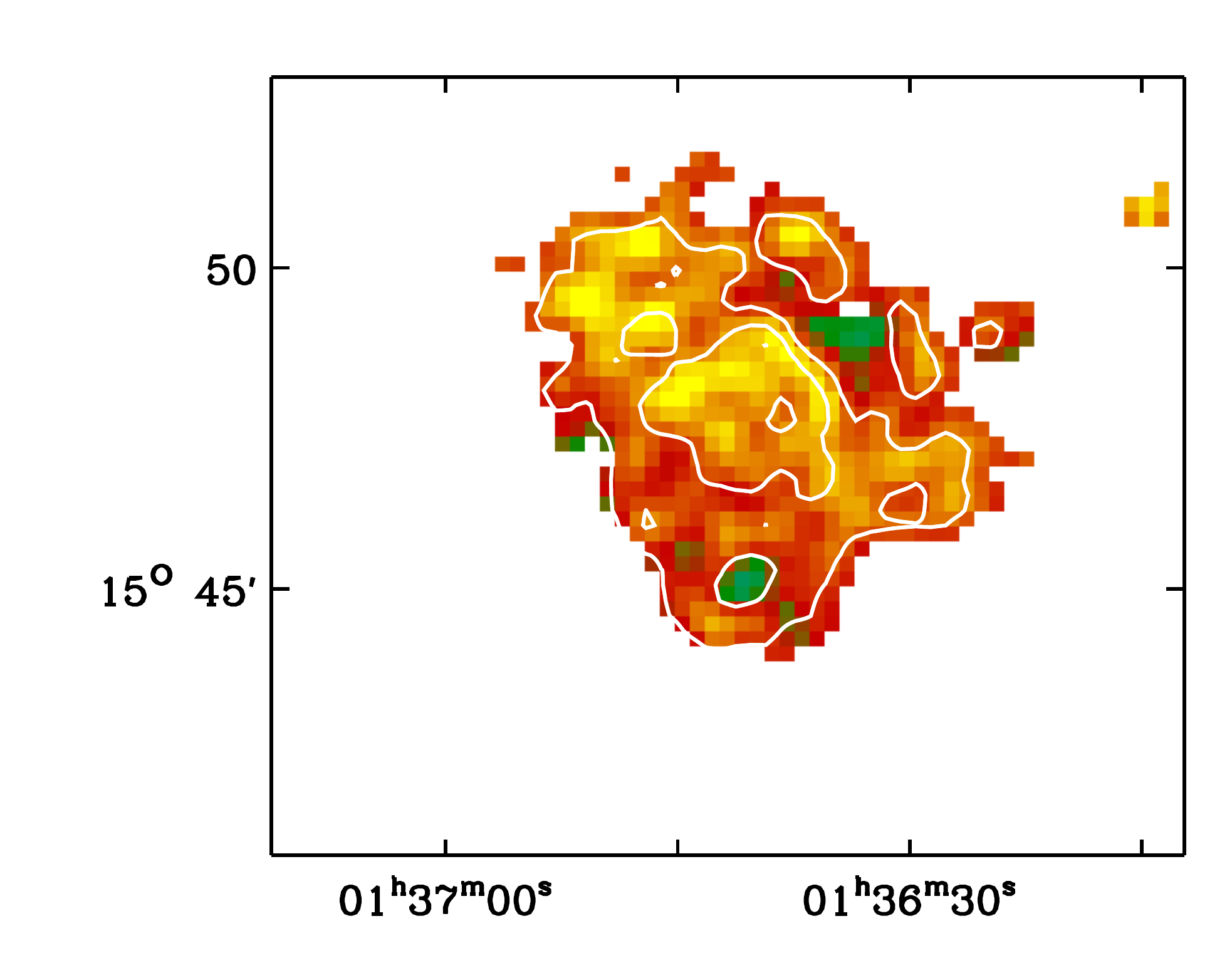} & 
\includegraphics[width=5.7cm]{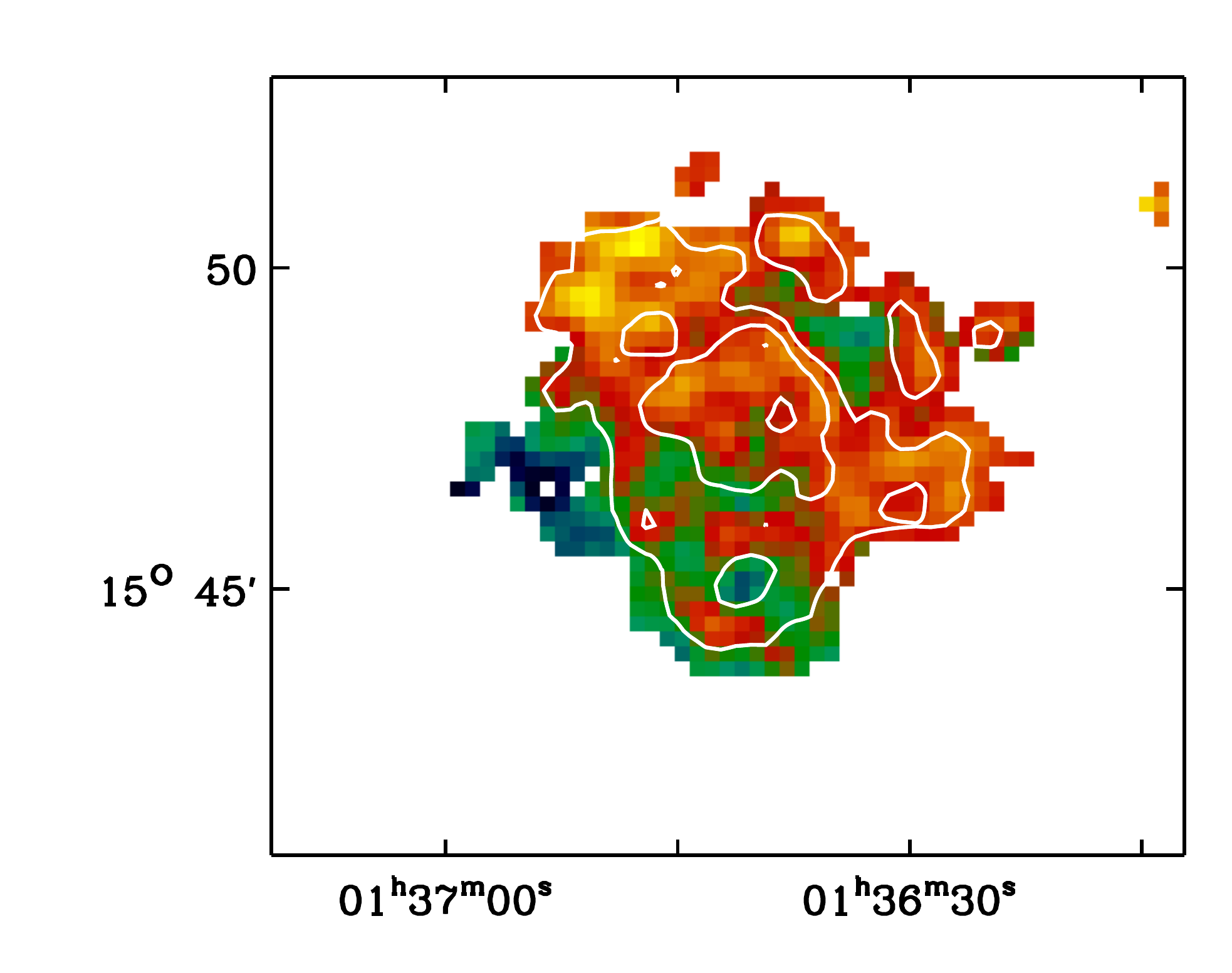} &
\includegraphics[width=5.7cm]{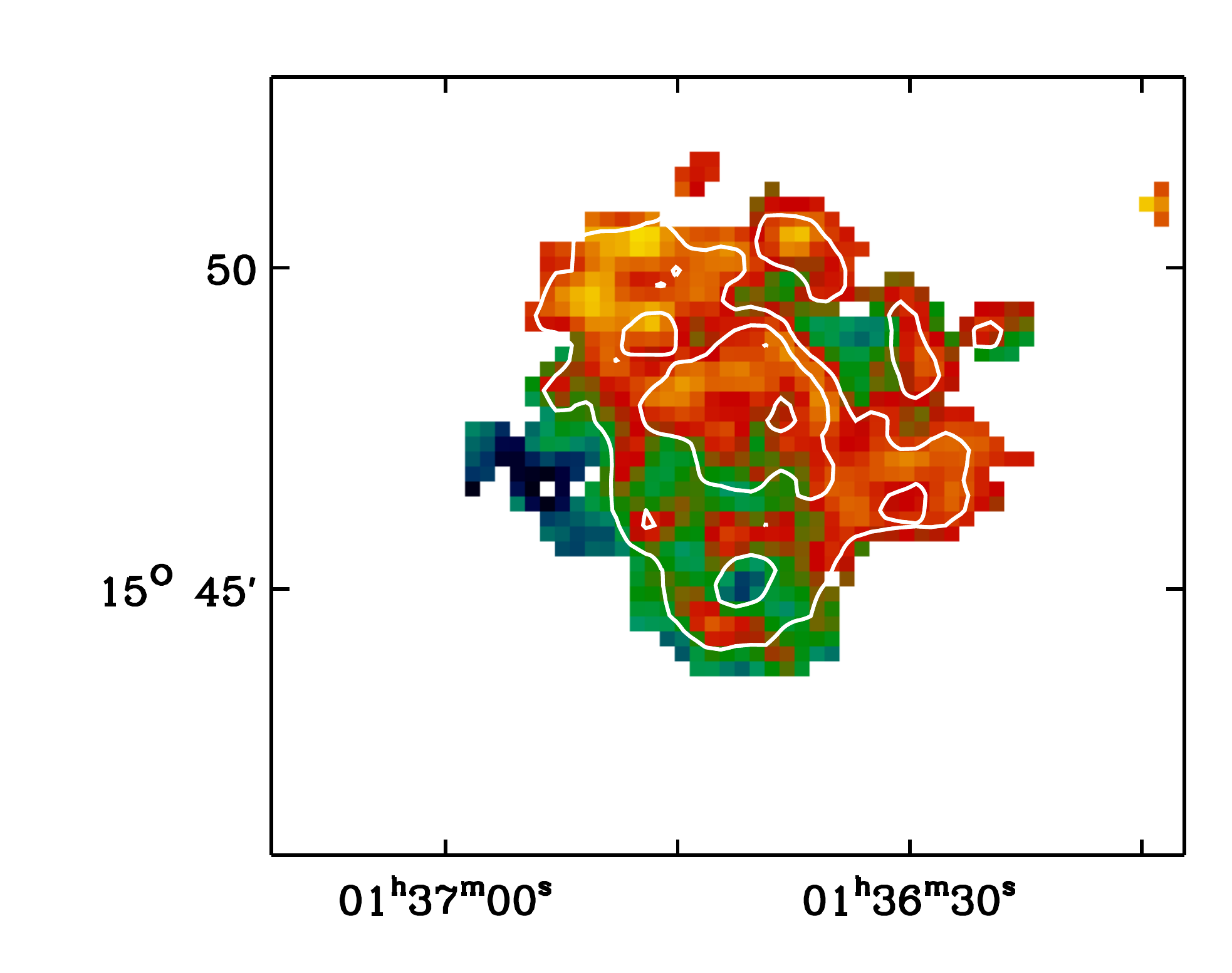}  &
\rotatebox{90}{\includegraphics[width=4cm, height=0.9cm]{NGC0628_Excess_ColorBars}}  \\
	 
\rotatebox{90}{\Large Relative Difference} &
\includegraphics[width=5.7cm]{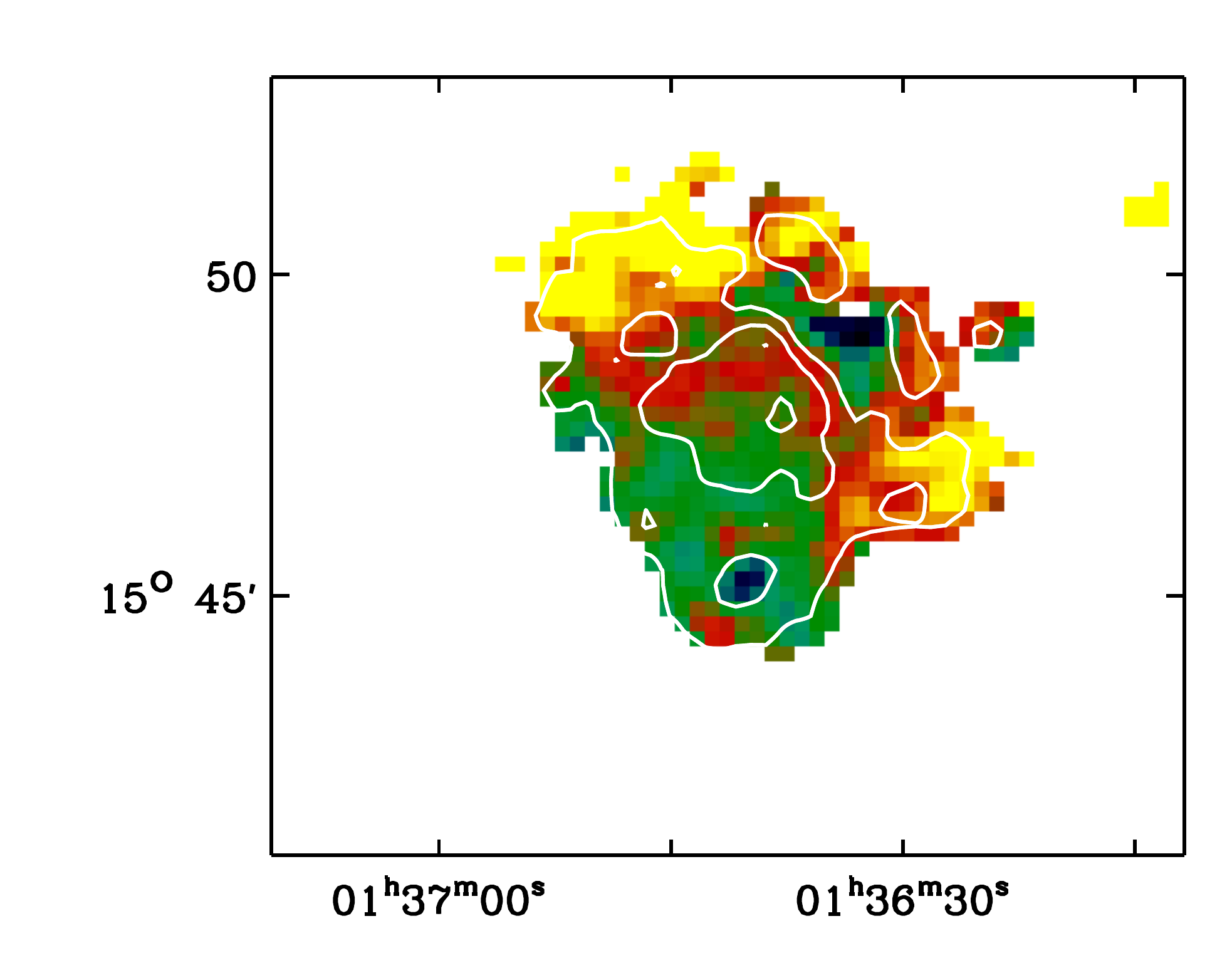} &
\includegraphics[width=5.7cm]{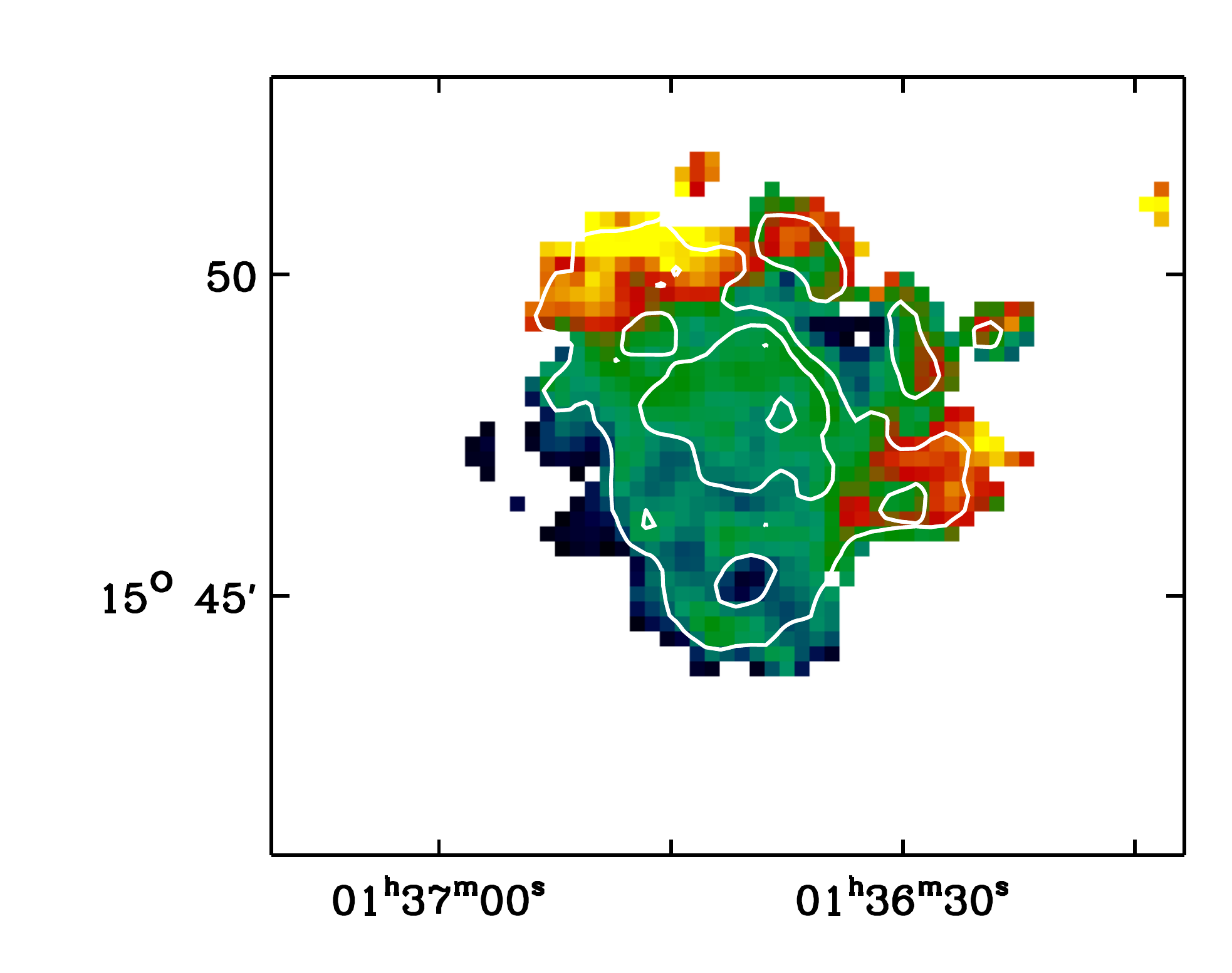} &
\includegraphics[width=5.7cm]{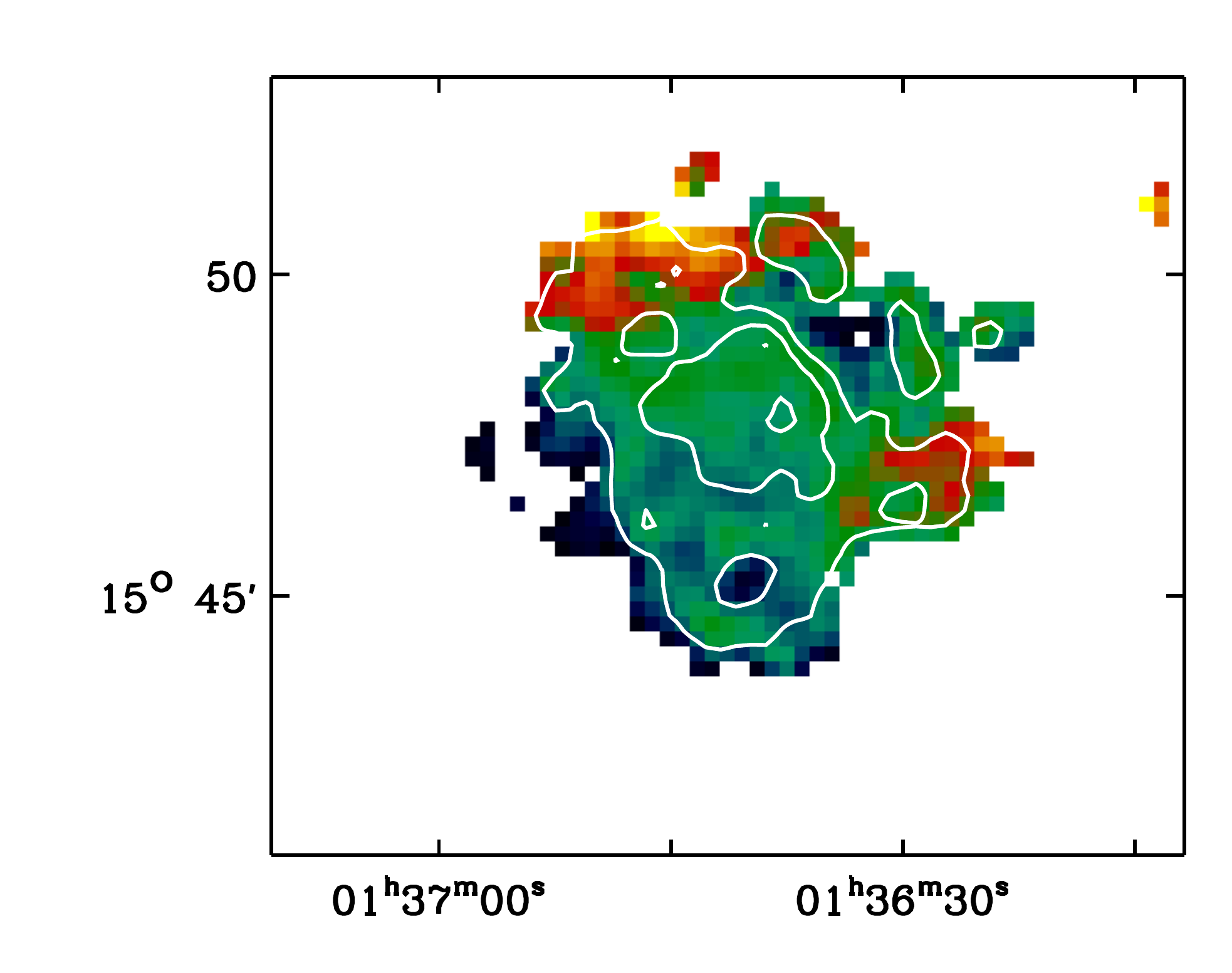}  &
\rotatebox{90}{\includegraphics[width=4cm, height=0.9cm]{RelativeExcess_ColorBars}}  \\	   
\end{tabular}  
    \caption{continued. }
\end{figure*}

\newpage
\addtocounter {figure}{-1}
\begin{figure*}
\centering
\begin{tabular}  { m{0cm} m{5.1cm} m{5.1cm} m{5.1cm}  m{0.7cm}}    
{\Large \bf~~~~~~~~~~NGC1097} &&&\\  
&\hspace{5cm}\rotatebox{90}{\Large 870 \mic\ Observed} & 
\includegraphics[width=5.7cm]{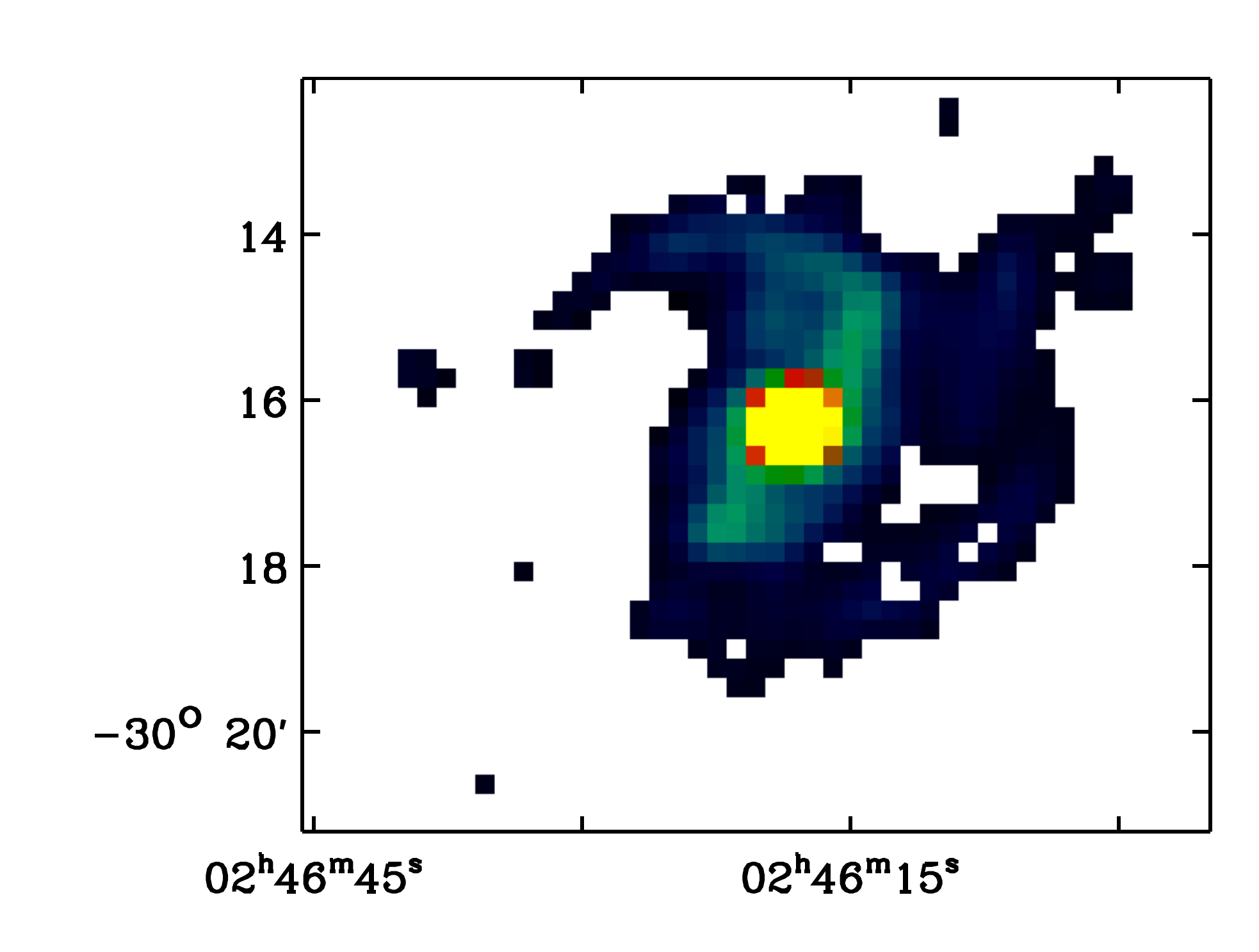} &&
\rotatebox{90}{\includegraphics[width=4cm, height=0.9cm]{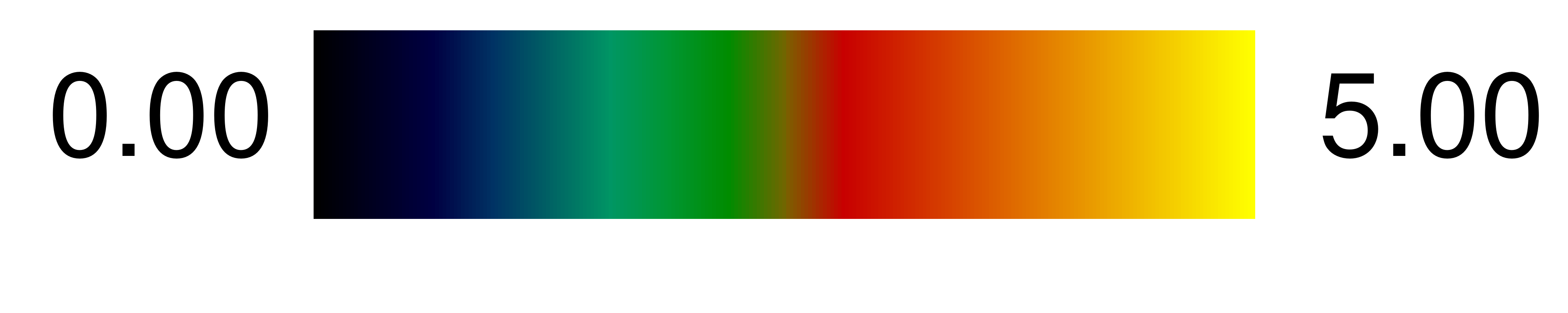}}  \\
&&\\
& {\Large \hspace{2.2cm}$\beta$$_c$ = 2.0 model} & {\Large \hspace{2.2cm}$\beta$$_c$ = 1.5 model}  & {\Large \hspace{2.2cm}[DL07] model} & \\

\rotatebox{90}{\Large 870 \mic\ Modelled} & 
\includegraphics[width=5.7cm]{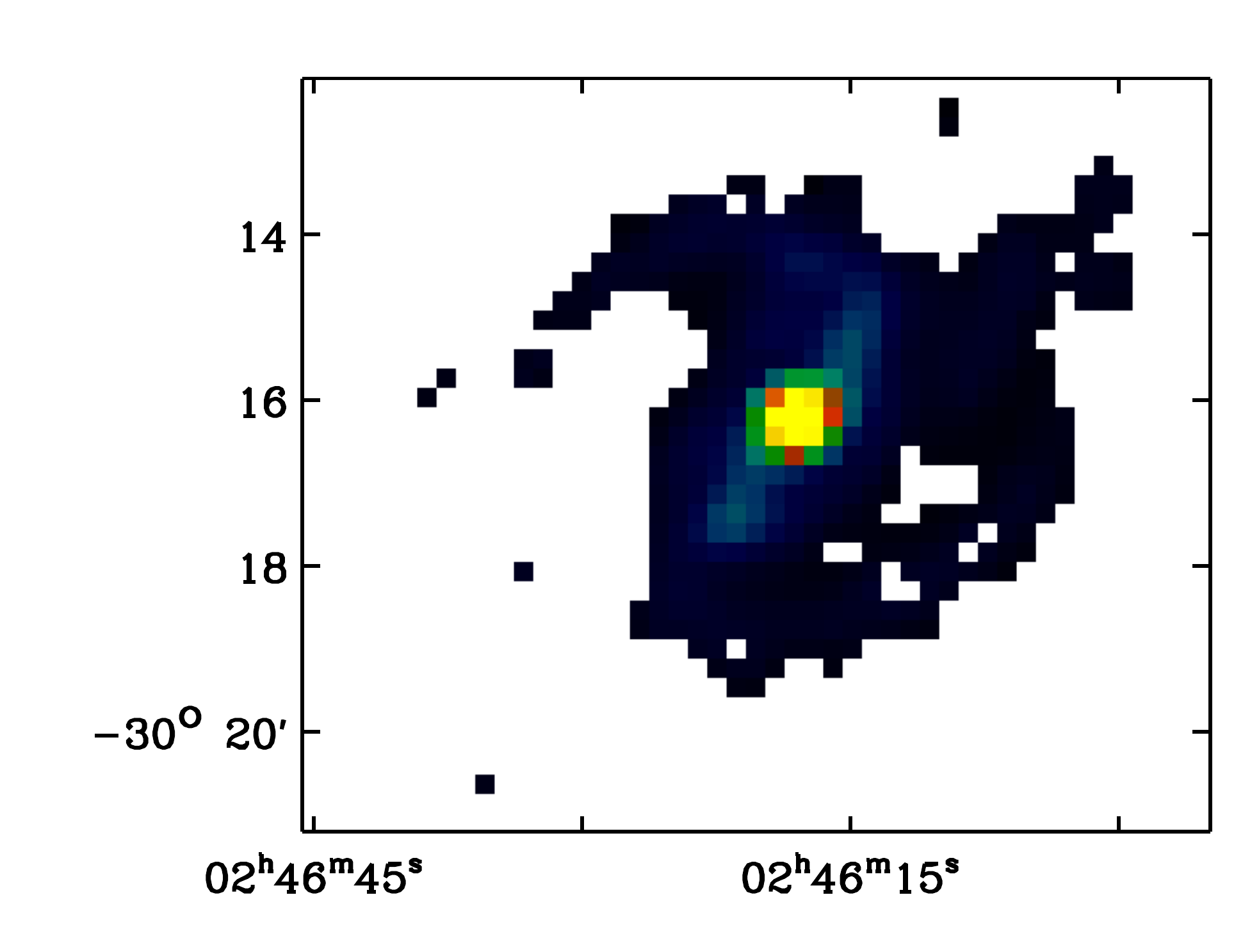} &
\includegraphics[width=5.7cm]{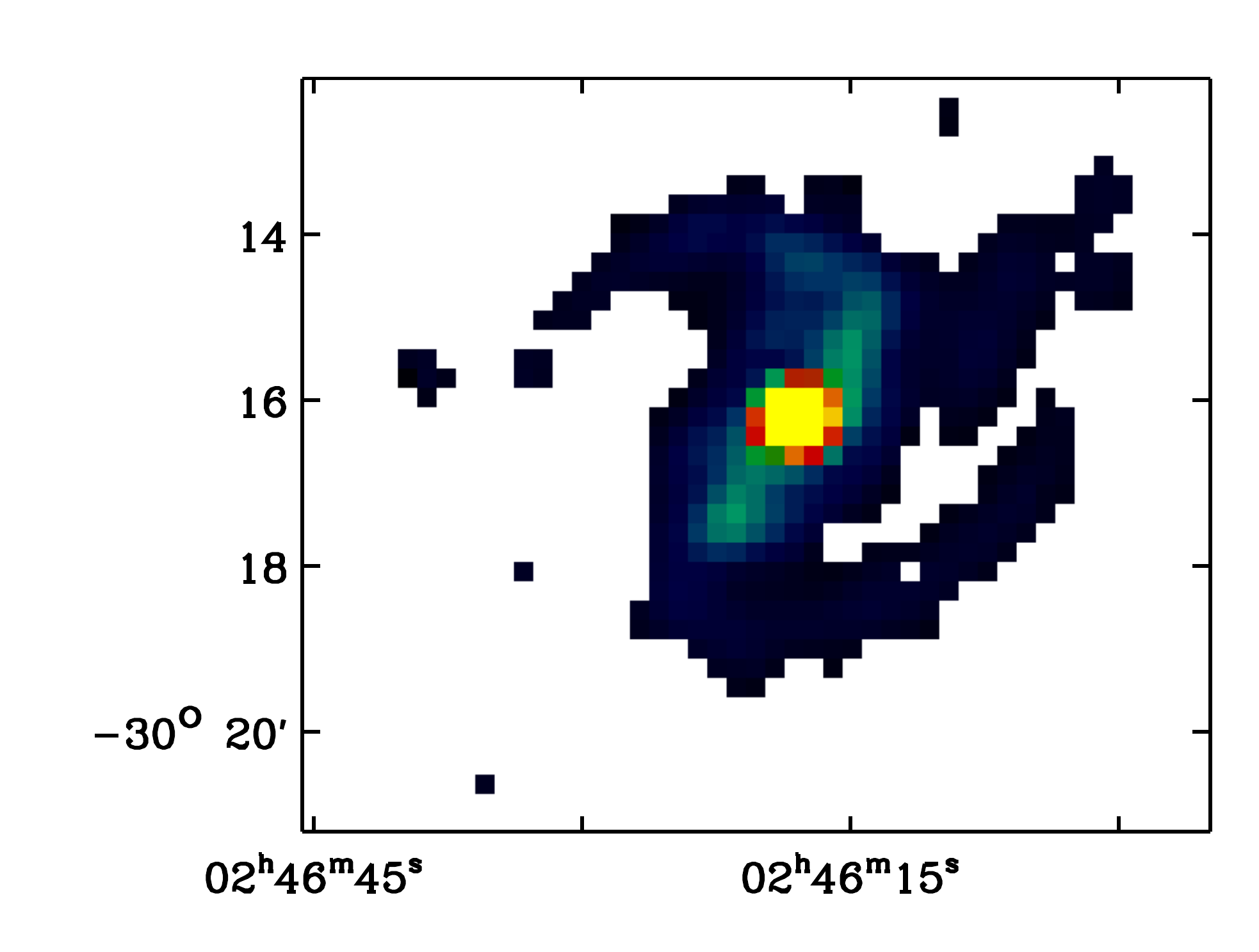} &
\includegraphics[width=5.7cm]{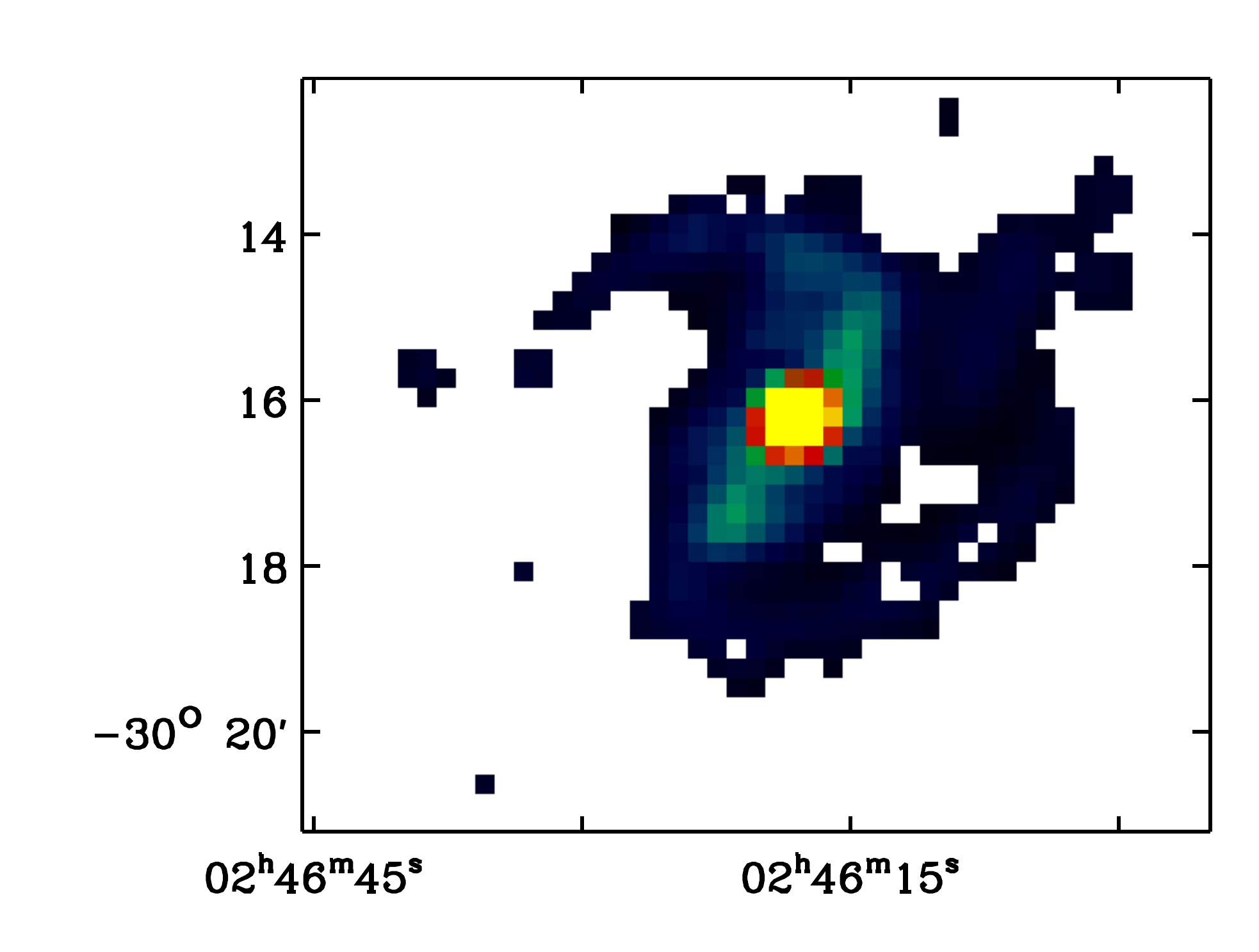}  &
\rotatebox{90}{\includegraphics[width=4cm, height=0.9cm]{NGC1097_Extrap870_ColorBars}}  \\
	
\rotatebox{90}{\Large Absolute Difference} &
\includegraphics[width=5.7cm]{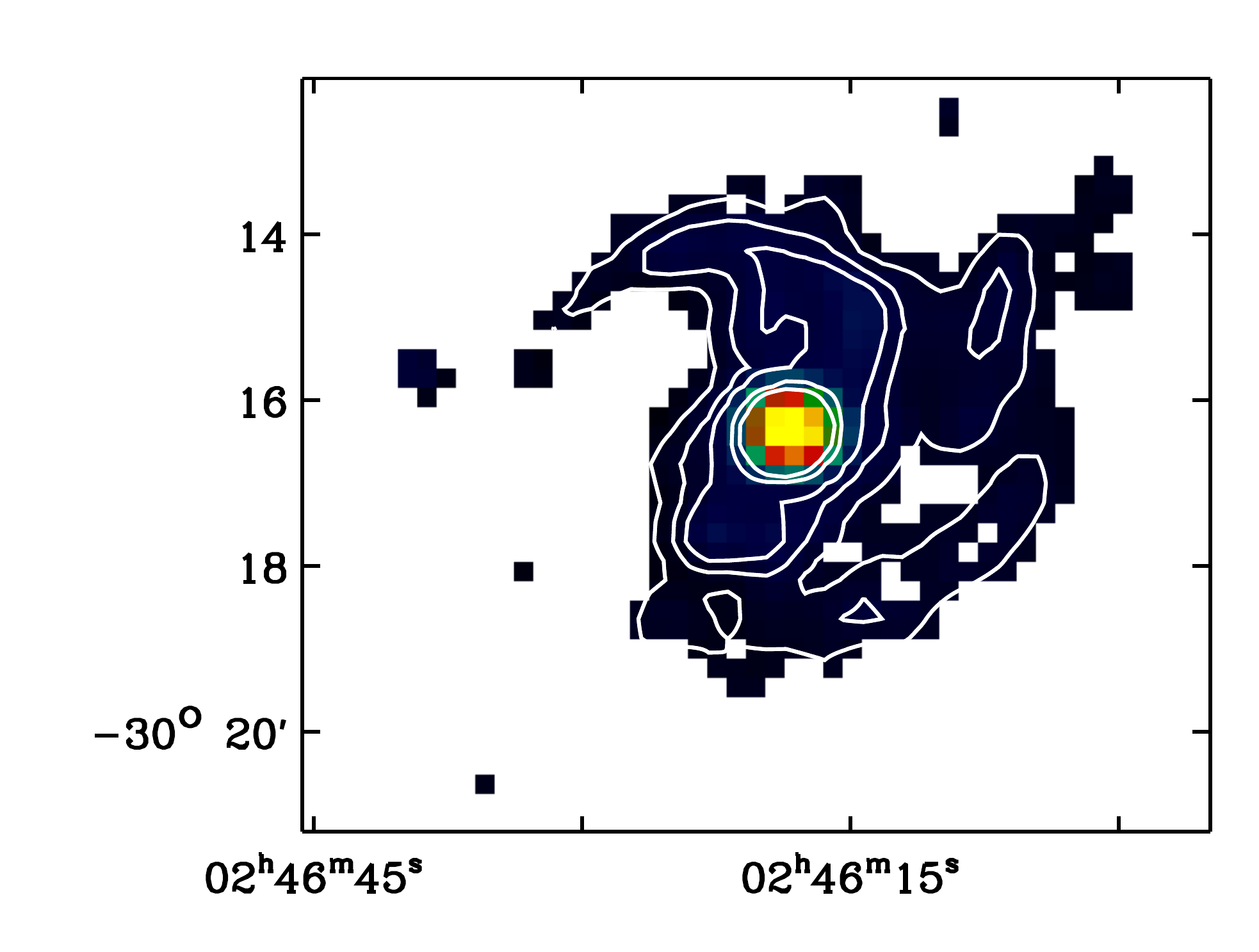} & 
\includegraphics[width=5.7cm]{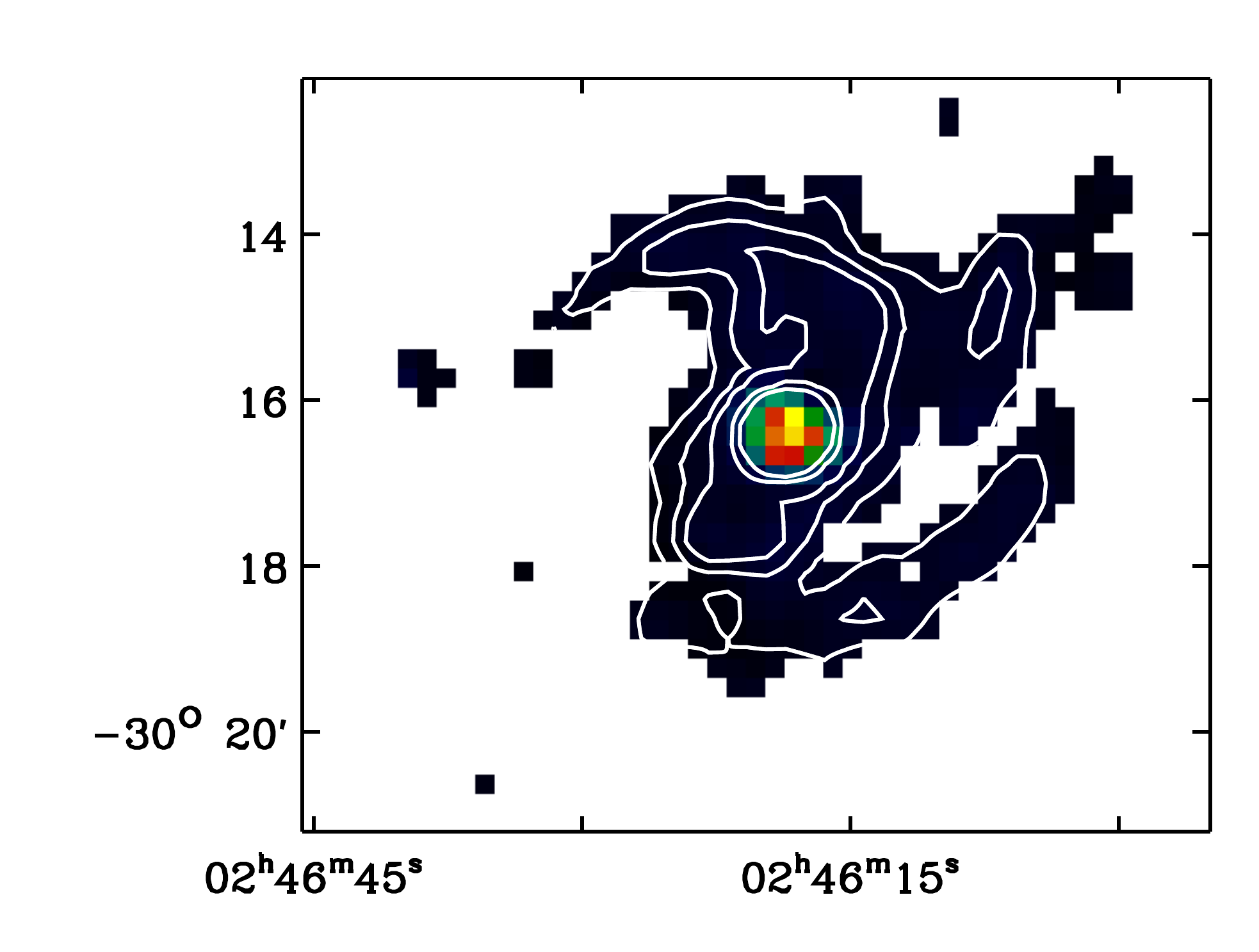} &
\includegraphics[width=5.7cm]{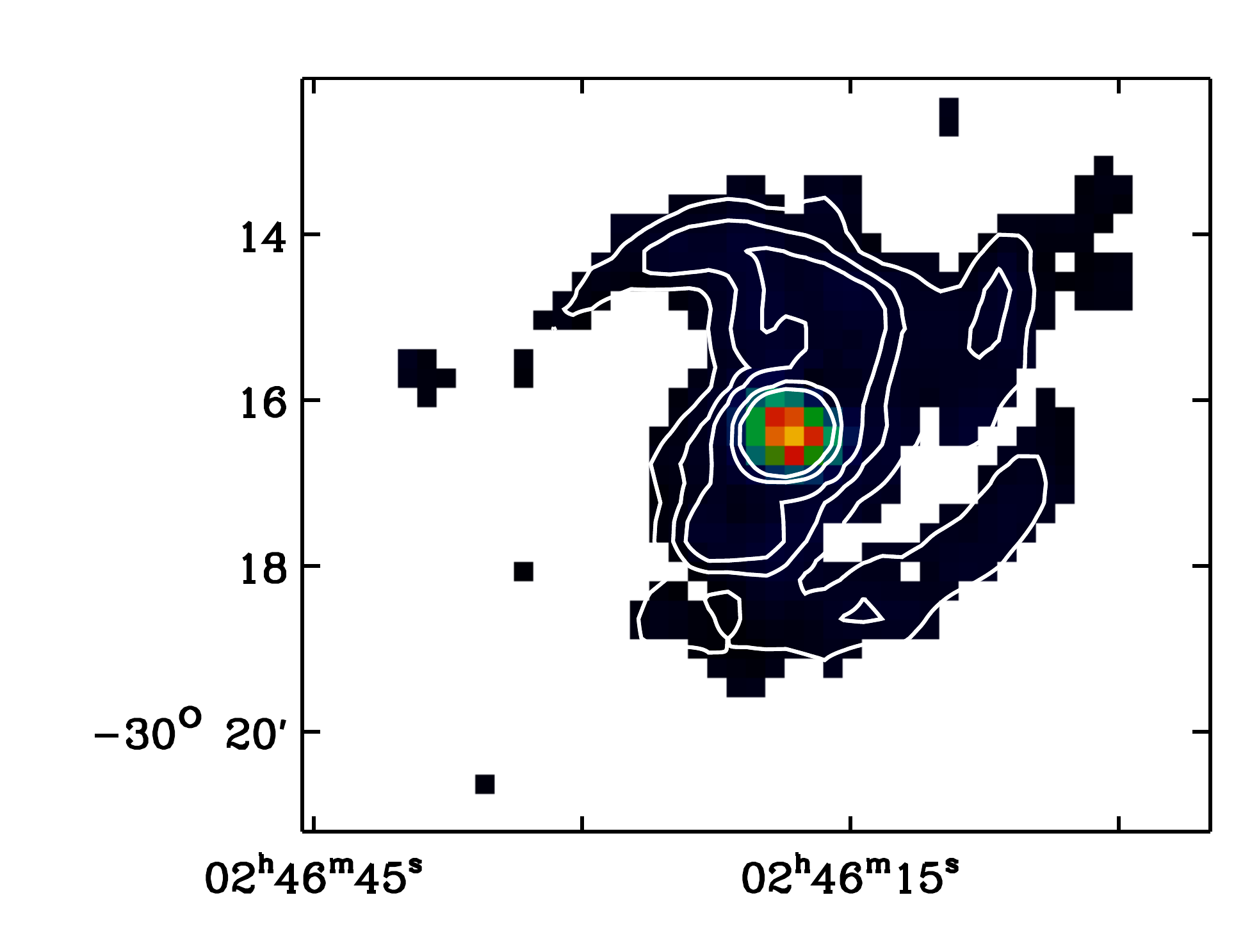}  &
\rotatebox{90}{\includegraphics[width=4cm, height=0.9cm]{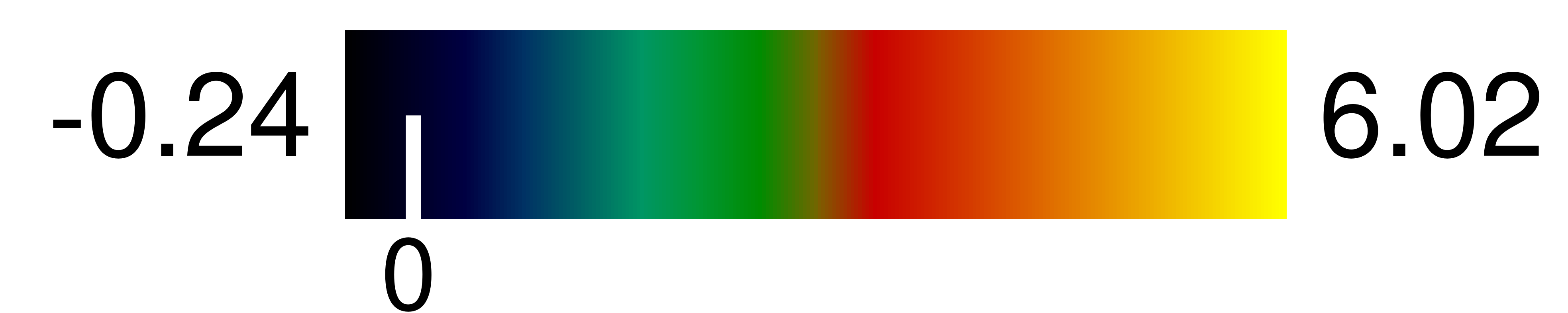}}  \\
	 
\rotatebox{90}{\Large Relative Difference} &
\includegraphics[width=5.7cm]{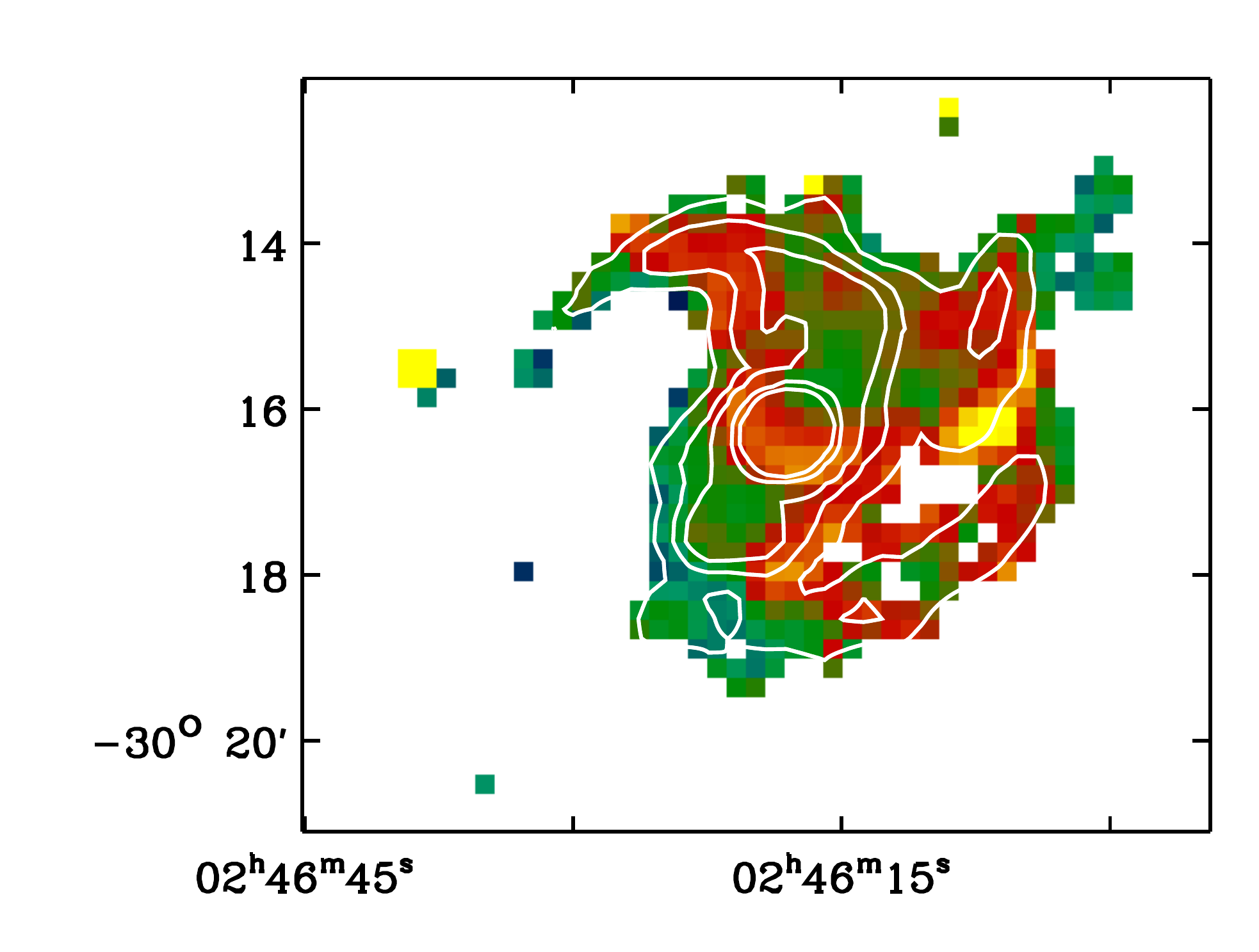} &
\includegraphics[width=5.7cm]{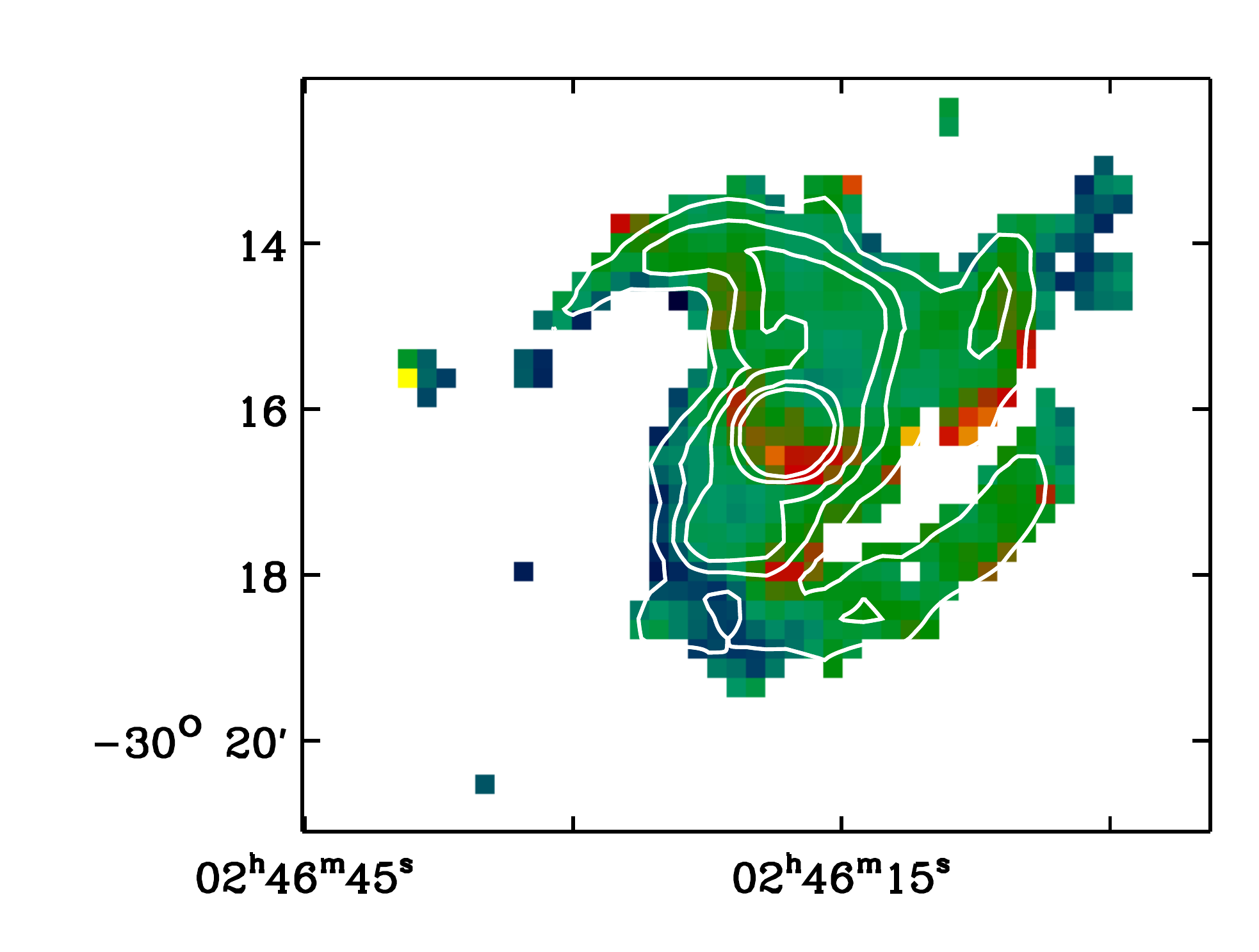} &
\includegraphics[width=5.7cm]{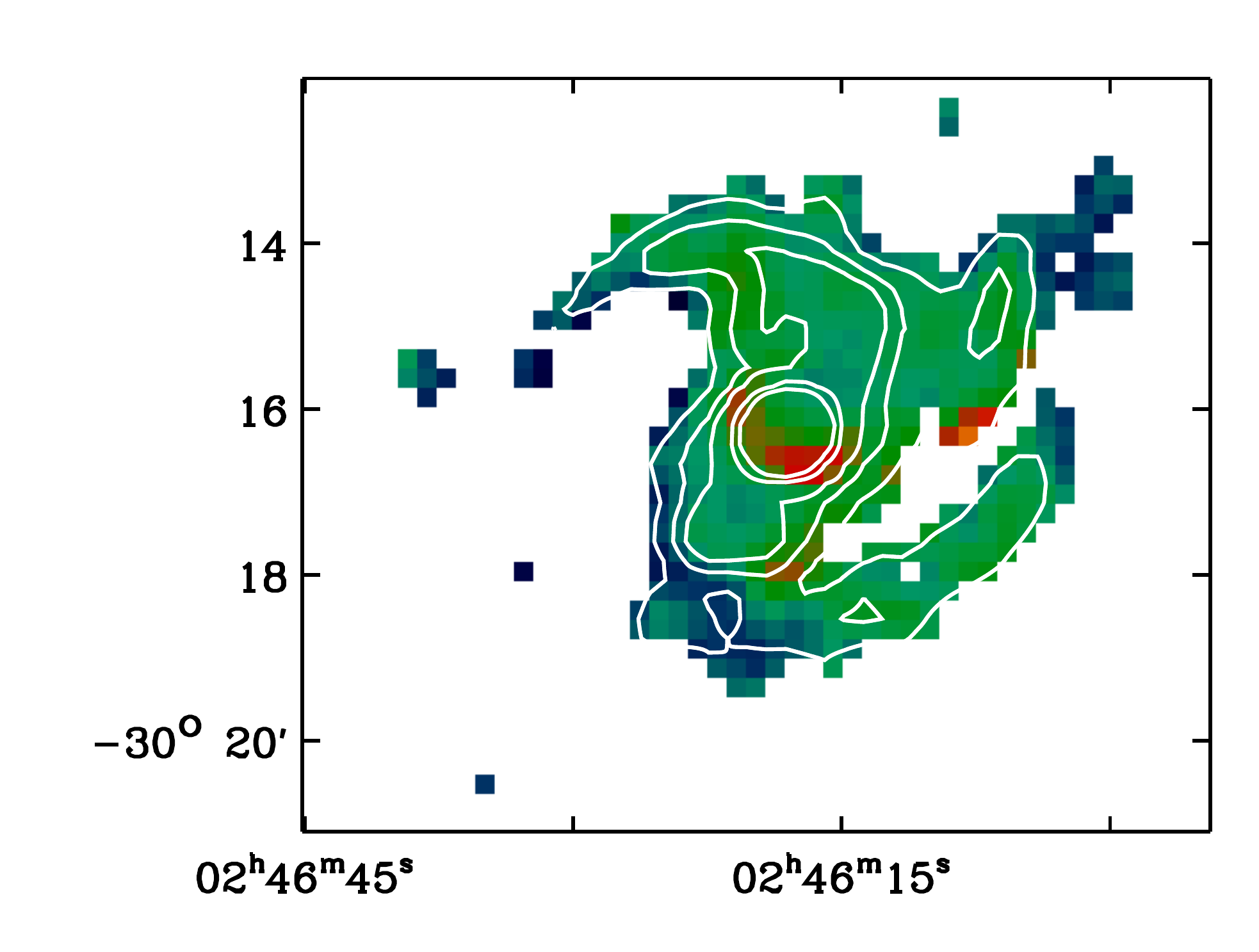}  &
\rotatebox{90}{\includegraphics[width=4cm, height=0.9cm]{RelativeExcess_ColorBars}}  \\	   
\end{tabular}  
    \caption{continued. }
    \end{figure*}

\newpage
\addtocounter {figure}{-1}
\begin{figure*}
\centering
\begin{tabular}  { m{0cm} m{5.1cm} m{5.1cm} m{5.1cm}  m{0.7cm}}    
{\Large \bf~~~~~~~~~~NGC1291} &&&\\  
&\hspace{5cm}\rotatebox{90}{\Large 870 \mic\ Observed} & 
\includegraphics[width=5.7cm]{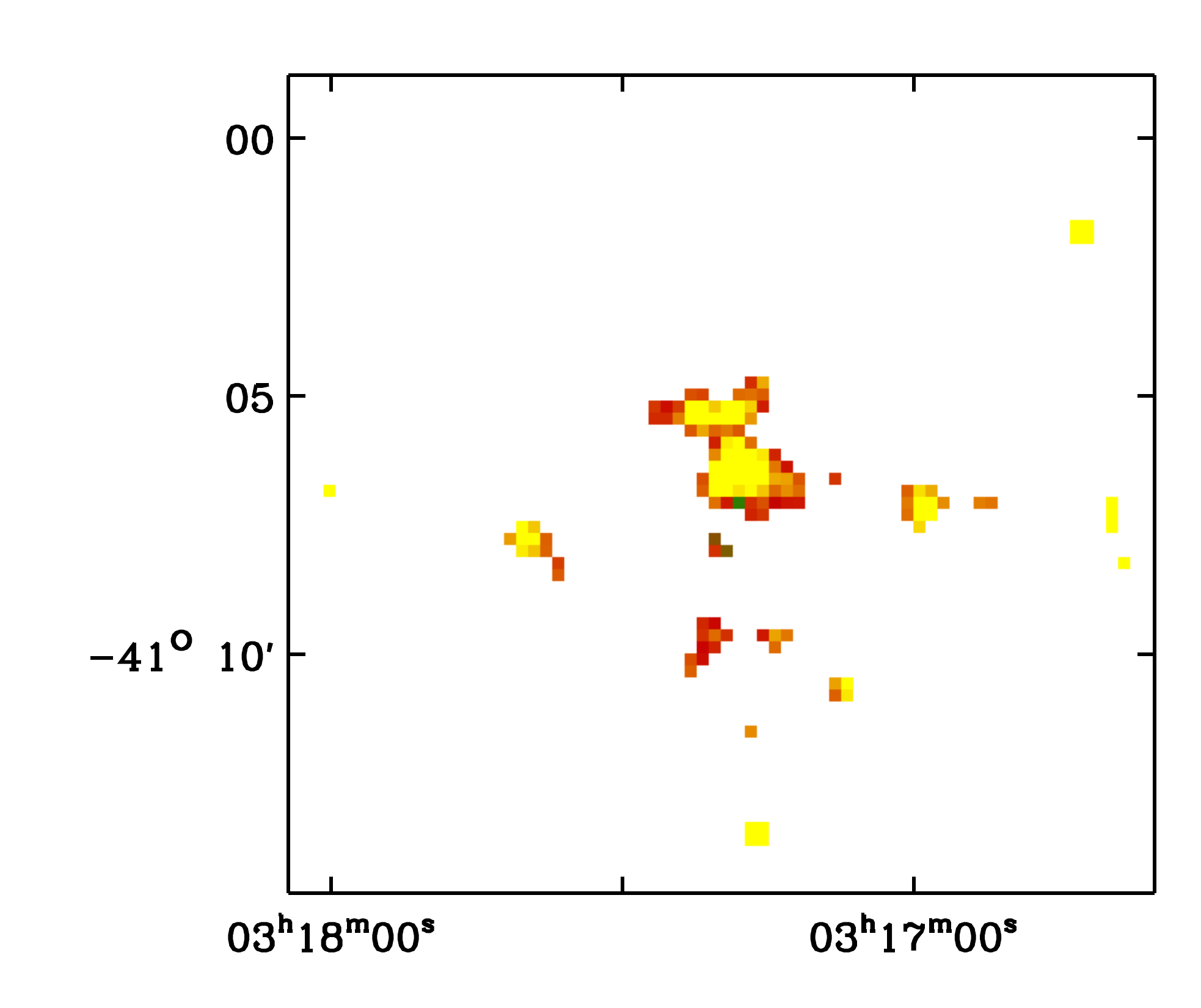} &&
\rotatebox{90}{\includegraphics[width=4cm, height=0.9cm]{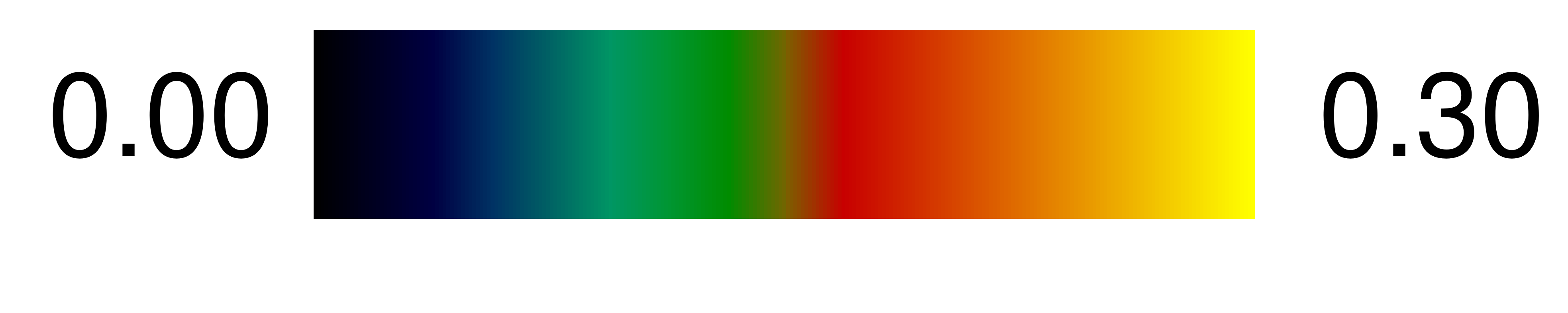}}  \\
&&\\
& {\Large \hspace{2.2cm}$\beta$$_c$ = 2.0 model} & {\Large \hspace{2.2cm}$\beta$$_c$ = 1.5 model}  & {\Large \hspace{2.2cm}[DL07] model} & \\

\rotatebox{90}{\Large 870 \mic\ Modelled} & 
\includegraphics[width=5.7cm]{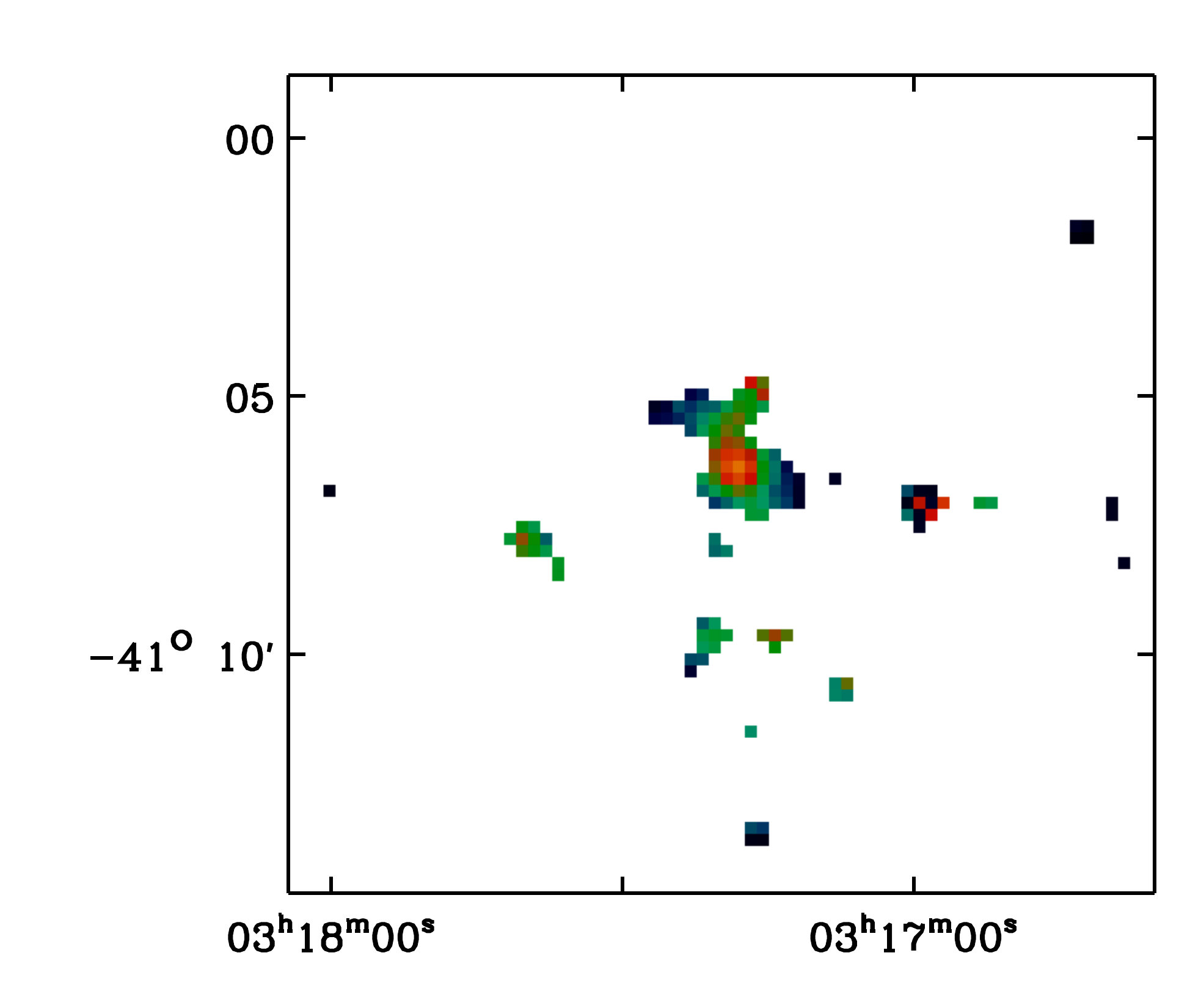} &
\includegraphics[width=5.7cm]{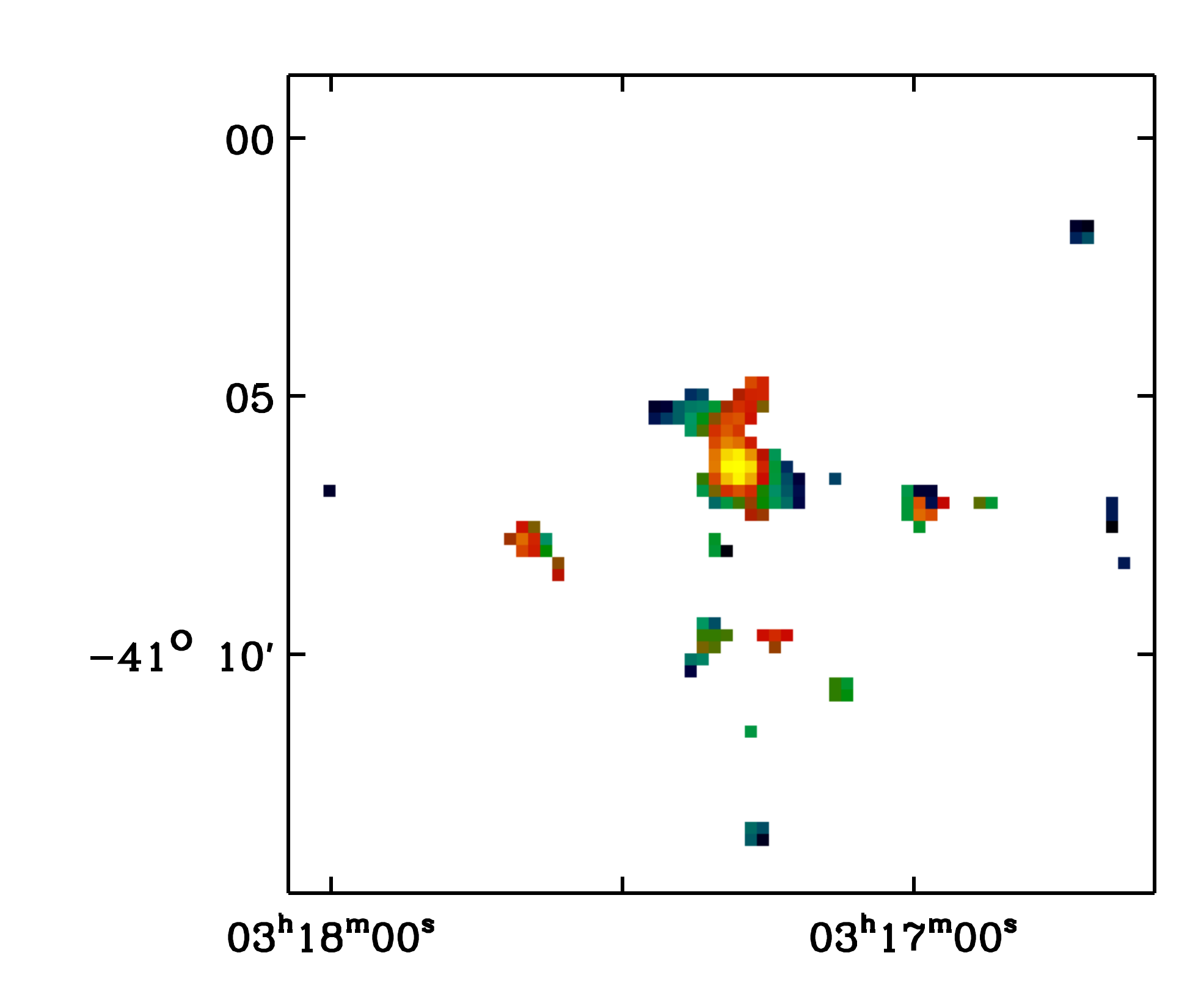} &
\includegraphics[width=5.7cm]{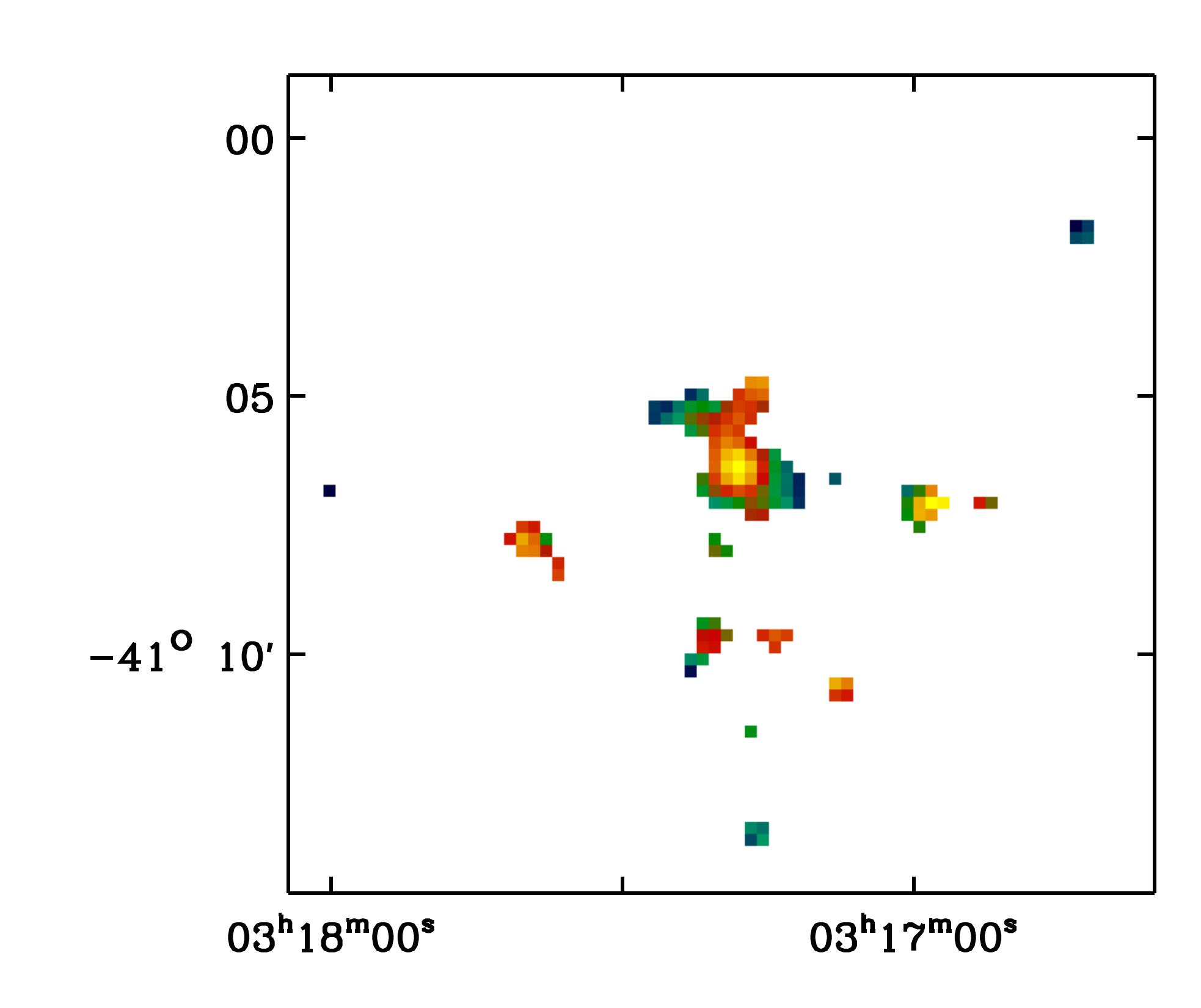}  &
\rotatebox{90}{\includegraphics[width=4cm, height=0.9cm]{NGC1291_Extrap870_ColorBars}}  \\
	
\rotatebox{90}{\Large Absolute Difference} &
\includegraphics[width=5.7cm]{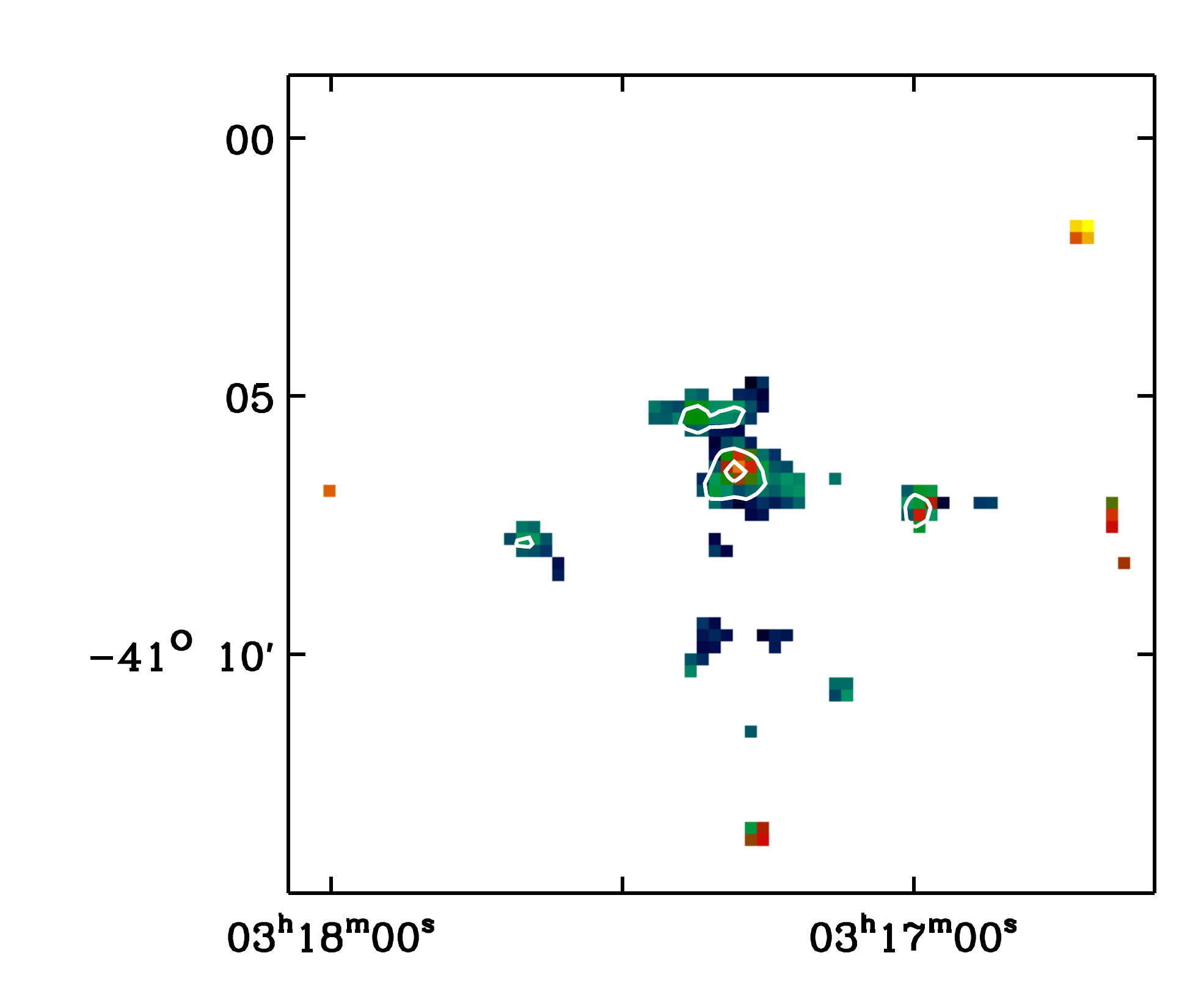} & 
\includegraphics[width=5.7cm]{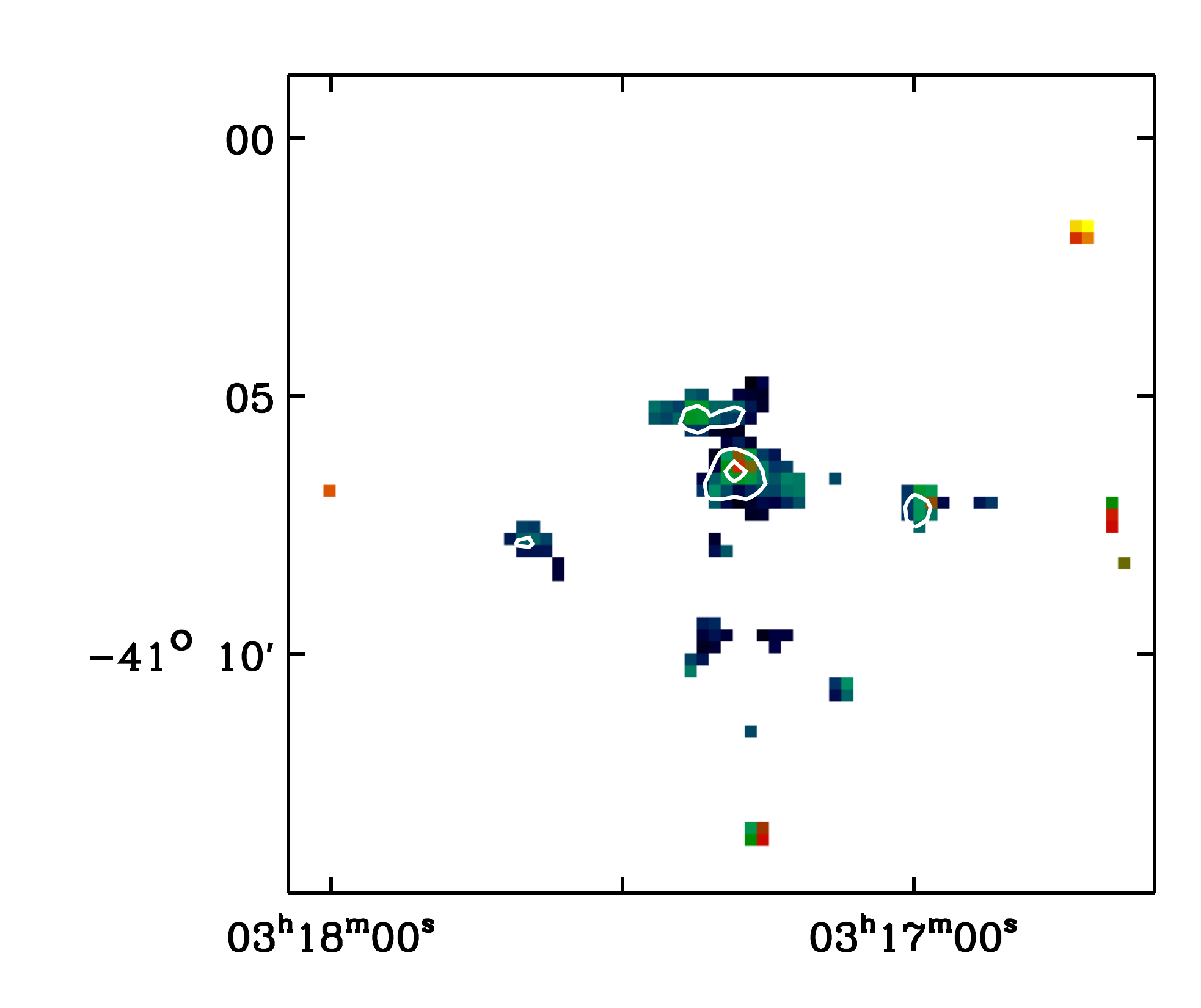} &
\includegraphics[width=5.7cm]{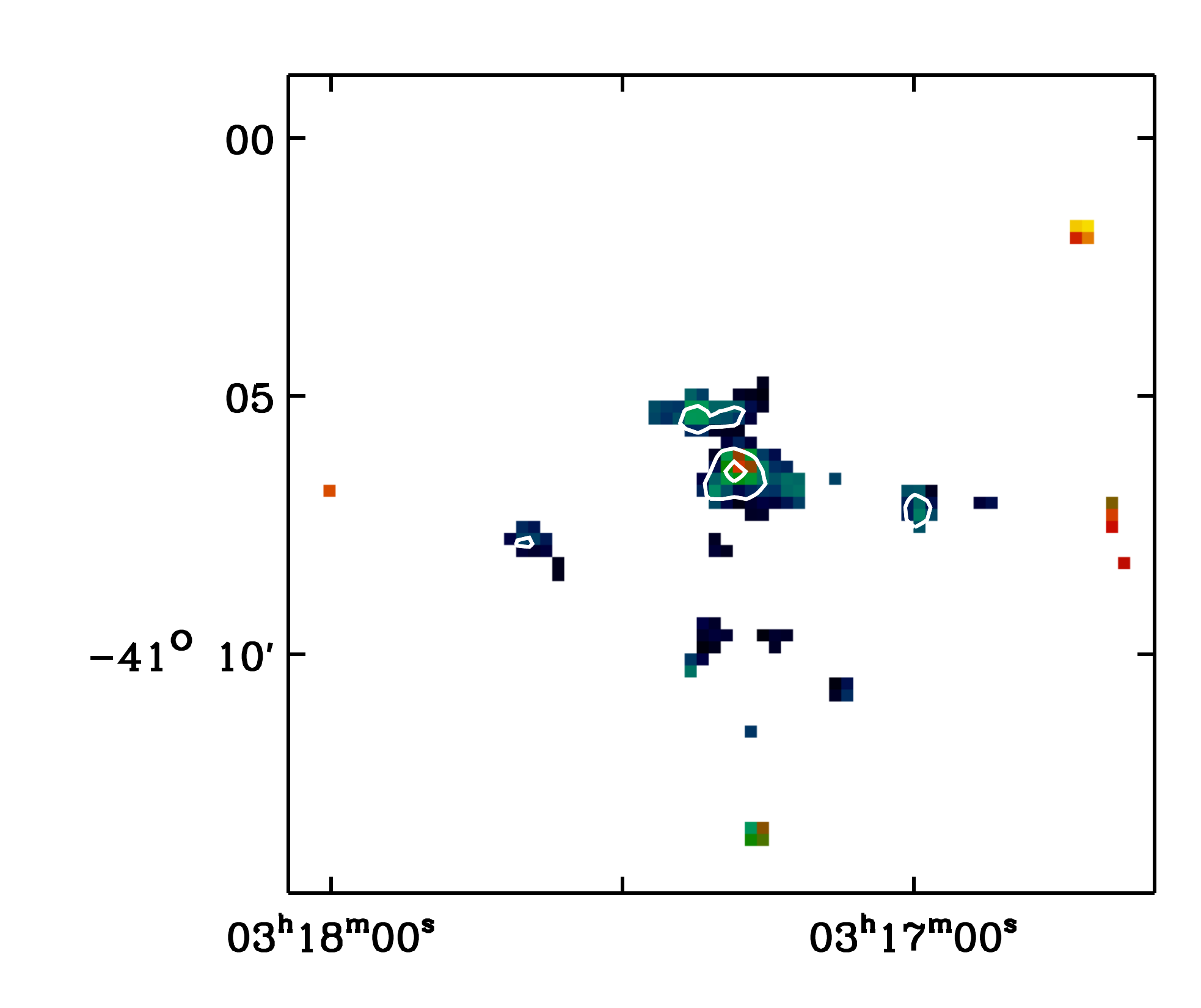}  &
\rotatebox{90}{\includegraphics[width=4cm, height=0.9cm]{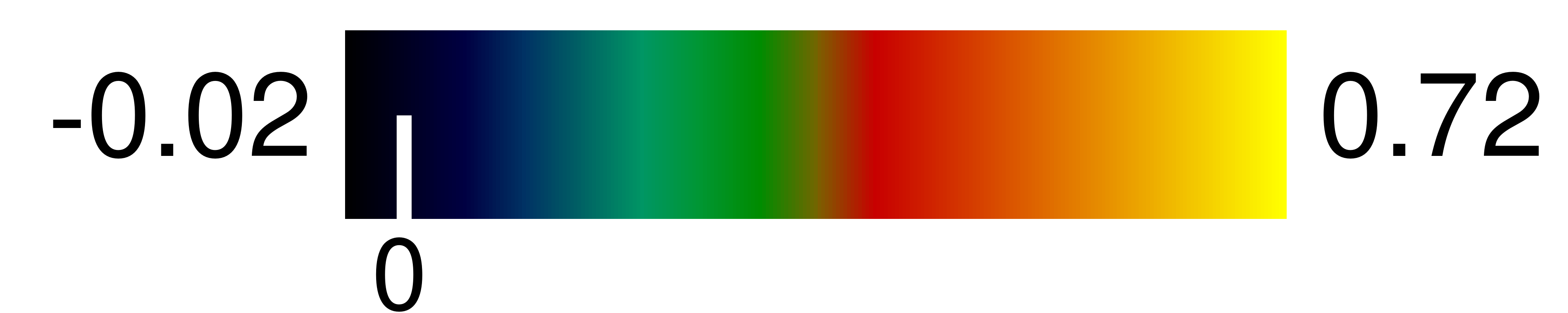}}  \\
	 
\rotatebox{90}{\Large Relative Difference} &
\includegraphics[width=5.7cm]{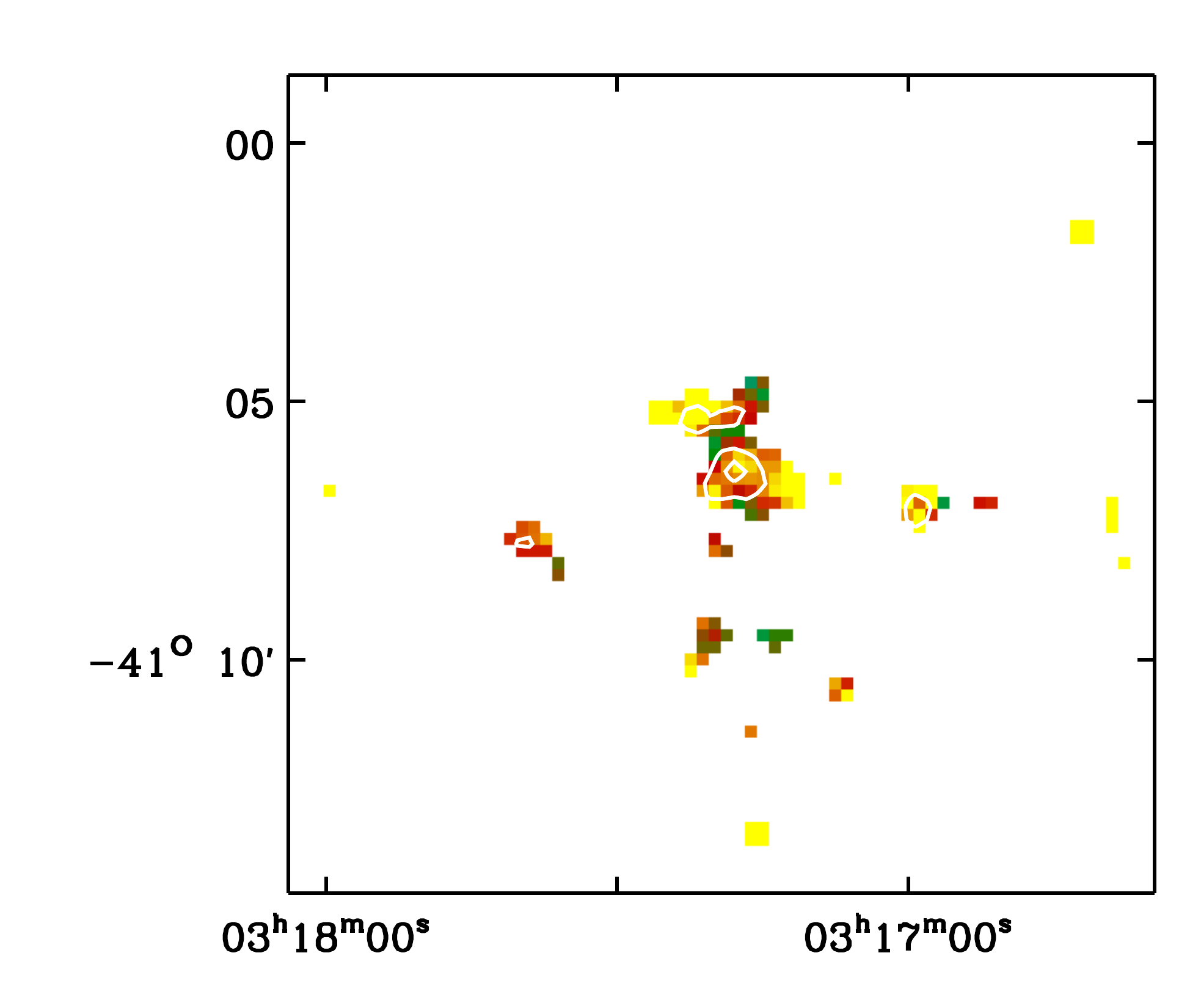} &
\includegraphics[width=5.7cm]{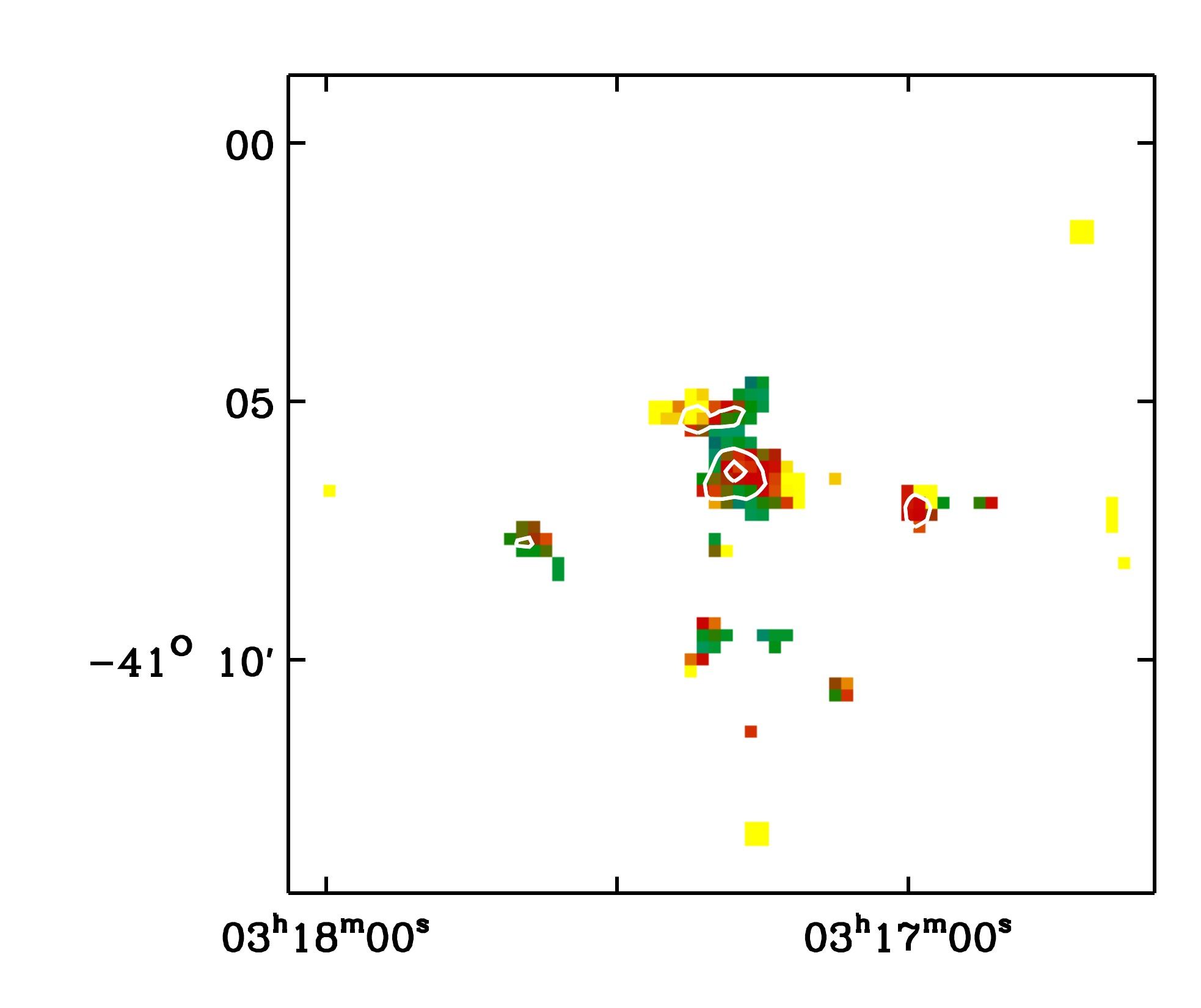} &
\includegraphics[width=5.7cm]{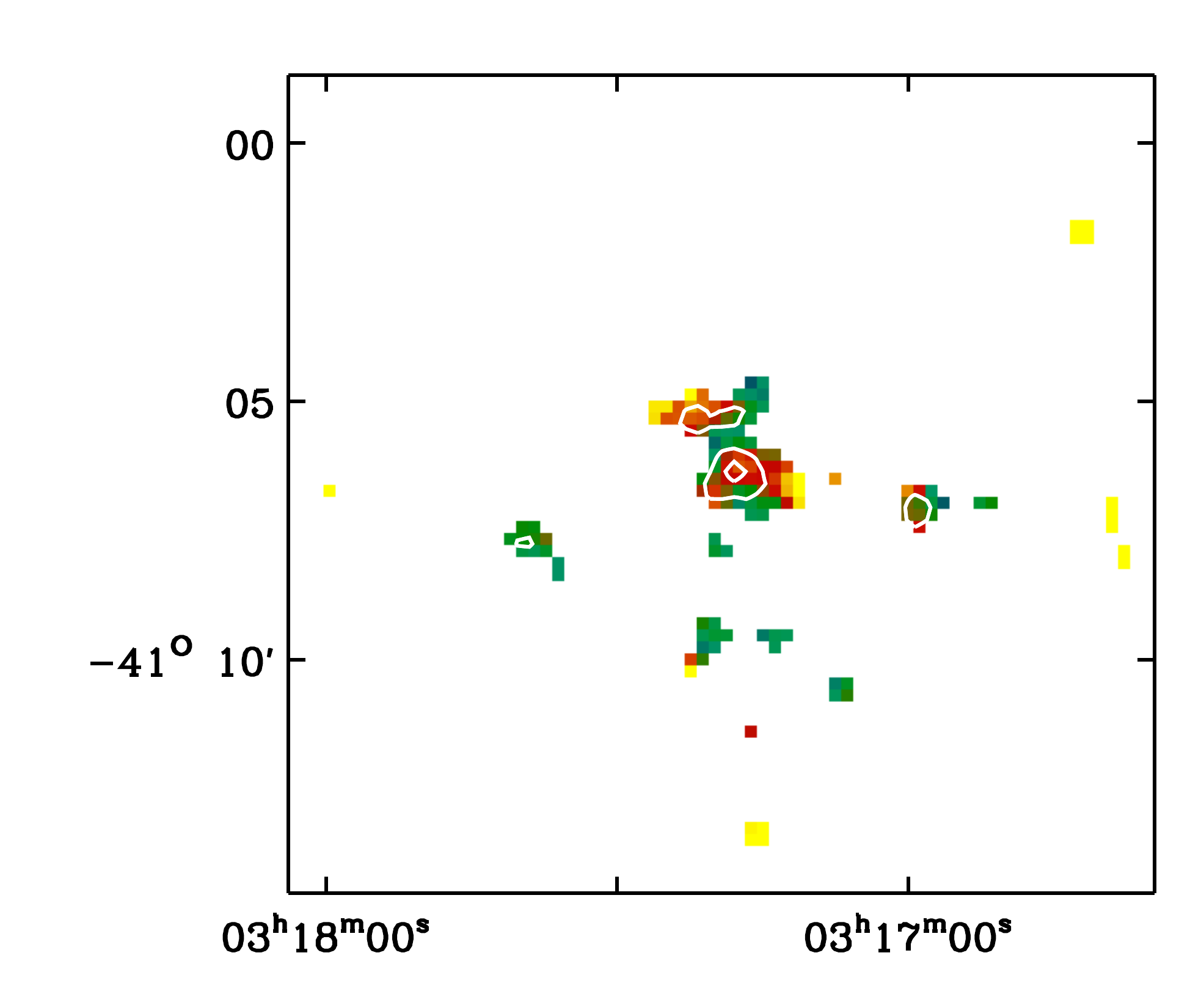}  &
\rotatebox{90}{\includegraphics[width=4cm, height=0.9cm]{RelativeExcess_ColorBars}}  \\	   
\end{tabular}  
    \caption{continued. }
    \end{figure*}

\newpage
\addtocounter {figure}{-1}
\begin{figure*}
\centering
\begin{tabular}  { m{0cm} m{5.1cm} m{5.1cm} m{5.1cm}  m{0.7cm}}    
{\Large \bf~~~~~~~~~~NGC1316} &&&\\  
&\hspace{5cm}\rotatebox{90}{\Large 870 \mic\ Observed} & 
\includegraphics[width=5.7cm]{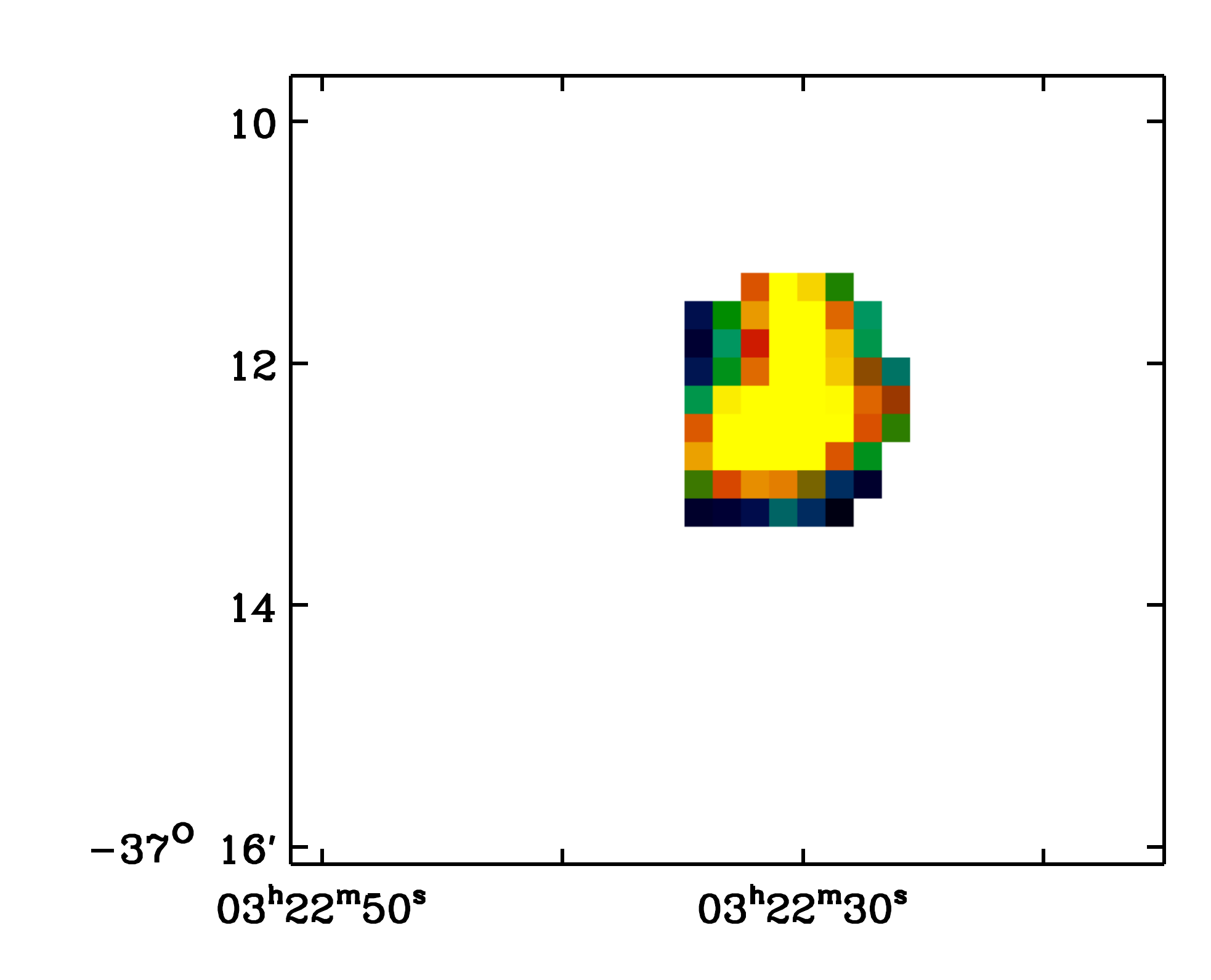} &&
\rotatebox{90}{\includegraphics[width=4cm, height=0.9cm]{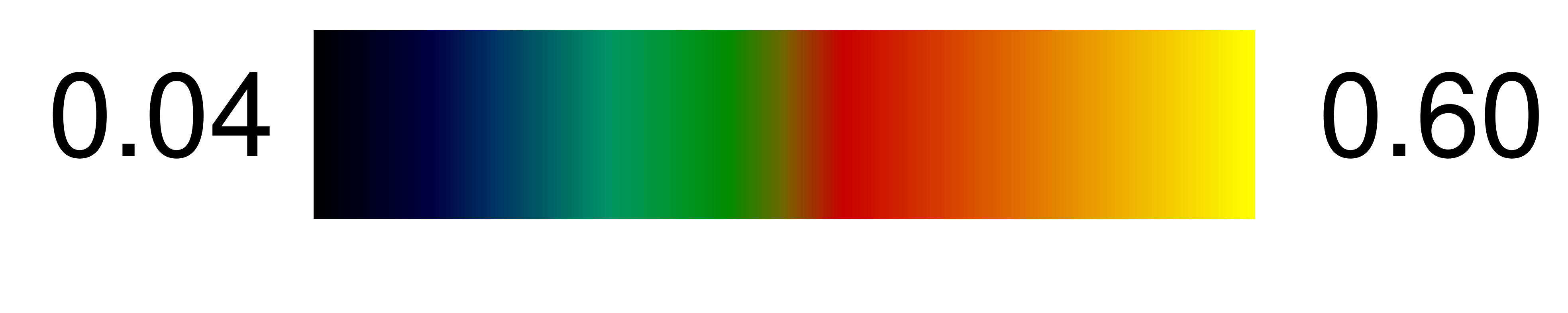}}  \\
&&\\
& {\Large \hspace{2.2cm}$\beta$$_c$ = 2.0 model} & {\Large \hspace{2.2cm}$\beta$$_c$ = 1.5 model}  & {\Large \hspace{2.2cm}[DL07] model} & \\

\rotatebox{90}{\Large 870 \mic\ Modelled} & 
\includegraphics[width=5.7cm]{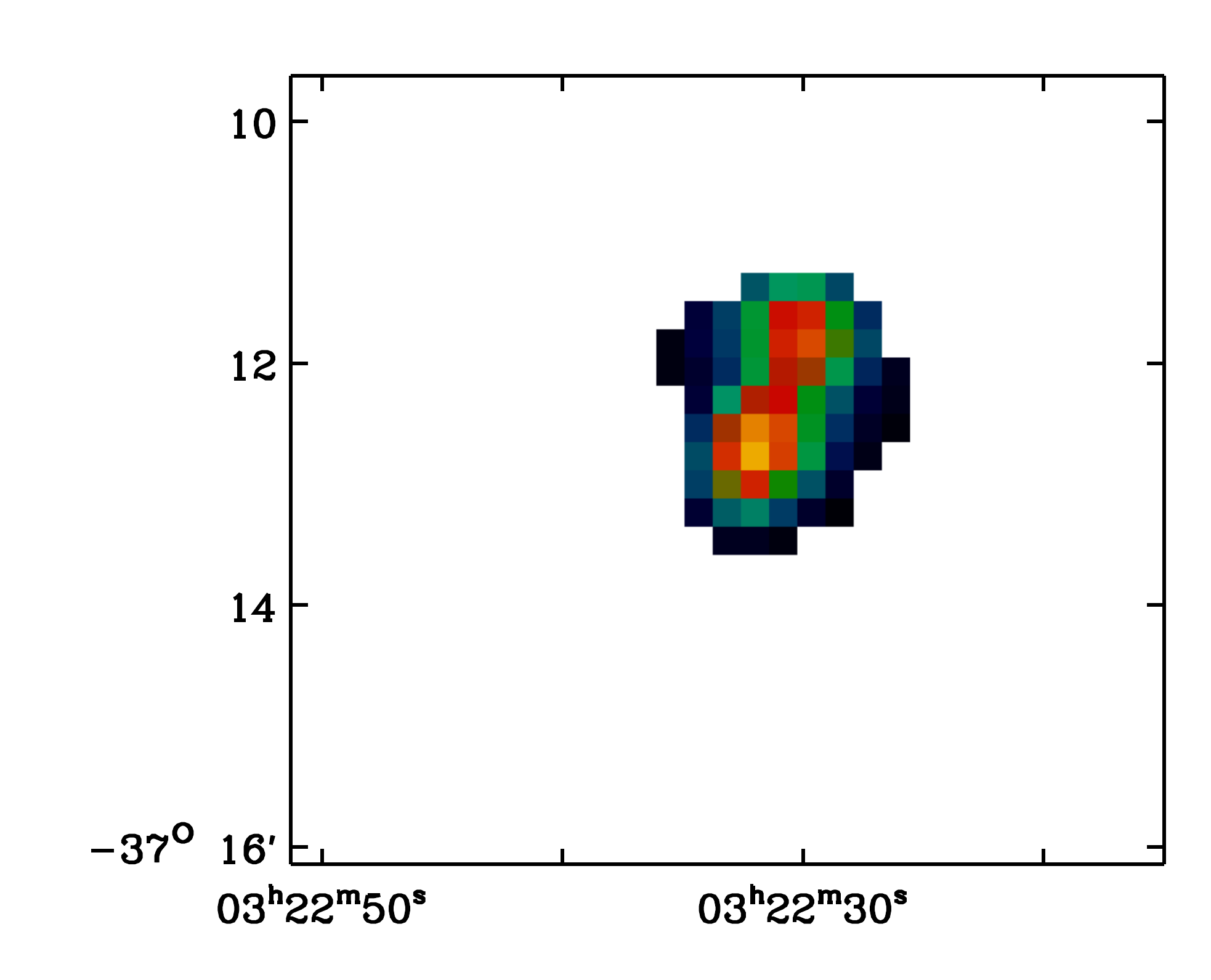} &
\includegraphics[width=5.7cm]{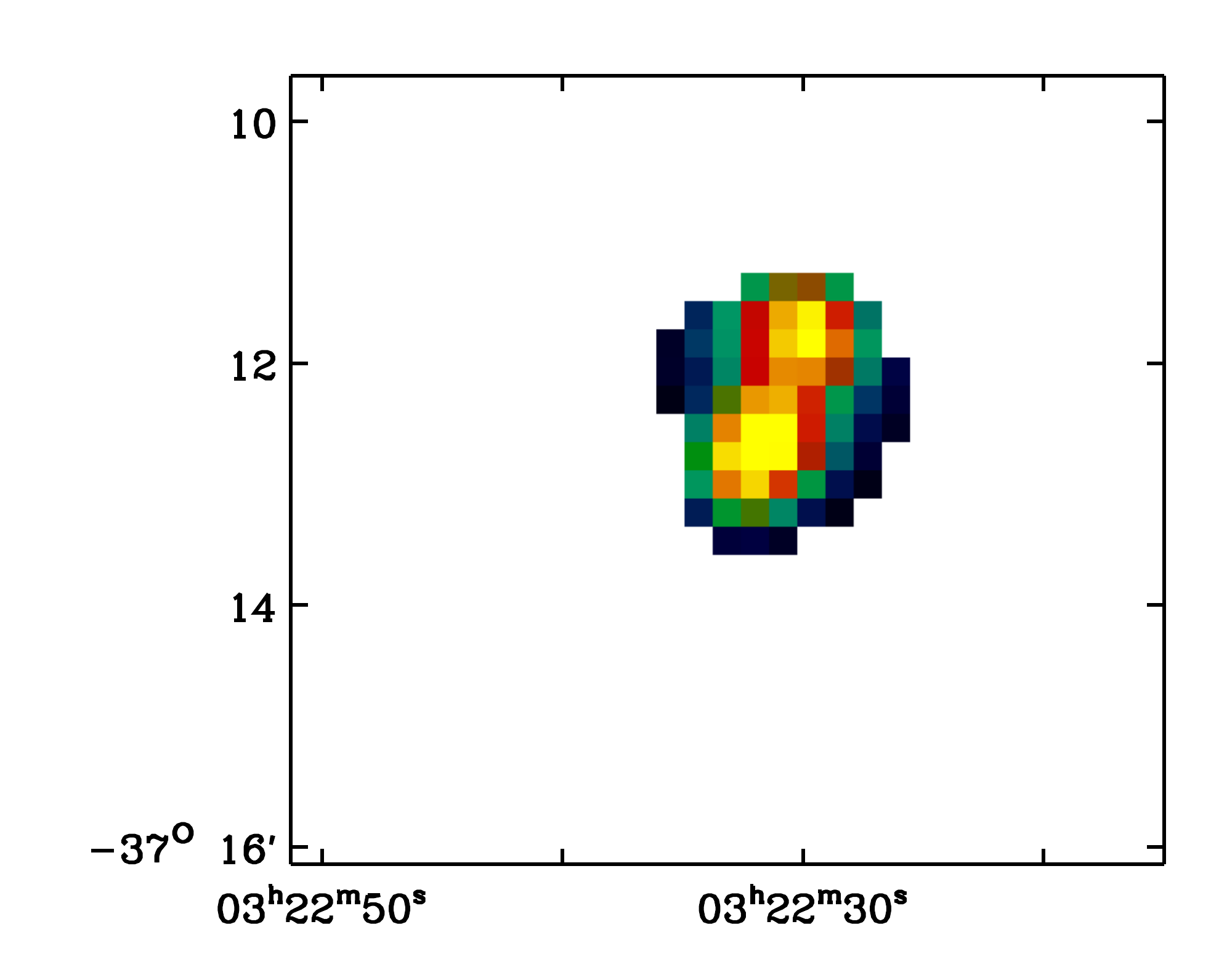} &
\includegraphics[width=5.7cm]{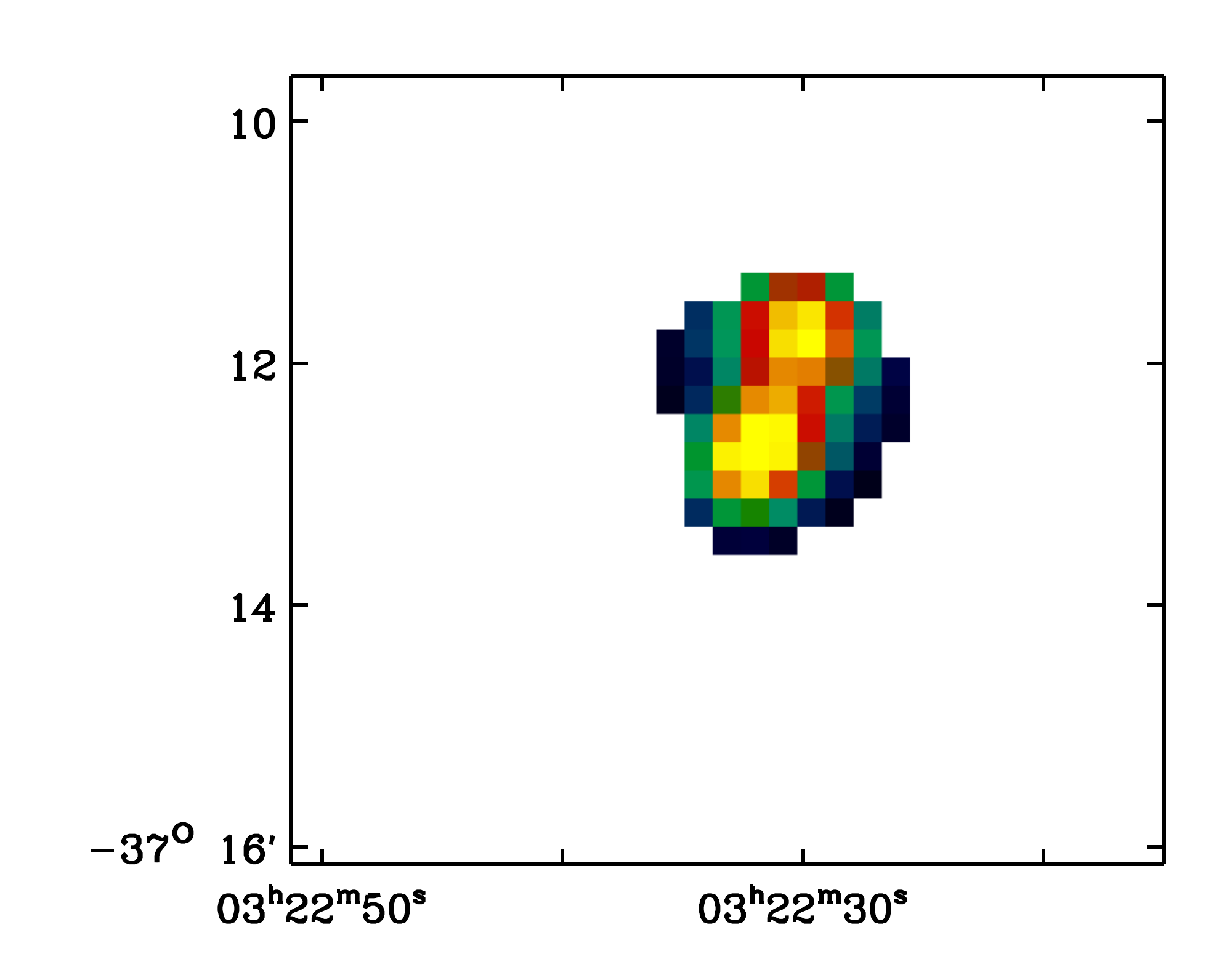}  &
\rotatebox{90}{\includegraphics[width=4cm, height=0.9cm]{NGC1316_Extrap870_ColorBars}}  \\
	
\rotatebox{90}{\Large Absolute Difference} &
\includegraphics[width=5.7cm]{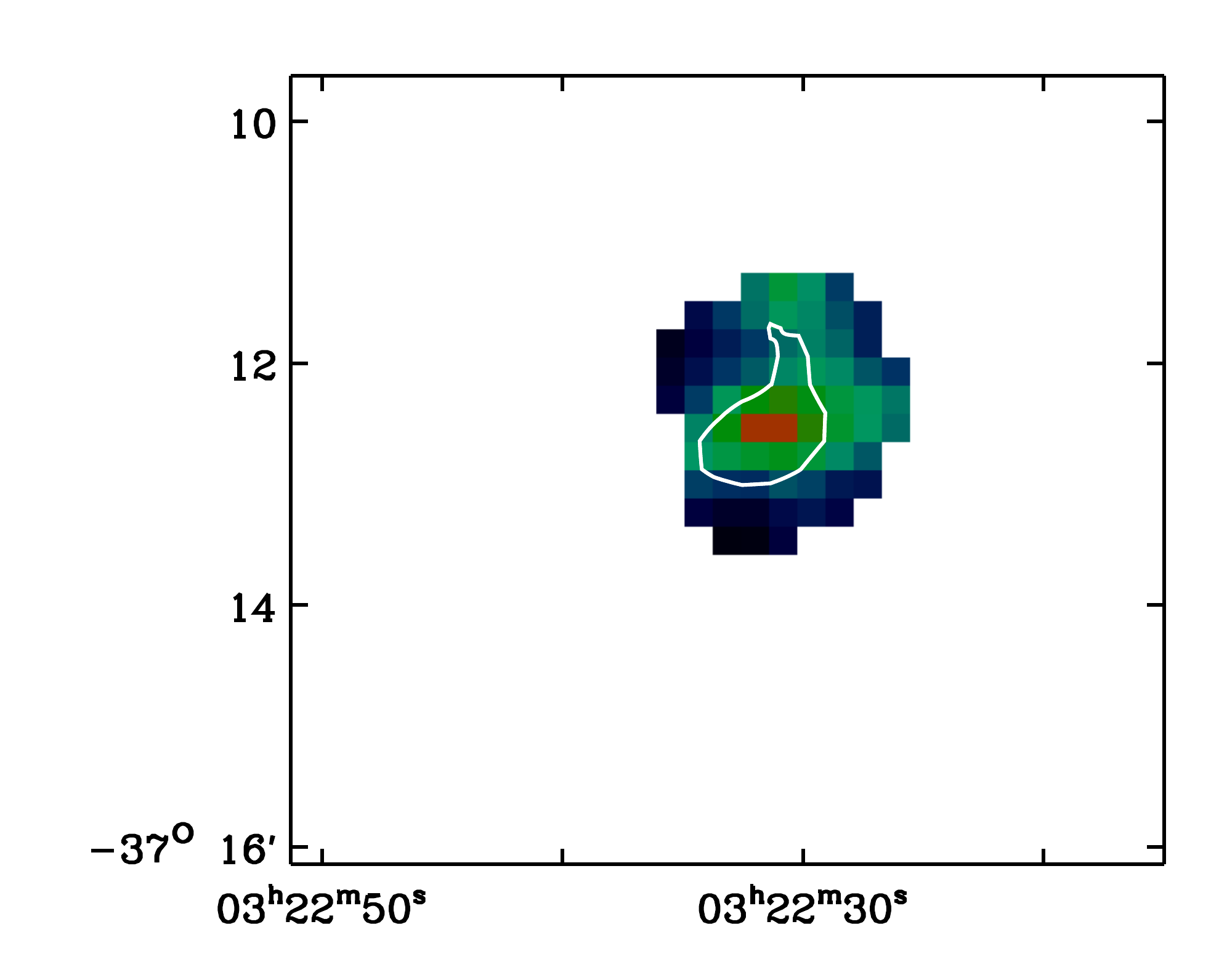} & 
\includegraphics[width=5.7cm]{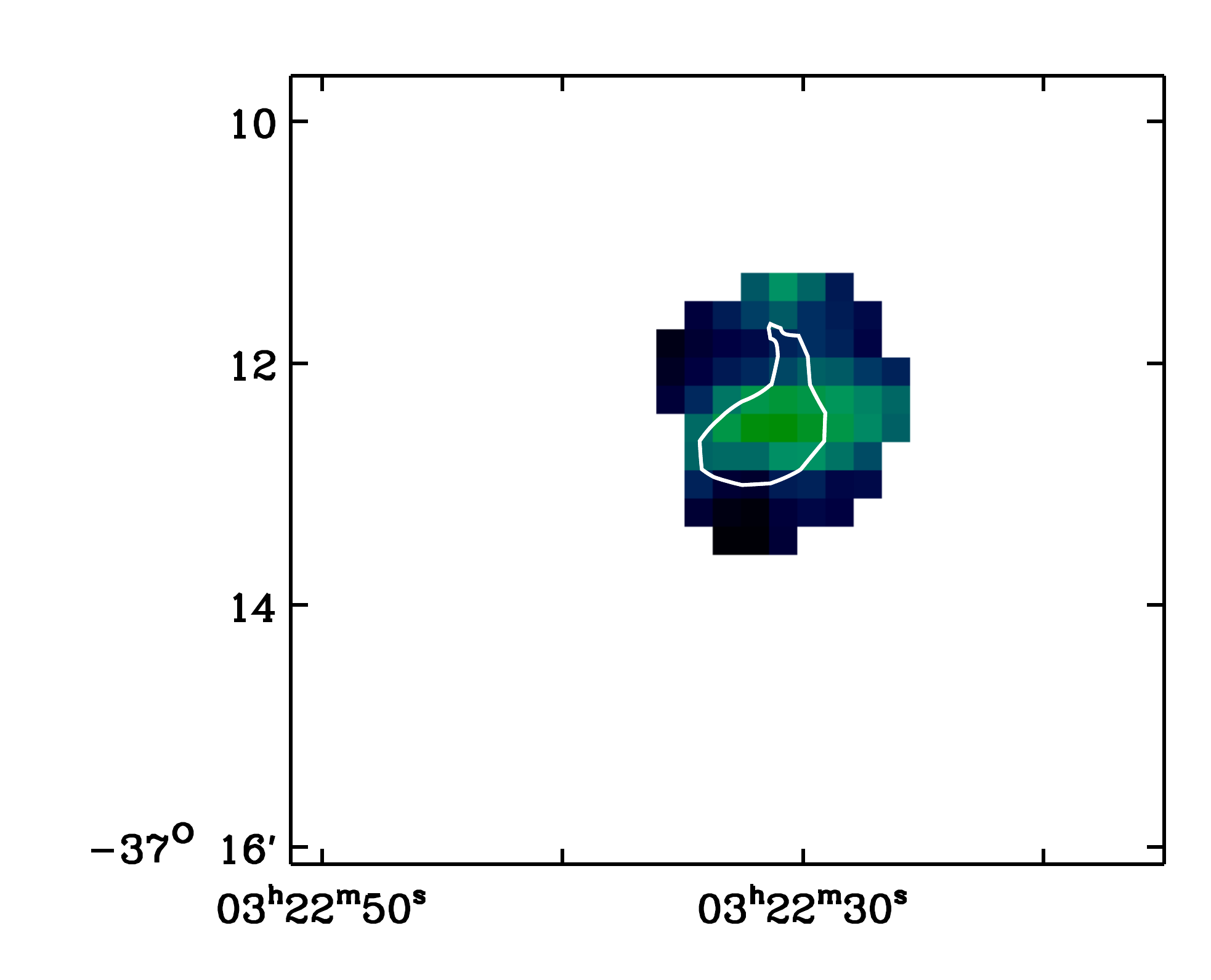} &
\includegraphics[width=5.7cm]{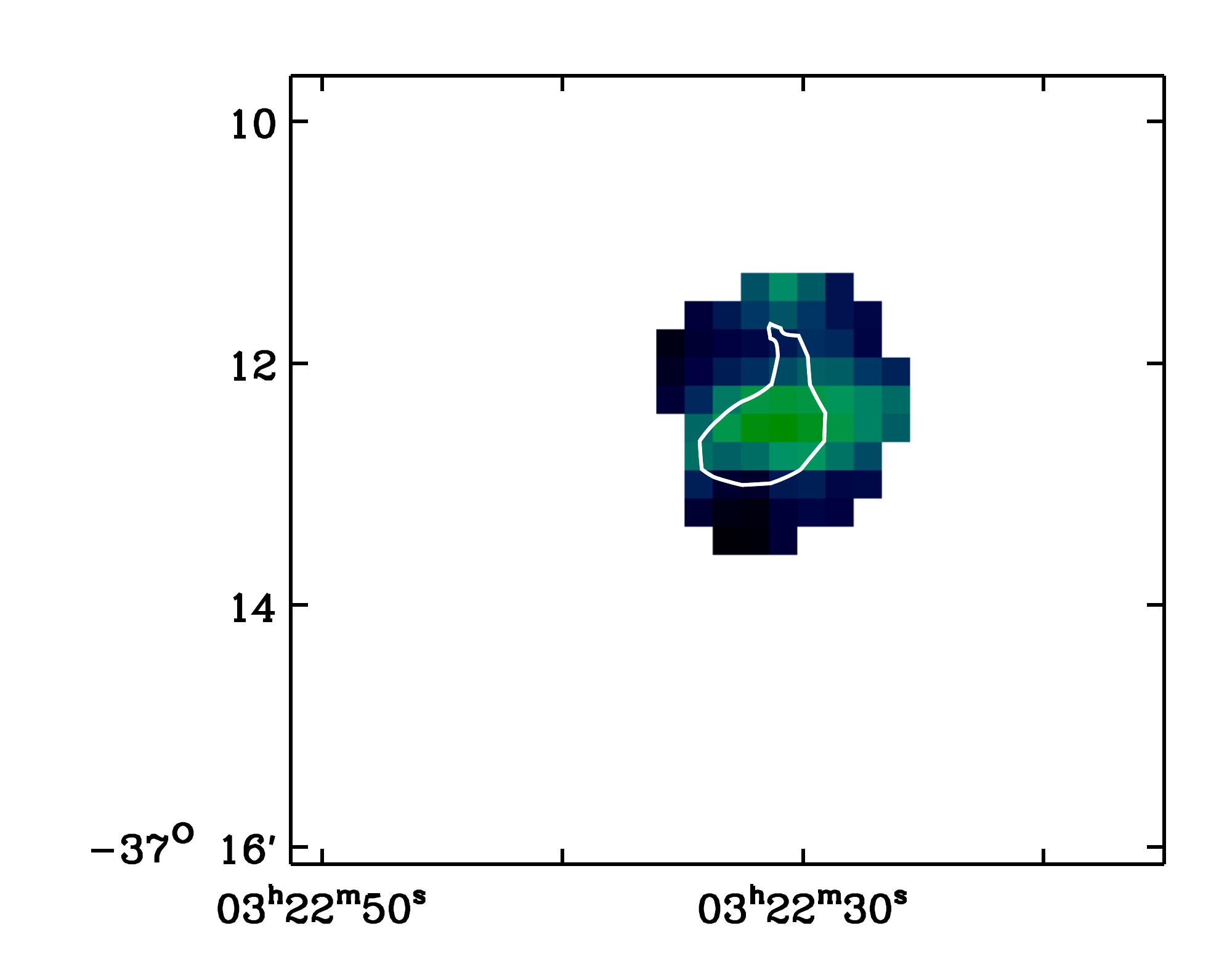}  &
\rotatebox{90}{\includegraphics[width=4cm, height=0.9cm]{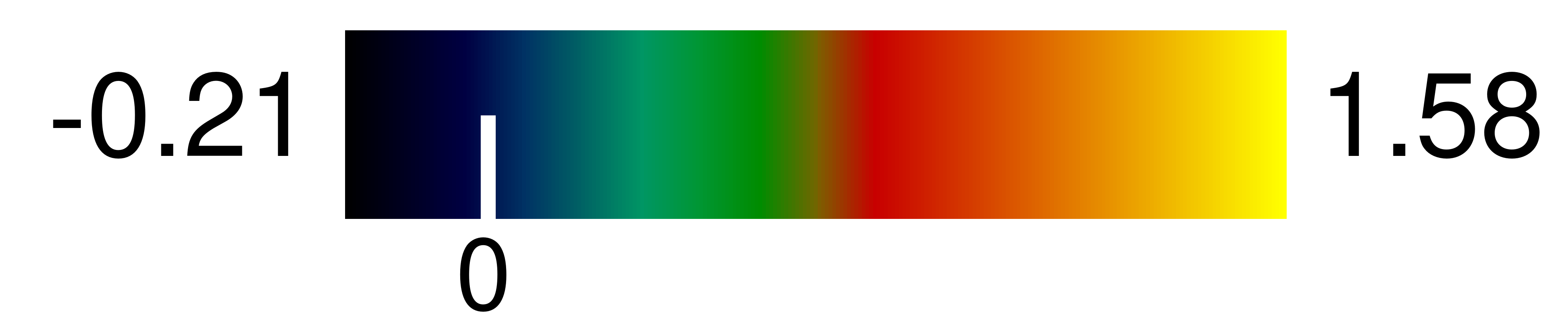}}  \\
	 
\rotatebox{90}{\Large Relative Difference} &
\includegraphics[width=5.7cm]{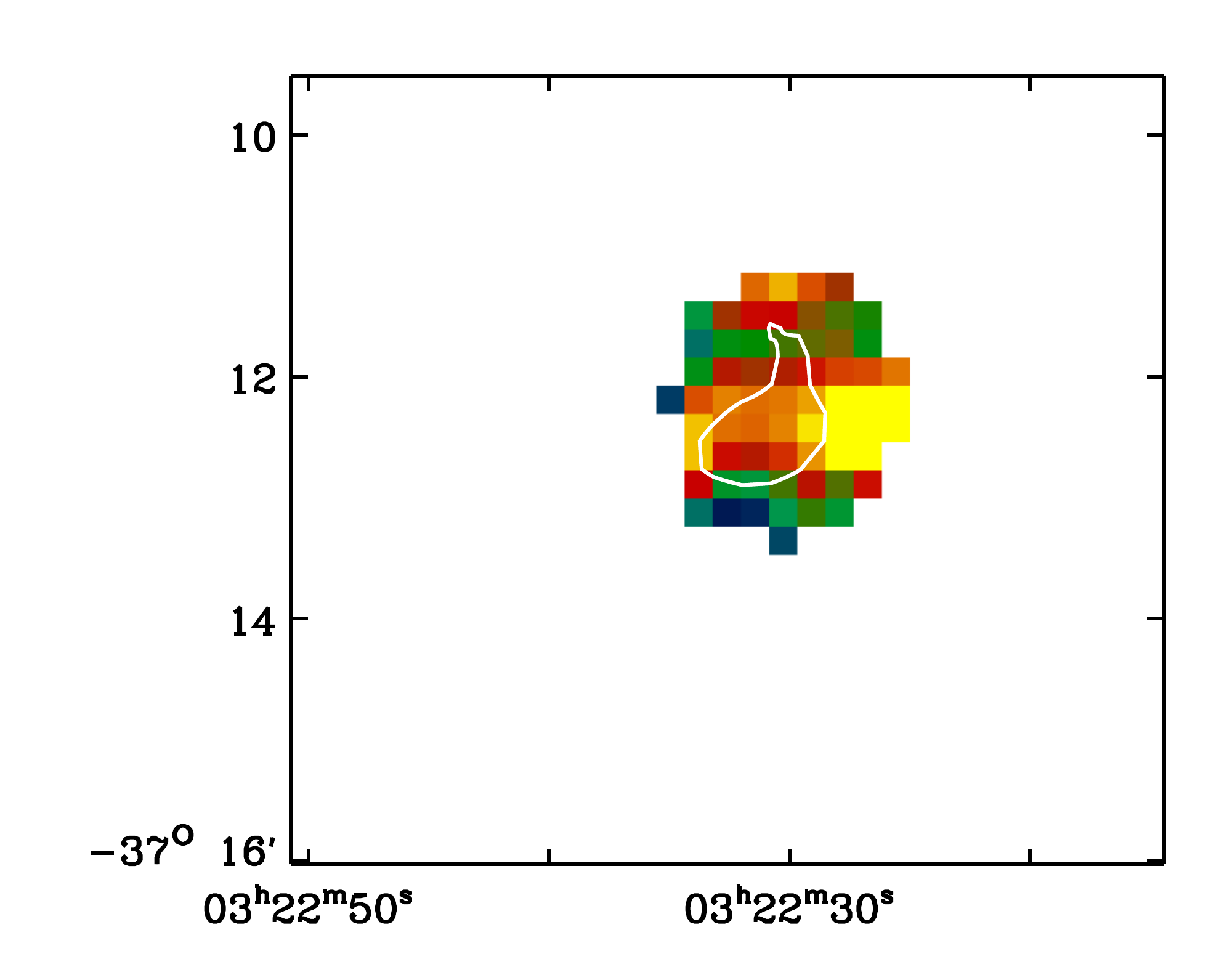} &
\includegraphics[width=5.7cm]{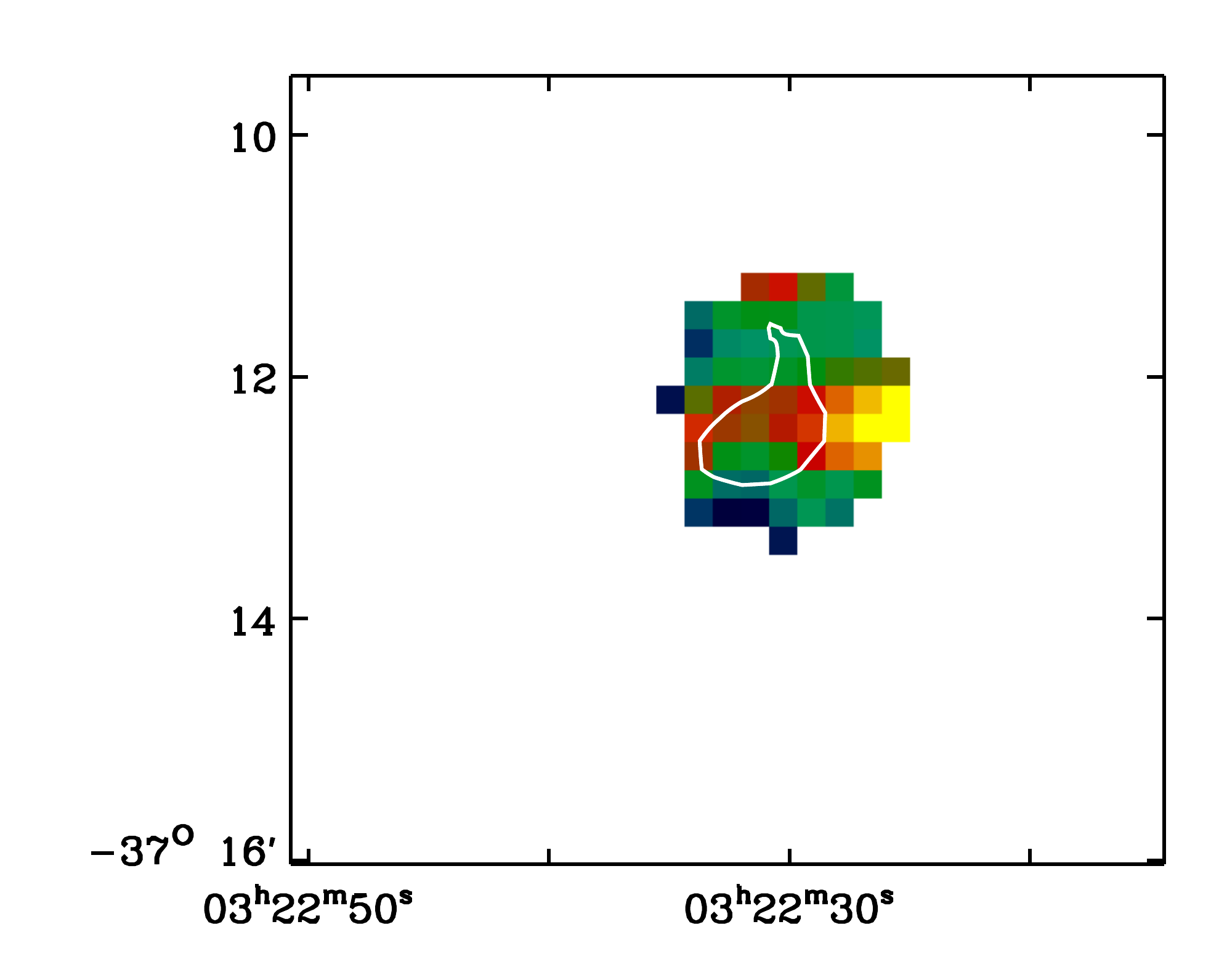} &
\includegraphics[width=5.7cm]{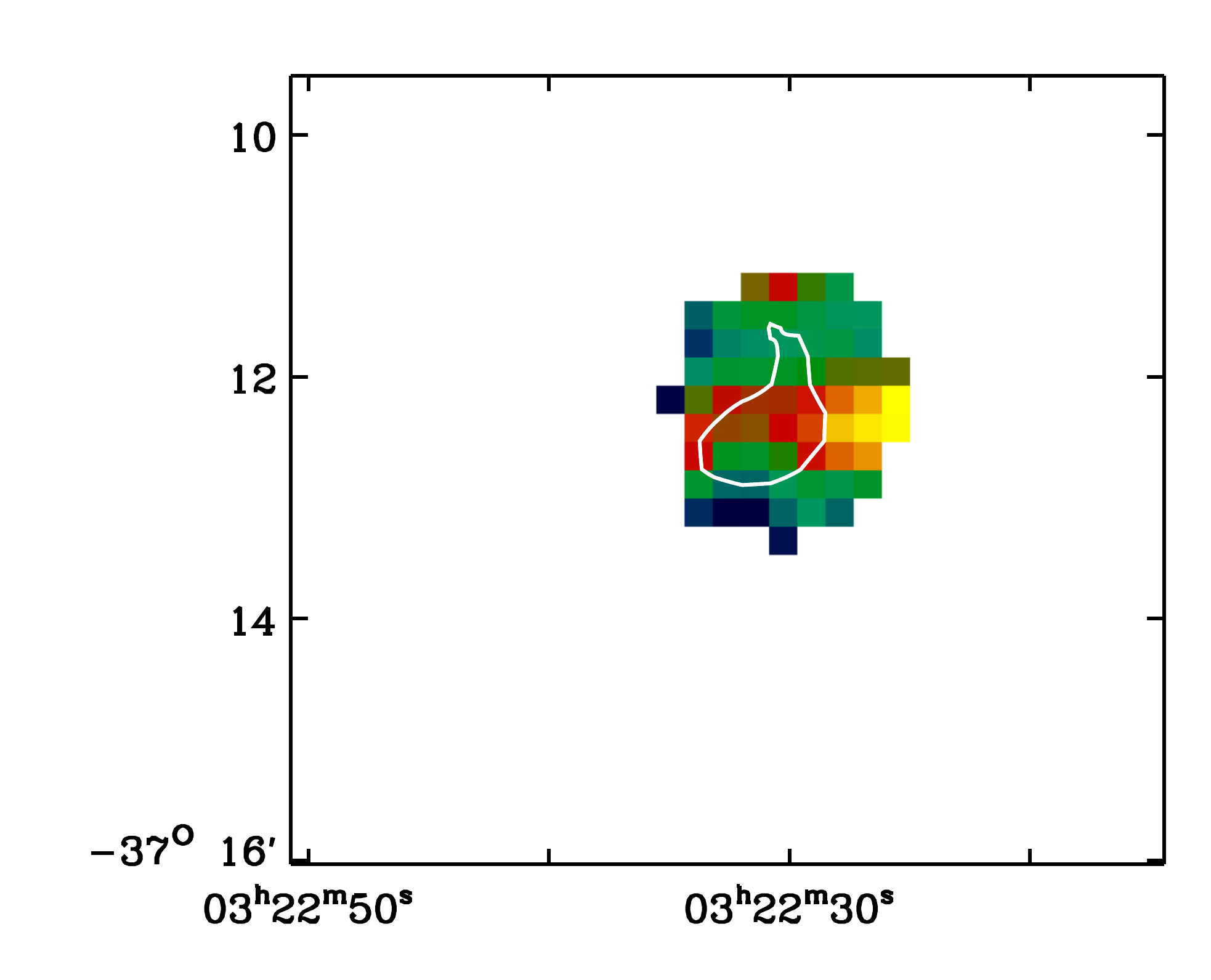}  &
\rotatebox{90}{\includegraphics[width=4cm, height=0.9cm]{RelativeExcess_ColorBars}}  \\	   
\end{tabular}  
\caption{continued. }
\end{figure*}

\newpage
\addtocounter {figure}{-1}
\begin{figure*}
\centering
\begin{tabular}  { m{0cm} m{5.1cm} m{5.1cm} m{5.1cm}  m{0.7cm}}    
{\Large \bf~~~~~~~~~~NGC1512} &&&\\  
&\hspace{5cm}\rotatebox{90}{\Large 870 \mic\ Observed} & 
\includegraphics[width=5.7cm]{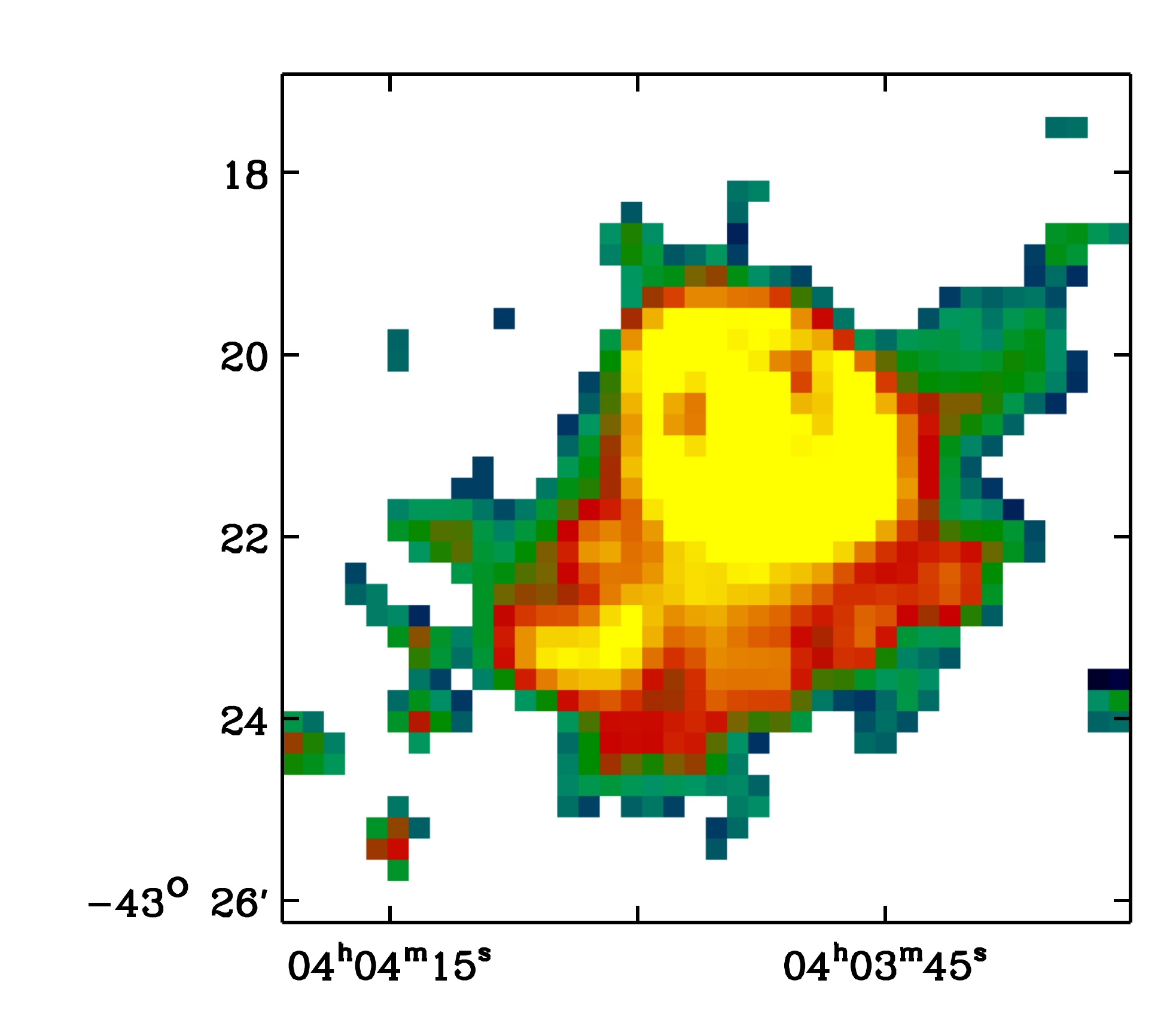} &&
\rotatebox{90}{\includegraphics[width=4cm, height=0.9cm]{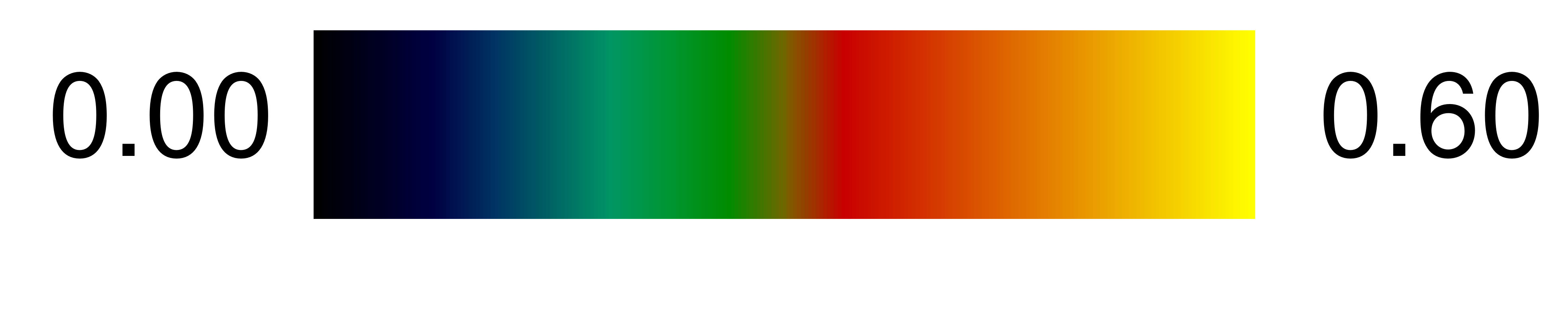}}  \\
&&\\
& {\Large \hspace{2.2cm}$\beta$$_c$ = 2.0 model} & {\Large \hspace{2.2cm}$\beta$$_c$ = 1.5 model}  & {\Large \hspace{2.2cm}[DL07] model} & \\

\rotatebox{90}{\Large 870 \mic\ Modelled} & 
\includegraphics[width=5.7cm]{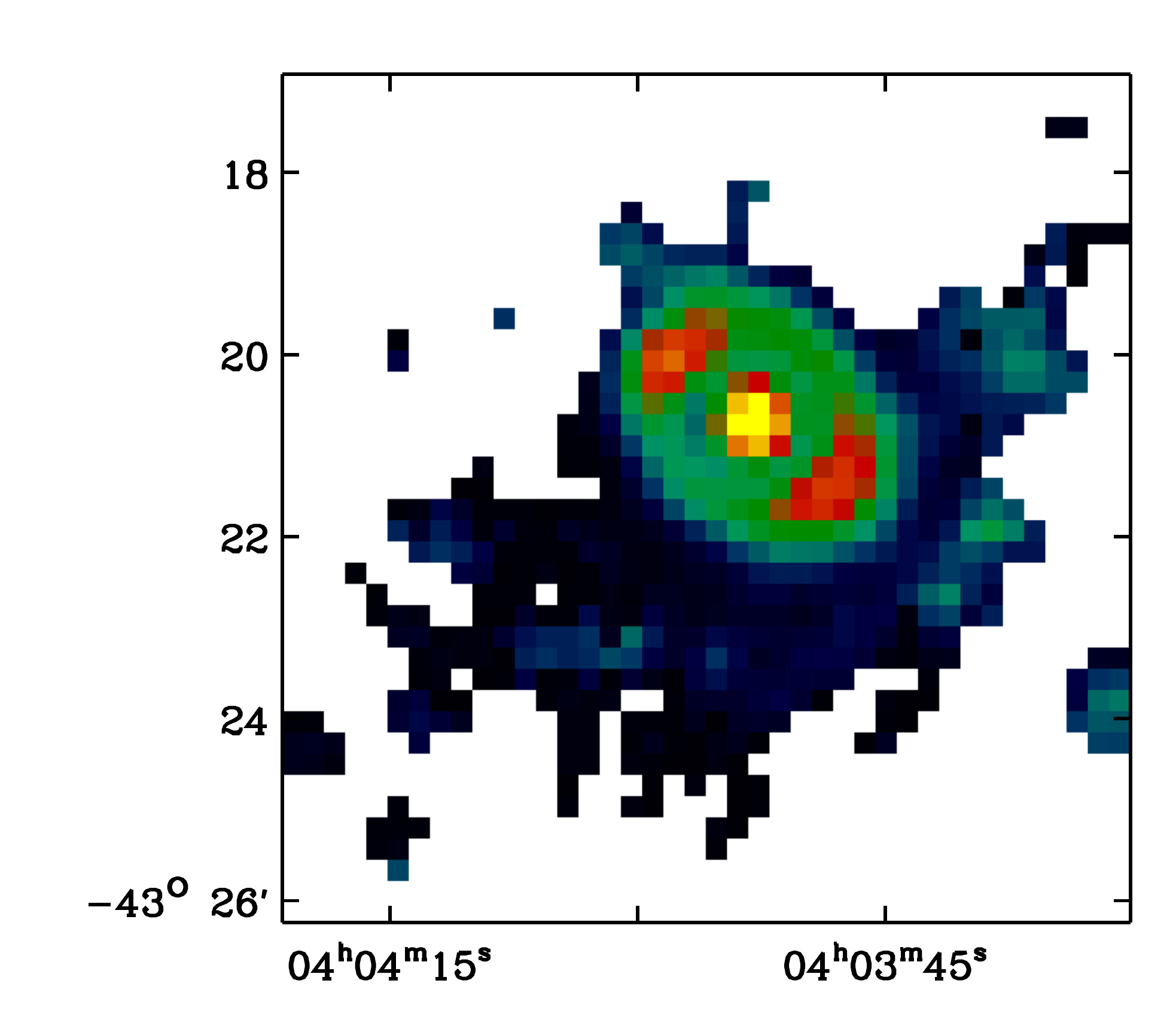} &
\includegraphics[width=5.7cm]{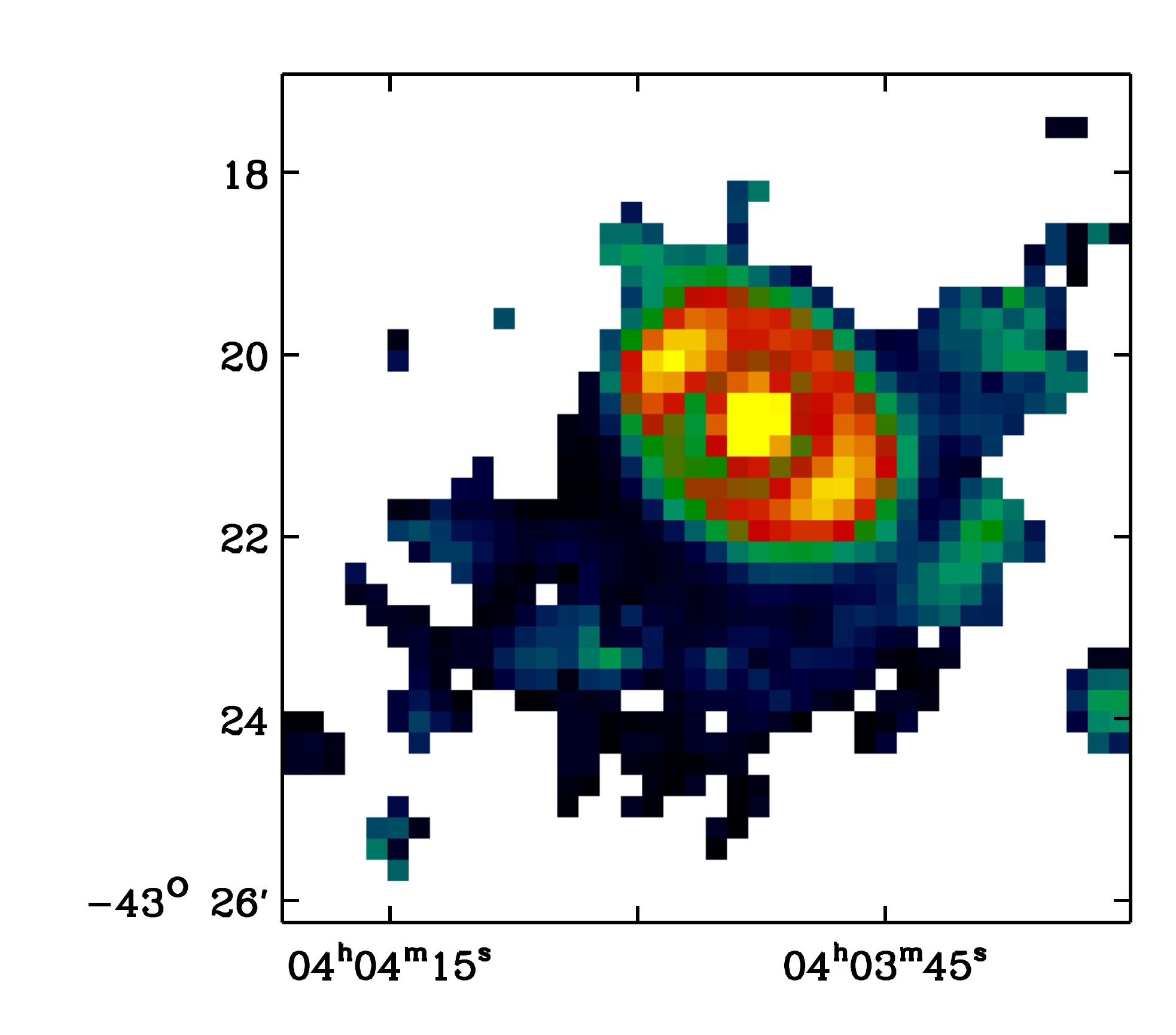} &
\includegraphics[width=5.7cm]{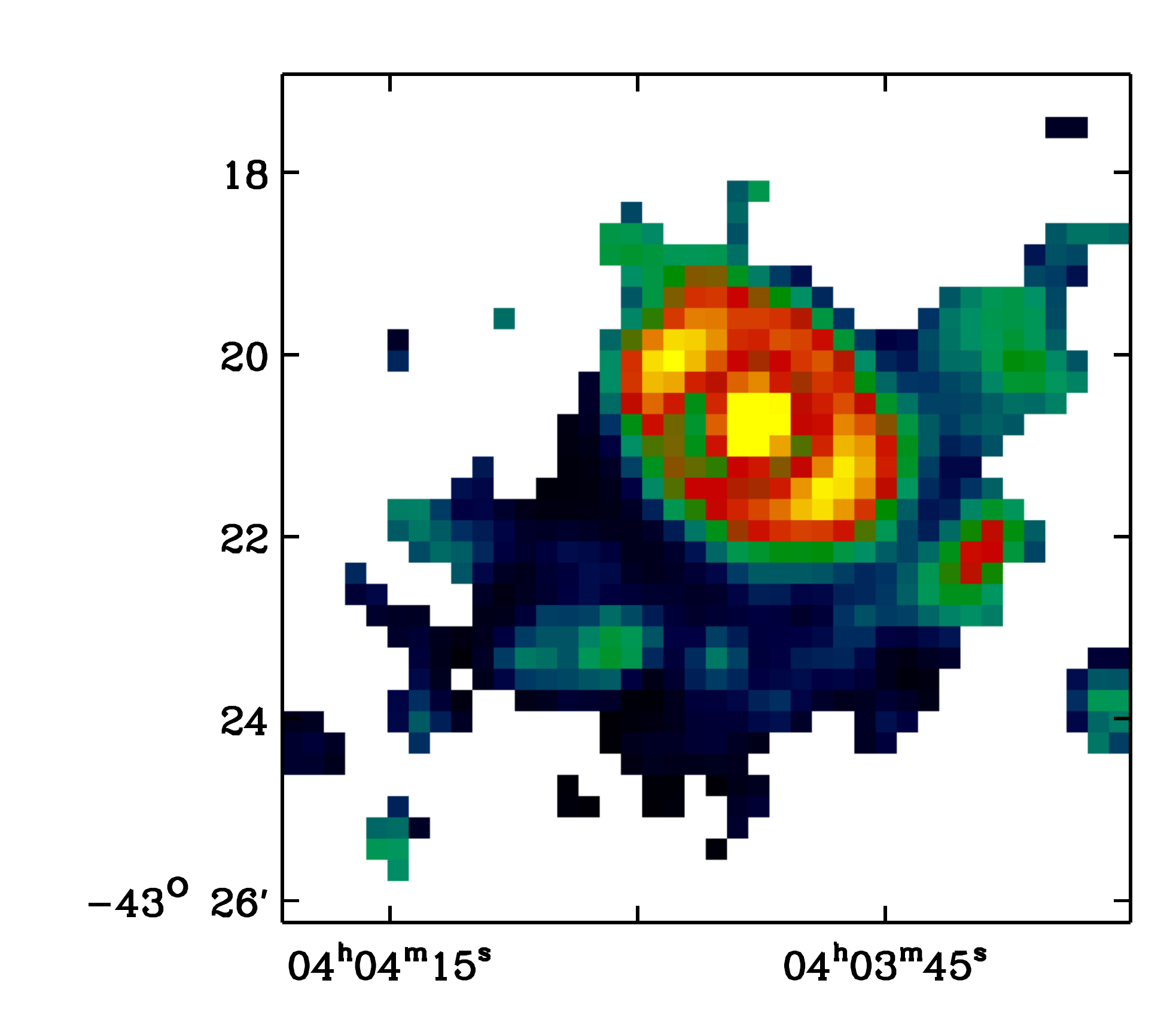}  &
\rotatebox{90}{\includegraphics[width=4cm, height=0.9cm]{NGC1512_Extrap870_ColorBars}}  \\
	
\rotatebox{90}{\Large Absolute Difference} &
\includegraphics[width=5.7cm]{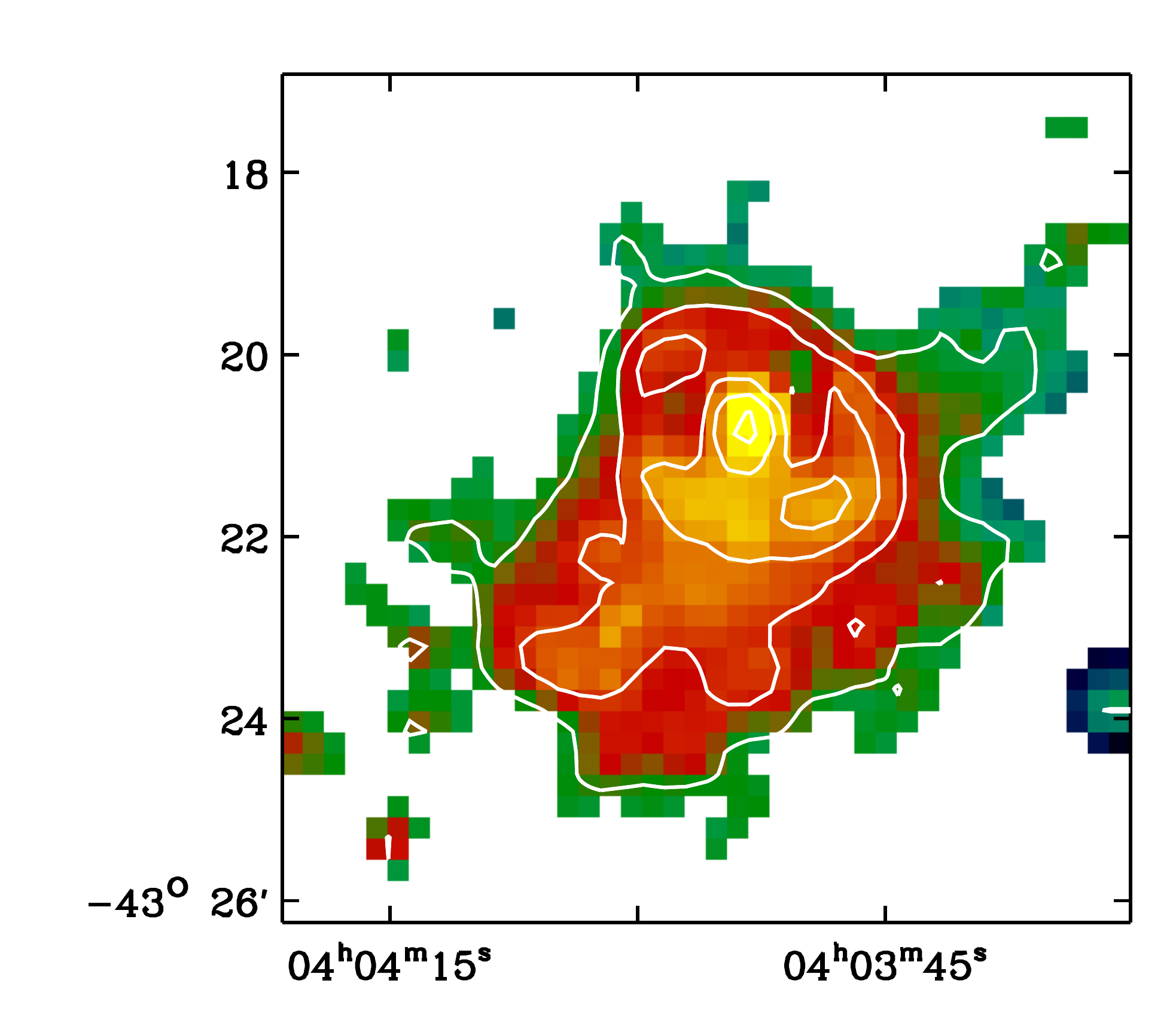} & 
\includegraphics[width=5.7cm]{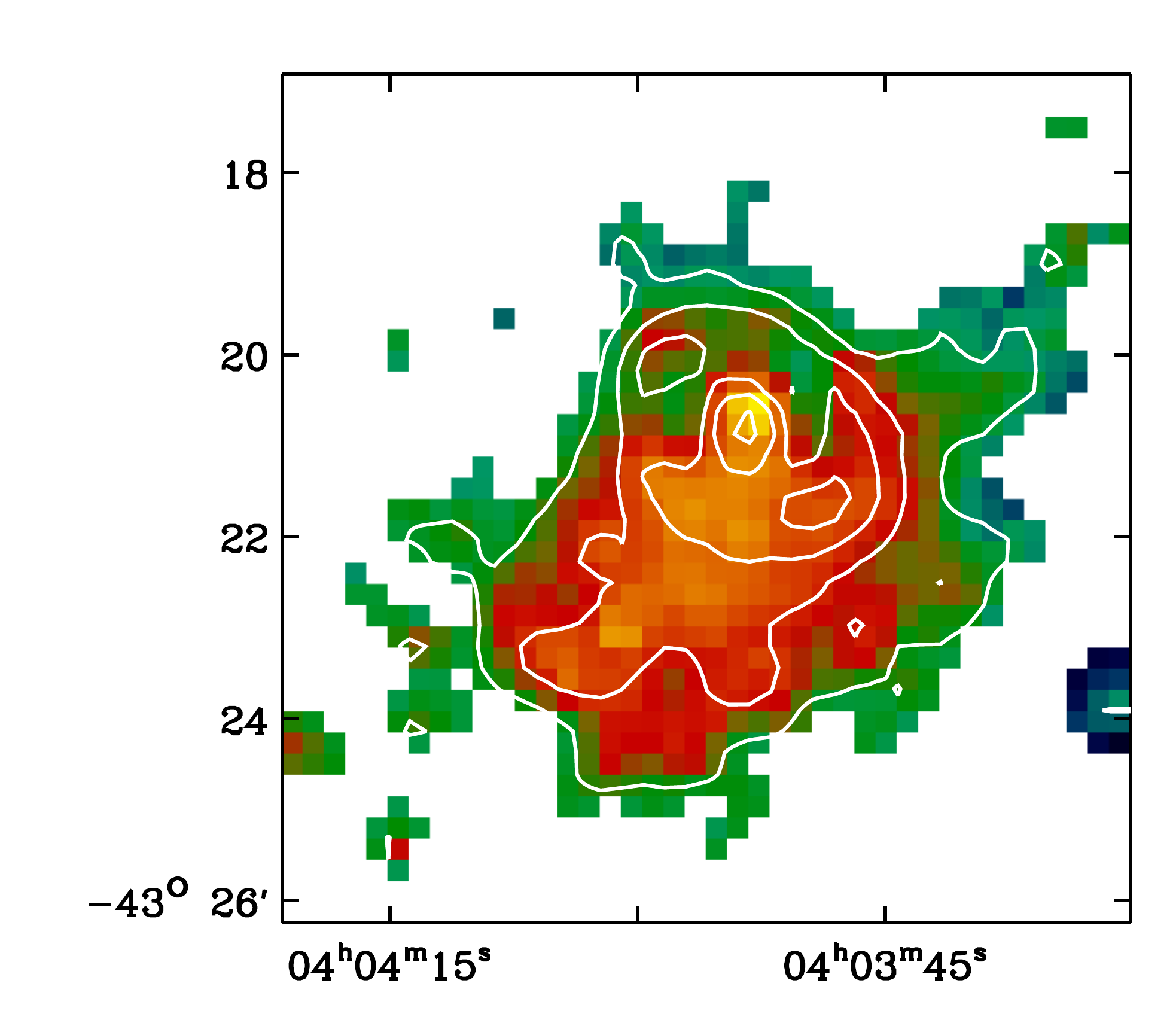} &
\includegraphics[width=5.7cm]{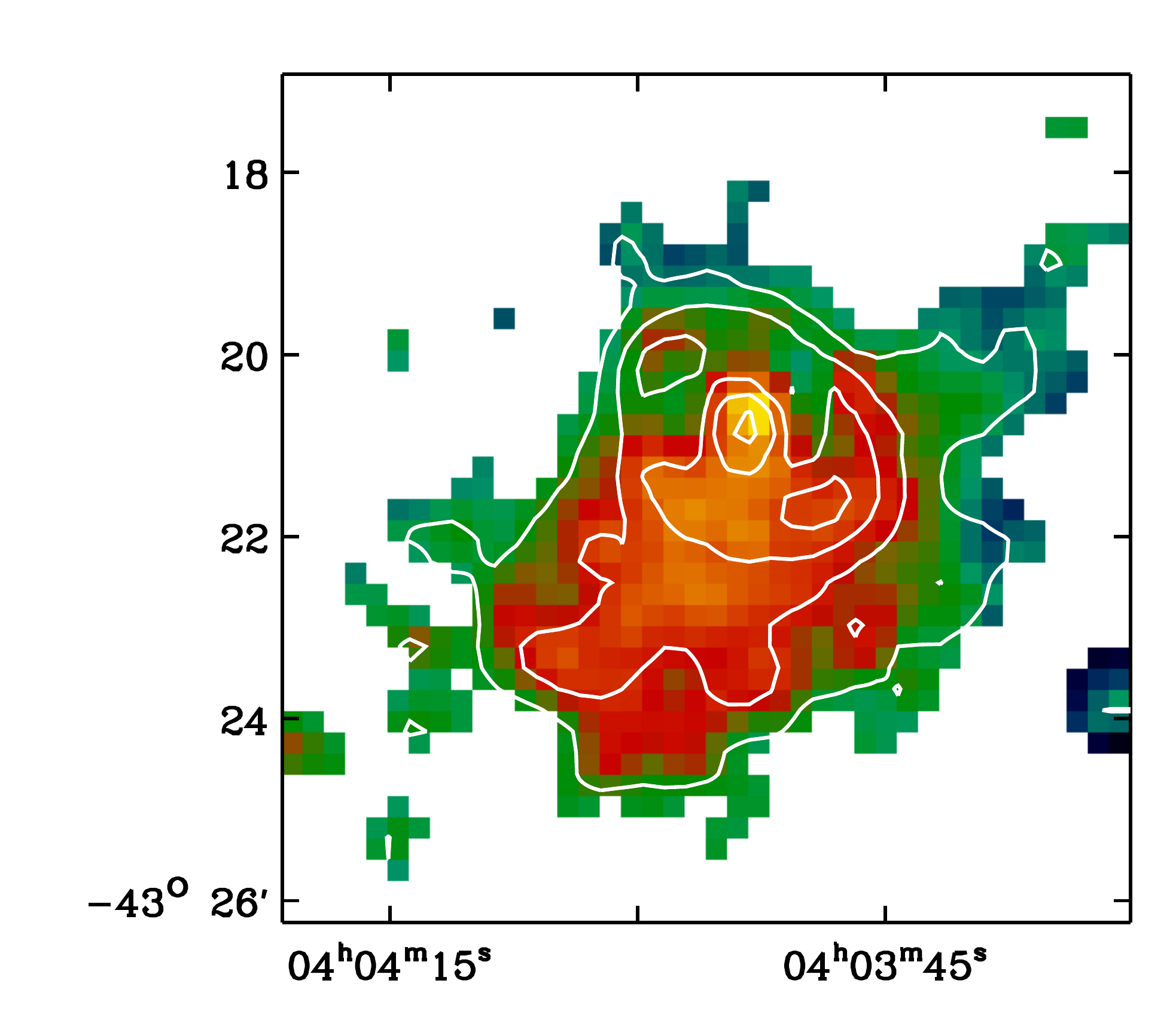}  &
\rotatebox{90}{\includegraphics[width=4cm, height=0.9cm]{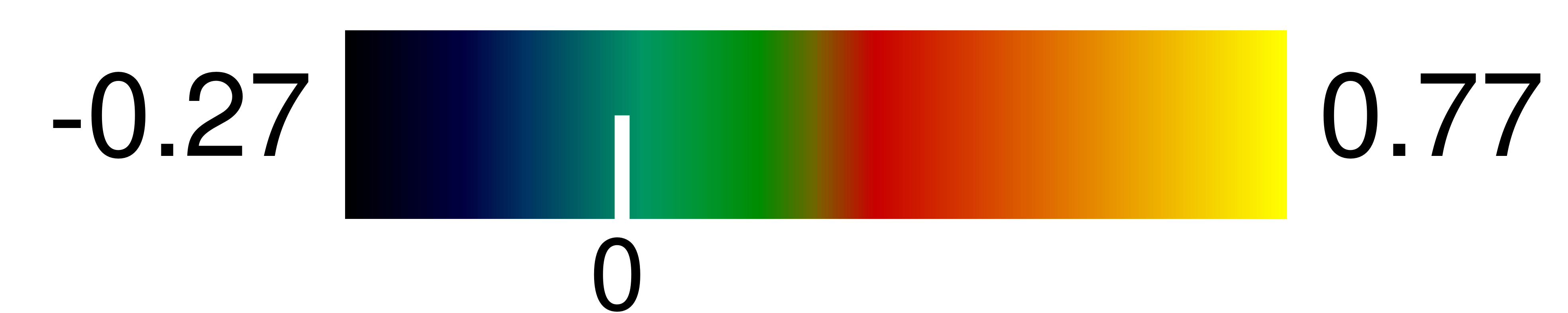}}  \\
	 
\rotatebox{90}{\Large Relative Difference} &
\includegraphics[width=5.7cm]{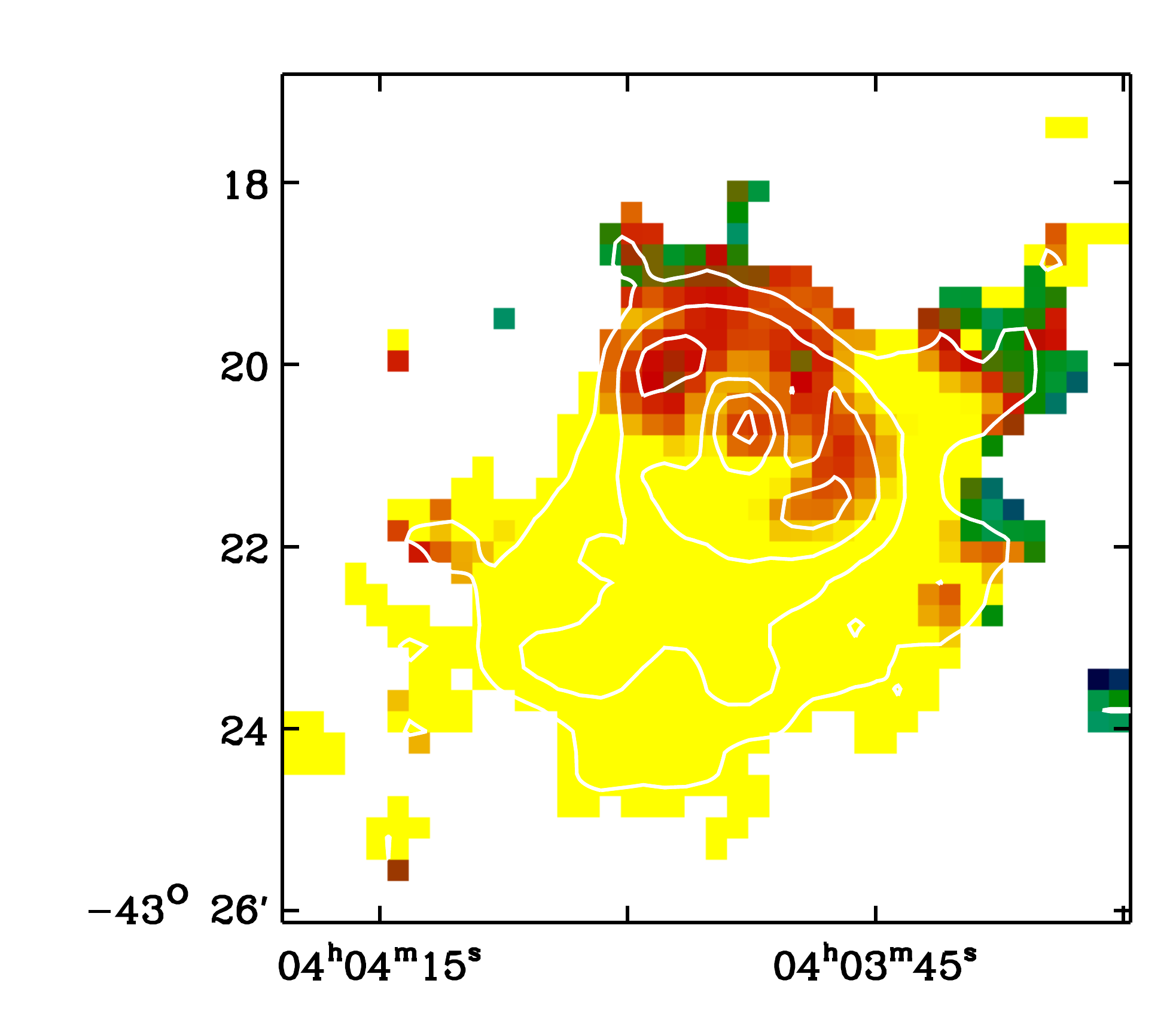} &
\includegraphics[width=5.7cm]{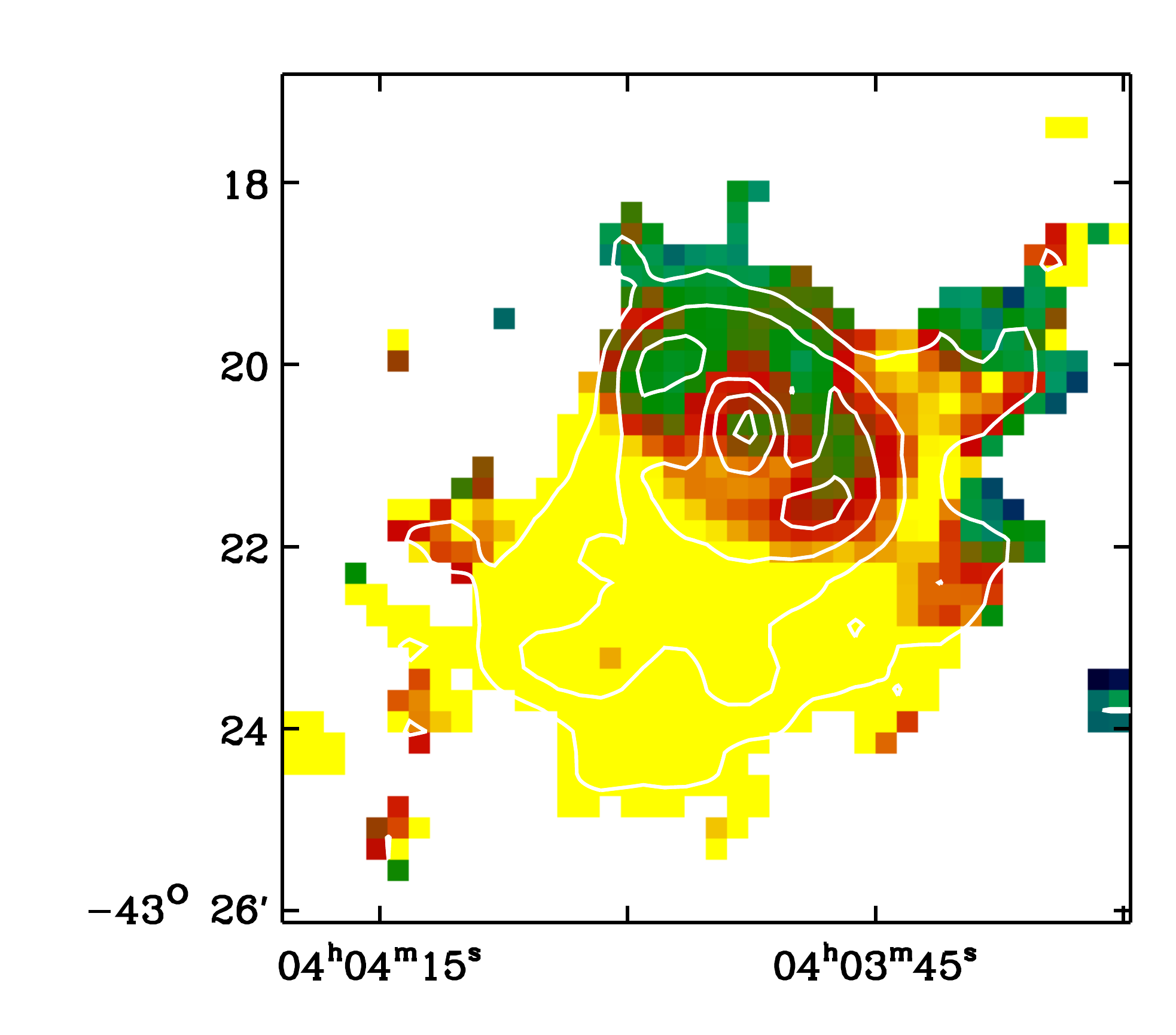} &
\includegraphics[width=5.7cm]{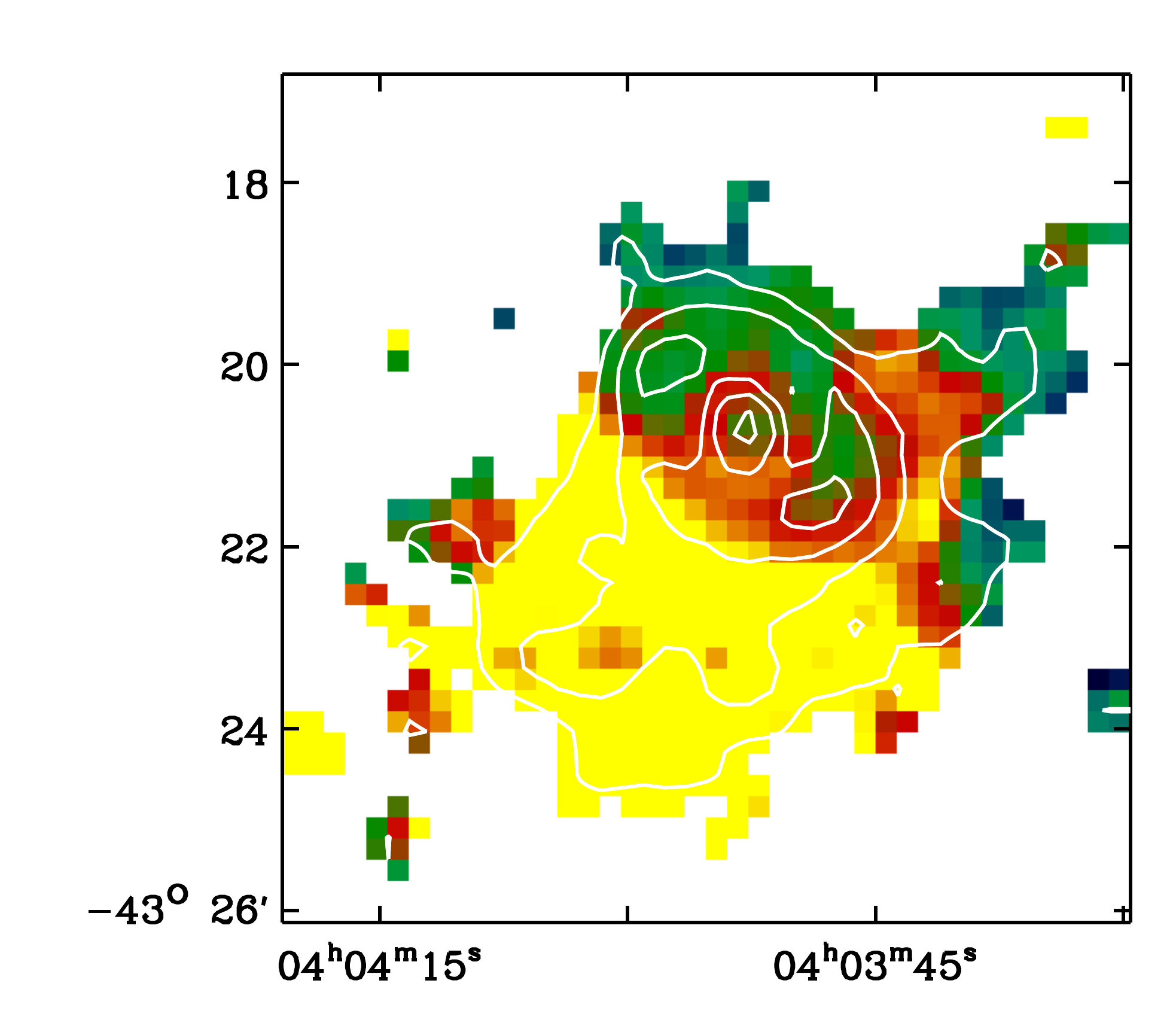}  &
\rotatebox{90}{\includegraphics[width=4cm, height=0.9cm]{RelativeExcess_ColorBars}}  \\	   
\end{tabular}  
\caption{continued. }
\end{figure*}

\newpage
\addtocounter {figure}{-1}
\begin{figure*}
\centering
\begin{tabular}  { m{0cm} m{5.1cm} m{5.1cm} m{5.1cm}  m{0.7cm}}    
{\Large \bf~~~~~~~~~~NGC3351} &&&\\  
&\hspace{5cm}\rotatebox{90}{\Large 870 \mic\ Observed} & 
\includegraphics[width=5.7cm]{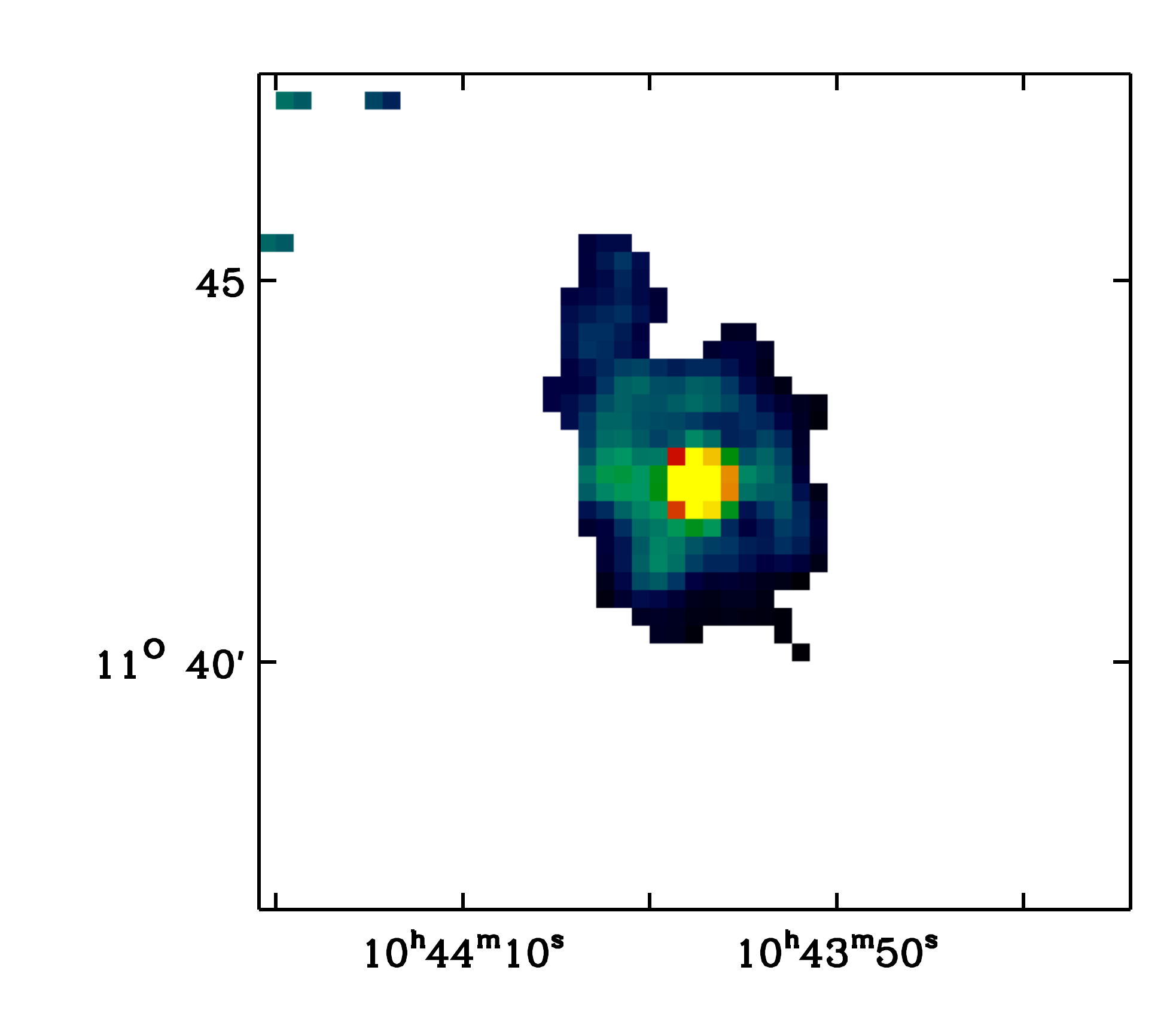} &&
\rotatebox{90}{\includegraphics[width=4cm, height=0.9cm]{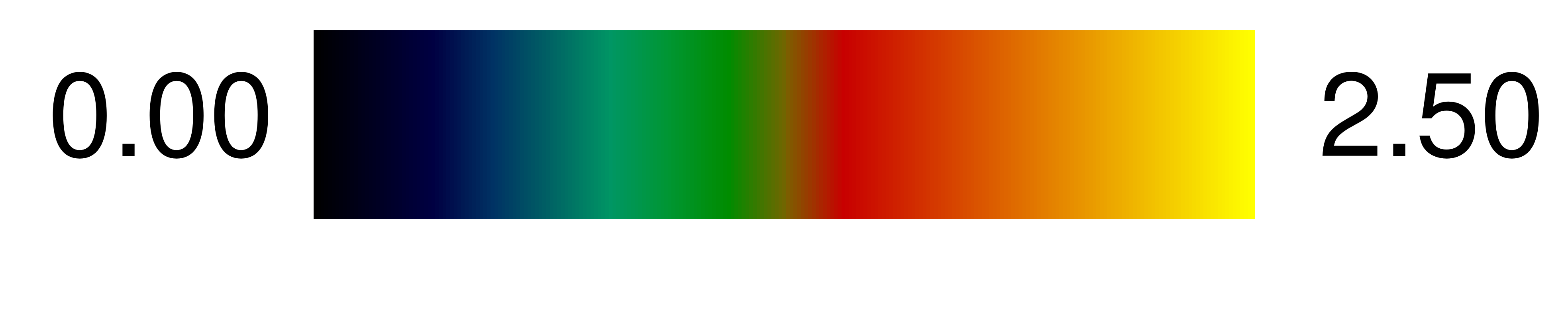}}  \\
&&\\
& {\Large \hspace{2.2cm}$\beta$$_c$ = 2.0 model} & {\Large \hspace{2.2cm}$\beta$$_c$ = 1.5 model}  & {\Large \hspace{2.2cm}[DL07] model} & \\

\rotatebox{90}{\Large 870 \mic\ Modelled} & 
\includegraphics[width=5.7cm]{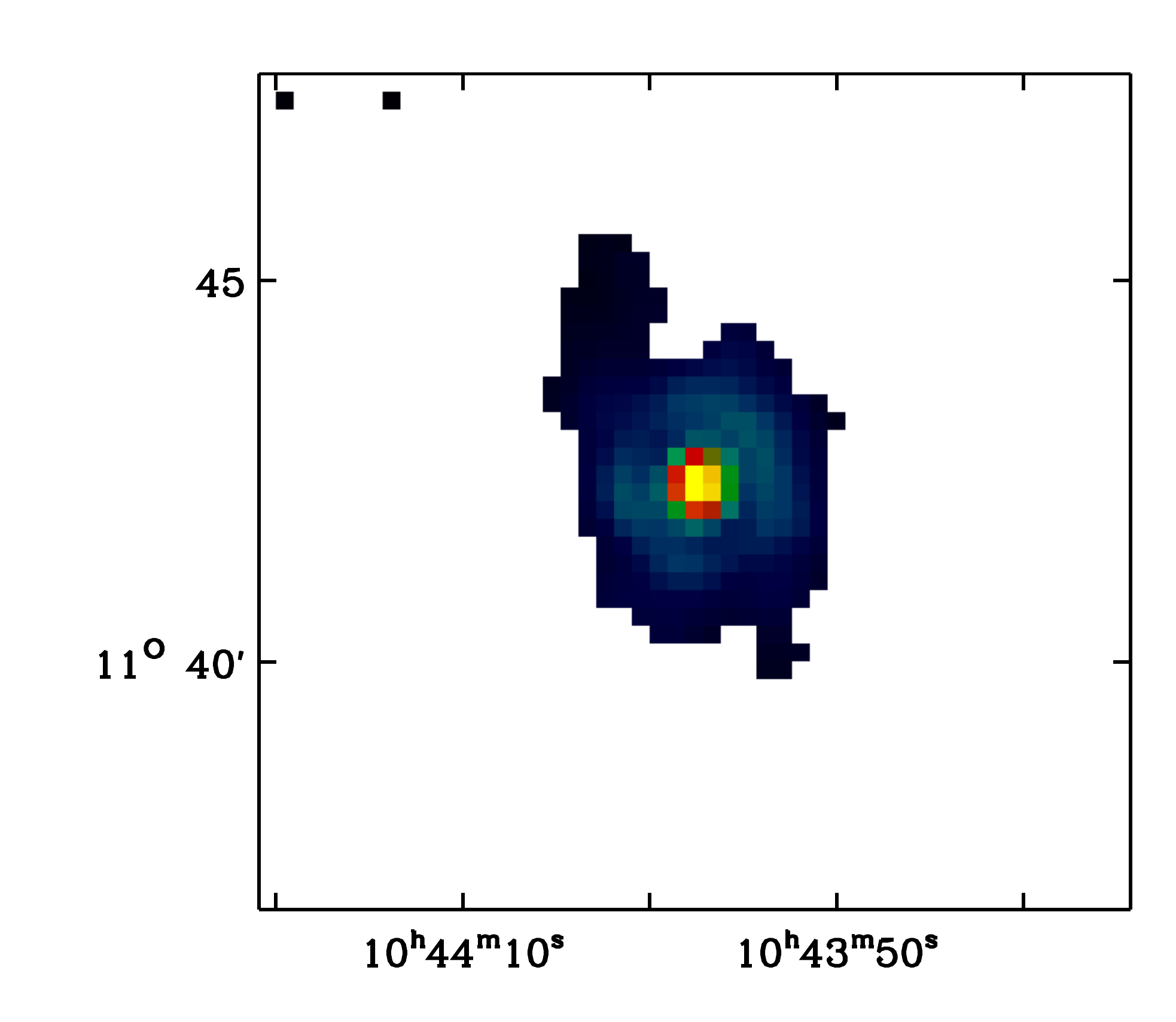} &
\includegraphics[width=5.7cm]{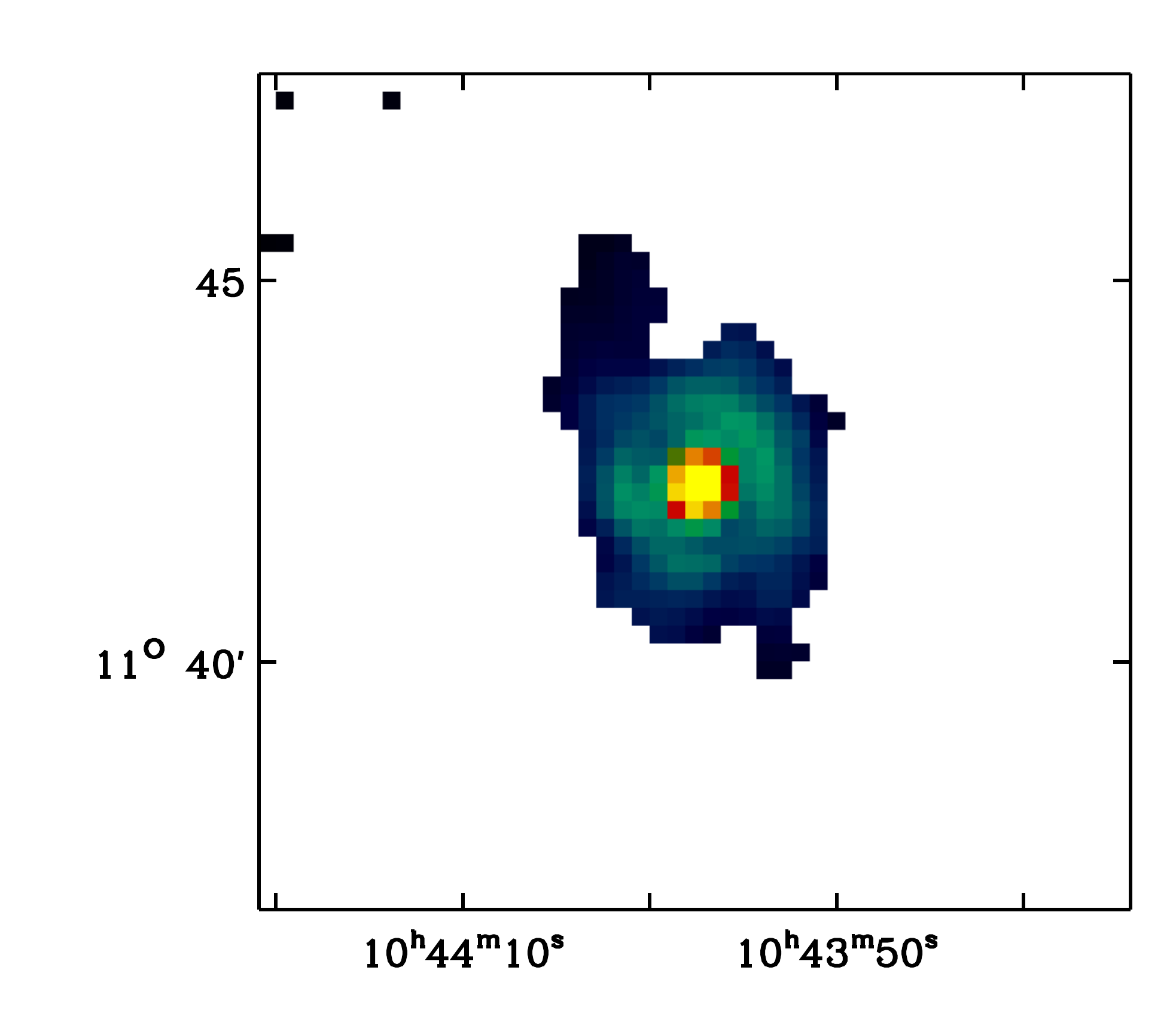} &
\includegraphics[width=5.7cm]{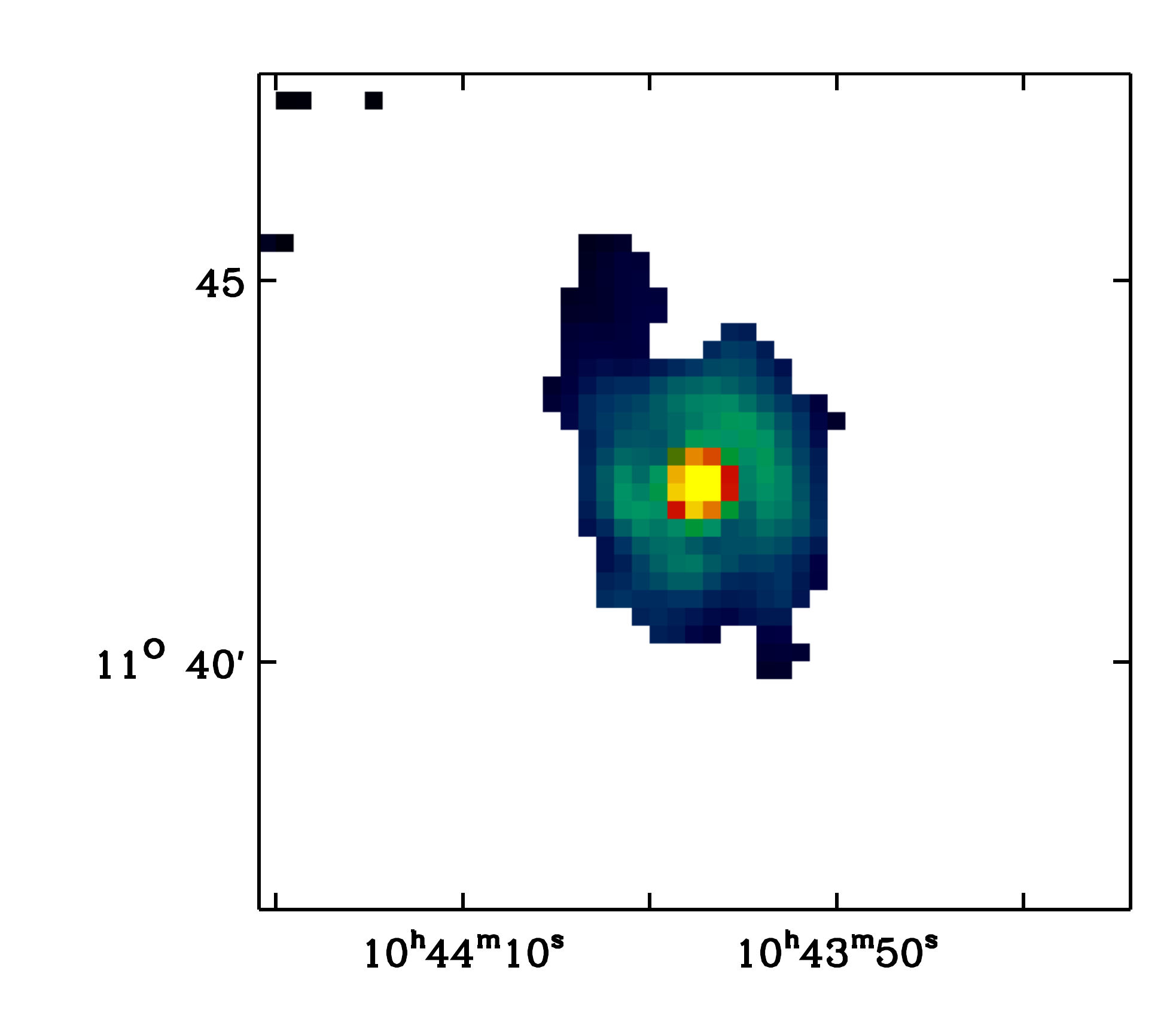}  &
\rotatebox{90}{\includegraphics[width=4cm, height=0.9cm]{NGC3351_Extrap870_ColorBars}}  \\
	
\rotatebox{90}{\Large Absolute Difference} &
\includegraphics[width=5.7cm]{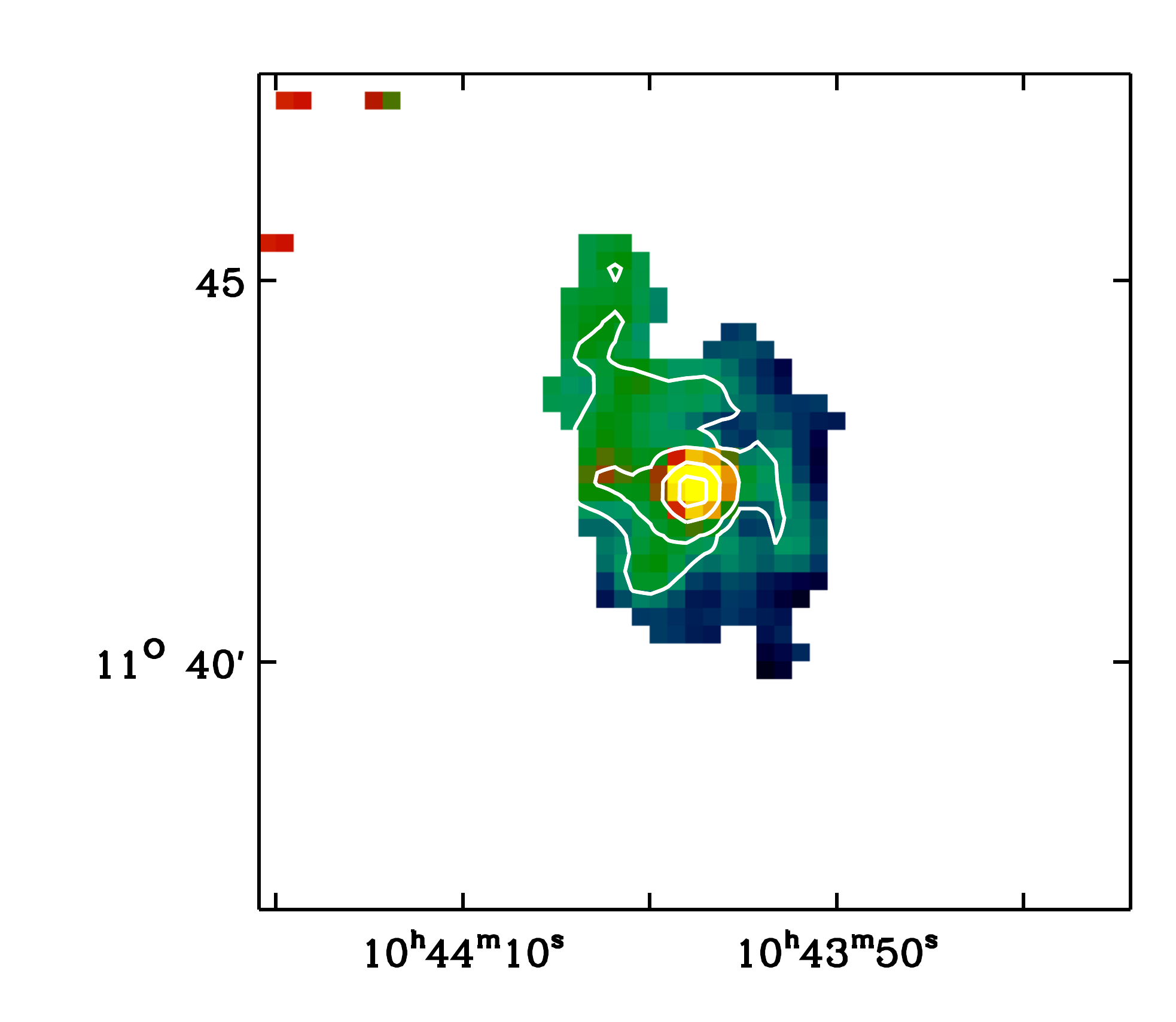} & 
\includegraphics[width=5.7cm]{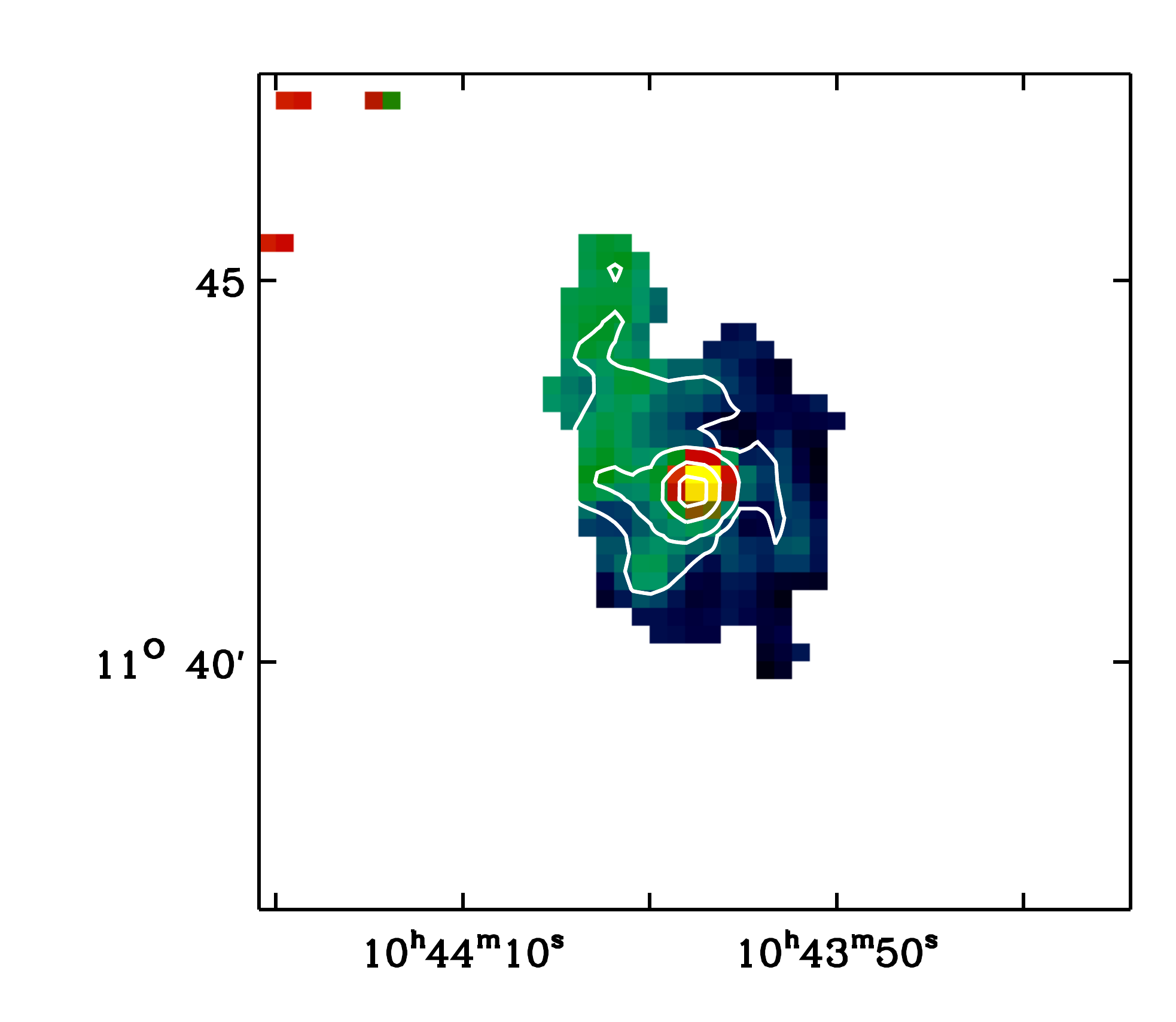} &
\includegraphics[width=5.7cm]{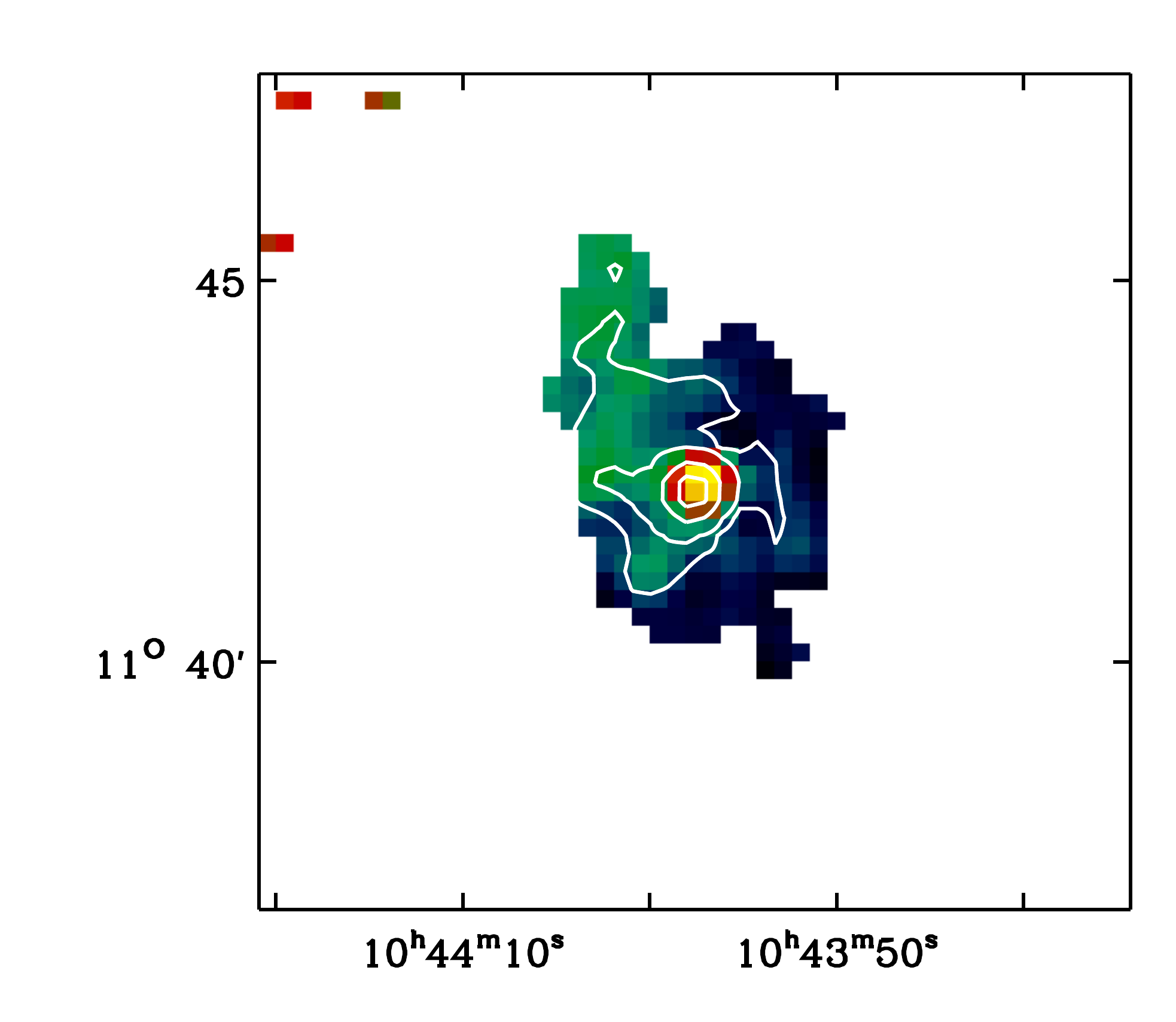}  &
\rotatebox{90}{\includegraphics[width=4cm, height=0.9cm]{NGC3351_Excess_ColorBars}}  \\
	 
\rotatebox{90}{\Large Relative Difference} &
\includegraphics[width=5.7cm]{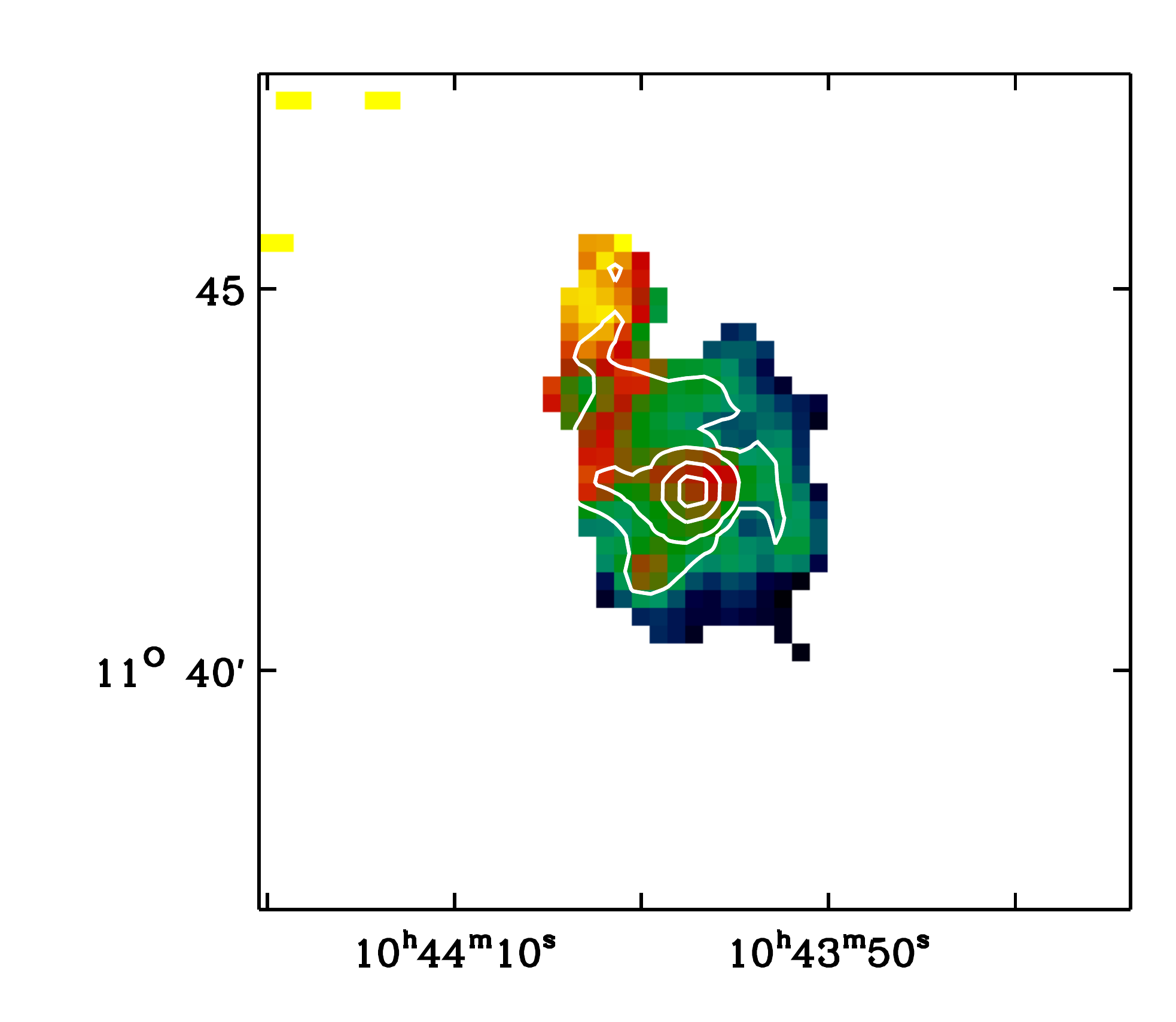} &
\includegraphics[width=5.7cm]{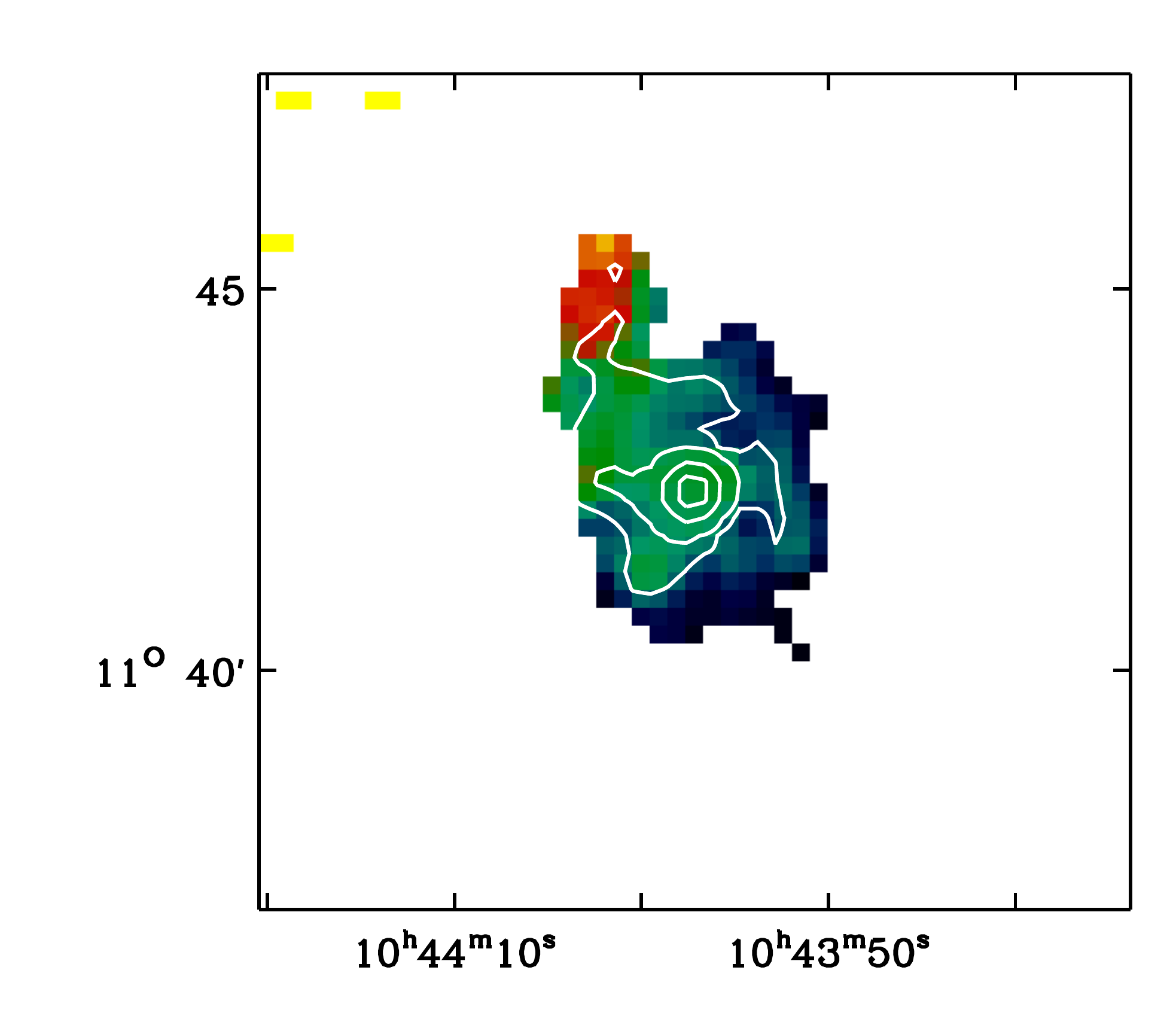} &
\includegraphics[width=5.7cm]{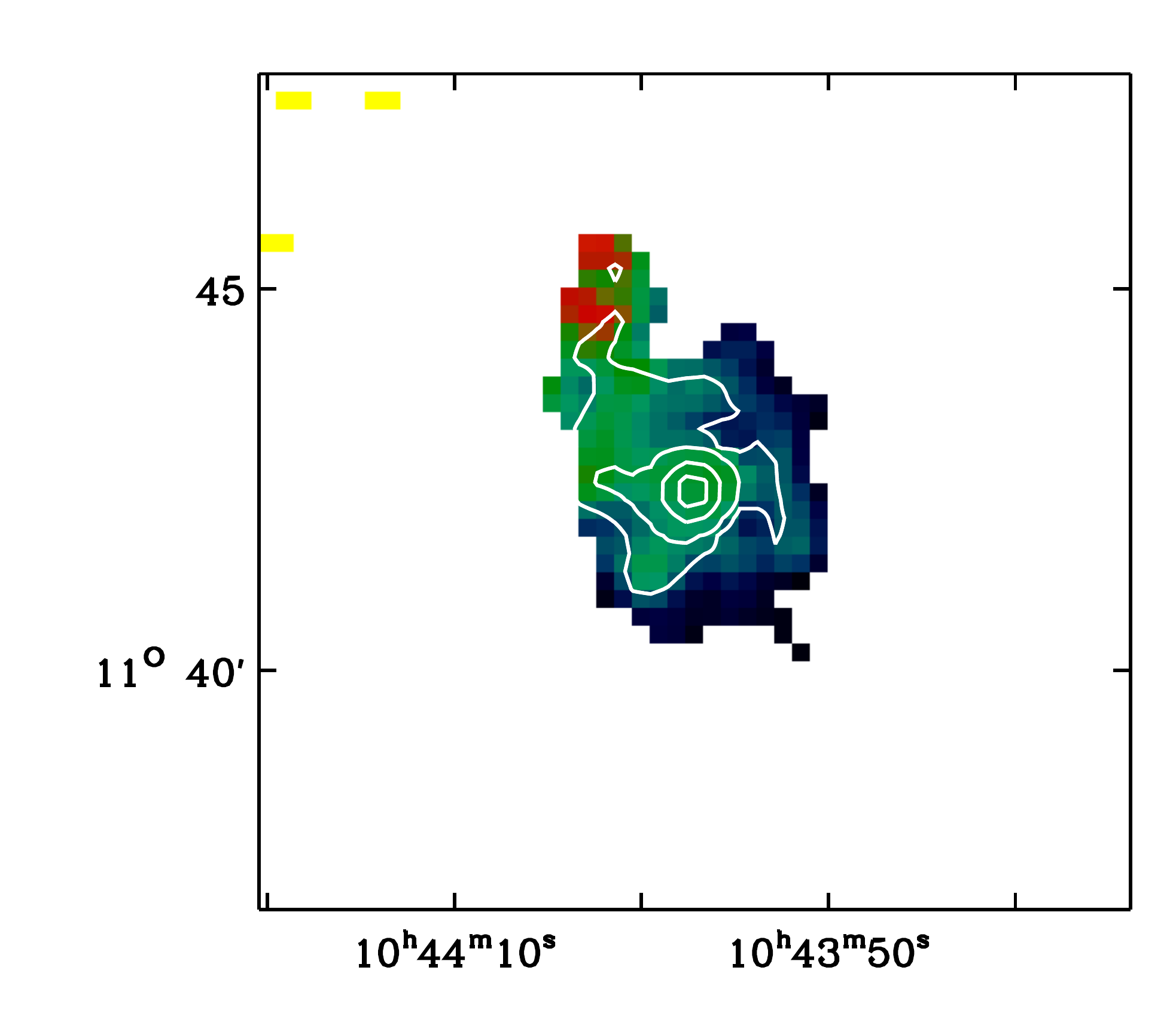}  &
\rotatebox{90}{\includegraphics[width=4cm, height=0.9cm]{RelativeExcess_ColorBars}}  \\	   
\end{tabular}  
\caption{continued. }
\end{figure*}

\newpage
\addtocounter {figure}{-1}
\begin{figure*}
\centering
\begin{tabular}  { m{0cm} m{5.1cm} m{5.1cm} m{5.1cm}  m{0.7cm}}    
{\Large \bf~~~~~~~~~~NGC3621} &&&\\  
&\hspace{5cm}\rotatebox{90}{\Large 870 \mic\ Observed} & 
\includegraphics[width=5.7cm]{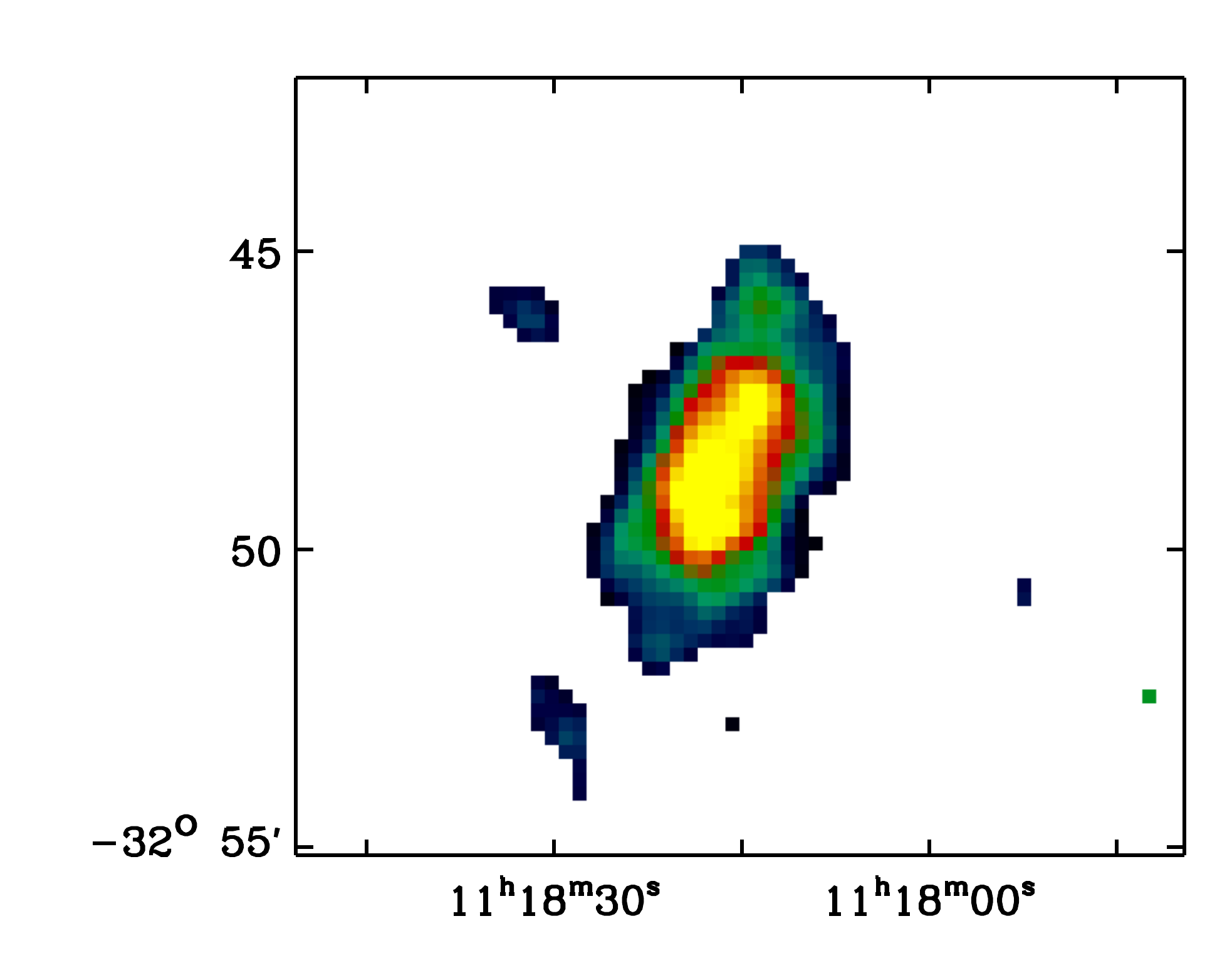} &&
\rotatebox{90}{\includegraphics[width=4cm, height=0.9cm]{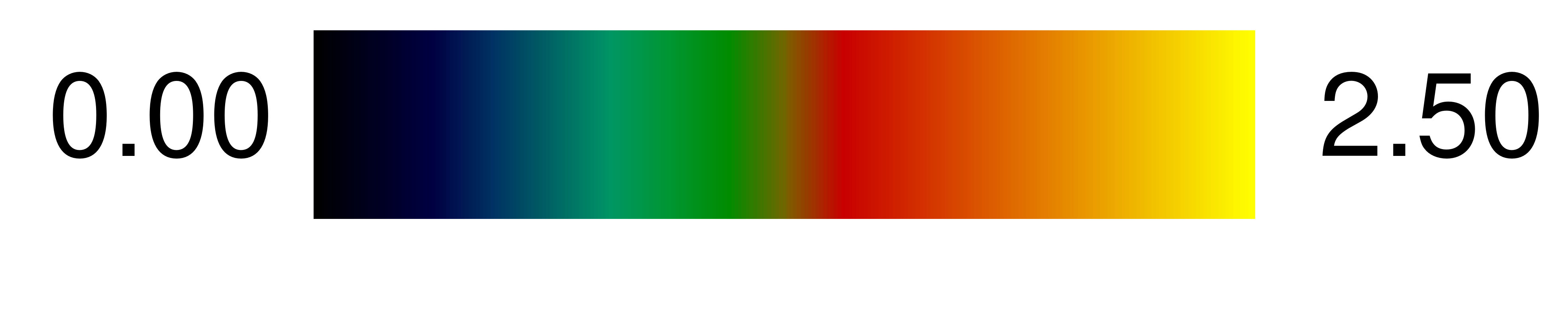}}  \\
&&\\
& {\Large \hspace{2.2cm}$\beta$$_c$ = 2.0 model} & {\Large \hspace{2.2cm}$\beta$$_c$ = 1.5 model}  & {\Large \hspace{2.2cm}[DL07] model} & \\

\rotatebox{90}{\Large 870 \mic\ Modelled} & 
\includegraphics[width=5.7cm]{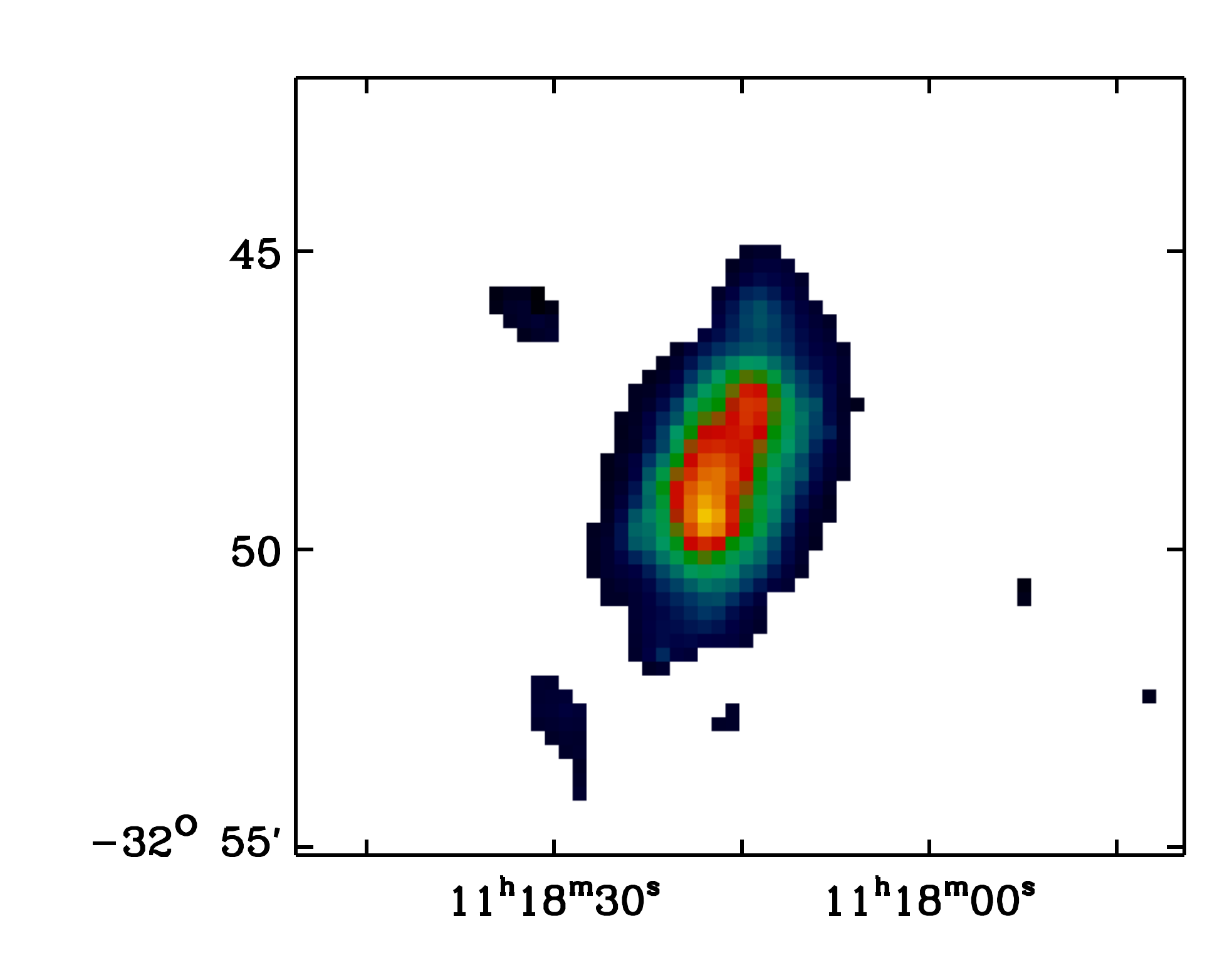} &
\includegraphics[width=5.7cm]{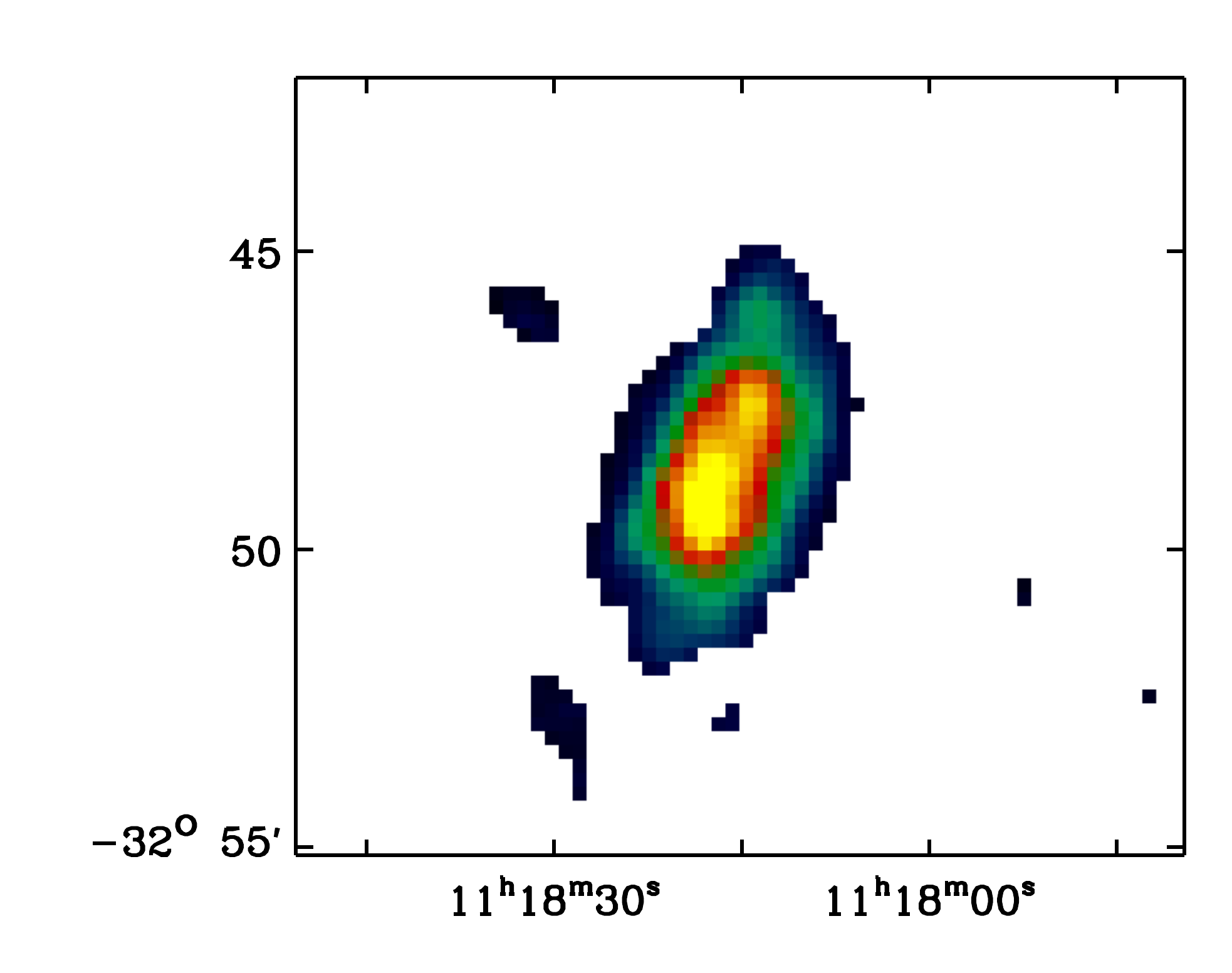} &
\includegraphics[width=5.7cm]{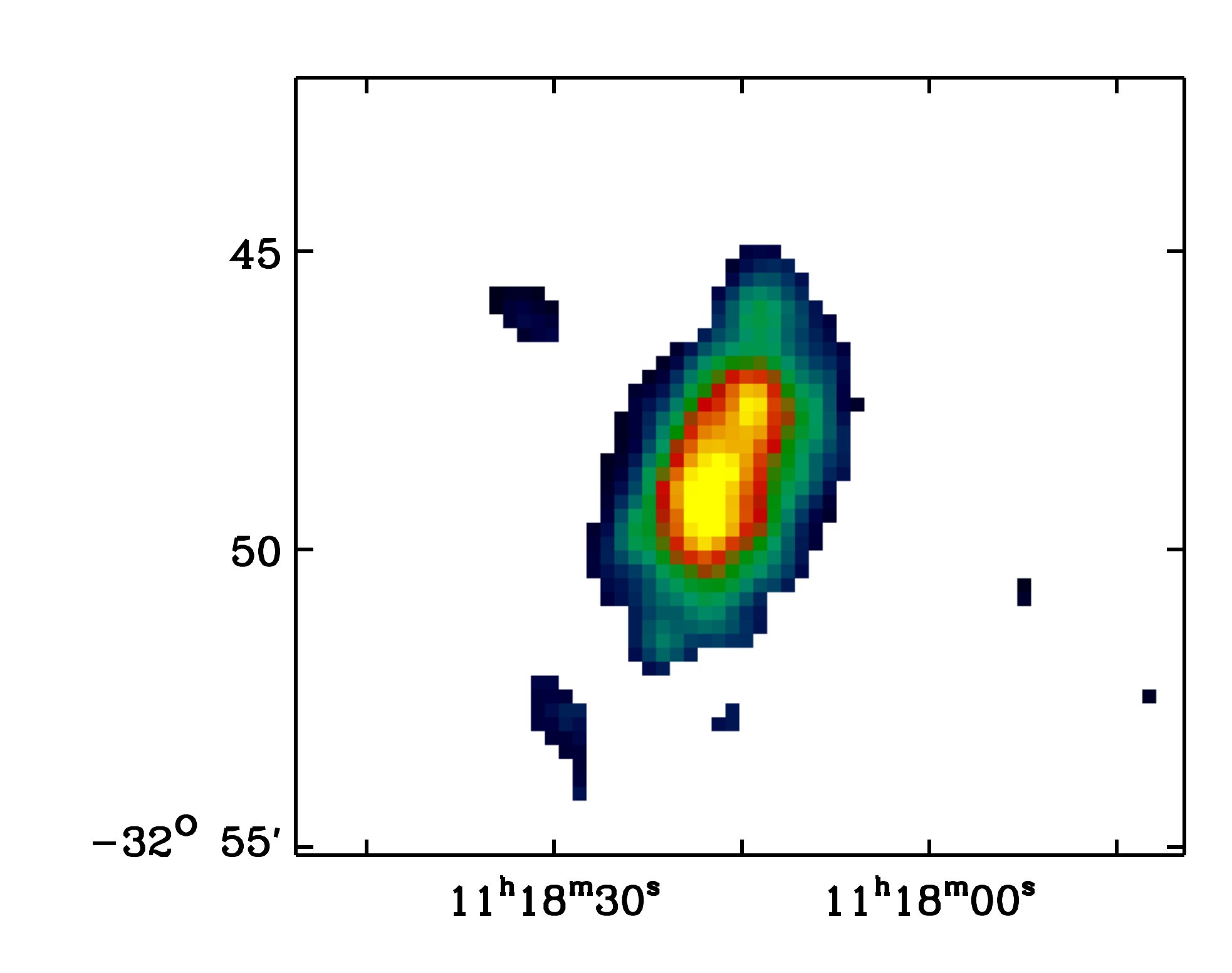}  &
\rotatebox{90}{\includegraphics[width=4cm, height=0.9cm]{NGC3621_Extrap870_ColorBars}}  \\
	
\rotatebox{90}{\Large Absolute Difference} &
\includegraphics[width=5.7cm]{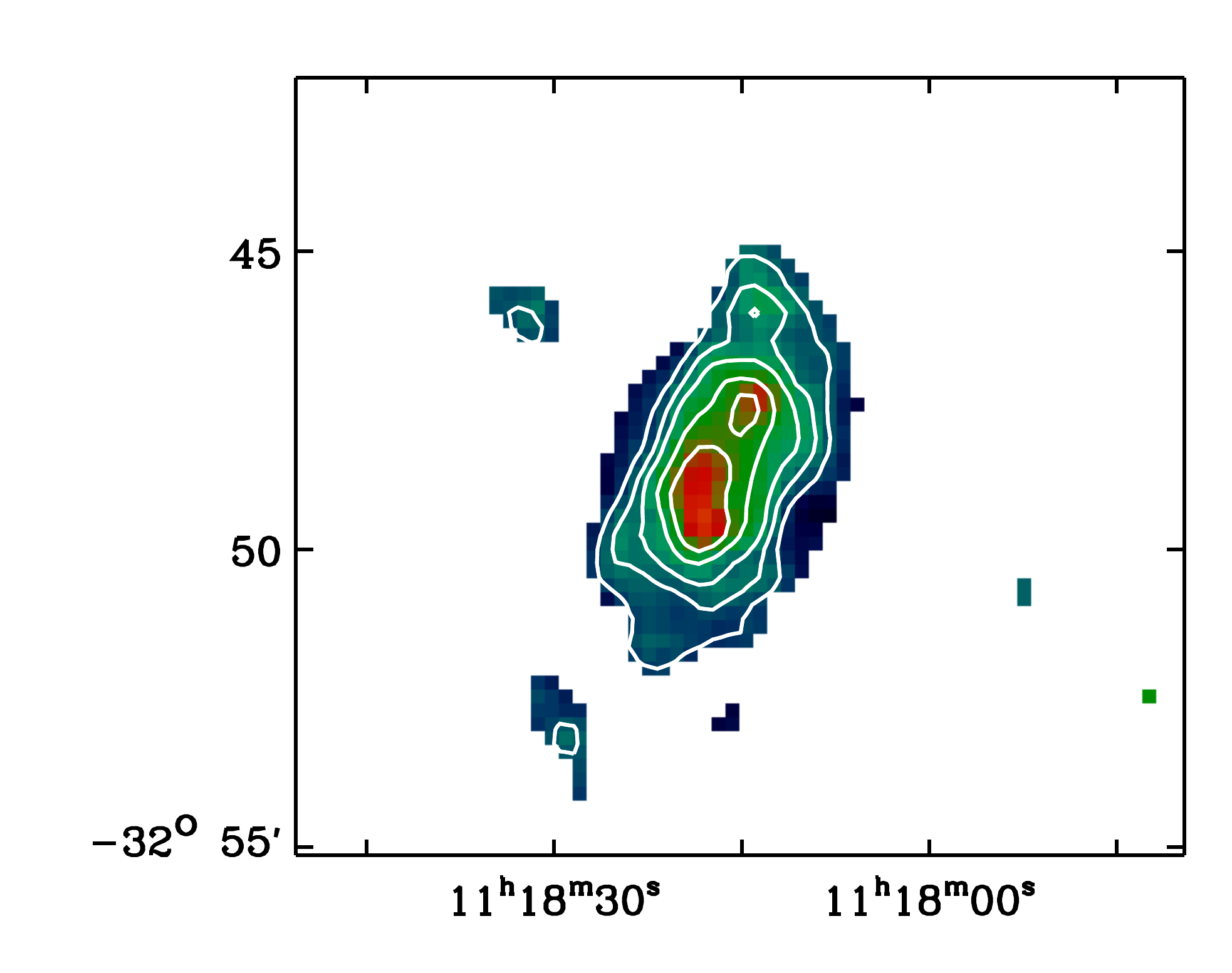} & 
\includegraphics[width=5.7cm]{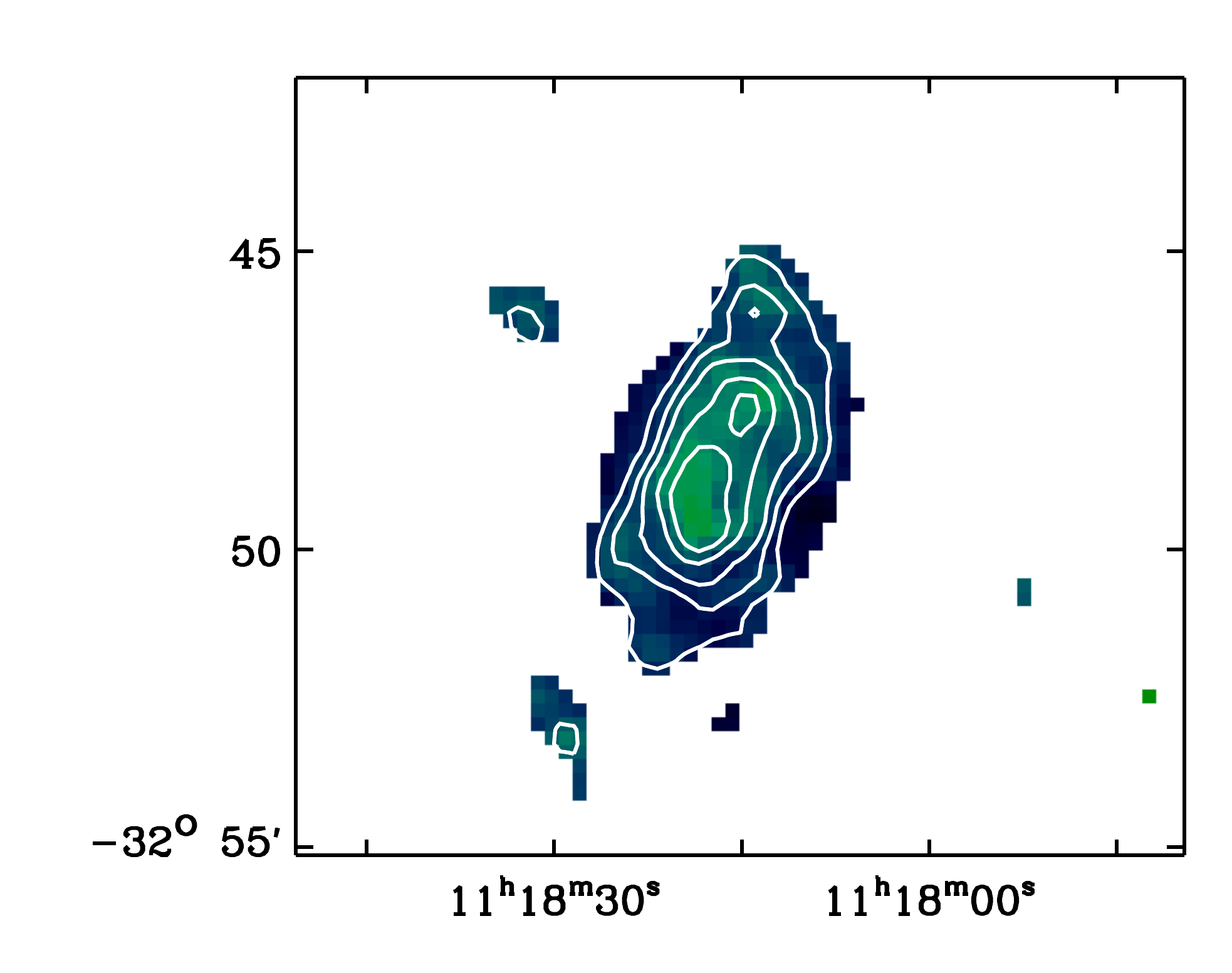} &
\includegraphics[width=5.7cm]{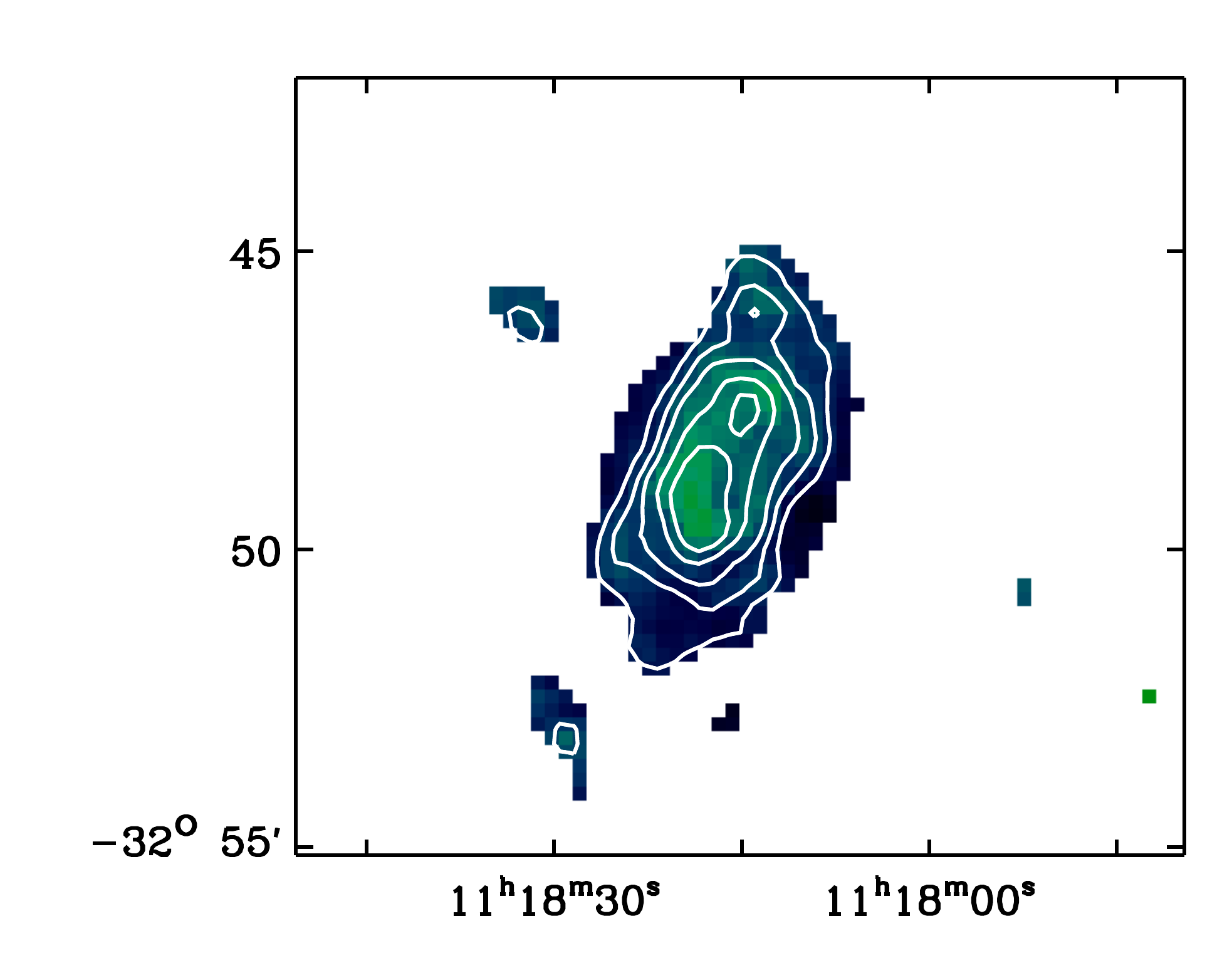}  &
\rotatebox{90}{\includegraphics[width=4cm, height=0.9cm]{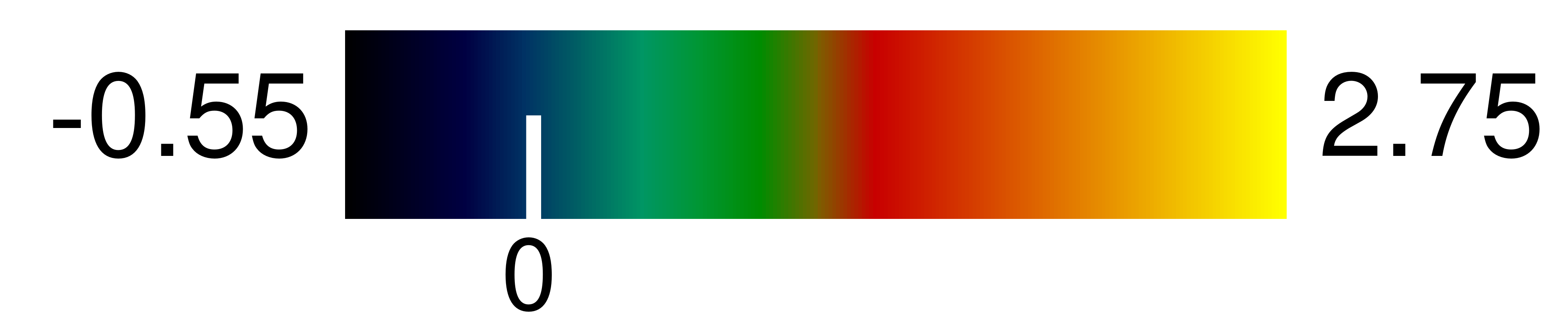}}  \\
	 
\rotatebox{90}{\Large Relative Difference} &
\includegraphics[width=5.7cm]{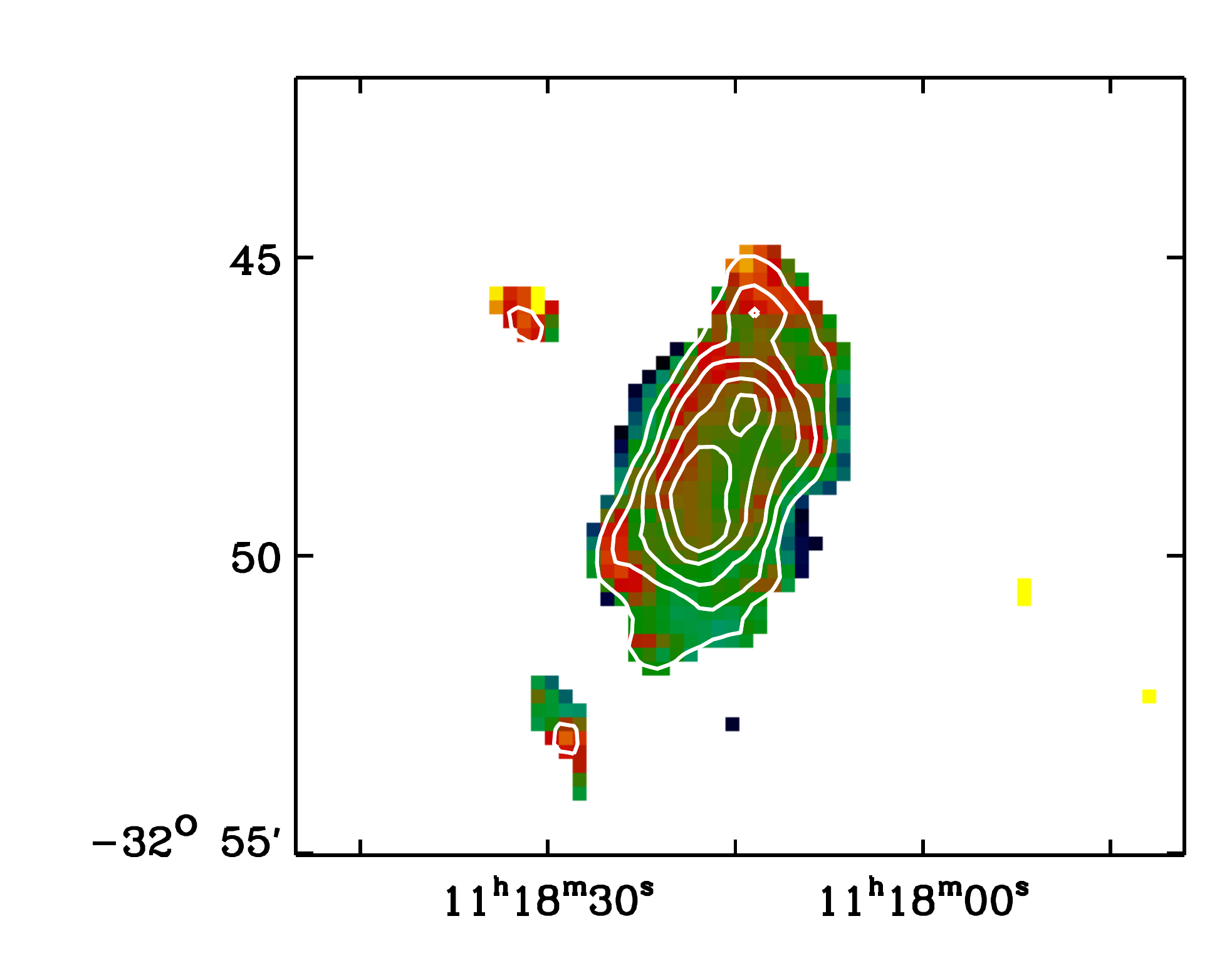} &
\includegraphics[width=5.7cm]{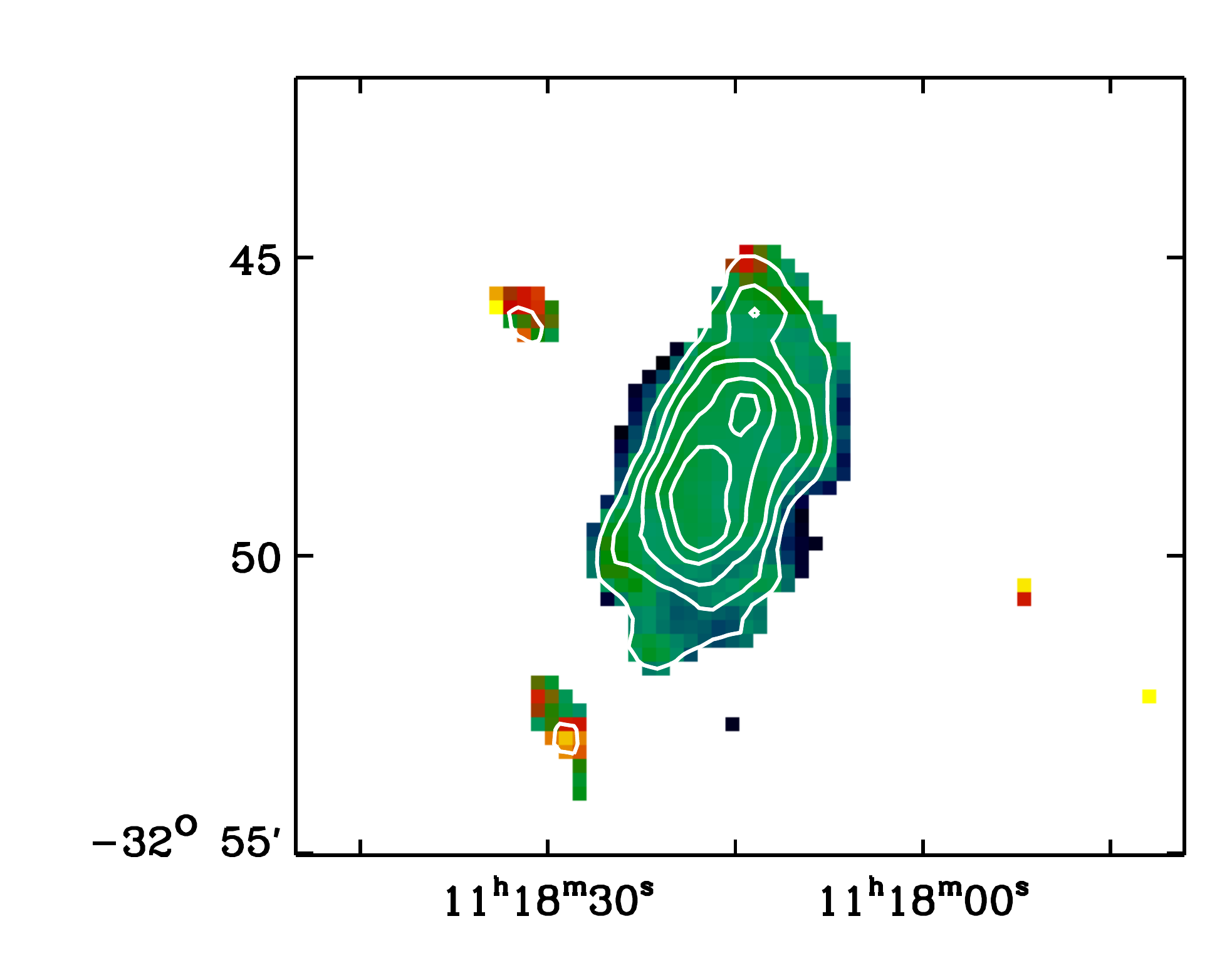} &
\includegraphics[width=5.7cm]{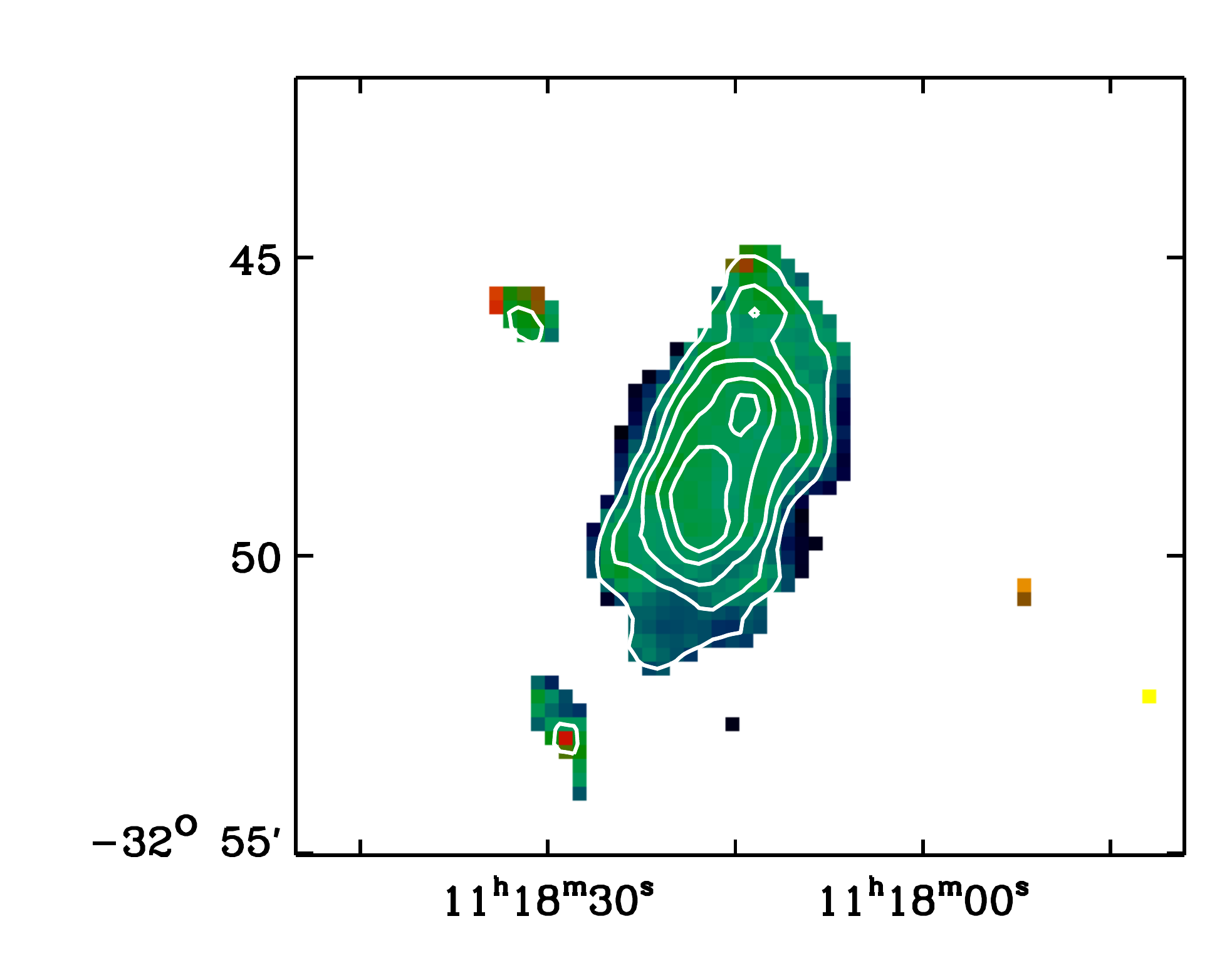}  &
\rotatebox{90}{\includegraphics[width=4cm, height=0.9cm]{RelativeExcess_ColorBars}}  \\	   
\end{tabular}  
\caption{continued. }
\end{figure*}

\newpage
\addtocounter {figure}{-1}
\begin{figure*}
\centering
\begin{tabular}  { m{0cm} m{5.1cm} m{5.1cm} m{5.1cm}  m{0.7cm}}    
{\Large \bf~~~~~~~~~~NGC4826} &&&\\  
&\hspace{5cm}\rotatebox{90}{\Large 870 \mic\ Observed} & 
\includegraphics[width=5.7cm]{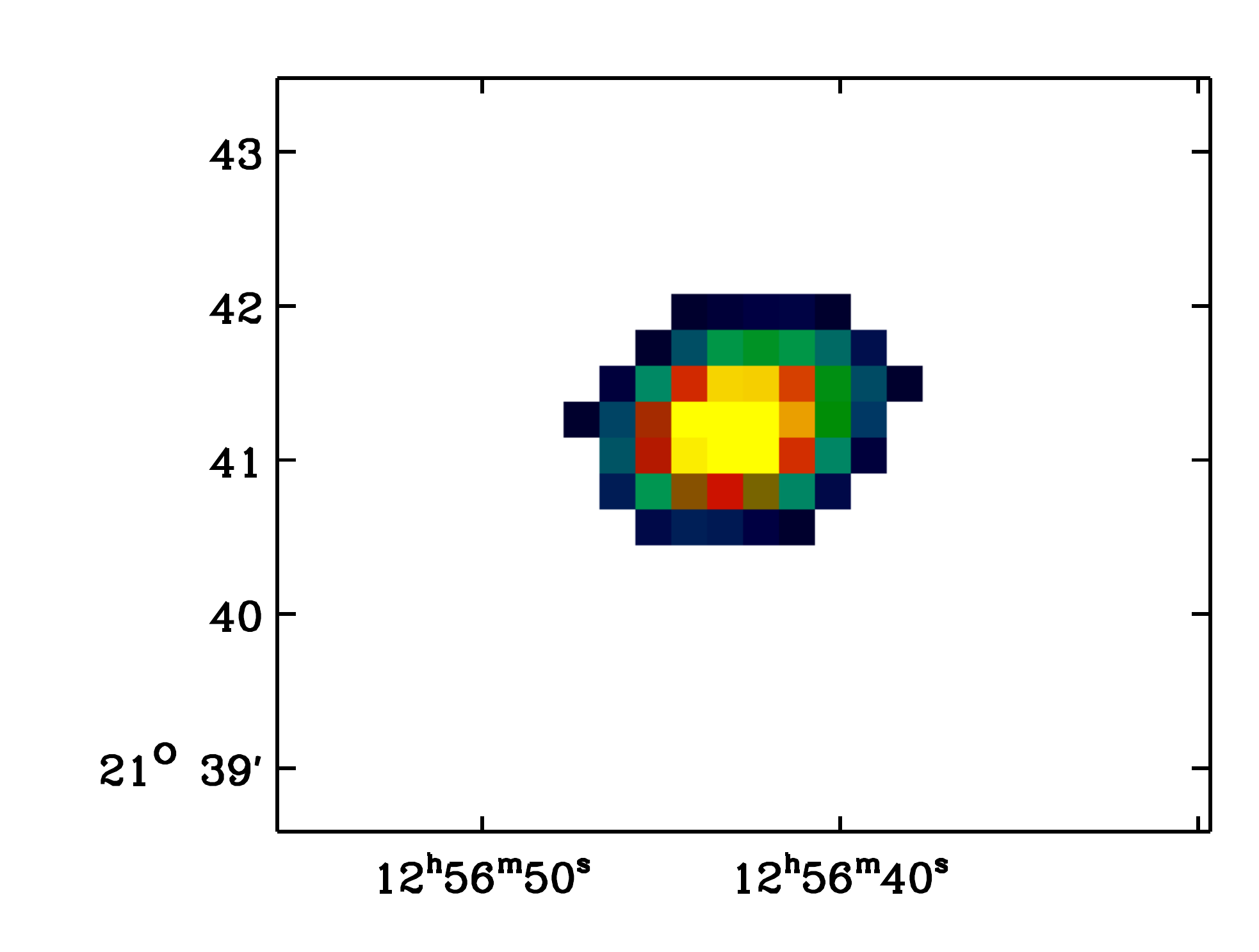} &&
\rotatebox{90}{\includegraphics[width=4cm, height=0.9cm]{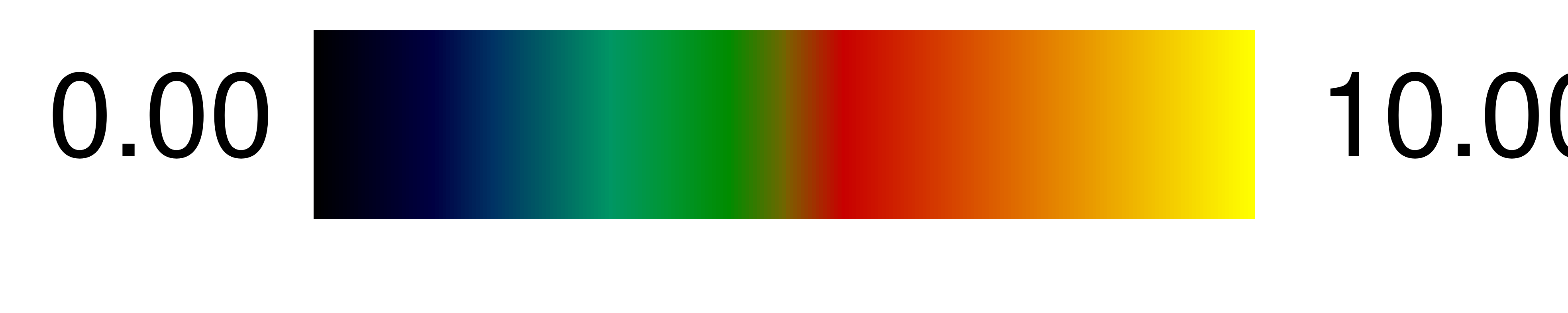}}  \\
&&\\
& {\Large \hspace{2.2cm}$\beta$$_c$ = 2.0 model} & {\Large \hspace{2.2cm}$\beta$$_c$ = 1.5 model}  & {\Large \hspace{2.2cm}[DL07] model} & \\

\rotatebox{90}{\Large 870 \mic\ Modelled} & 
\includegraphics[width=5.7cm]{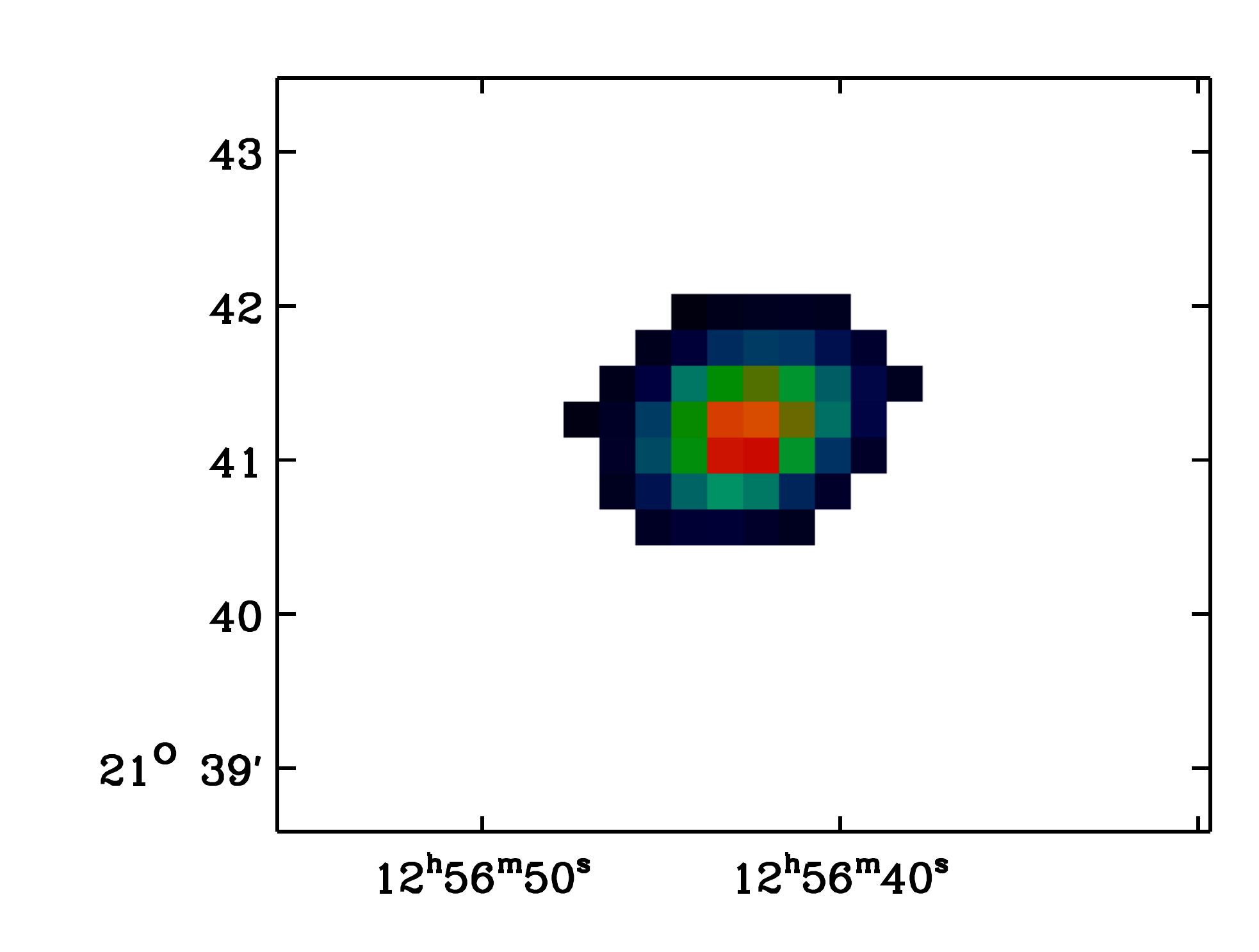} &
\includegraphics[width=5.7cm]{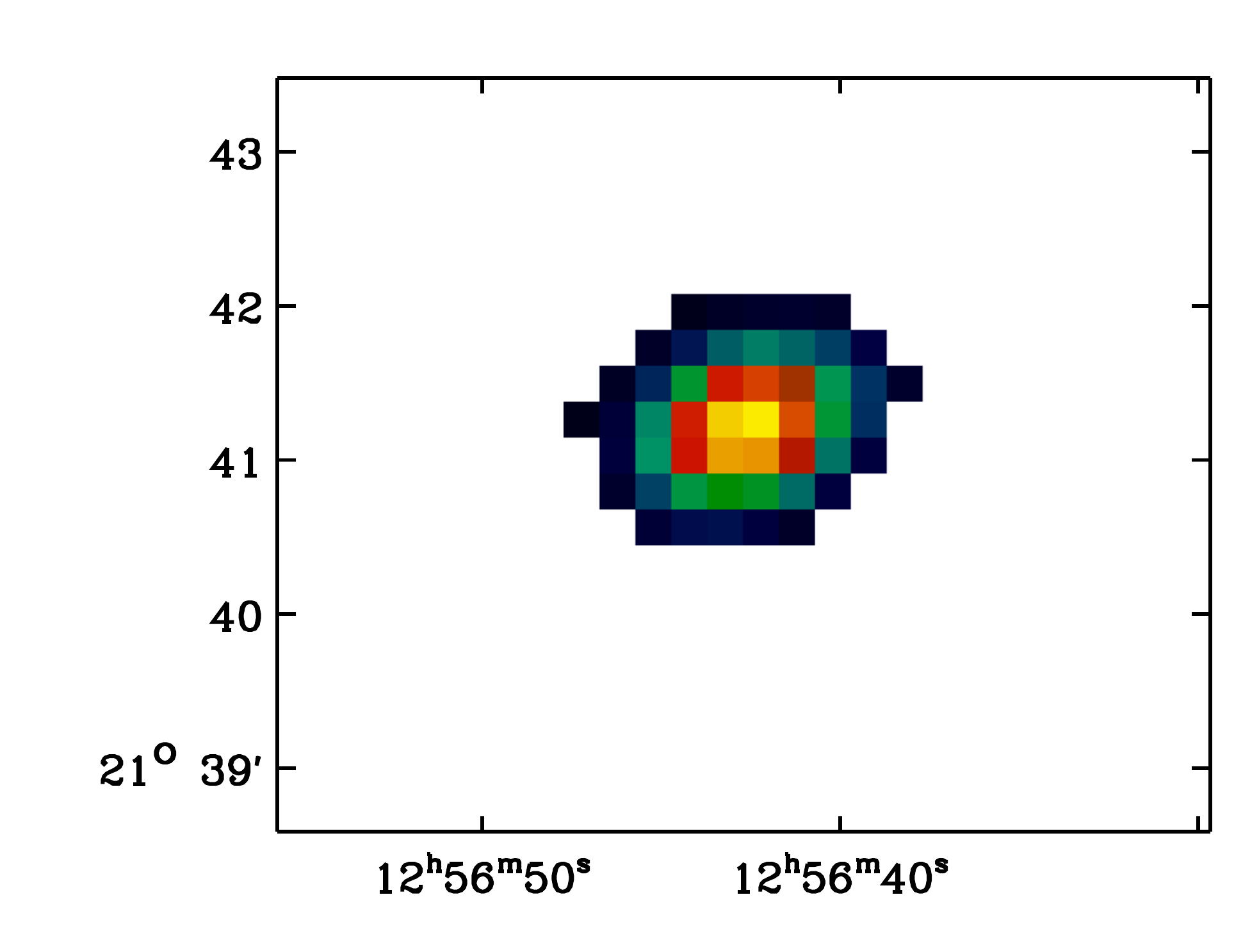} &
\includegraphics[width=5.7cm]{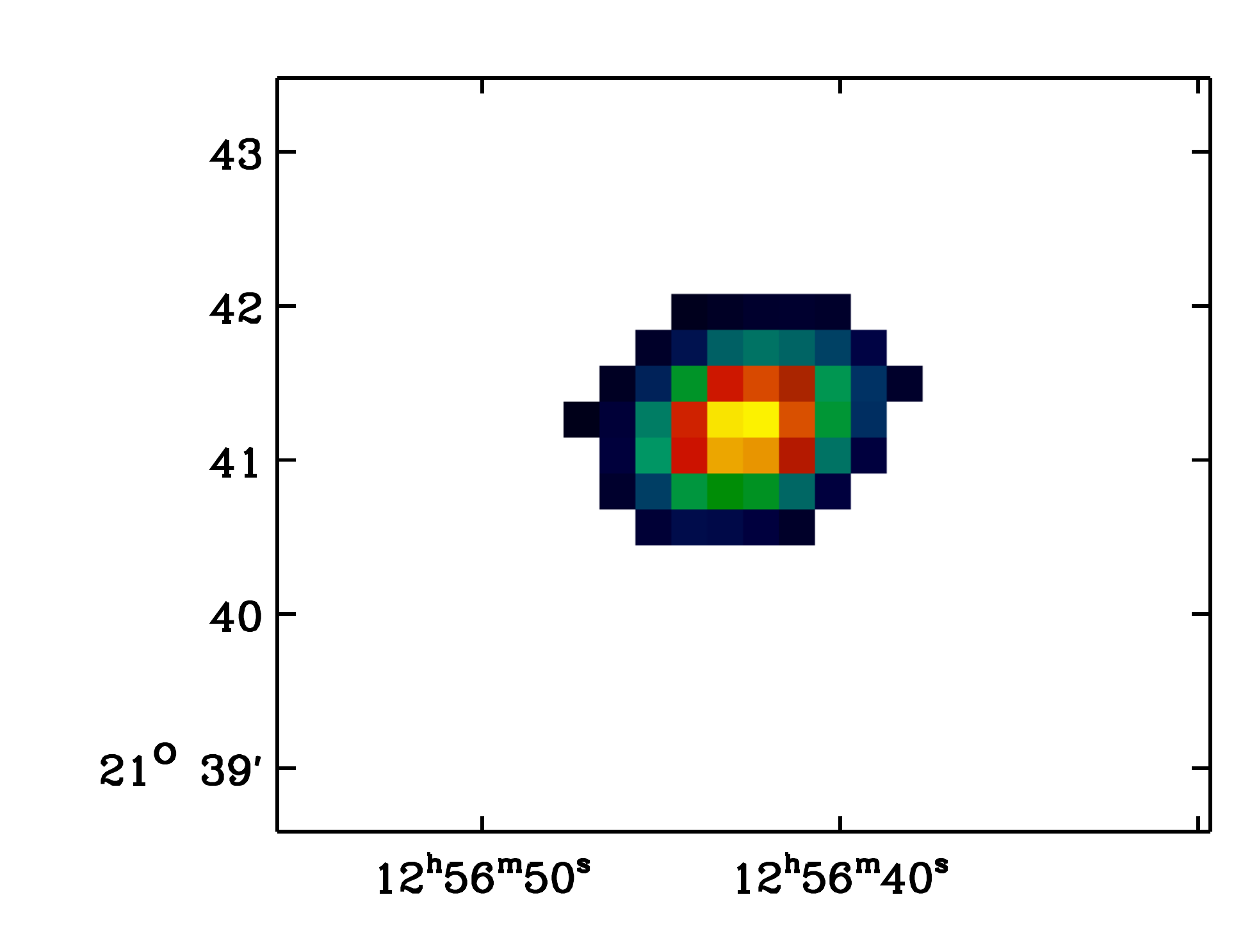}  &
\rotatebox{90}{\includegraphics[width=4cm, height=0.9cm]{NGC4826_Extrap870_ColorBars}}  \\
	
\rotatebox{90}{\Large Absolute Difference} &
\includegraphics[width=5.7cm]{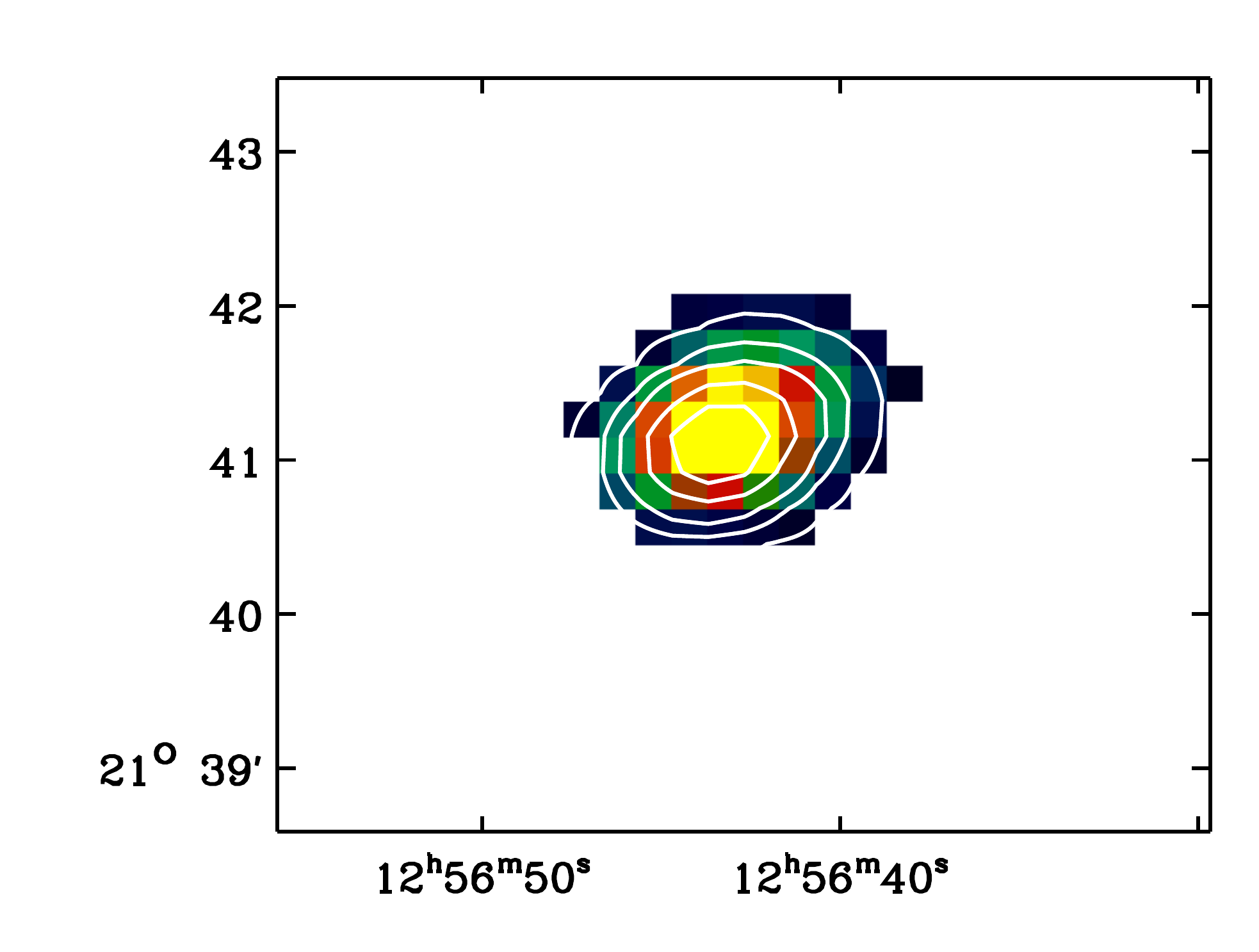} & 
\includegraphics[width=5.7cm]{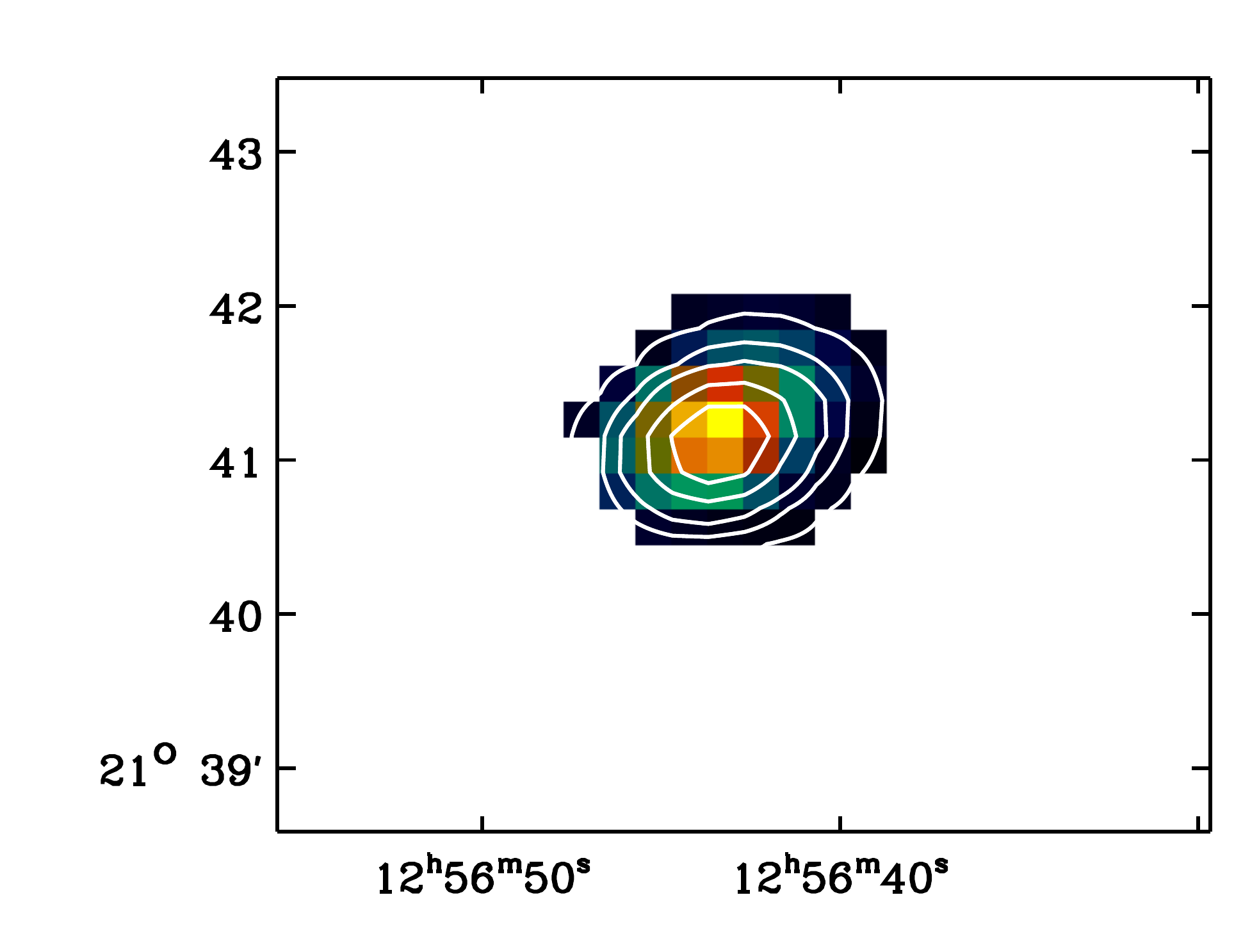} &
\includegraphics[width=5.7cm]{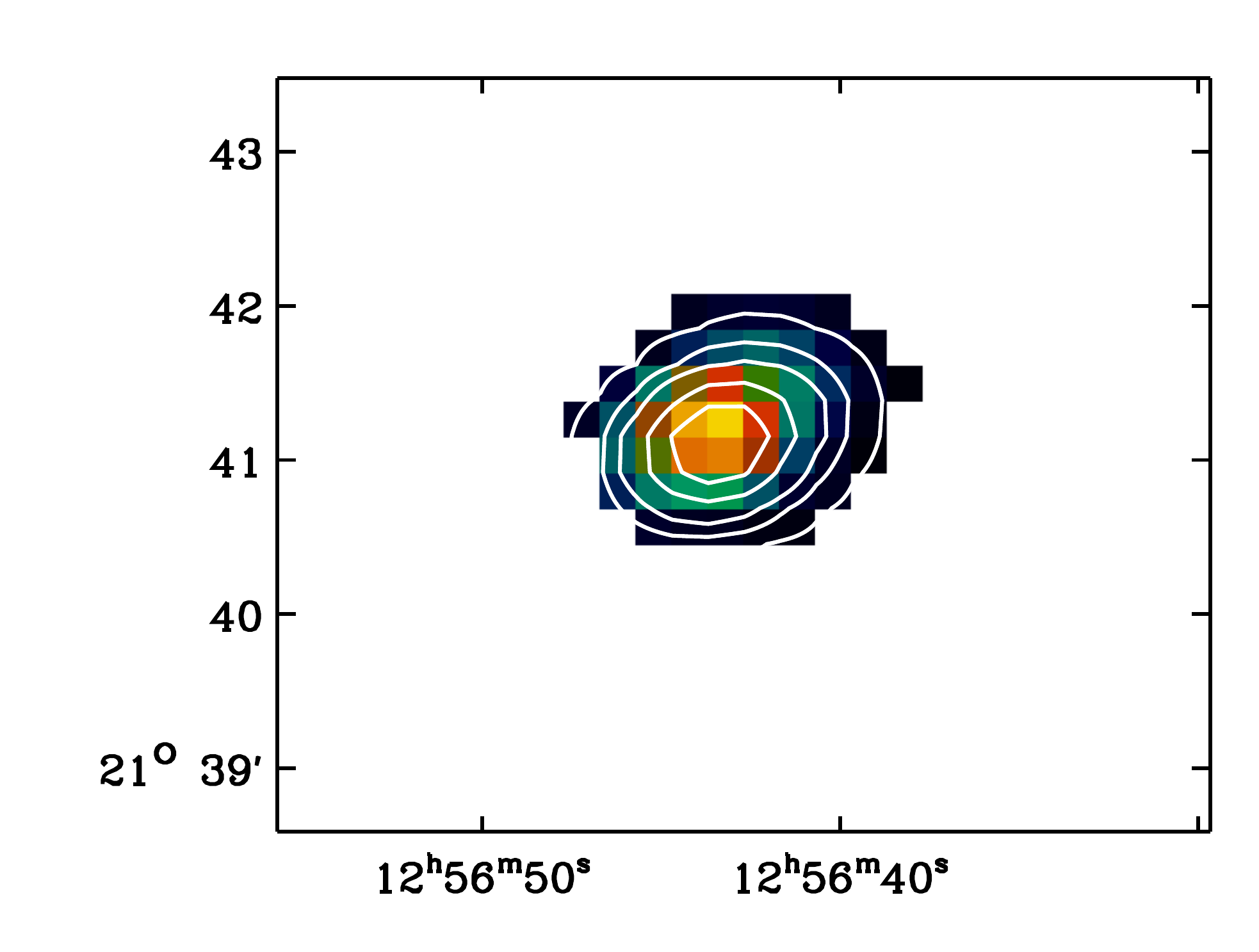}  &
\rotatebox{90}{\includegraphics[width=4cm, height=0.9cm]{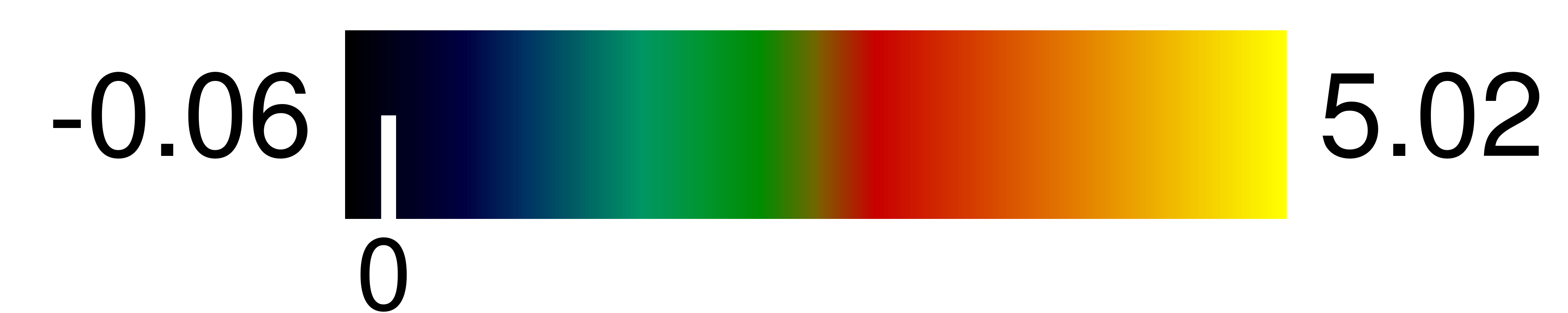}}  \\
	 
\rotatebox{90}{\Large Relative Difference} &
\includegraphics[width=5.7cm]{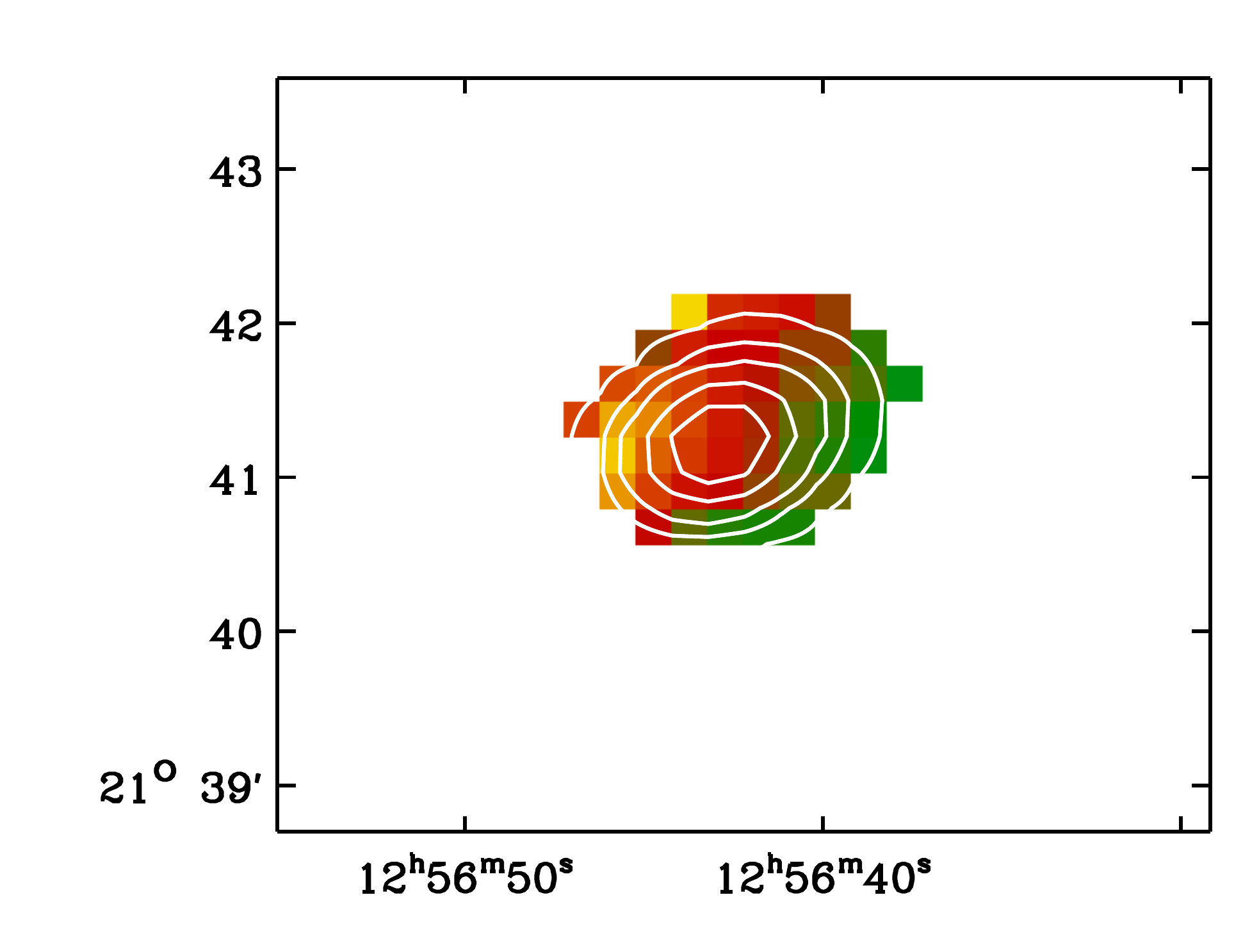} &
\includegraphics[width=5.7cm]{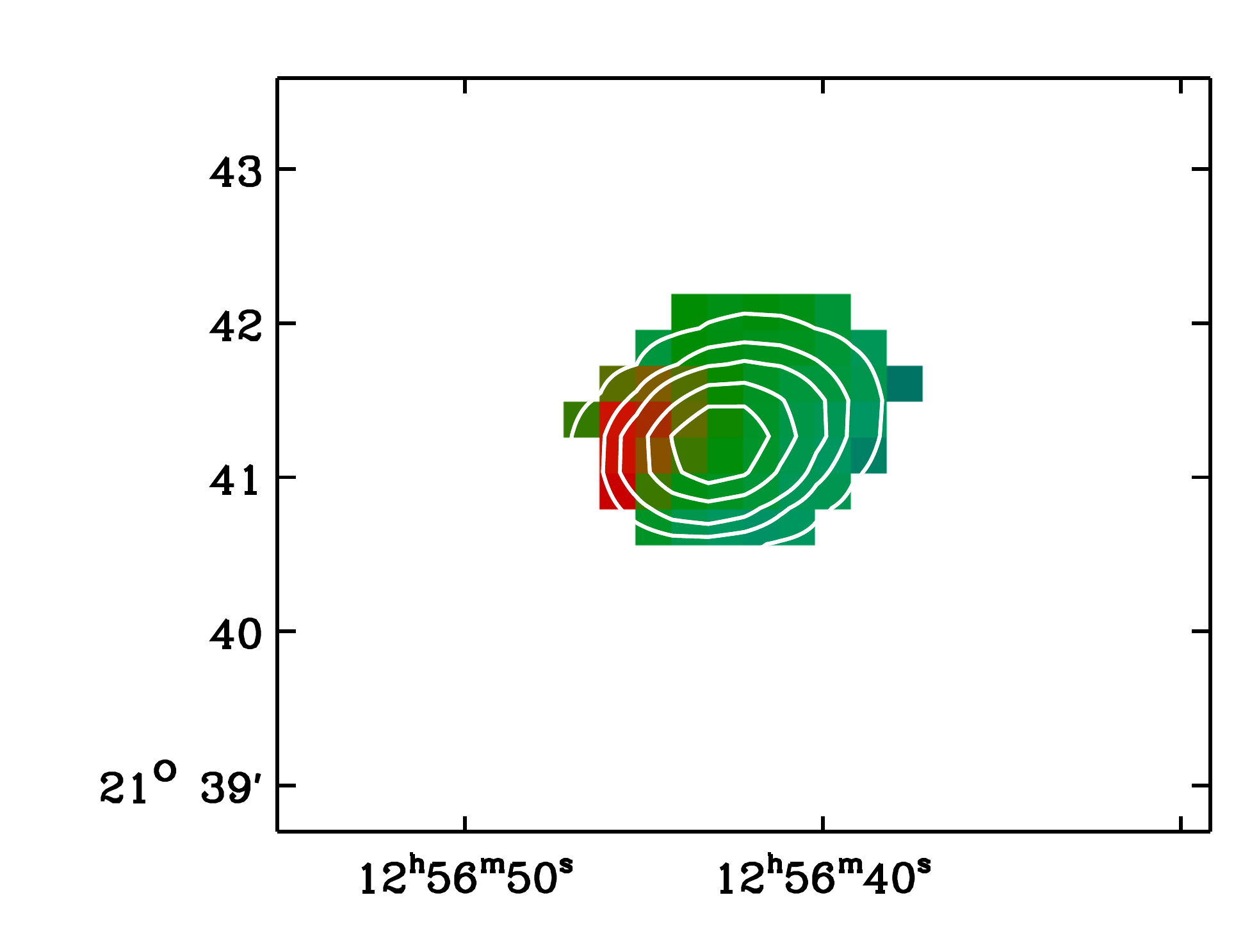} &
\includegraphics[width=5.7cm]{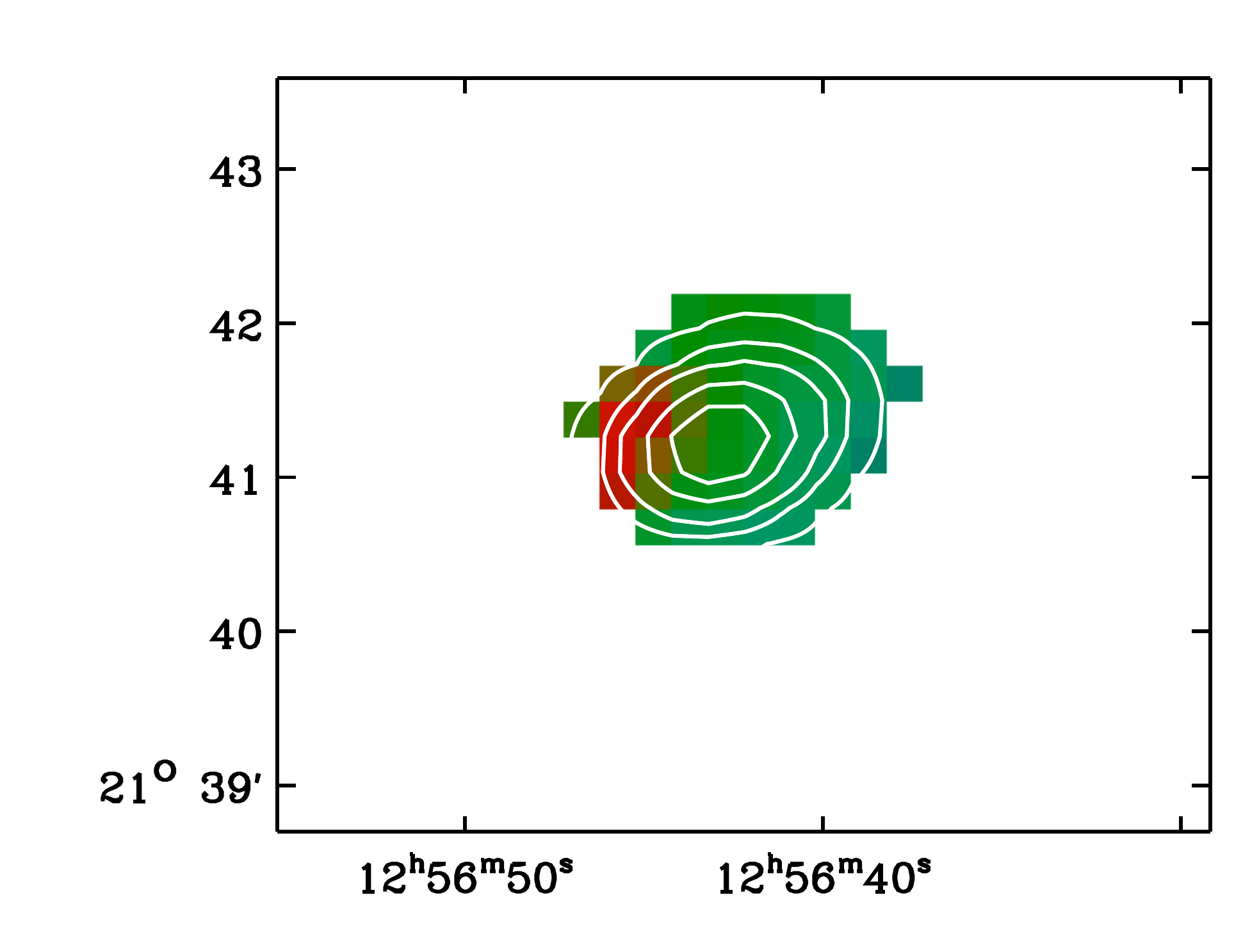}  &
\rotatebox{90}{\includegraphics[width=4cm, height=0.9cm]{RelativeExcess_ColorBars}}  \\	   
\end{tabular}  
\caption{continued. }
\end{figure*}

\newpage
\addtocounter {figure}{-1}
\begin{figure*}
\centering
\begin{tabular}  { m{0cm} m{5.1cm} m{5.1cm} m{5.1cm}  m{0.7cm}}    
{\Large \bf~~~~~~~~~~NGC7793} &&&\\  
&\hspace{5cm}\rotatebox{90}{\Large 870 \mic\ Observed} & 
\includegraphics[width=5.7cm]{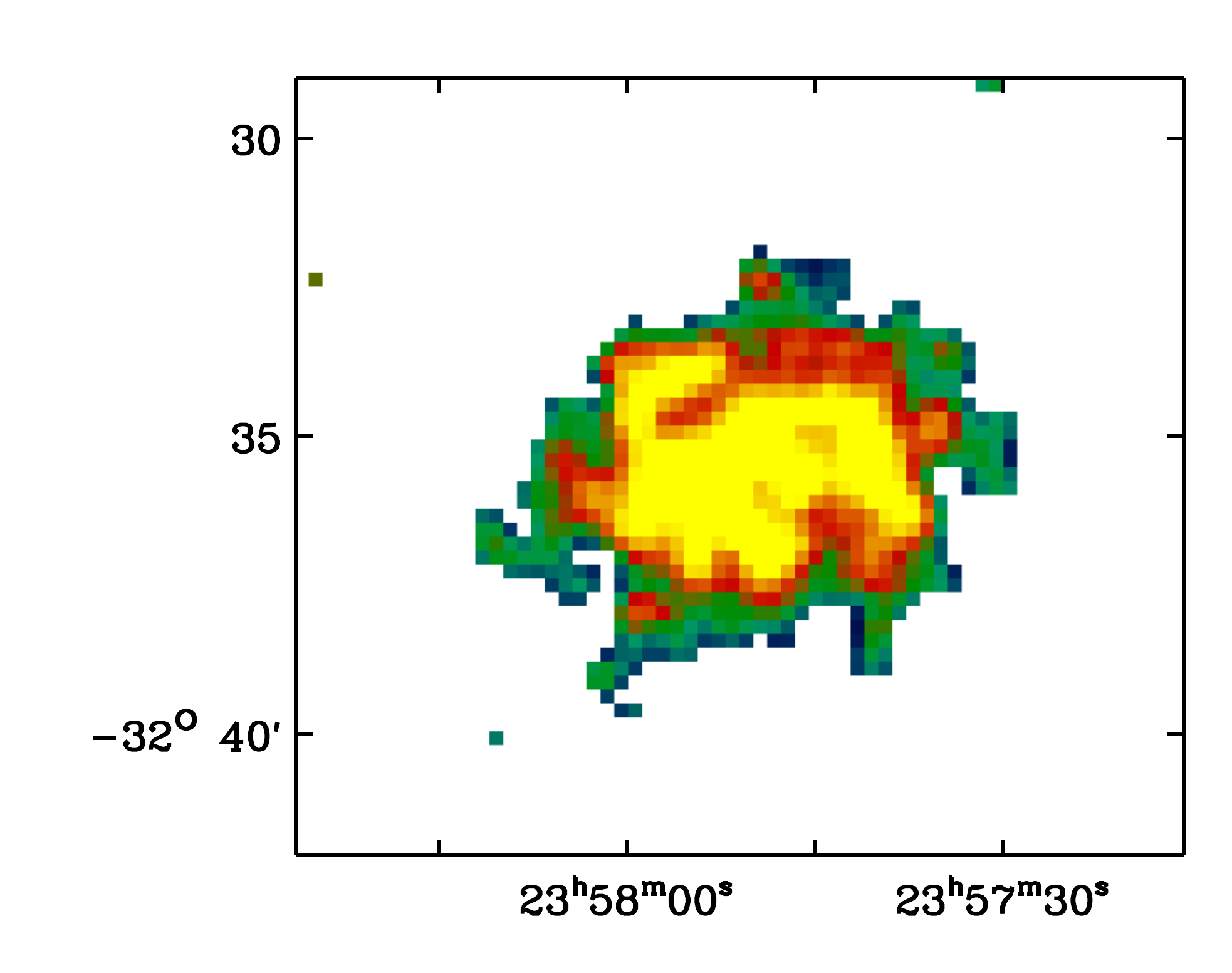} &&
\rotatebox{90}{\includegraphics[width=4cm, height=0.9cm]{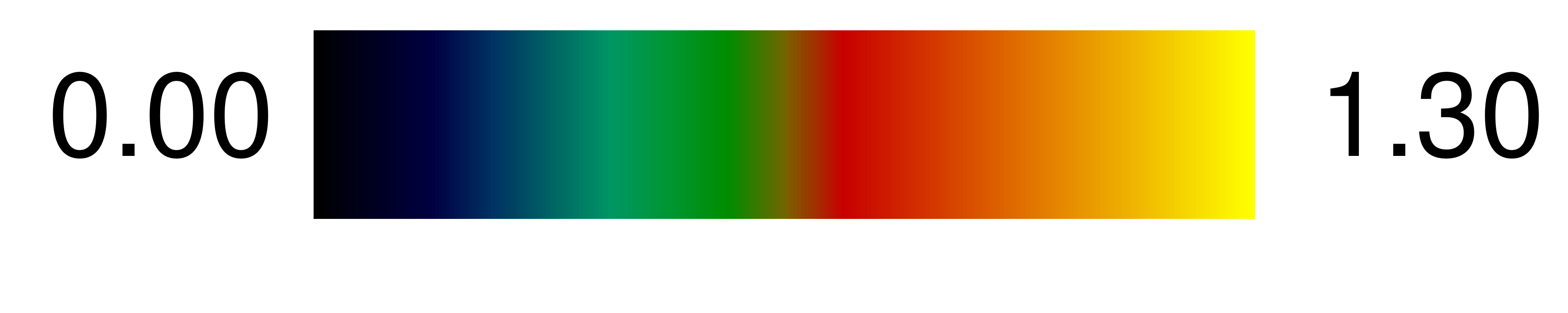}}  \\
&&\\
& {\Large \hspace{2.2cm}$\beta$$_c$ = 2.0 model} & {\Large \hspace{2.2cm}$\beta$$_c$ = 1.5 model}  & {\Large \hspace{2.2cm}[DL07] model} & \\

\rotatebox{90}{\Large 870 \mic\ Modelled} & 
\includegraphics[width=5.7cm]{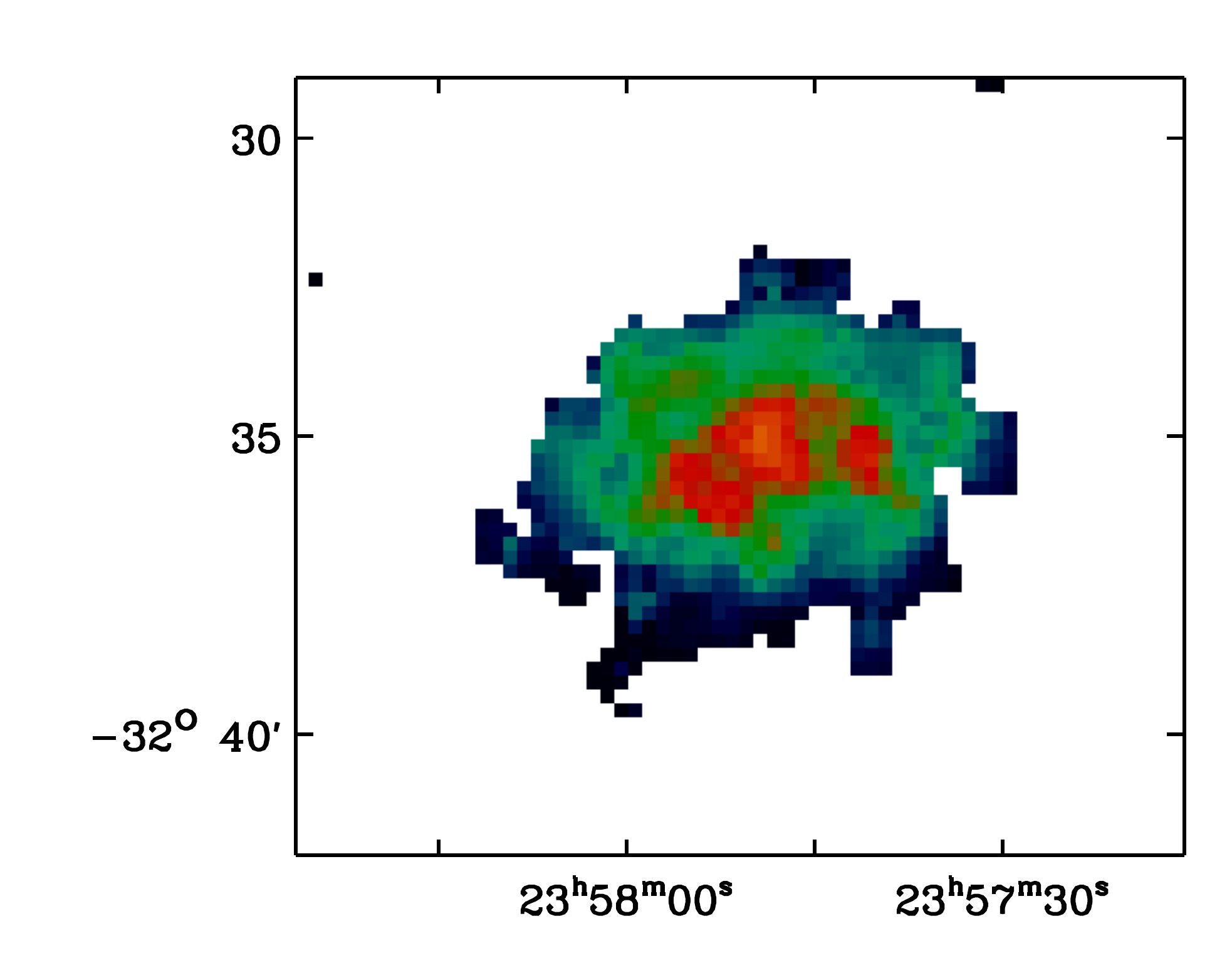} &
\includegraphics[width=5.7cm]{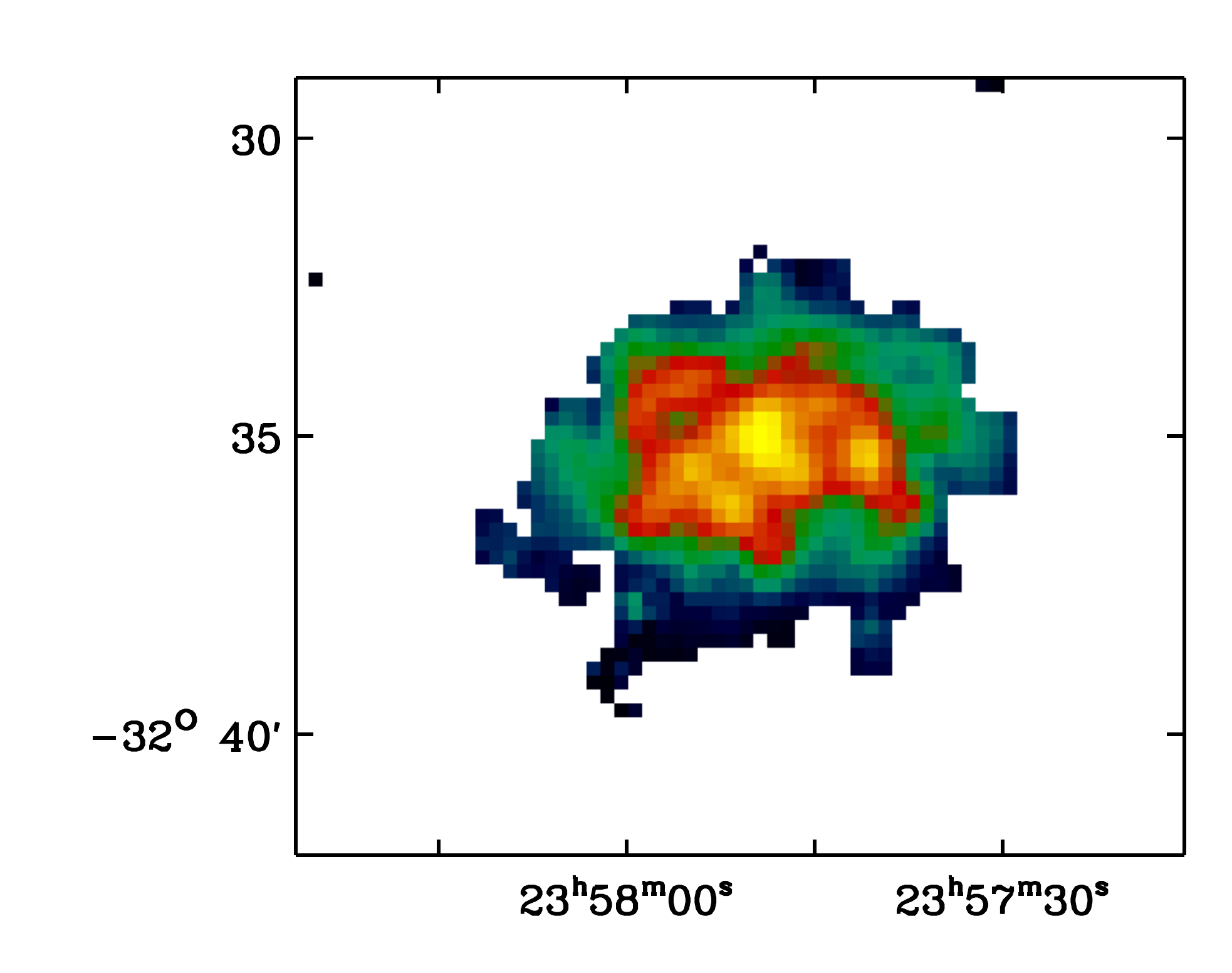} &
\includegraphics[width=5.7cm]{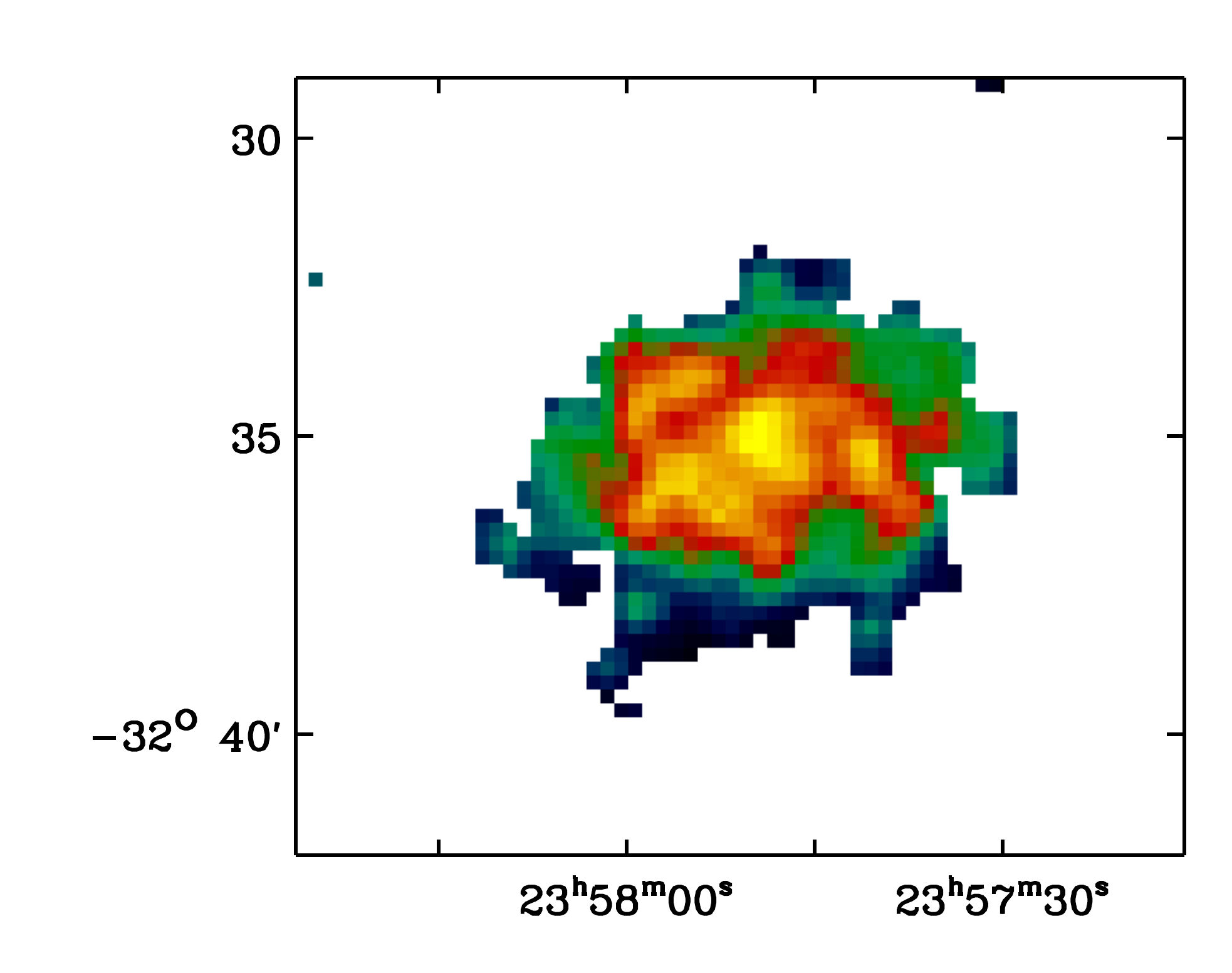}  &
\rotatebox{90}{\includegraphics[width=4cm, height=0.9cm]{NGC7793_Extrap870_ColorBars}}  \\
	
\rotatebox{90}{\Large Absolute Difference} &
\includegraphics[width=5.7cm]{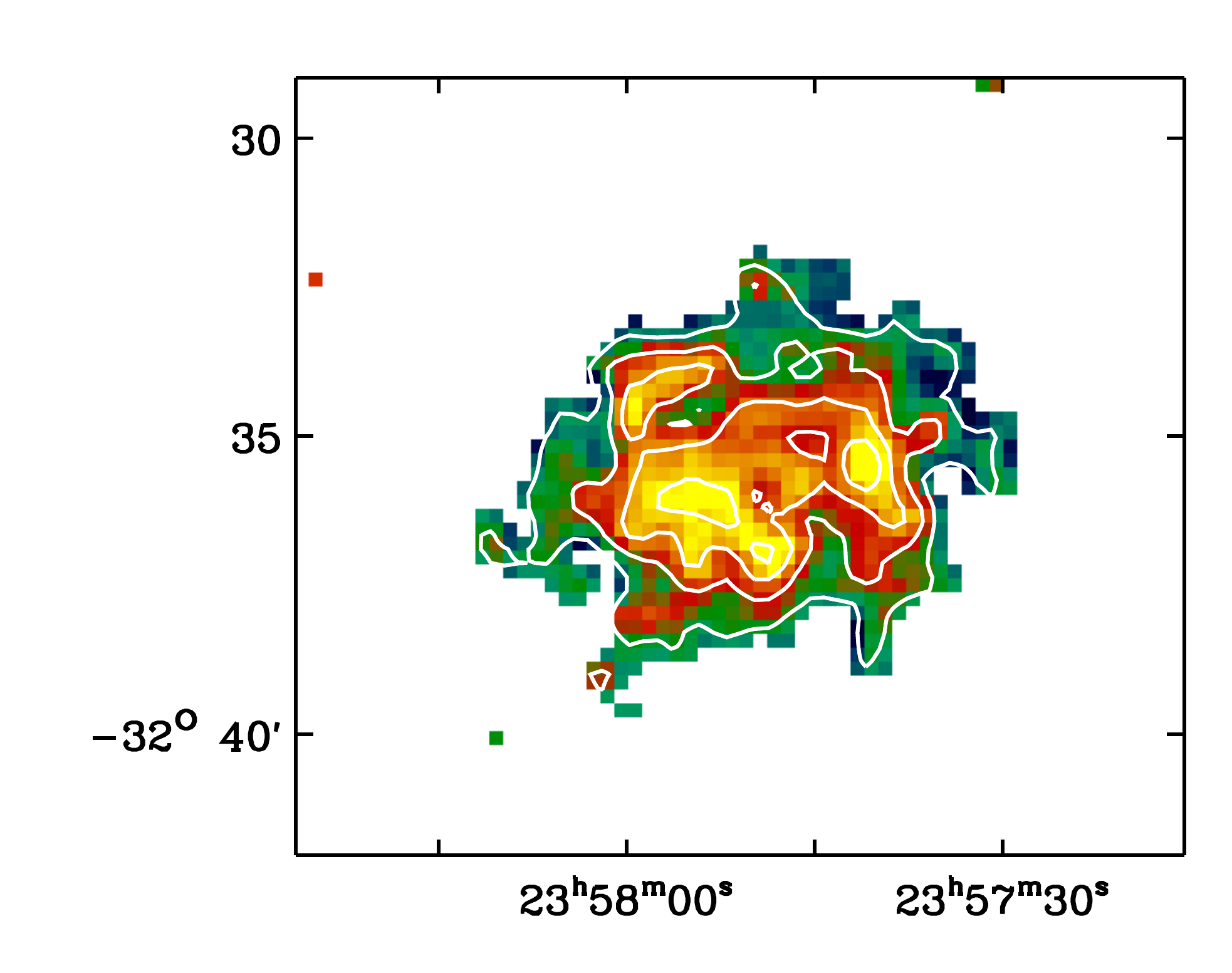} & 
\includegraphics[width=5.7cm]{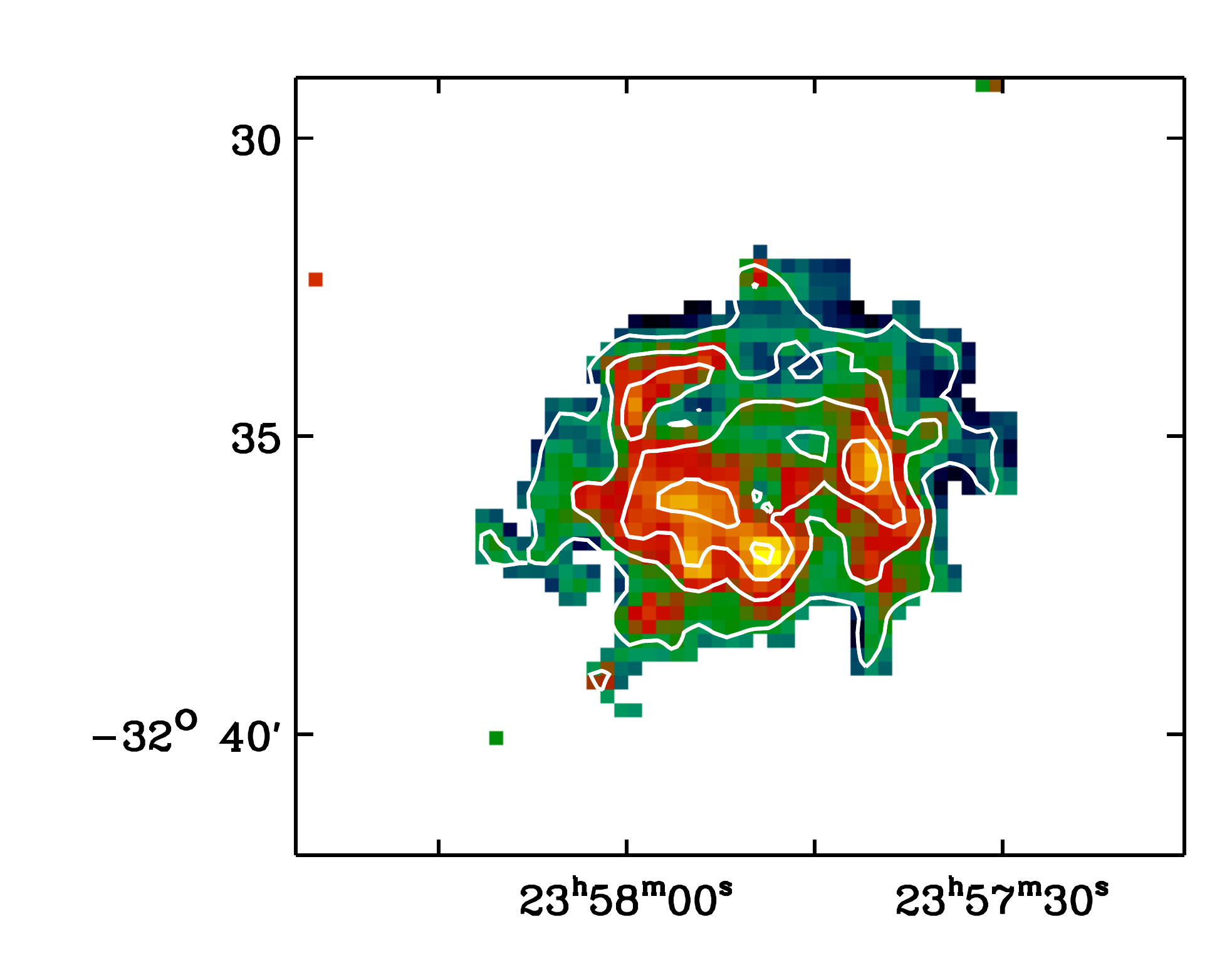} &
\includegraphics[width=5.7cm]{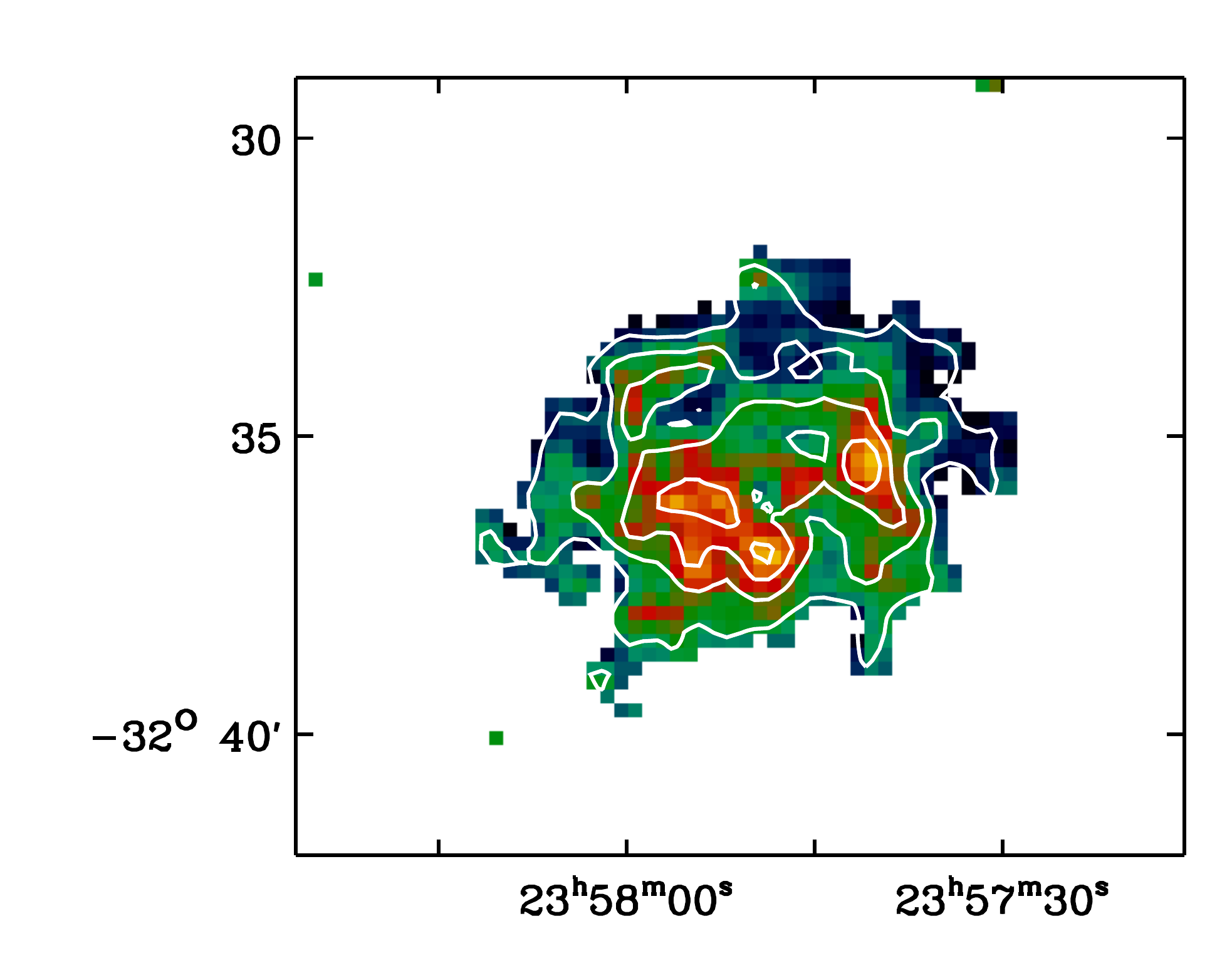}  &
\rotatebox{90}{\includegraphics[width=4cm, height=0.9cm]{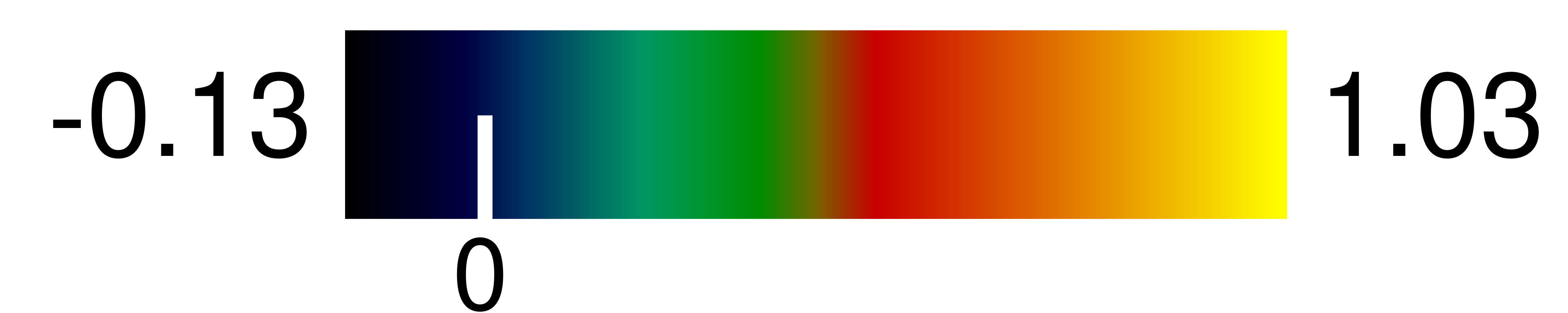}}  \\
	 
\rotatebox{90}{\Large Relative Difference} &
\includegraphics[width=5.7cm]{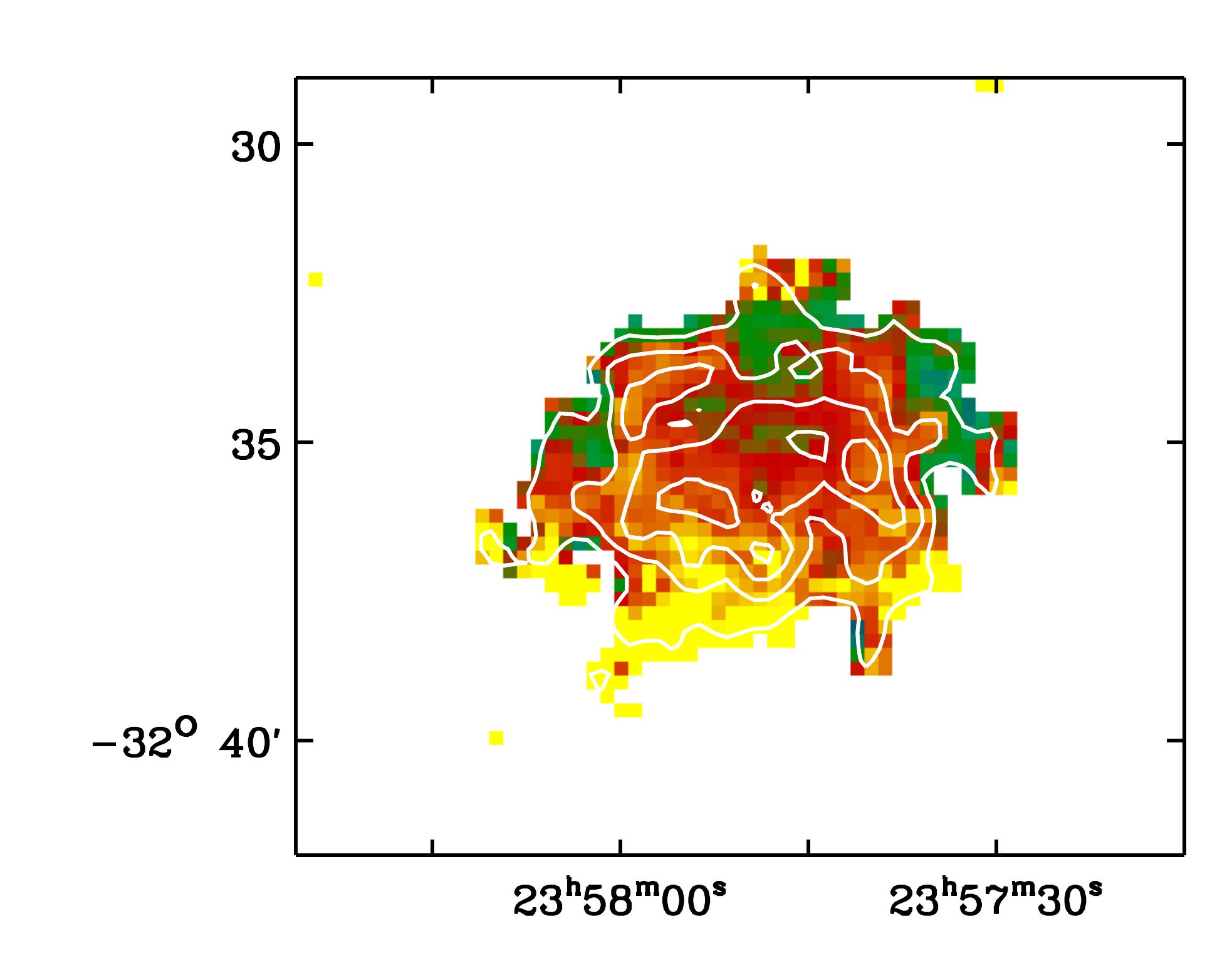} &
\includegraphics[width=5.7cm]{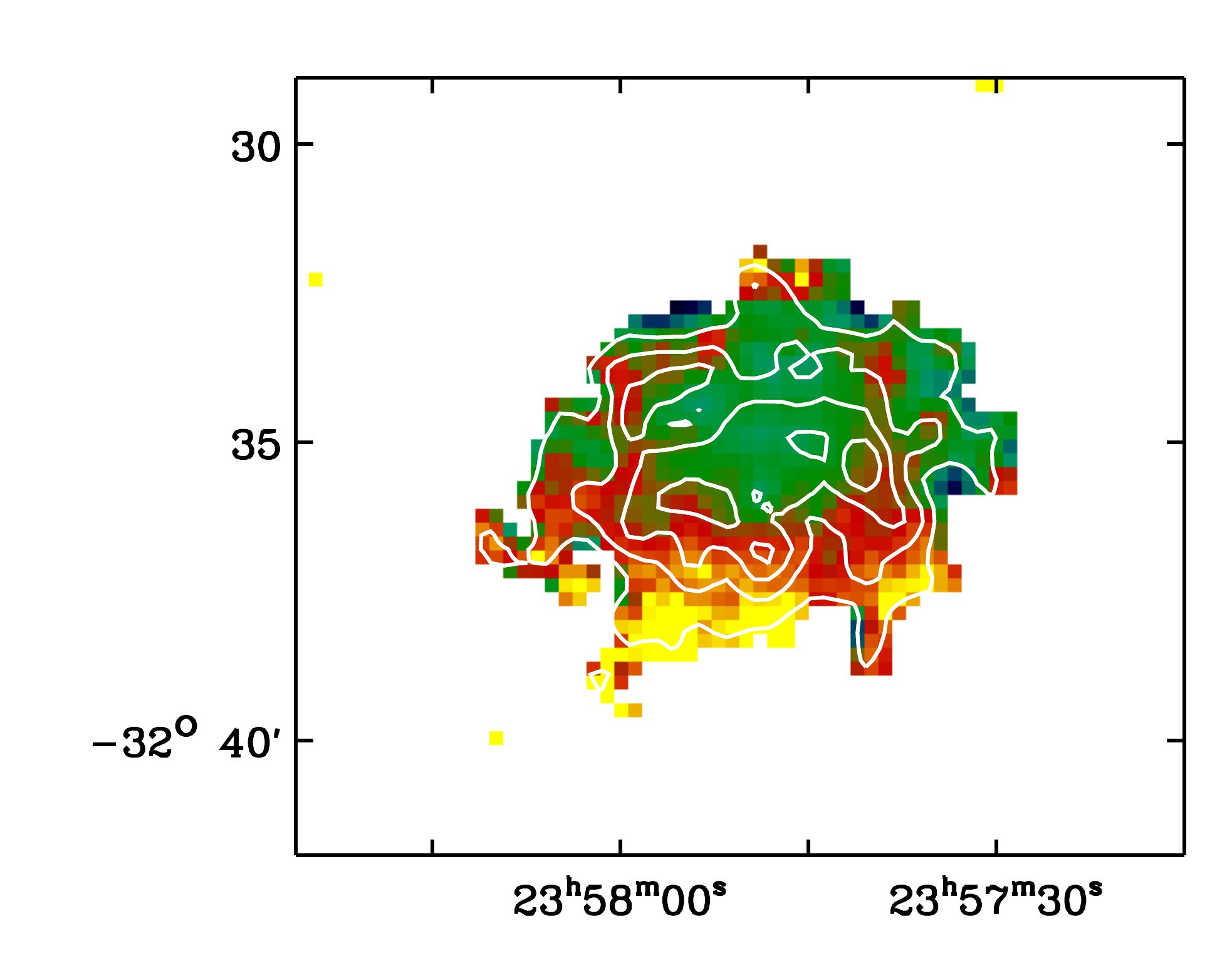} &
\includegraphics[width=5.7cm]{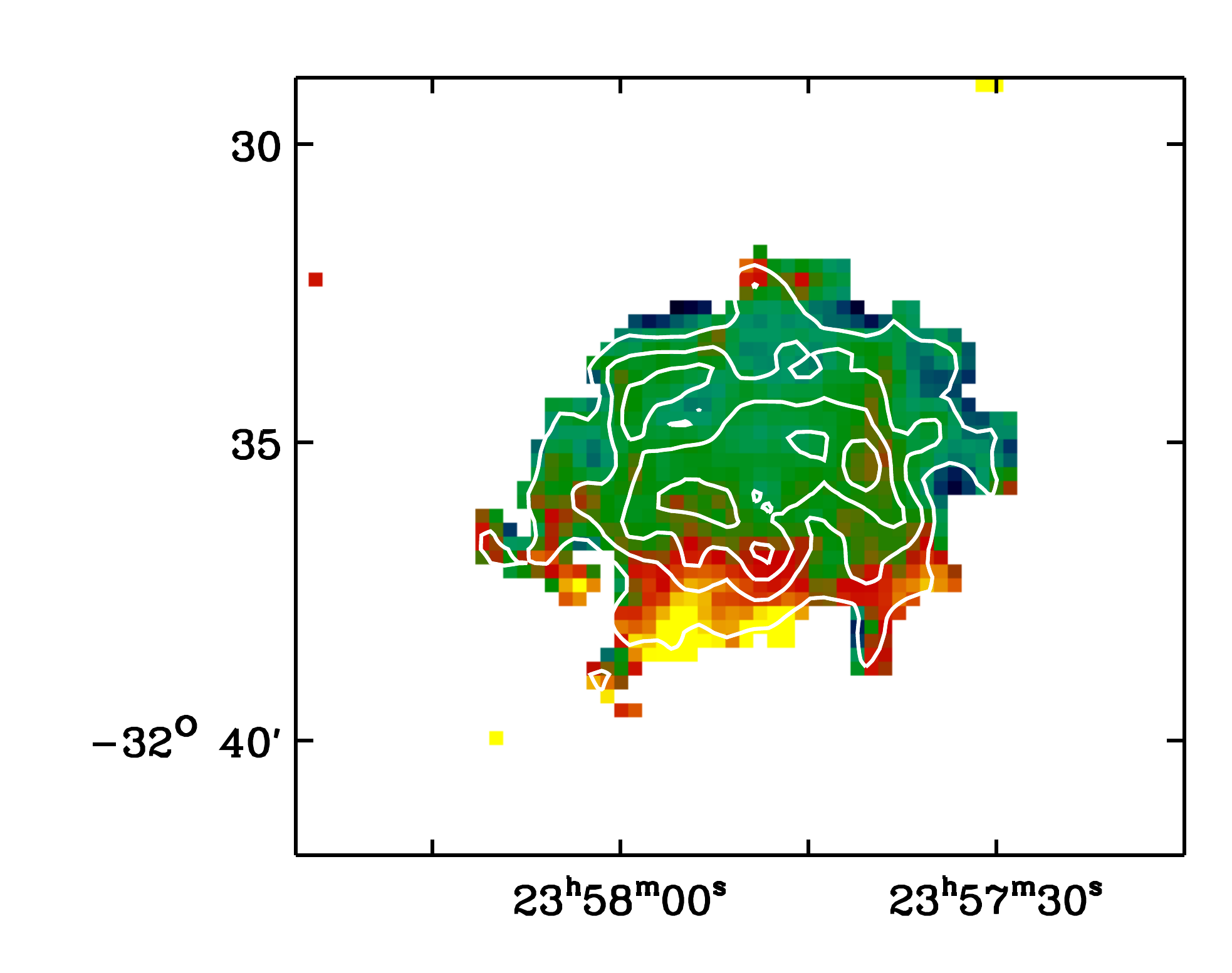}  &
\rotatebox{90}{\includegraphics[width=4cm, height=0.9cm]{RelativeExcess_ColorBars}}  \\	   
\end{tabular}  
\caption{continued.} 
\end{figure*}


\end{document}